%% file: PhDThesis.tex
\documentclass[a4paper,12pt,twoside]{book}

\usepackage[english]{babel}
\usepackage{minitoc}              
\usepackage{fancyhdr}             
\usepackage{graphicx}
\usepackage{colordvi}
\usepackage{array}
\usepackage{pstcol}
\usepackage{amsmath}
\usepackage{amsfonts}
\usepackage{amssymb}
\usepackage{pstricks}

\def \be{\begin{equation}}
\def \ee{\end{equation}}
\def \bea{\begin{eqnarray}}
\def \eea{\end{eqnarray}}
\def \ban{\begin{align}}
\def \ean{\end{align}}
\def \ben{\begin{enumerate}}
\def \een{\end{enumerate}}
\def \bit{\begin{itemize}}
\def \eit{\end{itemize}}
\def \paren#1{\left(#1\right)}

\def \ufm {\mathrm{fm}}
\def \umev {\mathrm{MeV}}
\def \ugev {\mathrm{GeV}}

\def \eq#1{Eq.~(\ref{#1})}

\def \fig#1{Fig.~\ref{#1}}

\def \rf{Ref.~\cite}
\def \rfs{Refs.~\cite}
\def \sec#1{Sec.~\ref{#1}}
\def \tb#1{Table~\ref{#1}}
\def \apx#1{Appendix~\ref{#1}}
\def \chp#1{Chapter~\ref{#1}}
\def \sp{\vspace{0.5cm}}

\def \mp{\hspace{0.4cm}}
\def \lneq#1{~(\ref{#1})}
\def \sprg{\hspace{0.65cm}}
\def  \figtk{Figure taken from Ref.~\cite}
\def \bkarr{\end{eqnarray}
            \begin{eqnarray}}

\date{\today}
 \large\normalsize
\newcommand{\clearemptydoublepage}{\newpage{\pagestyle{empty}\cleardoublepage}}
\newcommand{\captions}{\sf\caption}
\begin{document}
%

\include{covers}
\clearemptydoublepage
\renewcommand{\thepage}{\roman{page}}
\clearemptydoublepage \setcounter{page}{1} \cfoot{\thepage}
\include{resumo}
\clearemptydoublepage
\include{abstract}
\clearemptydoublepage
\include{acknowledgments}
\clearemptydoublepage
\include{fct}
%
%
\clearemptydoublepage
\dominitoc
\lhead[]{\fancyplain{}{\bfseries Table of Contents}}
\rhead[\fancyplain{}{\bfseries Table of Contents}]{}
\cfoot{\thepage}
\tableofcontents
\clearemptydoublepage
\lhead[]{\fancyplain{}{\bfseries List of Figures}}
\rhead[\fancyplain{}{\bfseries List of Figures}]{}
\listoffigures

\clearemptydoublepage
\lhead[]{\fancyplain{}{\bfseries List of Tables}}
\rhead[\fancyplain{}{\bfseries List of Tables }]{} \listoftables

\clearemptydoublepage
\include{abreviations}
%
%
\clearemptydoublepage
%
\renewcommand{\thepage}{\arabic{page}}
\setcounter{page}{1}
\addcontentsline{toc}{section}{\numberline{}{\bf \hskip -0.8cm
Preface}}

\lhead[]{\fancyplain{}{\bfseries Preface}}
\rhead[\fancyplain{}{\bfseries Preface}]{} \cfoot{\thepage}
\include{Preface}
%
%
\clearemptydoublepage
\lhead[\fancyplain{}{\bfseries\thepage}]{\fancyplain{}{\bfseries\rightmark}}
\rhead[\fancyplain{}{\bfseries\leftmark}]{\fancyplain{}{\bfseries\thepage}}

\include{Chapter1}
\clearemptydoublepage
\include{Chapter2}
\clearemptydoublepage
\include{Chapter3}
\clearemptydoublepage
\include{Chapter4}
\clearemptydoublepage
\include{Chapter5}
\clearemptydoublepage
\include{Chapter6}
\clearemptydoublepage
\begin{appendix}
\include{Appendix_A}
\clearemptydoublepage
\include{Appendix_B}
\clearemptydoublepage
\include{Appendix_C}
\clearemptydoublepage
\include{Appendix_D}
\clearemptydoublepage
\include{Appendix_E}
\clearemptydoublepage
\include{Appendix_F}
\clearemptydoublepage
\include{Appendix_G}
\clearemptydoublepage
\include{Appendix_H}
\clearemptydoublepage

\end{appendix}


\clearemptydoublepage

\addcontentsline{toc}{chapter}{\numberline{}{\bf \hskip -0.8cm
Bibliography}} \clearemptydoublepage

\lhead[\fancyplain{}{\bfseries\thepage}]{\fancyplain{}{\bfseries
Bibliography}} \rhead[\fancyplain{}{\bfseries
Bibliography}]{\fancyplain{}{\bfseries\thepage}}

\include{References}

\end{document}

%% file: covers.tex
\thispagestyle{empty}

\newpsobject{showgrid}{psgrid}{subgriddiv=1,griddots=10,gridlabels=6pt}

\psset{unit=1cm}

\begin{pspicture}(1.0,0)(11,24)
\vspace{5cm}
\includegraphics[width=0.90\textwidth]{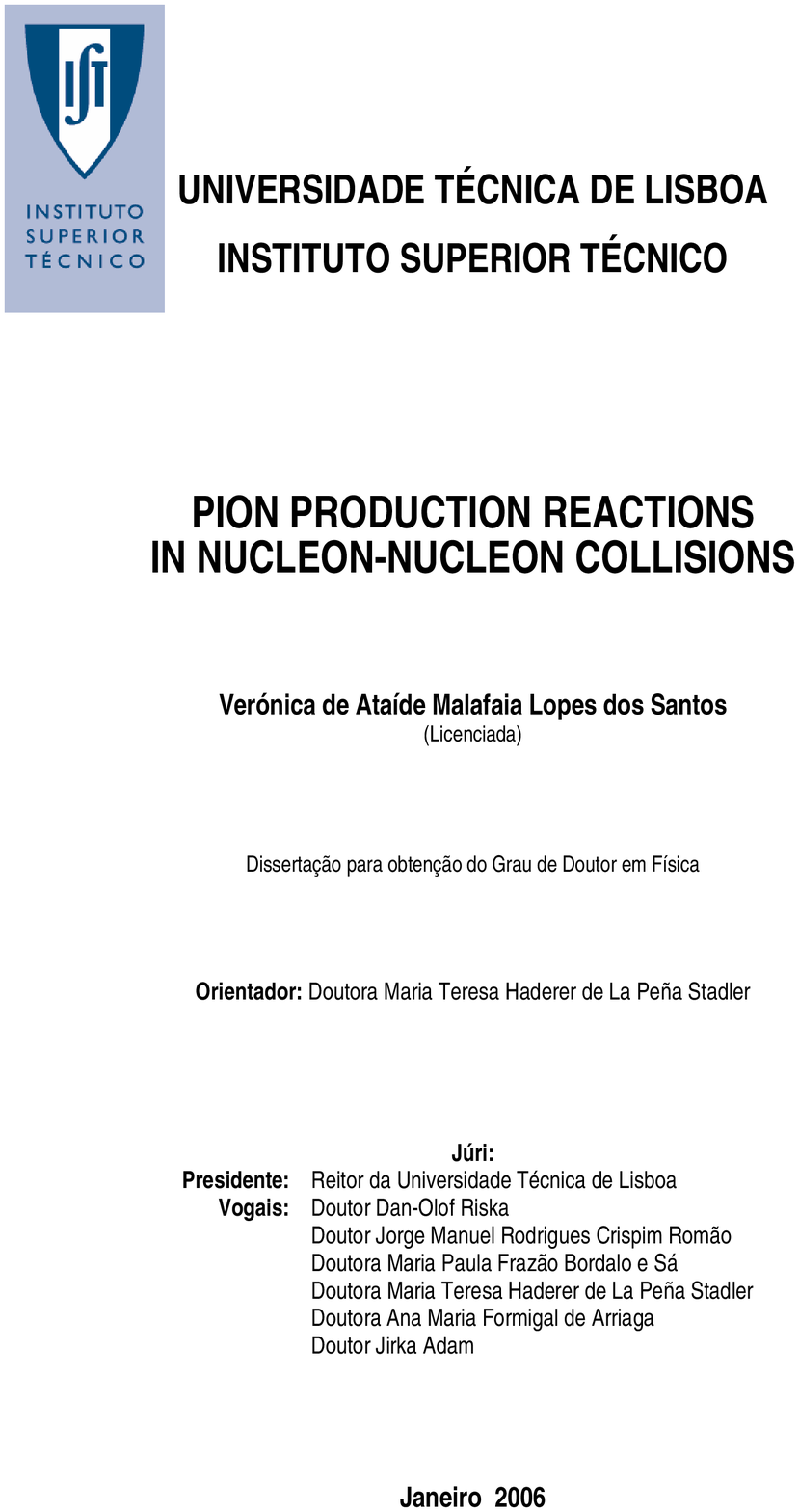}
\end{pspicture}

%% file: resumo.tex
\chapter*{Resumo}

\sprg
A produ\c c\~ao de pi\~oes em colis\~oes nucle\~ao-nucle\~ao,
junto ao limiar, tem sido um desafio nas \'ultimas d\'ecadas. A
reac\c c\~ao $pp \rightarrow pp \pi^{0}$ em particular, \'e muito
sens\'\i vel a mecanismos de curto alcance porque a conserva\c
c\~ao do isospin suprime o termo de troca de pi\~oes que de outra
forma seria dominante. Assim, tem sido muito dif\'\i cil
estabelecer a relativa import\^ancia dos v\'arios processos de
reac\c c\~ao.

Ap\'os rever o estado-da-arte da teoria, abordamos a validade da
\textit{distorted-wave Born approximation} (DWBA), atrav\'es da
sua liga\c c\~ao aos diagramas da teoria das perturba\c c\~oes
ordenadas no tempo (TOPT). Analisamos igualmente as escolhas
poss\'\i veis para a energia do pi\~ao trocado, inevit\'aveis no
formalismo n\~ao-relativista subjacente a DWBA.

O operador de re-dispers\~ao resultante de TOPT \'e comparado com
o obtido pela abordagem mais simples de matriz-S, que tem sido
usada eficazmente abaixo do limiar. A t\'ecnica de matriz-S, tendo
reproduzido os resultados de TOPT para a re-dispers\~ao em $pp
\rightarrow pp \pi^{0}$ \'e aplicada \`a produ\c c\~ao de pi\~oes
neutros e carregados, descrevendo-se com sucesso as sec\c c\~oes
eficazes nestes diferentes canais. Os principais mecanismos de
produ\c c\~ao e ondas parciais correspondentes a momento angular
mais elevado s\~ao inclu\'\i dos. Finalmente, discutimos o efeito
na sec\c c\~ao eficaz das aproxima\c c\~oes usuais para a energia
do pi\~ao trocado.

\vspace{2cm}

\noindent \textbf{PALAVRAS-CHAVE:}\\
Produ\c c\~ao de pi\~oes, prescri\c c\~oes de energia, matriz-S,
DWBA, teoria das perturba\c c\~oes ordenadas no tempo,
re-dispers\~ao.

%% file: abstract.tex
\chapter*{Abstract}

\sprg
Understanding pion production in nucleon-nucleon collisions near
threshold has been a challenge for the last decades. In
particular, the reaction $pp \rightarrow pp\pi^{0}$ is highly
sensitive to short-range mechanisms, because isospin conservation
suppresses the otherwise dominant pion exchange term. However, the
relative importance of the various reaction processes has been
very difficult to establish.

After reviewing the state-of-the-art of the theoretical
approaches, we address the validity of the distorted-wave Born
approximation (DWBA) through its link to the time-ordered
perturbation theory (TOPT) diagrams.
As the energy of the exchanged pion is not determined
unambiguously within the non-relativistic formalism underlying
DWBA, we analyse several options to determine which one is closer
to TOPT.

%
The S-matrix technique, successfully used below threshold, is
shown to reproduce the results of TOPT for the re-scattering
mechanism in $\pi^{0}$ production. It is afterwards applied to
full calculations of both charged and neutral pion production
reactions, the cross sections of which are described successfully.
The main production mechanisms and partial waves corresponding to
high angular momentum are included in the calculations. Finally we
discuss the effect on the cross section of the frequent
prescriptions for the energy of the exchanged pion.

\vspace{2cm}

\noindent \textbf{KEYWORDS:}\\
Pion production, energy-prescriptions, S-matrix, DWBA,
time-ordered perturbation theory, re-scattering.

%% file: acknowledgments.tex
\chapter*{Acknowledgments}

\sprg
I am deeply grateful to my advisor, Professor Teresa Pe\~na, who
first motivated me to study Nuclear Physics. I wish to thank her
for her constant support, encouragement, and extensive advice
throughout my Ph.D., and for all the important remarks in the
writing of this manuscript. I am also very grateful for the
possibility to establish valuable international collaborations and
for participating in international conferences.

I wish to thank my collaborators, Professor Jiri Adam and
Professor Charlotte Elster, for their advice. I cannot forget
their kind hospitality during my stays in \v{R}e\v{z} and in
Athens, OH.

My gratitude goes to CFTP (Centro de F\'\i sica Te\'orica de
Part\'\i culas), and all its members, for the excellent working
conditions and stimulating environment during the preparation of
my Ph.D. I also wish to thank CFIF (Centro de F\'\i sica das
Interac\c c\~oes Fundamentais) for hosting me during the first two
years of my investigations, before CFTP was created in 2004. The
Physics Department secretaries, Sandra Oliveira and F\'atima
Casquilho, offered constant and valuable help, and I thank them
for all their kindness and assistance.

I am indebted to FCT (Funda\c c\~ao para a Ci\^encia e Tecnologia)
for the financial support of my research and participation in
several international conferences.

Regarding computer support, I would like to thank Doctor Juan
Aguilar-Saavedra, for helping me to install Linux, and Doctor
Ricardo Gonz\'alez Felipe and Doctor Filipe Joaquim for their
precious assistance in many computer-crisis. I am also grateful to
Eduardo Lopes, especially for his readily help concerning the
cluster of computers.

I am indebted to Professor Alfred Stadler for his valuable
comments during my investigations. I wish to thank Doctor Zoltan
Batiz, my office colleague during 2002, and Doctor Gilberto
Ramalho, for their suggestions.

I would like to express my gratitude to my teachers during my
graduate studies, Professor Pedro Bicudo, Professor L\'\i dia
Ferreira, Professor Jorge Rom\~ao and Professor V\'\i tor Vieira,
for their encouragement.

\sp

My gratitude goes to my colleagues and friends at Instituto
Superior T\'ecnico, Ana Margarida Teixeira, Bruno Nobre, David
Costa, Filipe Joaquim, Gon\c calo Marques, Juan Aguilar-Saavedra,
Miguel Lopes, Nuno Santos and Ricardo Gonz\'alez Felipe, for their
friendship and immense support, and for creating an excellent
working environment. I will never forget their help and advice, as
well as the many stimulating discussions we had.

I wish to thank my friends outside Instituto Superior T\'ecnico,
Patr\'\i cia, Rita and Ana Filipa, for our weekly lunches and for
their incessant encouragement, and my movie-companions Miguel and
In\^es. I am very grateful to my dear friends Ana Filipa and Ana
Margarida for our leisure trips together and for their precious
suggestions in the writing of this thesis.

Without the continuous support of my family I could not have
completed this work. I wish to thank my grandparents for all their
kindness and unceasing support. I am deeply grateful to my
parents, for having always supported my options, and for their
unconditional love and constant encouragement. Finally, I wish to
thank Gon\c calo for his endless patience and unfailing love
during so many years.

%% file: fct.tex
\thispagestyle{plain}

\begin{tabular}{l}
  \vspace{15cm} \\
\end{tabular}

Este trabalho foi financiado pela Funda\c c\~ao para a Ci\^encia e
Tecnologia, sob o contrato SFRD/BD/4876/2001.\\

\vspace{2cm}

This work was supported by Funda\c c\~ao para a Ci\^encia e
Tecnologia, under the grant SFRD/BD/4876/2001.\\

%% file: Preface.tex
\chapter*{Preface}

\sprg
The study of pion production processes close to threshold was
originally initiated to explore the application of fundamental
symmetries to near-threshold phenomena. These investigations on
meson production reactions from hadron-hadron scattering began in
the fifties, when high energy beams of protons became available. A
strong interdependence between developments in accelerator
physics, detector performance and theoretical understanding led to
an unique vivid field of physics.

Triggered by the unprecedented high precision data for
proton-proton induced reactions (in new cooler rings), the
interest on pion production studies was revitalised in the last
decade. The (large) deviations from the predictions of one-meson
exchange models controlled by the available phase-space, are
indications of new and exciting physics.

The reaction $NN \rightarrow NN \pi$ is the basic inelastic
process related to the nucleon-nucleon ($NN$) interaction. It
sheds light on the $NN$ and $\pi N$ interactions and is key to
understanding pion production in more complex systems. Close to
the threshold the process is simpler because it is characterised
only by a small number of combinations of initial and exit
channels. Moreover, at these reduced energies, meson production
occurs with large momentum transfers, making it a powerful tool to
study short range phenomena.

Pion production occurs when the mutual interaction between the two
nucleons causes a real pion to be emitted.
The other contribution comes from a virtual pion being produced by
one nucleon and knocked on to its mass shell by an interaction
with the second nucleon. This is the so-called re-scattering
diagram, which is found to be highly sensitive to the details of
the calculations, namely the treatment of the exchanged pion
energy.

This pion re-scattering mechanism is suppressed in $pp \rightarrow
pp \pi^{0}$ due to isospin conservation. The transition amplitude
then results from a delicate interference between various
additional contributions of shorter range.  A treatment of these
mechanisms consistent with the $NN$ interactions employed in the
distortion of the initial and final state is essential to clarify
this large model dependence. In this work, a consistent
description of not only neutral, but also charged pion production,
is shown to be possible.

\sp

\newpage

In this thesis we will address the problem of charged and neutral
pion production in nucleon-nucleon collisions. The main steps of
this investigations are:
\begin{enumerate}
\item Starting with relativistic field theory, time-ordered
perturbation theory (TOPT) is used to justify the distorted-wave
Born approximation (DWBA) approach for pion production;
\item Using the DWBA amplitude fixed by TOPT as a reference
result, the effect of the traditional prescriptions for the energy
of the exchanged pion in the re-scattering operator is analysed;
\item Defining a single effective production operator within the
S-matrix technique, its relation to the DWBA amplitude yielded by
TOPT is established;
\item Employing the S-matrix approach, charged and neutral pion
production reactions are described consistently.
\end{enumerate}

This work is based on the following publications:
\begin{itemize}
\item   V.~Malafaia and M.~T.~Pe\~na, \textit{Pion re-scattering
in $\pi^{0}$ production}, Phys.\ Rev.\ C {\bf 69} (2004) 024001
[nucl-th/0312017].
\item  V.~Malafaia, J.~Adam and M.~T.~Pe\~na,
  \textit{Pion re-scattering operator in the S-matrix approach},
  Phys.\ Rev.\ C {\bf 71} (2005) 034002
  [nucl-th/0411049].
\item  V.~Malafaia, M.~T.~Pe\~na, Ch.~Elster and J.~Adam,
     \textit{Charged and neutral pion production in the S-matrix
                  approach}, [nucl-th/0511038], submitted to Phys.
                  Lett. B.
\item  V.~Malafaia, M.~T.~Pe\~na, Ch.~Elster and J.~Adam,
     \textit{Neutral and charged pion production with realistic
      $NN$ interactions}, to submit to Phys. Rev. C.
%
\end{itemize}

\sp

\chp{MesonProdTh} introduces the physics accessible through the
study of the meson production reactions. It focuses on the
specific aspects of hadronic meson production reactions close to
threshold, namely the rapidly varying phase-space, the (high)
initial and (low) final relative momenta, and the general energy
dependence of the production operator.

\sp

\chp{StateArt} is a historical review of the main theoretical
approaches to pion production developed so far. The first part is
dedicated to the distorted-wave Born approximation (DWBA), the
frequently used approach in all the calculations near the
threshold energy, in which the nucleon-nucleon interaction is
treated non-perturbatively, whereas the transition amplitude $NN
\rightarrow NN \pi$ is treated perturbatively. The second part
focuses on the coupled-channel phenomenological approaches and the
third part aims to present the actual status of chiral
perturbation theory ($\chi$PT) calculations.

DWBA calculations apply a three-dimensional formulation for the
initial- and final-$NN$ distortion, which is not obtained from the
Feynman diagrams. As a consequence, in calculations performed so
far within the DWBA approximation, the energy of the exchanged
pion has been treated approximately and under different
prescriptions.
%
%
%
A clarification of these formal issues is thus needed before one
can draw conclusions about the physics of the pion production
processes.
The re-scattering mechanism, being highly sensitive to these
energy prescriptions, is the ideal starting point for this
clarification.

\vspace{0.45cm}

\chp{FTtoDWBA} aims to obtain a three-dimensional formulation from
the general Feynman diagrams. This chapter discusses the validity
of the DWBA approach by linking it to the time-ordered
perturbation theory (TOPT) diagrams which result from the
decomposition of the corresponding Feynman diagram.
Since in the time-ordered diagrams energy is not conserved at
individual vertices, each of the re-scattering diagrams for the
initial- and final-state distortion defines a different off-energy
shell extension of the pion re-scattering amplitude.
This imposes the evaluation of the matrix elements between quantum
mechanical wave functions, of two different operators.
Henceforth, in \chp{Smatrixapp} we are lead to an alternative
approach to the TOPT formalism (the S-matrix approach), which in
contrast to it avoids this problem.

The S-matrix approach provides a consistent theoretical framework
for the two diagrams, as well as for the $NN$ distortion. Besides,
from the practical point of view, it simplifies tremendously the
numerical effort demanded in TOPT by the presence of logarithmic
singularities in the pion propagator for ISI.
The S-matrix construction has already been successfully used below
pion production threshold and in particular for $NN$ interactions
and electroweak meson exchange currents.
In \chp{Smatrixapp} also, the effective DWBA amplitude obtained in
\chp{FTtoDWBA} is employed to re-examine the nature and extent  of
the uncertainty resulting from the approximations made in the
evaluation of the effective operators.

\vspace{0.45cm}

In \chp{Chargedandneutral},  the S-matrix technique, shown in
\chp{Smatrixapp} to reproduce the results of time-ordered
perturbation theory for the re-scattering mechanism in $\pi^{0}$
production, is applied to charged and neutral pion production
reactions. The major production mechanisms, namely the
contributions from the direct-production, re-scattering,
Z-diagrams and $\Delta$-isobar excitation are considered. Higher
angular momentum partial waves, which are not included in
traditional calculations, are also considered. The last part is
dedicated to the effect on the cross section of the usual
prescriptions for the energy of the exchanged pion.
For all the charge channels, the S-matrix approach for the
description of the pion production operators reproduces well the
DWBA result coming from TOPT.
Importantly, the effect of some approximations usually employed is
also assessed. The $\pi^{+}$ reaction is seen to be especially
sensitive to those. Previous failures in its description are
overcome and clarified.

\sp

Finally, in \chp{Conclusions} we outline and summarise the most
relevant aspects discussed in this thesis, and mention the future
prospects of pion production in nucleon-nucleon collisions.

%% file: Chapter1.tex
\setcounter{minitocdepth}{2}

\chapter{Meson production close to threshold} \label{MesonProdTh}

\minitoc

\noindent
{\bfseries Abstract:} Meson production reactions in nucleonic
collisions near threshold are a powerful tool to investigate
short-range phenomena. Pion production reactions play a very
special role, since they yield the lowest hadronic inelasticity
for the nucleon-nucleon interaction and thus they are an important
test of the phenomenology of the nucleon-nucleon interaction
at intermediate energies. \\
\newpage
\section{What physics can we learn from meson production reactions?}
%
\sprg
Quantum Chromodynamics (QCD), the fundamental theory for strong
interactions, unfolds an impressive predictive power mainly at
high energies. However, at low energies the perturbative expansion
no longer converges.

Although a large amount of data on hadronic structure and dynamics
is available from measurements with electromagnetic probes (for
instance, from MAMI at Mainz, ELSA at Bonn and JLAB at Newport
News), there is still much to be learned about the physics with
hadronic probes at intermediate energies, comprising the
investigation of production, decay and interaction of hadrons. In
particular, meson production reactions (close to threshold) in
nucleon-nucleon, nucleon-nucleus and nucleus-nucleus collisions
constitute a very important class of experiments in this field.

With the advent of strong focusing synchrotrons having
high-quality beams (for instance, the IUCF Cooler in Bloomington,
CELSIUS in Upsala and COSY in J\"{u}lich), a new class of
experiments could be performed in the last decade, differing from
the previous ones with respect to the unprecedented quality of the
data (polarised as well as unpolarised) for several $NN
\rightarrow NN x$ reaction channels (a recent review on the
experimental and theoretical aspects of meson production can be
found in \rfs{Moskal:2002jm,Machner:1999ky} and in
\rf{Hanhart:2003pg}, respectively).

The study of meson production close to threshold has several
attractive key features, in particular concerning,
\begin{itemize}
\item[(i)] \textbf{Large momentum transfer in the entrance channels} \\
Meson production near threshold occurs at large momentum transfers
and therefore is a powerful  tool to study short range phenomena
in the entrance channel;
\item[(ii)] \textbf{Simplicity of the entrance and exit channels} \\
The analysis of the reaction data is straightforward allowing one
to study the underlying reaction mechanisms;
\item[(iii)] \textbf{Small phase-space} \\
Only a very limited part of the phase-space is available for the
reaction products and hence, only a few partial waves contribute
to the observables, especially very near the threshold.
\end{itemize}
Although a small number of partial waves is clearly an advantage
for calculations, on the other hand, as a result of the small
phase-space, the cross sections are also small. There is a
delicate interplay between the several possible reaction
mechanisms and thus it is essential that all the technical aspects
are under control for a meaningful interpretation of the data.

The large momentum transfer can also look like an disadvantage
since it is difficult to reliably construct the production
operator. However, in the regime of small invariant masses, the
production operator is largely independent of the relative energy
of a particular particle pair in the final state\footnote{If there
are resonances near by, this statement is no longer true.}.
Consequently, dispersion relations can be used to extract
low-energy scattering parameters of the final state interaction
and at the same time, to estimate the error in a model independent
way.

\sp

In the overall, meson production reactions in nucleonic collisions
have a huge potential to give insight into the strong interaction
physics at intermediate energies, namely concerning the following
aspects:
\begin{itemize}
    \item {\bfseries Final state interactions} \\
          Scattering of unstable particles off nucleons is
          experimentally very problematic since it is difficult to
          prepare intense beams of these particles with the
          required accuracy. Production reactions where such particles
          emerge in the final state are an attractive alternative.
    \item {\bfseries Baryon resonances in a nuclear environment}\\
          The systems studied in nucleon-nucleon and
          nucleon-nucleus collisions allow to investigate particular
          resonances in the presence of other baryons and excited by various
          exchanged particles. One example is the $N^{*}\left(1535\right)$,
          which is clearly visible as a bump in any $\eta$
          production cross section\cite{Hanhart:2003pg}.
    \item {\bfseries Charge symmetry breaking}\\
          The existence of available several possible initial
          isospin states (for instance $pp$, $pn$ and $dd$),
          with different possible spin states, allows experiments
          which enable the study of isospin symmetry breaking.
    \item {\bfseries Effective field theory in large momentum
          transfer reactions}\\
          The investigation of hadronic processes at low and
          intermediate momenta are essential to test the convergence of
          the low-energy expansion of chiral perturbation theory
          ($\chi$PT).
    \item {\bfseries Three-nucleon forces}\\
          The information on the short-range mechanisms that can be deduced
          from pion production in nucleon-nucleon ($NN$) collisions is
          relevant for constraining the three-nucleon forces\cite{Epelbaum:2002ji}.
\end{itemize}
\sp

This work will focus on pion production reactions in
nucleon-nucleon collisions. It is the lowest hadronic inelasticity
for the nucleon-nucleon interaction and thus an important test of
the understanding of the phenomenology of the $NN$ interaction.
Secondly, as pions are the Goldstone bosons of chiral symmetry, it
is possible to study pion production also using $\chi$PT. This
provides the opportunity to improve the phenomenological
approaches via matching to the chiral expansion, as well as to
constrain the chiral contact terms via resonance saturation. As
mentioned before, a large number of (un)polarised data is
available\footnote{
From the experimental point of view, it is important to notice
that since the vector mesons have much larger widths compared with
the pseudoscalar mesons, their detection is very difficult on top
of a large physical background of multi-pion production events.
Secondly, as there seems to be a general trend that the larger the
mass, the smaller the cross section, the generally heavier vector
mesons have smaller production cross sections, are thus much
harder to investigate than the lighter ones\cite{Machner:1999ky}.}
to be used as constraints.
Moreover, meson-exchange models of the nucleon-nucleon interaction
above the pion threshold rely on detailed information about the
strongly coupled inelastic channels which must be treated together
with the elastic interaction. Also, information on pion production
in the $NN$ system is required in models of pion production or
absorption in nuclei.
%
%
%
\section{Specific aspects of hadronic meson production close to
threshold\label{SpMeson}}
%
\subsection{Rapidly varying phase-space}

\sprg
Threshold production reactions are characterised by excess
energies which are small compared to the produced masses. In the
near threshold regime the available phase-space changes very
quickly (although remaining small). Therefore, to compare
different reactions, one needs an appropriate measure of the
energy relative to threshold. For pion production, the
traditionally used variable is the maximum pion momentum $\eta$
(in units of the pion mass). For all heavier mesons, the so-called
excess energy $Q$, defined as
\begin{equation}
Q=\sqrt{s}-\sqrt{s^{threshold}},
\end{equation}
is used instead. In \apx{Apkinematics}, a compilation of useful
kinematic relations is presented, and the importance of
relativistic kinematics for the near threshold reactions is
stressed.

\begin{figure}[t!]
\begin{center}
  \includegraphics[width=.60\textwidth,keepaspectratio]{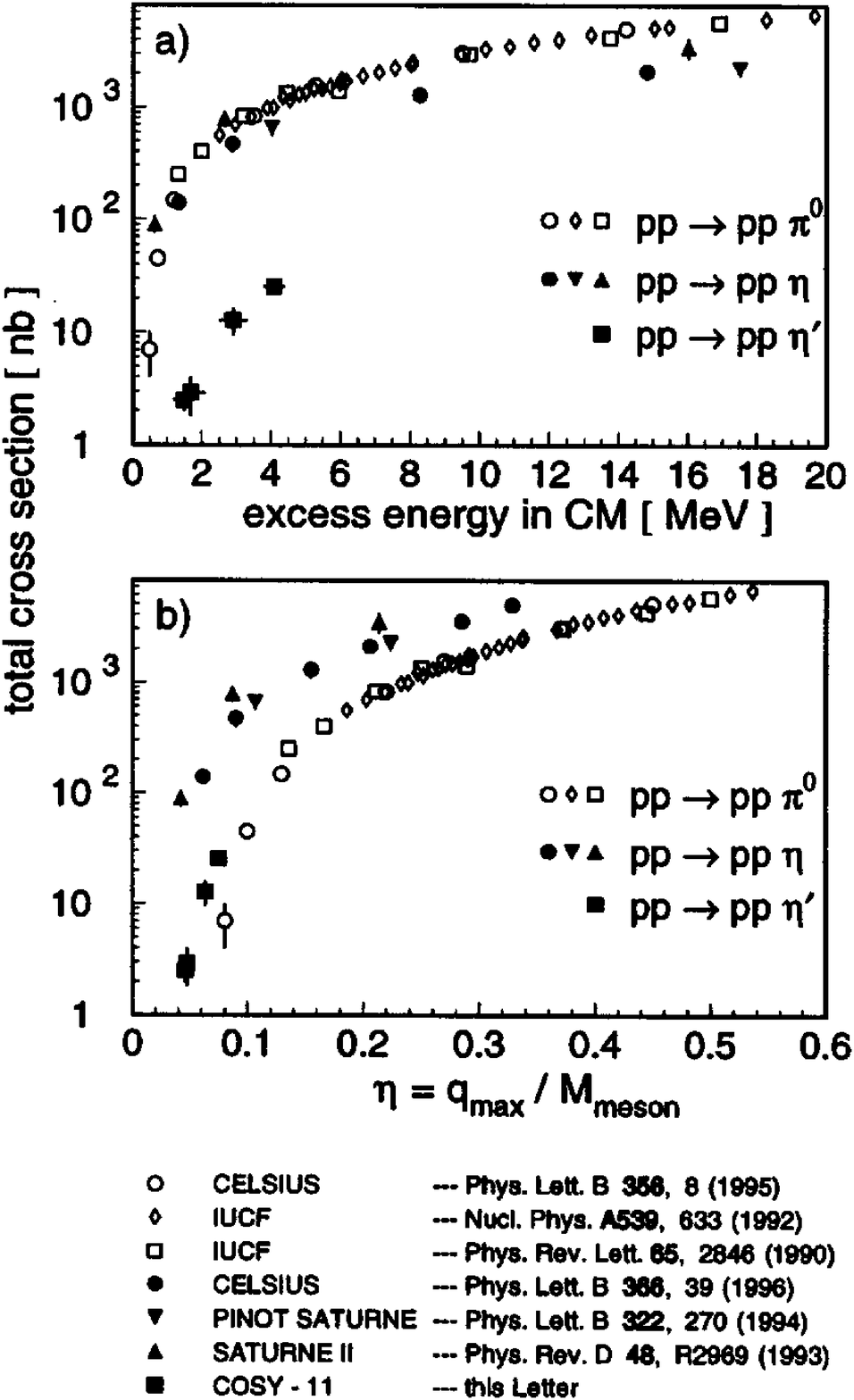}
  \captions[Comparison for several meson production reactions of the variables
   $\eta$ and $Q$ used to represent the energy]{Total cross sections for the reactions $pp
   \rightarrow pp \pi^{0}$, $pp \rightarrow pp \eta$ and $pp
   \rightarrow pp \eta^{\prime}$ as a function of a)the excess energy
   $Q$ and b)the maximum pion momentum
   $\eta$ in units of the mass of the produced meson.
   \figtk{Moskal:1998pc}.}\label{compEtaQ}
\end{center}
\end{figure}

If $Q$ gives the available energy for the final state, the
interpretation of $\eta$ is somewhat more involved. In a
non-relativistic, semiclassical picture, the maximum angular
momentum allowed can be estimated via
\begin{equation}
l_{max} \simeq R q^{\prime},
\end{equation}
where $R$ is a measure of the force range and $q^{\prime}$ is the
typical momentum of the corresponding particle. Identifying $R$
with the Compton wavelength of the meson of mass $m_{x}$, $\eta$
can be interpreted as the maximum angular momentum
allowed\cite{Rosenfeld:1954pr}:
\begin{equation}
l_{max} \simeq \frac{q^{\prime}}{m_{x}} \simeq \eta \hspace{1cm}
(\mathrm{with} \hspace{0.2cm} \hbar=c=1) \label{lmaxRosen}.
\end{equation}
To compare the cross sections for reactions with different final
states in order to extract information about the reaction
mechanisms, one has to choose carefully the variable that is used
to represent the energy. Indeed, as it is shown in \fig{compEtaQ},
the total cross sections for $pp \rightarrow pp \pi^{0}$, $pp
\rightarrow pp \eta$ and $pp \rightarrow pp \eta^{\prime}$ are
different when compared at equal $\eta$ (panel (b)) or at equal
$Q$ (panel (a)).
When the dominant final state interaction is the $pp$ interaction,
which is the case for those reactions,
it appears thus to be more appropriate to compare the cross
sections at equal $Q$, since then at any given energy, the impact
of the final state interaction is equal for all the reactions.
This is not the case for equal values of $\eta$, as $\eta$ depends
on the mass of the produced meson.
%

\subsection{Initial and final relative momenta}

\sprg
Meson production in nucleon-nucleon collisions requires that the
kinetic energy of the initial particles is sufficiently high to
put the outgoing meson on its mass shell. In other words, the
relative momentum of the initial nucleons must exceed the
threshold value,
\begin{equation}
p_{initial}^{th.} = \sqrt{m_{x} M +\frac{m_{x}^{2}}{4}},
\end{equation}
where $M$ is the nucleon mass and $m_{x}$ is the mass of the
produced meson. For a close-to-threshold regime, the particles in
the final state have small momenta and thus $p_{initial}^{th.}$
also sets the scale for the typical momentum transfer. In a
non-relativistic picture, this large momentum transfer translates
into a small reaction volume, characterised by a size
parameter\footnote{As in \eq{lmaxRosen}, $\hbar=c=1$ is assumed.},
\begin{equation}
R \sim \frac{1}{p_{initial}^{th.}} \simeq 0.5 \ufm \hspace{0.5cm}
\text{for pion production}.
\end{equation}
The two nucleons in the initial state have thus to approach each
other very closely before the production of a meson can happen.
For this reason it is important to understand not only the elastic
but also the inelastic $NN$ interaction to obtain quantitative
predictions.
\subsection{Remarks on the production operator and nucleonic distortions}

\begin{figure}[h!]
\begin{center}
  \includegraphics[width=.78\textwidth,keepaspectratio]{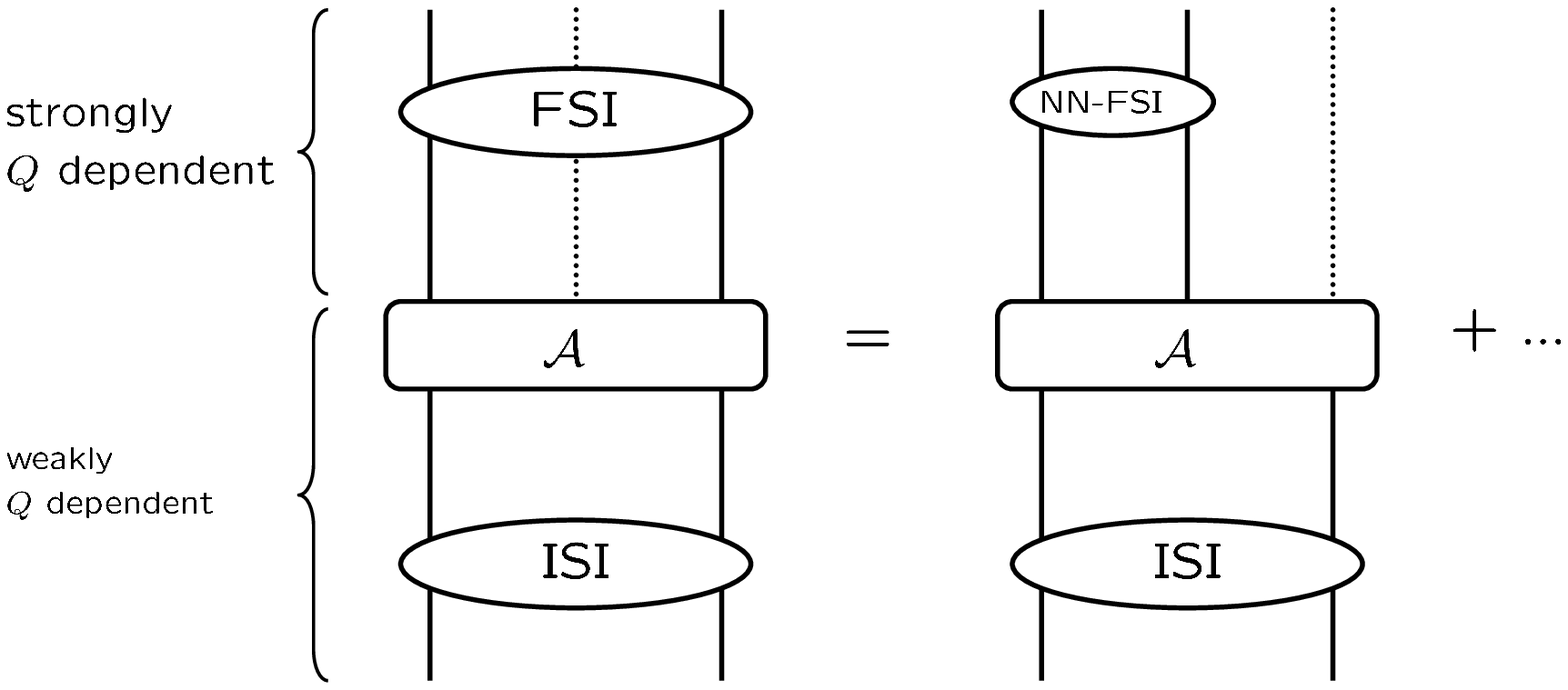}
  \captions[Sketch of the excess energy $Q$-dependence of the meson
   production reaction]
  {Sketch of the excess energy $Q$-dependence of the meson
  production reaction. $\mathcal{A}$ is the
   production operator. The left diagram shows the
   complexity possible, with interaction between the three final particles,
   whereas the right diagram shows the first and potentially dominant term (with
   FSI only for the two final nucleons).}\label{IntQdep}
\end{center}
\end{figure}

\sprg
%
%
In the near threshold regime, all the particles in the final state
have low relative momenta and thus can potentially undergo strong
final state interactions (FSI) that can induce strong energy
dependencies.
On the other hand, close to the threshold, the initial energy is
significantly larger than the excess energy $Q$, and consequently
the initial state interaction (ISI) should at most mildly
influence the energy dependence.
The dependence of the production operator on the excess energy
should also be weak, since it is controlled by the typical
momentum transfer, which is significantly larger than the typical
outgoing momenta.
\fig{IntQdep} illustrates the energy dependence of the production
operator both FSI and ISI cases.

The large momentum transfer characteristic of meson production
reactions at threshold leads to a large momentum mismatch for any
one-body operator that might contribute to the production
reaction.
%
\section{Theoretical considerations on $NN \rightarrow NNx$ reactions}
%

\begin{figure}
\begin{center}
  \includegraphics[width=.64\textwidth,keepaspectratio]{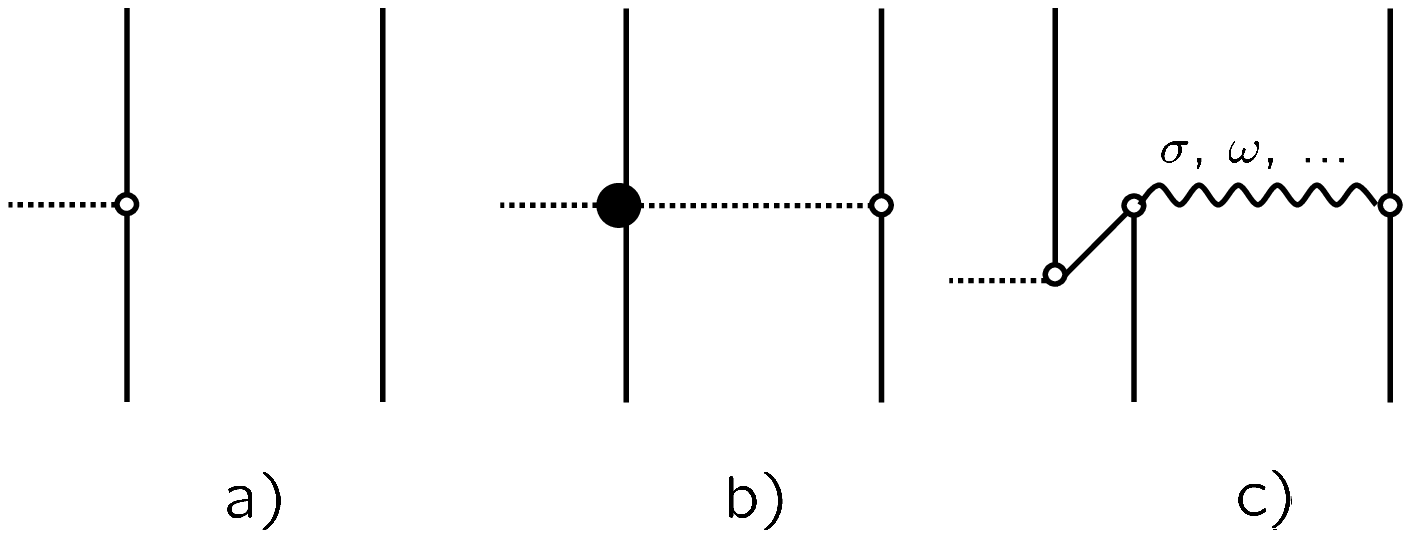}
  \captions[Contributions to the production
  operator in pion production in $NN$ collisions which in the DWBA
  formalism are convoluted with $NN$ wave functions:
  a) direct-production, b) re-scattering and c) short-range
  contributions ]
  {Contributions to the production
  operator in pion production in $NN$ collisions which in the DWBA
  formalism are convoluted with $NN$ wave functions:
  a) direct-production, b) re-scattering and c) short-range
  contributions. The solid and dotted lines represent the nucleons and
  pions, respectively.}
  \label{diagDwba}
\end{center}
\end{figure}

\sprg
Most of the theoretical models\footnote{The development of
theoretical approaches for the reactions $NN \rightarrow NNx$ has
a long history. A review of earlier works can be found in
\rfs{{Blankleider:1991yp},{Garcilazo:1990ws}}.} for the reactions
$NN \rightarrow NN x$ can be grouped in two distinct classes:
\begin{itemize}
\item{\textbf {Distorted wave Born approximation (DWBA)}} \\
A production operator is constructed within some perturbative
scheme approximation and then is convoluted with the nucleon wave
functions.

\item{\textbf{(Truly) Non-perturbative approaches}} \\
 Integral equations are solved for the full $ \paren {NN,NNx}$
 coupled-channel problem, describing multiple re-scattering and preserving
 three-body unitarity.
\end{itemize}
The great majority of theoretical studies of pion production in
nucleon-nucleon collisions in the threshold region have been done
within the DWBA formalism\cite{Woodruff:1960pr}.
%
%
This approach is motivated by the fact that close to pion
production threshold, where the kinetic energy
of the particles in the final state is practically zero,
the forces between the nucleons are much stronger than the
interaction between the pion and the nucleon. Consequently, only
the interaction between the nucleons is taken into account up to
all orders, for instance, by employing wave functions that are
solutions of a scattering (Lippmann-Schwinger) equation, whereas
the pion production process is treated perturbatively and the pion
is assumed to propagate freely after its production. Typically,
the diagrams that contribute are of the type of those of
\fig{diagDwba}.

\sp

On the other hand, all calculations performed so far for pion
production within the full $ \paren {NN,NNx}$ coupled-channel
approach were within the framework of time-ordered perturbation
theory\cite{Weinberg:1966tp} (TOPT), or its extension to the
$N\Delta$ sector done by the
Helsinki\cite{{Green:1975dm},{Niskanen:1978vm},{Niskanen:1984di}},
Argonne\cite{{Betz:1981rt},{Lee:1985jq},{Matsuyama:1986sh},{Lee:1987hd}}
and Hannover\cite{{Popping:1987ex},{Pena:1992ui},{Pena:1993pd}}
groups, which have a reasonable predictive power at higher
energies but cannot describe the physics very near threshold, as
discussed in \sec{Hannovermodel}.

%
%
\section{The cross section}
\begin{figure}[h!]
\begin{center}
  \includegraphics[width=.58\textwidth,keepaspectratio,angle=0.581]{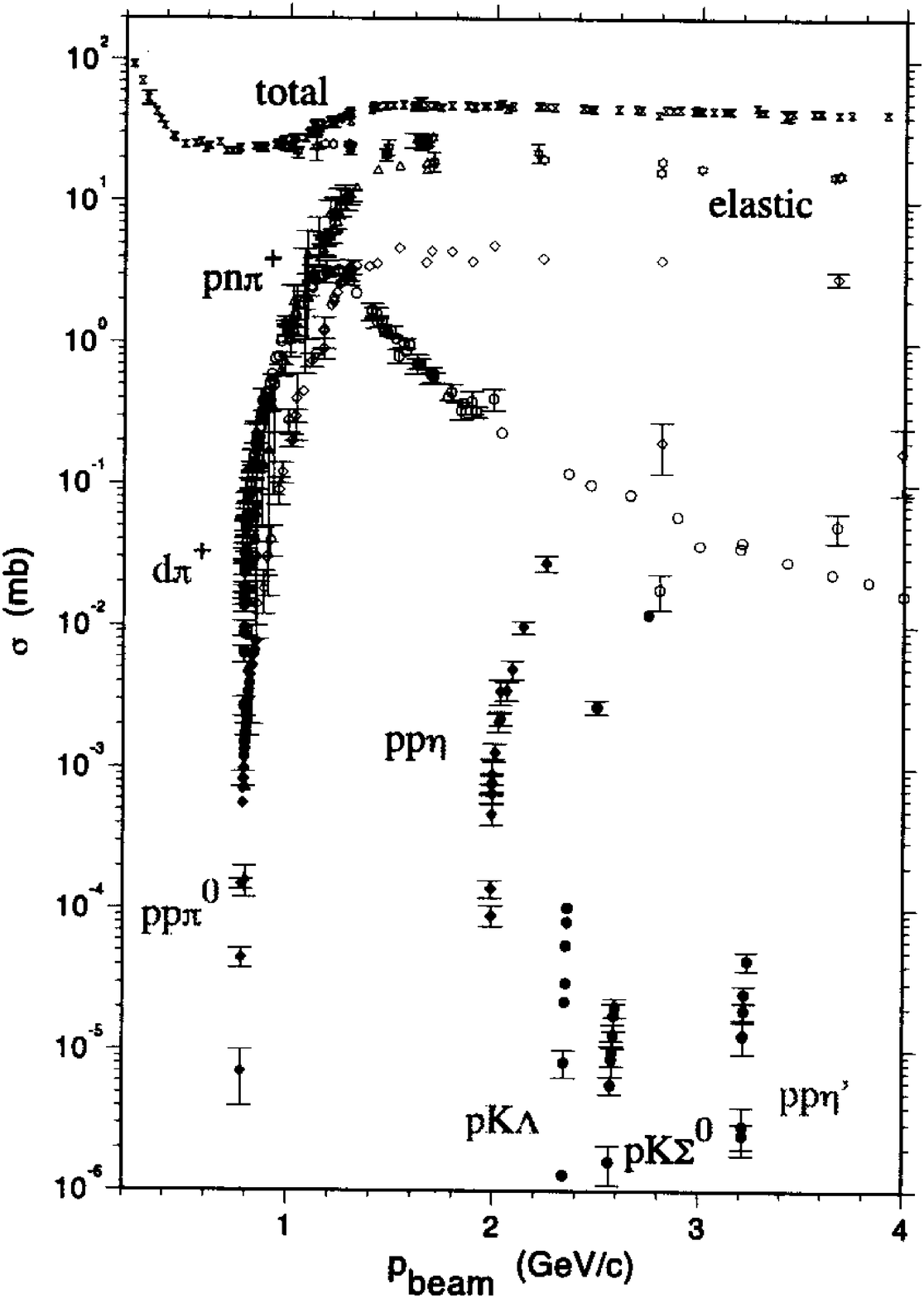}
  \captions[Cross sections for $pp$ interactions as a function of
  the beam momentum.]{Cross sections for $pp$ interactions as a function of
  the beam momentum. \figtk{Machner:1999ky}.}
  \label{crossMachner}
\end{center}
\end{figure}

\sprg In \fig{crossMachner}, there is an overview of the total
cross sections in $pp$ interactions below $4\ugev$ beam momentum.
In \tb{thmom} we list the threshold momenta $p^{thr.}$ and
threshold laboratory energies $T_{lab}^{thr.}$, for the $NN
\rightarrow NN \pi$ reactions considered in this work. The right
column is a compilation of references with the experimental
determination of $\sigma_{tot}$.

The first threshold which opens with increasing beam energy is
$\pi^{0}$ production followed very soon after by $\pi^{\pm}$
production (see \tb{thmom}). Pion production exhausts all
inelasticity in this momentum range (see \fig{crossMachner}) and
therefore is fundamental to understand the nucleon-nucleon
interaction.

\begin{table}[b!]
\begin{center}
\begin{tabular}
[c]{cccc}\hline\hline
reaction & $p^{\text{thr.}}\left[ \umev \right]  $ & $T_{lab}^{\text{thr.}%
}\left[ \umev \right]  $ & $\sigma_{tot}$\\\hline $pp\rightarrow
pp\pi^{0}$ & $724.4$ & $279.7$ &
\cite{{Meyer:1990yf},{Stanislaus:1991sy},{Meyer:1992jt},{Bondar:1995zv},{Bilger:2001qz},{AbdElSamad:2002fj}}  \\
$pp\rightarrow pn\pi^{+}$ & $737.3$ & $289.5$ &
 \cite{{Daehnick:1995gw},{Hardie:1997mg},{Brinkmann:1998ux},{Flammang:1998bb}}\\
$pn\rightarrow pp\pi^{-}$ & $737.3$ & $289.5$ &
\cite{{VerWest:1981dt},{Bachman:1995gn},{Daum:2001yh}}
\\\hline\hline
\end{tabular}
\end{center}
\captions[Threshold momenta, laboratory energies and
$\sigma_{tot}$ data for the pion production reactions considered
in this work]{Threshold momenta and threshold laboratory energies
for pion production reactions in the $NN$ collisions considered in
this work. The right column refers to the experimental
determination of the corresponding cross sections. } \label{thmom}
\end{table}

%% file: Chapter2.tex
\setcounter{minitocdepth}{2}
\chapter{State of the art of theoretical models for pion production \label{StateArt}}

\minitoc {\bfseries Abstract:} The significant experimental
progress in the last decade resulted in high-quality data on pion
production near threshold. For neutral pion production these new
and accurate data posed a theoretical challenge since they were
largely under-predicted by the existent calculations. This Chapter
is a historical review of the main theoretical approaches
developed.

\newpage
%
\section{DWBA in the Meson exchange approach}
%
\subsection{Problems of the earlier calculations}
\subsubsection{The work of Koltun and Reitan}

\sprg Pioneering work on pion production was done in the 1960s by
Woodruf\cite{Woodruff:1960pr} and by Koltun and
Reitan\cite{Koltun:1966pr}. All later investigations of meson
production, including the very recent efforts to analyse the
high-precision data from the new generation of accelerators, have
followed basically the same approach, if one excludes the Hannover
model\cite{{Popping:1987ex},{Pena:1992ui},{Pena:1993pd}} for
coupled $NN$, $N\Delta$-$NN \pi$ channels.

\begin{figure}
\begin{center}
  \includegraphics[width=.80\textwidth,keepaspectratio]{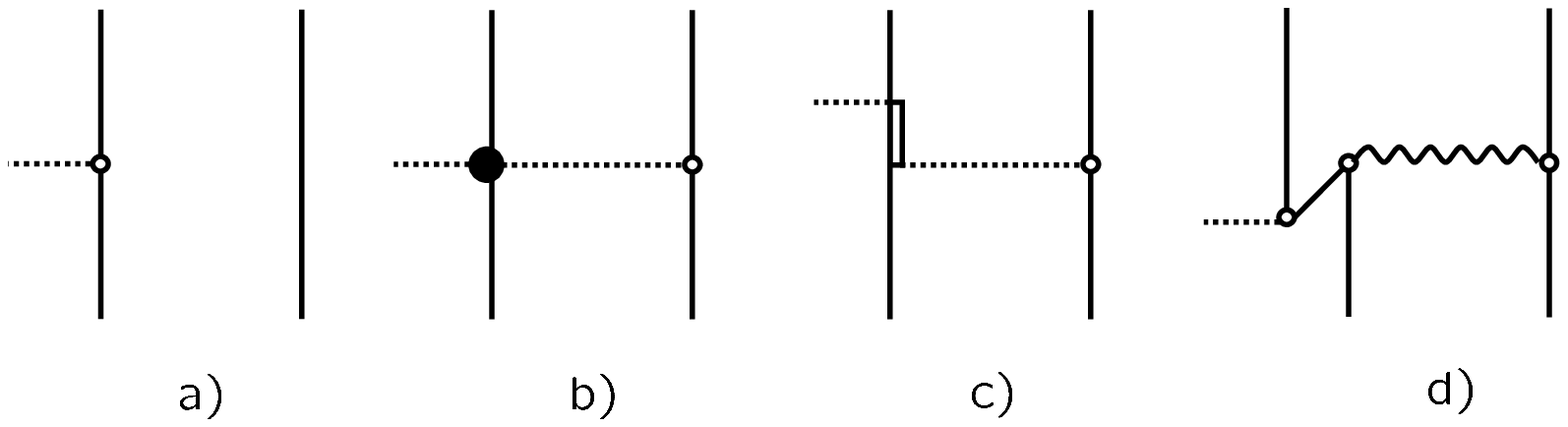}
  \captions[The main mechanisms considered in the works on pion production:
  direct production, re-scattering, pion
  re-scattering via a $\Delta$ and short-range contributions]
  {The main mechanisms considered in the works on pion production:
  a) direct production, b) re-scattering,  c) pion
  re-scattering via a $\Delta$ and d) short-range contributions.
  The solid and dashed lines represent the nucleon and pion fields,
  respectively. Heavy mesons ($\sigma$, $\omega$,...) are represented by the wavy line.}\label{histdiagrams}
\end{center}
\end{figure}

These works focused on the reactions $pp \rightarrow pp \pi^0$ and
$pp \rightarrow d\pi^+$. The processes considered were direct
production by either nucleon (diagram (a) of \fig{histdiagrams}),
the so-called impulse approximation, and production from
pion-nucleon scattering (diagram (b)), the so-called re-scattering
term.
The $\pi N \rightarrow \pi N$ transition amplitudes were
parameterised in terms of scattering lengths, through the
Hamiltonian\footnote{Actually, \eq{h1phen} and \eq{h2phen} are the
Lagrangian, but we kept here the term Hamiltonian for historical
reasons. } $\mathcal{H}=\mathcal{H}_{1}+\mathcal{H}_{2}$, with
\begin{eqnarray}
  \mathcal{H}_{1} &=& i\frac{f_{\pi NN}}{m_\pi}  \vec{\sigma} \cdot \left[ \vec{\nabla}_\pi \boldsymbol{\tau}
  \cdot \boldsymbol{\pi}+\frac{1}{2M}\left(\vec{p}\boldsymbol{\tau}\cdot \boldsymbol{\dot{\pi}}+
  \boldsymbol{\tau}\cdot\boldsymbol{\dot{\pi}}\vec{p} \right)\right ] \label{h1phen}\\
  \mathcal{H}_{2} &=& 4\pi \frac{\lambda_1}{m_\pi}\boldsymbol{\pi}\cdot\boldsymbol{\pi} +
  4\pi \frac{\lambda_2}{m_\pi^2} \boldsymbol{\tau}\cdot\boldsymbol{\pi}\times
  \boldsymbol{\dot{\pi}} \label{h2phen}
\end{eqnarray}
where $\vec{\sigma}$ and $\boldsymbol{\tau}$ are the usual nucleon
spin and isospin operators, and $\vec{p}$ is the nucleon momentum
operator. The pion(nucleon) mass is $m_{\pi}\left(M \right)$ and
the pion field is $\boldsymbol{\pi}$. The $\pi NN$ pseudovector
coupling constant is $f_{\pi NN}$.

$\mathcal{H}_1$ is obtained from a non relativistic reduction of
the pseudovector $NN\pi$ vertex\footnote{In the chiral limit for
vanishing momenta, the interaction of pions with nucleons has to
vanish. Thus the coupling of pions naturally occurs either as
derivative or as an even power of the pion mass. In general, the
pseudovector coupling for the $\pi NN$ vertex is preferred. The
pseudovector coupling automatically incorporates a strong(weak)
attractive $p$-wave($s$-wave) interaction between pions and
nucleons\cite{Iwamoto:1989mh}.} and gives diagram (a) in
\fig{histdiagrams}.
The first term of \eq{h1phen} represents $p$-wave $\pi N$
coupling, while the second term (``galilean" term) accounts for
the nucleon recoil effect\cite{Park:1995ku}. For $s$-wave pion
production, only the second term contributes. Since this second
term is smaller than the first term by a factor of $\sim
\frac{m_{\pi}}{M}$, the contribution of the Born term to $s$-wave
pion production is intrinsically suppressed, and as a consequence
the process becomes sensitive to two-body contributions,
\fig{histdiagrams}(b) and (d).

$\mathcal{H}_2$ is a phenomenological effective Hamiltonian
describing pion re-scattering. The isoscalar and isovector parts
of the $\pi N$ s-wave scattering amplitude, $\lambda_1$ and
$\lambda_2$, were obtained from the $s$-wave phase shifts
$\delta_{1}$ and $\delta_{3}$ for the pion-nucleon scattering,
through the Born-approximation relations\cite{Koltun:1966pr},
\begin{equation}
\lambda_{1}=-\frac{1}{6\eta}\left(\delta_{1}+2 \delta_{3}
\right)=0.005 \hspace{2.5cm}
\lambda_{2}=-\frac{1}{6\eta}\left(\delta_{1}- \delta_{3}
\right)=0.045,
\end{equation}
where $\eta$ is the maximum pion momentum (in units of $m_{\pi}$).
Although $\lambda_{2} \gg \lambda_{1}$, the isospin structure of
the $\lambda_{2}$ term is such that it cannot contribute to
$\pi^{0}$ production.

We note that without initial- or final-state distortions, the
diagram (a) of \fig{histdiagrams} vanishes because of
four-momentum conservation. The calculations of \rf{Koltun:1966pr}
were performed using the Hamada-Johnston phenomenological
potential for the $NN$ distortion, and neglecting the Coulomb
interaction between the two protons. The calculated cross section
for $pp \rightarrow pp \pi^{0}$ of $17 \eta^{2} \mu b$ was found
to be consistent with the measured cross section near threshold,
which was, however, not well determined\cite{Stallwood:1959pr} by
that time.
\subsubsection{The role of final-state and Coulomb interactions}
\begin{figure}
\begin{center}
  \includegraphics[width=.59\textwidth,keepaspectratio]{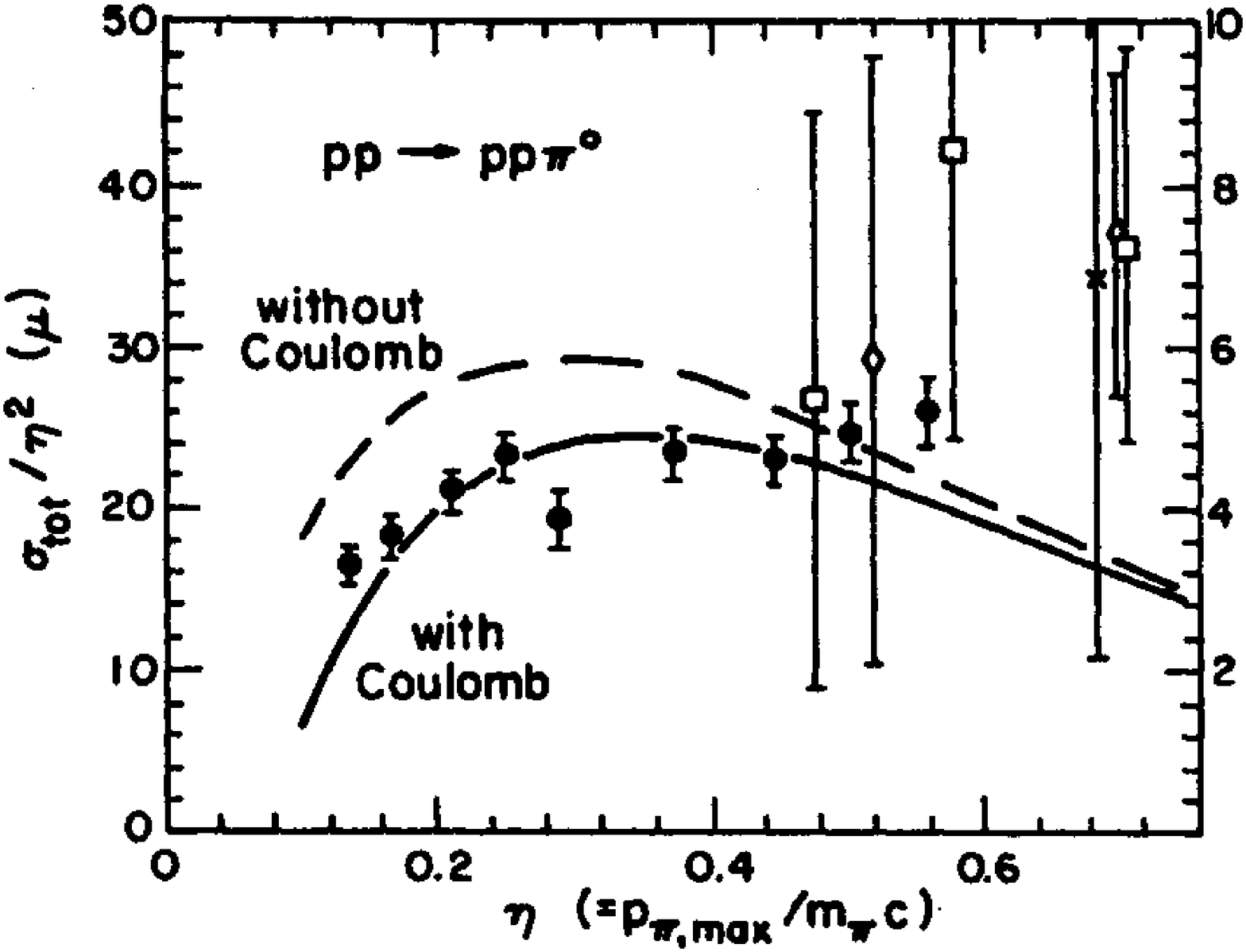}
  \captions[Energy dependence of the total cross section for
  the reaction $pp \rightarrow pp \pi^{0}$.]
  {Energy dependence of the total cross section for
  the reaction $pp \rightarrow pp \pi^{0}$. The solid dots are data from
  \rf{Meyer:1990yf}.
  The squares, cross and diamonds are older data. The solid curve is the full calculation
  and the dashed line shows the effect of omitting the Coulomb interaction. The
  data are to be evaluated using the scale on the left, while the theory used the scale
  on the right.
  Note that the calculations underestimate the data by a factor of
  $\sim 5$.
  \figtk{Miller:1991pi}.}\label{crosssauer}
\end{center}
\end{figure}

\sprg When the first high precision data\cite{Meyer:1990yf} on the
reaction $pp \rightarrow pp\pi^{0}$ appeared, they contradicted
the predicted $\eta^{2}$ dependence near threshold.
According to the work of \rfs{{Meyer:1992jt},{Miller:1991pi}} the
energy dependence of the $s$-wave cross section followed from the
phase space and a simple treatment of the final-state
interaction\cite{{Watson:1952ji},{Gell:1954an}} between the two
(charged) protons.
This was found sufficient to reproduce the shape of the measured
cross section up to $\eta \simeq 0.5$ (see \fig{crosssauer}),
where higher partial waves start to contribute\cite{Meyer:1992jt}.
%
%
Also, the inclusion of the Coulomb interaction was found to be
essential to describe the energy dependence of the total cross
section in particular for energies close to threshold. The
validity of the effective-range approximation employed in
\rf{Koltun:1966pr} for the energy dependence of the final state
turned out to be limited to energies rather close to threshold
($\eta \leq 0.4$).

However, the theory failed in describing quantitatively the cross
section for $pp \rightarrow pp \pi^{0}$ by a factor of $5$, which
was in contrast to the reaction $pp \rightarrow pn \pi^{+}$, where
the discrepancy was less than a factor of $2$, as reported in
\rf{Daehnick:1995gw}.
\rf{Miller:1991pi} suggested that the problem arose from the use
of an over-simplified pion nucleon interaction, namely in
considering the exchanged pion to be on-shell. The on-shell
$s$-wave pion nucleon interaction is constrained to be small by
the requirements of chiral symmetry but, for the production
reaction to proceed, either the re-scattered pion or a nucleon
must be off-shell. This means that the $\pi N$ amplitude relevant
for pion production must be larger than the theoretical one for
on-shell particles, whose investigation was to be subsequently
pursued.
\subsubsection{The inclusion of the $\Delta$}
\begin{figure}
\begin{center}
  \includegraphics[width=.58\textwidth,keepaspectratio]{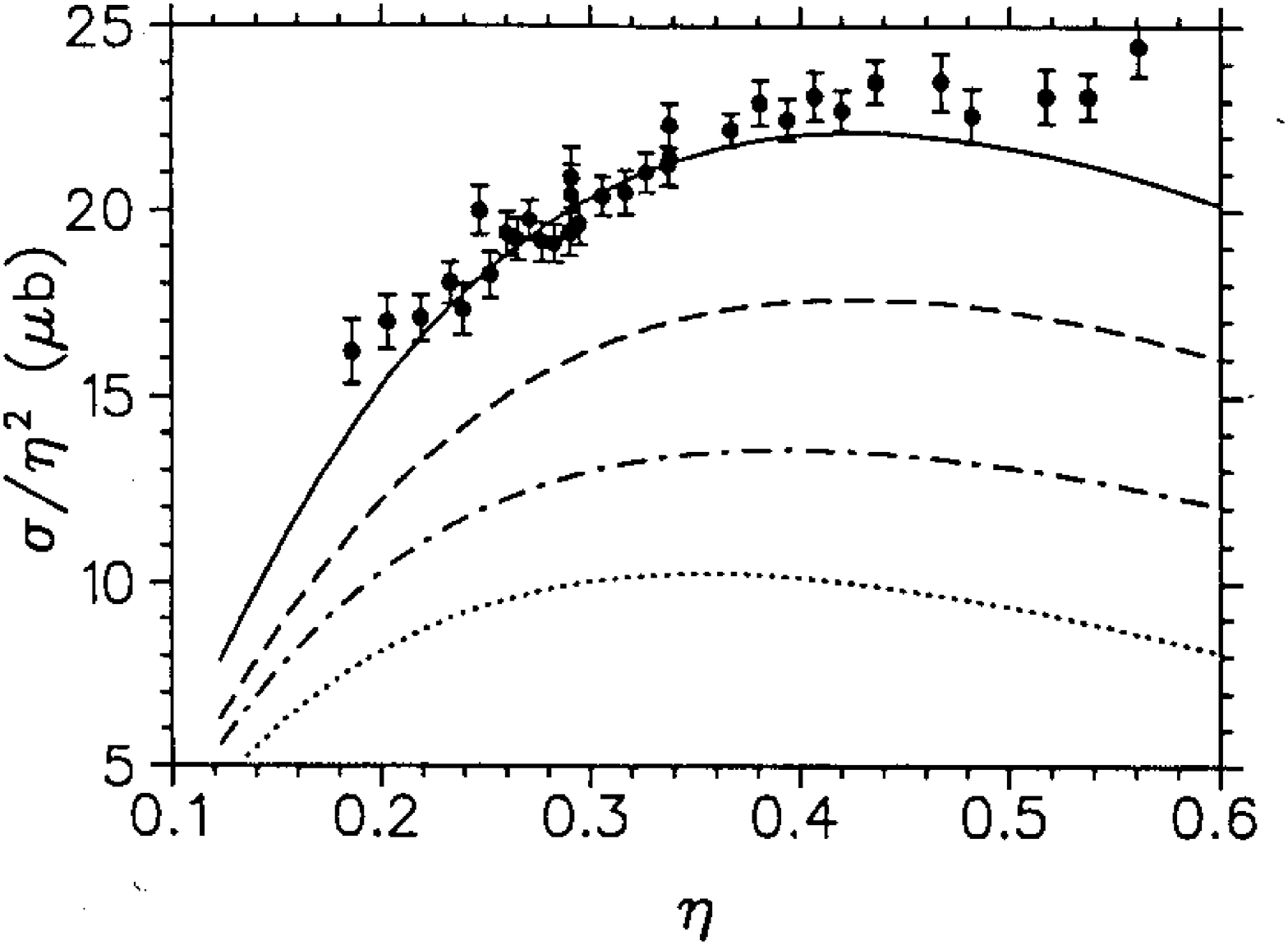}
  \captions[Cross section for $pp \rightarrow pp \pi^{0}$
  including also the $\Delta$ resonance mechanism]
  {Cross section for $pp \rightarrow pp \pi^{0}$. The solid
  line is the full calculation. The dotted line is the
  direct production due only to second term of \eq{h1phen} and going beyond
  the static approximation, the
  dash-dotted line includes also re-scattering through the $\Delta$. The
  dashed line is the full purely nucleonic production. All the
  calculations were multiplied by a common arbitrary factor of $3.6$.
  The data points are from \rf{Meyer:1990yf}. \figtk{Niskanen:1992yt}.}
  \label{crossniskanen1}
\end{center}
\end{figure}

\sprg The work of \rf{Niskanen:1992yt} considered the $p$-wave
re-scattering through a $\Delta \left(1232 \right)$ resonance
(diagram c) of \fig{histdiagrams}) by introducing finite-range $N
\Delta$ coupled-channel admixtures to the nucleonic wave
functions. The transition potential $NN \rightarrow N \Delta$
included both $\pi$ and $\rho$ exchanges.
%
%
The pion production vertex was taken from \eq{h2phen} and
including the relativistic effects arising from the use of the
pion total energy $E_{\pi}$ instead of the pion mass (static
approximation).
The isoscalar and isovector parameters $\lambda_{1}$ and
$\lambda_{2}$ of the phenomenological hamiltonian of \eq{h2phen}
were allowed an energy dependence through the
momentum of the pion.  The results are shown in
\fig{crossniskanen1}.

The small difference between the theoretical models was attributed
to the different value used for the $\pi N$ coupling constant and
to the relativistic kinematics. The inclusion of the non-galilean
term and the $s$-wave scattering gave an enhancement of over
$60\%$ (dashed curve). Another $25\%$ enhancement arose from the
inclusion of the re-scattering through the $\Delta$.
%
%
However, the cross section was still missed by almost a factor of
$\sim 4$.

\subsubsection{Charged pion production}

\sprg The first calculations on $\pi^{+}$ production where those
of Schillaci, Silbar and Young\cite{Schillaci:1969hq}. General
isospin and phase space
arguments\cite{{Rosenfeld:1954pr},{Gell:1954an}} were employed to
predict the spin, isospin and energy dependence of the total cross
section, including all partial waves for the $NN$ amplitudes, but
accounting only for $s$-wave pion-nucleon states. It fails if
contributions from the $\Delta$ resonance are significant.

\begin{figure}[h!]
\begin{center}
  \includegraphics[width=.58\textwidth,keepaspectratio]{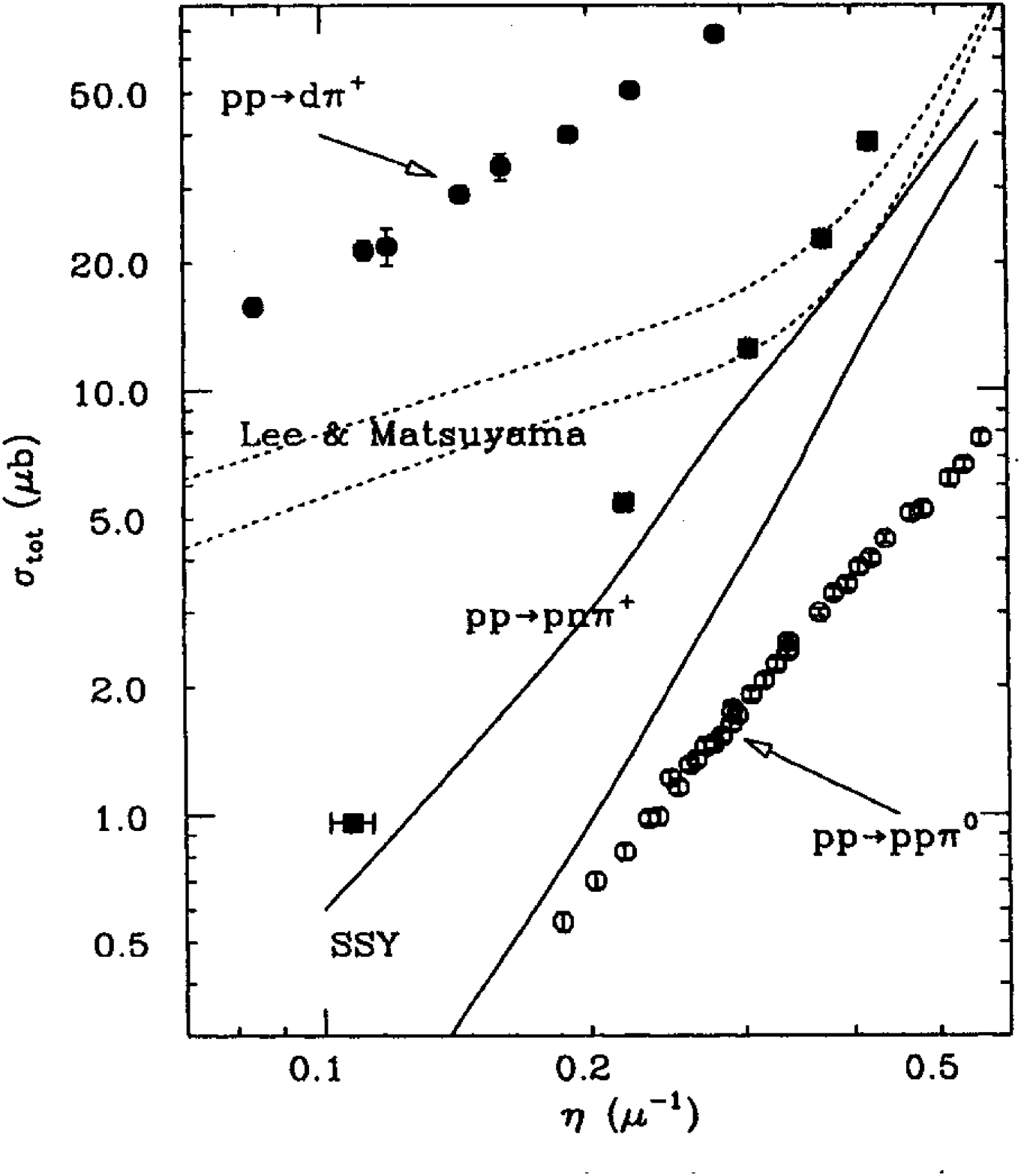}
  \captions[Comparison of measured total $pp \rightarrow pn \pi^{+}$
  cross section with theoretical predictions]
  {Comparison of measured total
  $pp \rightarrow pn \pi^{+}$ cross section with the theoretical
  predictions of \rfs{{Lee:1987hd},{Schillaci:1969hq}}. The solid lines are the
  calculations of Schillacci, Silbar and Young\cite{Schillaci:1969hq}.
  The dotted lines are the calculations of Lee and Matsyuama\cite{Lee:1987hd}.  \figtk{Daehnick:1995gw}. }
  \label{crossDaehnick}
\end{center}
\end{figure}

The second prediction was made by Lee and
Matsuyama\cite{Matsuyama:1986sh} with a coupled channel formalism
that focused on the effects of the $\Delta$ intermediate state. In
\rf{Matsuyama:1986sh,Lee:1987hd}, the $\Delta$ process is handled
rigorously while the non resonant pion production process is
introduced as a perturbation. The estimated  $\Delta$ contribution
was roughly $15 \%$ of the total cross section. Both calculations
were not able to describe the experimental data for $pp
\rightarrow pn\pi^{+}$ as it is shown in \fig{crossDaehnick}
(solid and dotted lines, respectively).

Calculations based on a relativistically covariant one-boson
exchange model\cite{Engel:1996ic}, also suggested that the
contribution of a $\Delta$-isobar is not important at energies
below $350 \umev$ (due to the fact that at lower energies pions
are predominantly produced in a $\pi N$ relative $s$-state and
thus the possibility of forming a $\Delta$-isobar is greatly
reduced), but dominate at higher beam energies. The cross sections
for $pp \rightarrow pn\pi^{+}$
near threshold were under-predicted a factor of $2$-$4$.

The work of \rf{Shyam:1996id} applied the Watson
theorem\cite{Watson:1952ji}\footnote{In 1952
Watson\cite{Watson:1952ji} showed that for a short-range strong
(attractive) $NN$ interaction and in the regime of low relative
energies of the interacting particles, the energy dependence of
the total $NN \rightarrow NN x$ cross section is determined only
by the phase space and by the on-shell $NN$ $T$-matrix,
\begin{equation}
\sigma_{NN\rightarrow NNx}\left(  \eta\right)
\propto\int_{0}^{m_{x}\eta }d\rho\left(  q\right)  \left\vert
T\left(  k_{0},k_{0}\right)  \right\vert
^{2}\propto\int_{0}^{m_{x}\eta}d\rho\left(  q\right)  \left[
\frac{\sin \delta\left(  k_{0}\right)  }{k_{0}}\right]  ^{2}. \label{crossWatson}%
\end{equation}
Here, the momentum of the outgoing meson is $q$ and $d \rho \left(
q\right)$ denotes the phase space. The relative momentum on the
final nucleons is $k_{0}$, and $\delta \left(k_{0}\right)$ are the
corresponding phase-shifts of the final $NN$ subsystem (restricted
to $s$-waves). Recently, the work of \rf{Hanhart:1998rn} concluded
that Watson's requirement of an attractive FSI is unnecessary to
obtain the energy dependence of the cross section given by
\eq{crossWatson}.}
to the final state interaction of \rf{Engel:1996ic}. The
calculated cross sections were within $25 \%$ of the $pp
\rightarrow pn \pi^{+}$ data\cite{Hardie:1997mg}, but were also
under-predicted near threshold.

Also, in the work of \rf{Lee:1995sg}, the isoscalar heavy meson
exchange  found to dominate in $pp \rightarrow pp \pi^{0}$ was
however shown to be less significant in $pp \rightarrow pn
\pi^{+}$, where the re-scattering diagram was the most important
one.
%
%
Further theoretical studies on these issues were then clearly
needed to clarify the role of the different production mechanisms.
\subsection{The first quantitative understandings}

\sprg
The first quantitative understanding of the $pp \rightarrow pp
\pi^{0}$ data was reported by Lee and Riska\cite{Lee:1993xh} and
later confirmed by Horowitz et al.\cite{Horowitz:1993hk}, where it
was demonstrated that short range mechanisms (diagram (d) of
\fig{histdiagrams}) can give a sizeable contribution.
In these works, the difficulty in describing the cross section for
$pp \rightarrow pp\pi^{0}$ was overcame by considering the pair
terms, positive and negative energy components of the nucleon
spinors, connected to the isoscalar part of the $NN$ interaction.
%
%
\subsubsection{The importance of short-range mechanisms \label{firstquant}}
In the work of \rf{Lee:1993xh}, the short range two-nucleon
mechanisms that are implied by the nucleon-nucleon interaction
were taken into account by describing the pion-nucleus interaction
by the extension of Weinberg's effective pion-nucleon interaction
to nuclei:
\begin{equation}
\mathcal{L}=-\frac{1}{f_{\pi}} \boldsymbol{A}^{\mu} \cdot
\partial_{\mu} \boldsymbol{\pi},
\end{equation}
where $\boldsymbol{A}^{\mu}$ is the isovector axial current of the
nuclear system\footnote{The relationship between the current and
the amplitude $\mathcal{M}$ is $Q_{\mu}A^{\mu}= f_{\pi}
\mathcal{M}$, where $Q=\left(E_{\pi},\vec{q} \right)$ is the
four-momentum of the emitted pion. Near threshold, the
interactions that involve $s$-pions should dominate, and thus the
amplitude is simply given by
$\mathcal{M}=-\frac{1}{f_{\pi}}A^{0}E_{\pi}$, which coincides with
the second term of \eq{h1phen}.}. This formulation reduces the
calculation of matrix elements for nuclear pion production to the
construction of the axial current operator, which is formed of
single-nucleon and two-nucleon (exchange) current operators. The
single-nucleon contribution is\footnote{The conventional
single-nucleon pion production operator (first term of
$\mathcal{H}_{1}$ of \eq{h1phen}) is recovered by using the
Goldberger-Treiman relation $ \frac{g_{A}}{2 f_{\pi}}=\frac{f_{\pi
NN}}{m_{\pi}}$.}
\begin{equation}
\mathbf{A}^{0}_{\mathrm{one-body}} =-\frac{g_{A}}{2} \sum_{i=1,2}
\left[\vec{\sigma}^{(i)} \cdot
\frac{\vec{p}^{\,\prime}_{i}+\vec{p}_{i}}{2M}\boldsymbol{\tau}^{(i)}
\right], \label{A0Iriska}
\end{equation}
where $\vec{p}_{i} \left(\vec{p}^{\,\prime}_{i}\right)$ is the
initial(final) nucleon momenta.
When the nucleon-nucleon interaction is expressed in terms of
Fermi invariants (scalar, vector, tensor and axial-vector) there
is an unique axial exchange charge operator that corresponds to
each invariant. The general two-body-exchange charge operator is
then
\begin{equation}
\mathbf{A}^{0}_{\mathrm{two-body}}=\frac{1}{\left(2 \pi
\right)^{3}}\left[\mathbf{A}^{0} \left( S \right)+\mathbf{A}^{0}
\left( V \right)+\mathbf{A}^{0} \left( T \right)+\mathbf{A}^{0}
\left( A \right) \right]. \label{A0IIriska}
\end{equation}
The axial exchange charge operators associated with the scalar
$\left( S \right)$ and vector $\left( V \right)$ components of the
$NN$ interaction are the most important. Dropping terms that
involve isospin flip and therefore do not contribute to $\pi^{0}$
production, $\mathbf{A}^{0}\left( S\right)$ and
$\mathbf{A}^{0}\left( V\right)$ read\cite{Kirchbach:1991af}
\begin{eqnarray}
\mathbf{A}^{0}\left(  S\right)    & =& \frac{g_{A}}{M^{2}}\left[  v_{S}%
^{+}(  \vec{k})  \boldsymbol{\tau}^{(1)}+v_{S}^{-}(  \vec{k})
\boldsymbol{\tau}^{(2)}\right]  \vec{\sigma}^{(1)}\cdot\vec{P}%
_{1}+\left(  1\leftrightarrow2\right)  ,\\
\mathbf{A}^{0}\left(  V\right)    & = & \frac{g_{A}}{M^{2}}\left\{
\left[ v_{V}^{+}(\vec{k})
\boldsymbol{\tau}^{(1)}+v_{V}^{-}(\vec{k})
\boldsymbol{\tau}^{(2)}\right]  \left[  \vec{\sigma}^{(1)}%
\cdot\vec{P}_{2}+\frac{1}{2}\vec{\sigma}^{(1)}\times
\vec{\sigma}^{(2)}\cdot\vec{k}\right]  \right.  \\
&& \left.  +\frac{1}{2}iv_{V}^{-}(  \vec{k}) \left(
\boldsymbol{\tau}^{(1)}\times\boldsymbol{\tau}^{(2)}\right)  \vec{\sigma}^{(1)}\cdot\vec{k}%
\right\}  +\left(  1\leftrightarrow2\right)  \nonumber .
\end{eqnarray}
The momentum operators are defined as
$\vec{P}_{i}=\frac{\vec{p}^{\,\prime}_{i}+\vec{p}_{i}}{2}$ and
$\vec{k}=\vec{p}^{\,\prime}_{2}-\vec{p}_{2}=\vec{p}_{1}-\vec{p}^{\,\prime}_{1}$
is the momentum transfer. The momentum dependent potential
functions $v_{j}^{\pm}$ are isospin independent $\left( + \right)$
and isospin dependent $\left( - \right)$ functions associated with
the corresponding Fermi invariants. These functions can be
constructed from the components of complete phenomenological
potential models\cite{Blunden:1992xx}, or alternatively, by
employing phenomenological meson exchange
models\cite{Horowitz:1993hk}.

The contribution of the $s$-wave pion re-scattering to the $pp
\rightarrow pp \pi^{0}$ amplitude can be included by adding to
$\mathbf{A}^{0}_{\mathrm{one-body}}+\mathbf{A}^{0}
_{\mathrm{two-body}}$ the following two-body axial charge operator
\begin{equation}
A^{0} \left( \pi \right)=-\frac{1}{\left( 2 \pi \right)^{3}}
\frac{8 \pi \lambda_{1}}{E_{\pi}} \frac{f_{\pi NN}}{m_{\pi}^{2}}
\frac{\vec{\sigma}^{(2)} \cdot
\vec{k}_{2}}{m_{\pi}^{2}+\vec{k}_{2}^{2}} f ( \vec{k}_{2} ) +
\left(1 \leftrightarrow 2 \right), \label{Arescriska}
\end{equation}
where $f ( \vec{k}_{2} )$ is a monopole form factor (to be
consistent with the Bonn boson exchange model for the
nucleon-nucleon potential) and $E_{\pi}$ the energy of the
produced pion. Note that $A^{0} \left( \pi \right)$ of
\eq{Arescriska} does not include the dependence on the energy of
the exchanged pion (i. e., the static approximation is employed
here).

\begin{figure}[h!]
\begin{center}
  \includegraphics[width=.69\textwidth,keepaspectratio,angle=0.69]{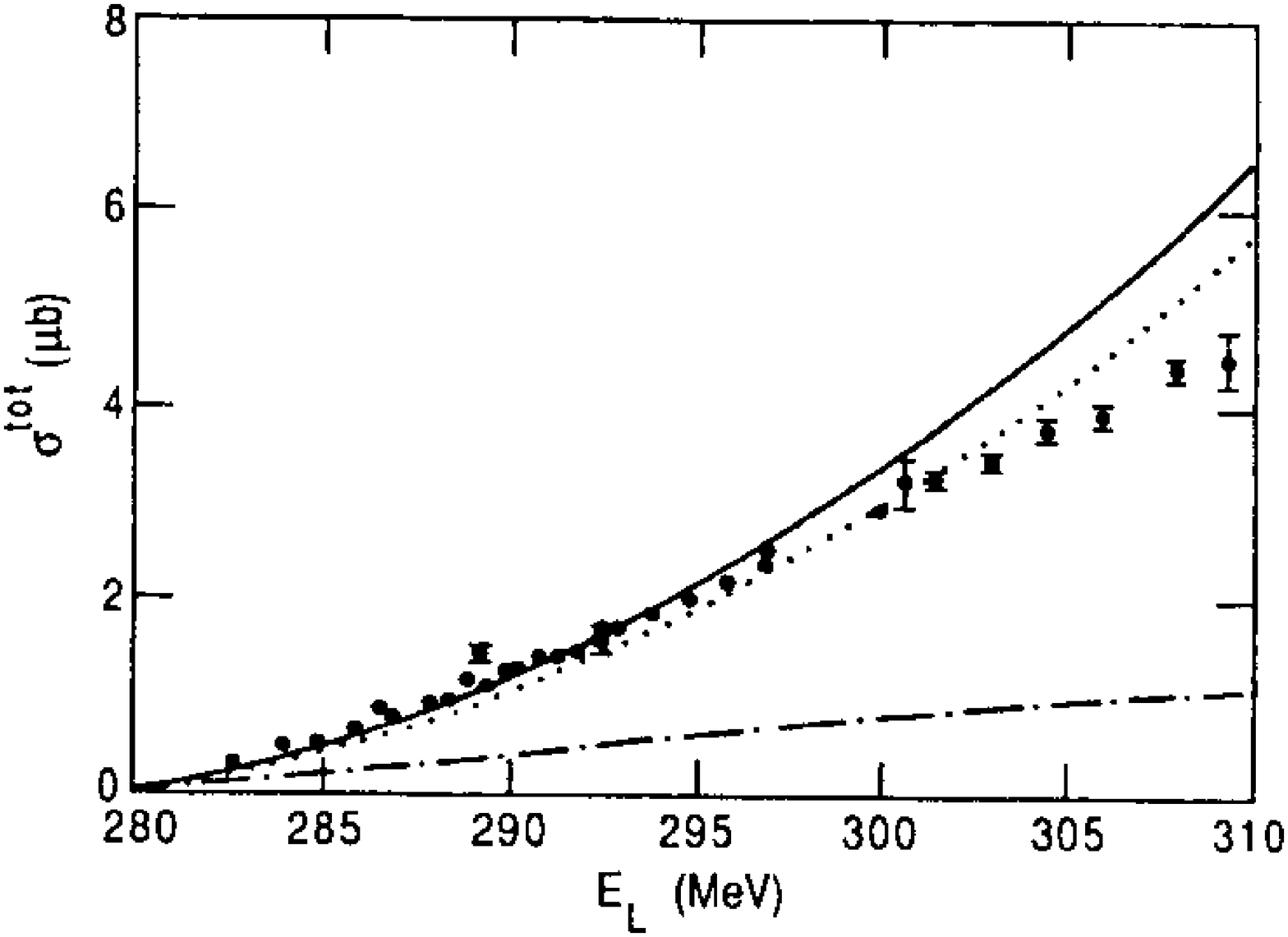}
  \captions[Cross section for $pp \rightarrow pp \pi^{0}$
  considering heavy-meson exchange mechanisms]
  {Cross section for $pp \rightarrow pp \pi^{0}$.
  The solid(dotted) line is the full calculation using the Bonn(Paris)
  potential to construct the axial exchange charge operator.
  The dot-dashed line is obtained by keeping only the one-body
  term of \eq{A0Iriska} and the pion re-scattering term of \eq{Arescriska} using the
  Paris potential. The data points are from \rfs{{Meyer:1990yf},{Meyer:1992jt}}.
  \figtk{Lee:1993xh}.}
  \label{crossriska}
\end{center}
\end{figure}

The results for the $pp \rightarrow pp\pi^{0}$ are shown in
\fig{crossriska}.
As already found in previous
works\cite{{Miller:1991pi},{Niskanen:1992yt}}, the impulse and
re-scattering mechanism  were not enough to describe the data
(dot-dashed line of \fig{crossriska}). The short-range axial
charge operators enhance largely the cross section and remove most
of the under prediction (solid and dotted lines of
\fig{crossriska}). However, it was also found that for both $NN$
potentials employed (Paris and Bonn), the energy dependence of the
data was not reproduced in detail. According to \rf{Lee:1993xh},
this might be due to the neglect of the energy dependence of the
parameter $\lambda_{1}$ in the effective amplitude of
\eq{Arescriska}, or to $p$-wave contributions and $\pi NN$
three-body scattering distortions in the final state. Although
these corrections were expected to be small, they could in
principle lead to significant contributions through the
interference with large amplitudes and thus have large significant
effects on the predicted energy dependence.

Shortly after the work of \rf{Lee:1993xh}, where the
meson-exchange contributions were calculated from phenomenological
potentials, \rf{Horowitz:1993hk} used an explicit one-boson
exchange model for the $NN$ interaction and for the calculation of
the MEC's.
As in \rf{Lee:1993xh} the $\pi N$ re-scattering vertex was
restricted however to the on-shell matrix element and to $s$-wave
pion production.
The largest contribution was found to come from the Z-diagrams
mediated by $\sigma$-exchange, which was of the order of the
one-body (impulse) term. The next important contribution was from
the $\omega$ meson Z-diagrams ($35-45 \%$ of the one-body term
contribution).
\subsubsection{The importance of off-shell effects}
\sprg Shortly after the discovery of the importance of the
short-range mechanisms, Hern\'{a}ndez and
Oset\cite{Hernandez:1995kj} demonstrated, using various
parameterisations for the $\pi N \rightarrow \pi N$ transition
amplitude, that its strong off-shell dependence could also be
sufficient to remove the discrepancy between the Koltun and Reitan
model\cite{Koltun:1966pr} and the data.

\begin{figure}
\begin{center}
  \includegraphics[width=.69\textwidth,keepaspectratio]{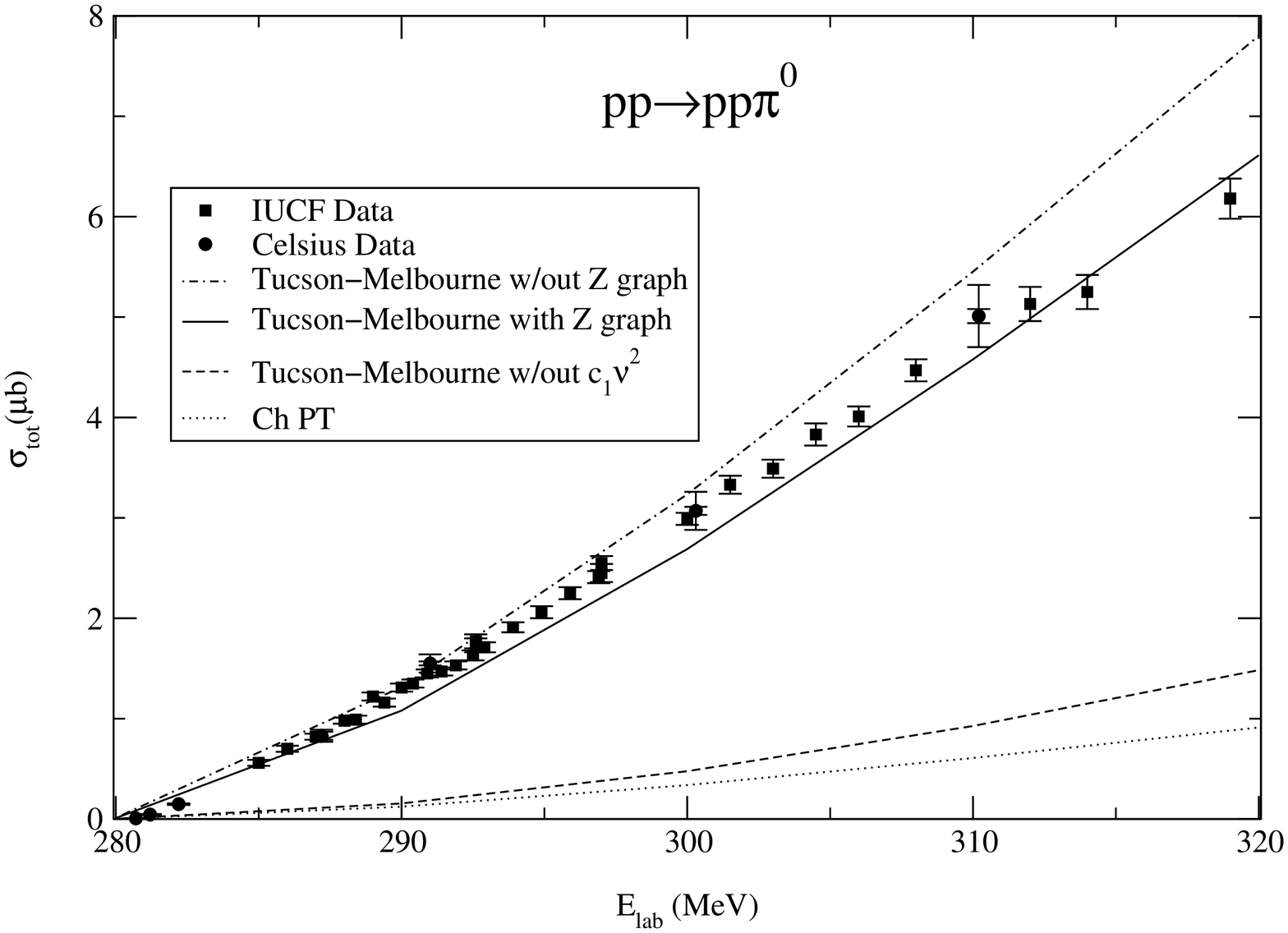}
  \captions[Cross section for $pp \rightarrow pp \pi^{0}$
  (re-scattering half-off shell amplitude adapted from the off-shell
  $\pi N$ amplitude of the Tucson-Melbourne three-body force)]
  {Cross section for $pp \rightarrow pp \pi^{0}$ using the Bonn B
  potential for the initial and final state interaction between the two protons.
  All calculations include both impulse and pion re-scattering
  diagrams. \figtk{Pena:1999pu}.}
  \label{crossCoon}
\end{center}
\end{figure}

The importance of the off-shell amplitudes was also seen in a
relativistic one boson exchange model\cite{Gedalin:1998mm}. In the
work of \rf{Pena:1999pu}, the model independent off-shell $\pi N$
amplitude obtained by current algebra (and used previously in the
Tucson-Melbourne three-nucleon force) was also considered as input
for the pion re-scattering contribution to $pp \rightarrow pp
\pi^{0}$ near threshold. It was found that this pion re-scattering
contribution, together with the direct-production term, provided a
good description of the $\pi^{0}$ production data, when the
current algebra $\pi N$ amplitude parameters were updated with the
phenomenological information obtained from the new meson factory
$\pi N$ scattering data (see \fig{crossCoon}).
%
%
\subsubsection{Z-diagrams: perturbative vs. non-perturbative}
\sprg In the succeeding years many theoretical efforts were made
for the calculation of the $pp \rightarrow pp \pi^{0}$ cross
section. In \rfs{{Engel:1996ic},{Bernard:1998sz}} covariant one
boson exchange models were used in combination with an approximate
treatment of the nucleon-nucleon interaction. Both models turned
out to be dominated by heavy meson exchanges, thus giving further
support to the picture proposed in
\rfs{{Lee:1993xh},{Horowitz:1993hk}}.

However, in \rf{Adam:1997pe} the negative energy nucleons were
re-examined using the covariant spectator description, for both
the production mechanism and for the initial and final state $pp$
interaction.
%
%
This approach differs crucially from earlier ones by including non
perturbatively the intermediate negative-energy states of the
nucleons\footnote{As mentioned before, in perturbative approaches
these contributions (often called \textit{Z-diagrams}) are
simulated by the inclusion of effective meson-exchange operators
acting in two-nucleon initial and final states.}.

The perturbative result for the direct-production diagram was
found to be about $3$ times larger that the non-perturbative one.
Although the calculation in \rf{Adam:1997pe} did not include the
re-scattering diagram contribution, it showed that the sensitive
cross section for $\pi^{0}$ production seemed to be an ideal place
to look for effects of relativistic dynamics.
\subsubsection{The role of the nucleon resonances}
\begin{figure}
\begin{center}
  \includegraphics[width=.69\textwidth,keepaspectratio]{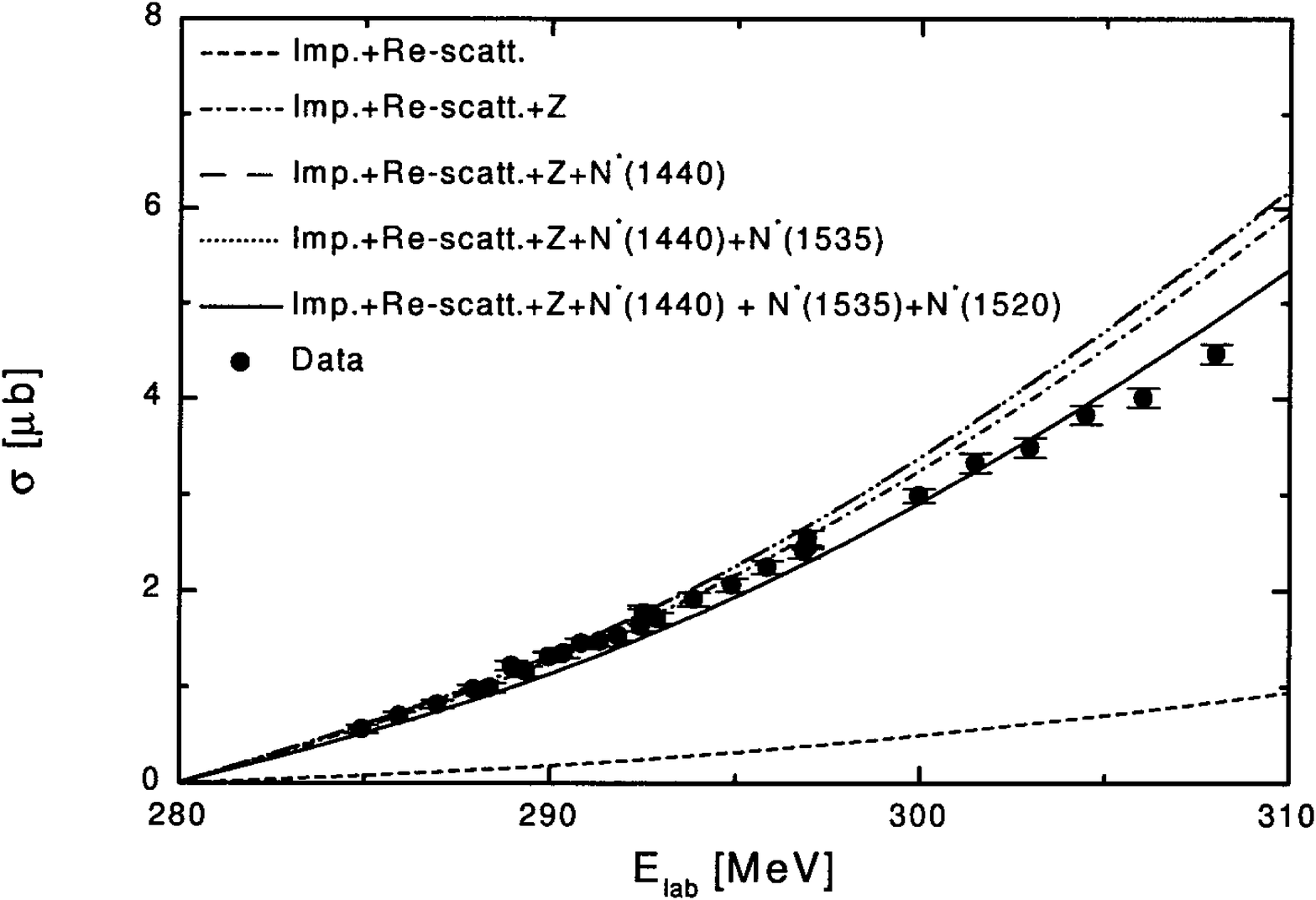}
  \captions[The effect of the nucleon resonances
   on the $pp \rightarrow pp \pi^{0}$ cross section]
  {Total cross section for $pp \rightarrow pp \pi^{0}$ as
  a function of the proton laboratory energy. The net effect of the resonances
  is small. \figtk{Pena:1999hq}.}
  \label{crossPena}
\end{center}
\end{figure}

\sprg Additional short-range contributions were also suggested,
namely the $\rho-\omega$ meson exchange
current\cite{vanKolck:1996dp}, resonance
contributions\cite{vanKolck:1996dp,Pena:1999hq,Hernandez:1998ra}
(see \fig{crossPena}) and loops that contain
reso\-nances\cite{Hernandez:1998ra}. All those, however, turned
out to be smaller when compared to the heavy meson exchanges and
the off-shell pion re-scattering.
%
\section{Coupled-channel phenomenological calculations \label{Phenapproach}}

\sprg
The starting point of the $NN$-$N \Delta$ approach is the
recognition that the nucleon is a composite system.
%
%
Since the $\Delta$ isobar is the most important mode of nucleonic
excitation at intermediate energies, a possible process
contributing (in second order) to nucleon-nucleon elastic
scattering is the transition from a pure nucleonic state into a
state into a nucleon plus a $\Delta$ (or two $\Delta$'s) with the
inverse transition taking the system back to a two-nucleon state
again. From this point of view, the nucleon-nucleon problem is a
coupled-channel system involving at least the $NN$ and the
$N\Delta$ channels\cite{Garcilazo:1990ws}.
%
\subsection{The Hannover model \label{Hannovermodel}}
\sprg The Hannover
model\cite{{Popping:1987ex},{Pena:1992ui},{Pena:1993pd}} for the
$NN$ system considers the $\Delta$ isobar and pion degrees of
freedom in addition to the nucleonic one. The model is based on a
hamiltonian approach within the framework of non covariant quantum
mechanics. In isospin-triplet partial waves, it extends the
traditional approach with purely nucleonic potentials. It is
constructed to remain valid up to $500 \umev$ CM energy. The
Hilbert space considered comprises $NN$ and $N\Delta$ basis
states, connected by transition potentials. Pion production and
pion absorption are mediated by the $\Delta$ isobar excited by
$\pi$ and $\rho$ exchange.
The model accounts with satisfactory accuracy for the experimental
data of elastic nucleon-nucleon scattering, of the inelastic
reactions $pp \rightarrow  \pi^{+} d$ and of elastic pion-deuteron
scattering.

Lee and Matsuyama also performed
calculations\cite{{Lee:1985jq},{Matsuyama:1986sh},{Lee:1987hd}}
for pion production within a coupled-channel approach, which
differed from that of the Hannover group mainly by the treatment
of the energy in the $\Delta$ propagator.
Both the theoretical
predictions\cite{{Matsuyama:1986sh},{Pena:1992ui}} from the
Hannover and the Lee and Matsuyama models for the $pp \rightarrow
pn \pi^{+}$ differential cross section were found to be quite
sensitive to the inclusion of the $N \Delta$ potential, but the
under-prediction of the data could not be completely removed.
The calculations of \rfs{{Matsuyama:1986sh},{Pena:1992ui}}
considered an energy region well above pion production threshold
($\sim 580 \umev - 800 \umev$), since pion production is assumed
to occur only via an intermediate $\Delta$ excitation and thus the
details of the $\pi N$ amplitude (related to chiral symmetry and
the chiral limit), which are important close to threshold, were
not included.

\subsection{The J\"{u}lich model}

\sprg The J\"{u}lich
model\cite{{Hanhart:1995ut},{Hanhart:1998za},{Hanhart:2000jj}}
attempts to treat consistently the $NN$ and the $\pi N$
interaction for meson production close to threshold, taking both
from microscopic models. Although not all parameters and
approximations used in the two systems are the same,  the same
effective Lagrangians consistent with the symmetries of the strong
interaction underlie the potentials to be used in the
Lippmann-Schwinger equations for $NN$ and $\pi N$, independently.
All single pion production channels including higher partial waves
are considered.

The model used for the $NN$ distortions in the initial and final
states is based on the Bonn
potential\cite{Machleidt:1987hj}.
%
%
The $\Delta$-isobar is treated in equal footing with the nucleons
through a coupled-channel framework including the $NN$ as well as
the $N\Delta$ and $\Delta \Delta$ channels. The model parameters
were adjusted to the phase shifts below the pion production
threshold.

Since all the short range mechanisms suggested in literature to
contribute to pion production in $NN$ collisions mainly influence
the production of $s$-wave pions, in the J\"{u}lich model only a
single diagram was included (heavy meson through the $\omega$ as
in diagram (d) of \fig{histdiagrams}) to parameterise these
various effects. The strength of this contribution was adjusted to
reproduce the total cross section of the reaction $pp \rightarrow
pp \pi^{0}$ close to threshold.
%
%
After this is done, all the parameters are fixed.

\begin{figure}[h!]
\begin{center}
  \includegraphics[width=.88\textwidth,keepaspectratio]{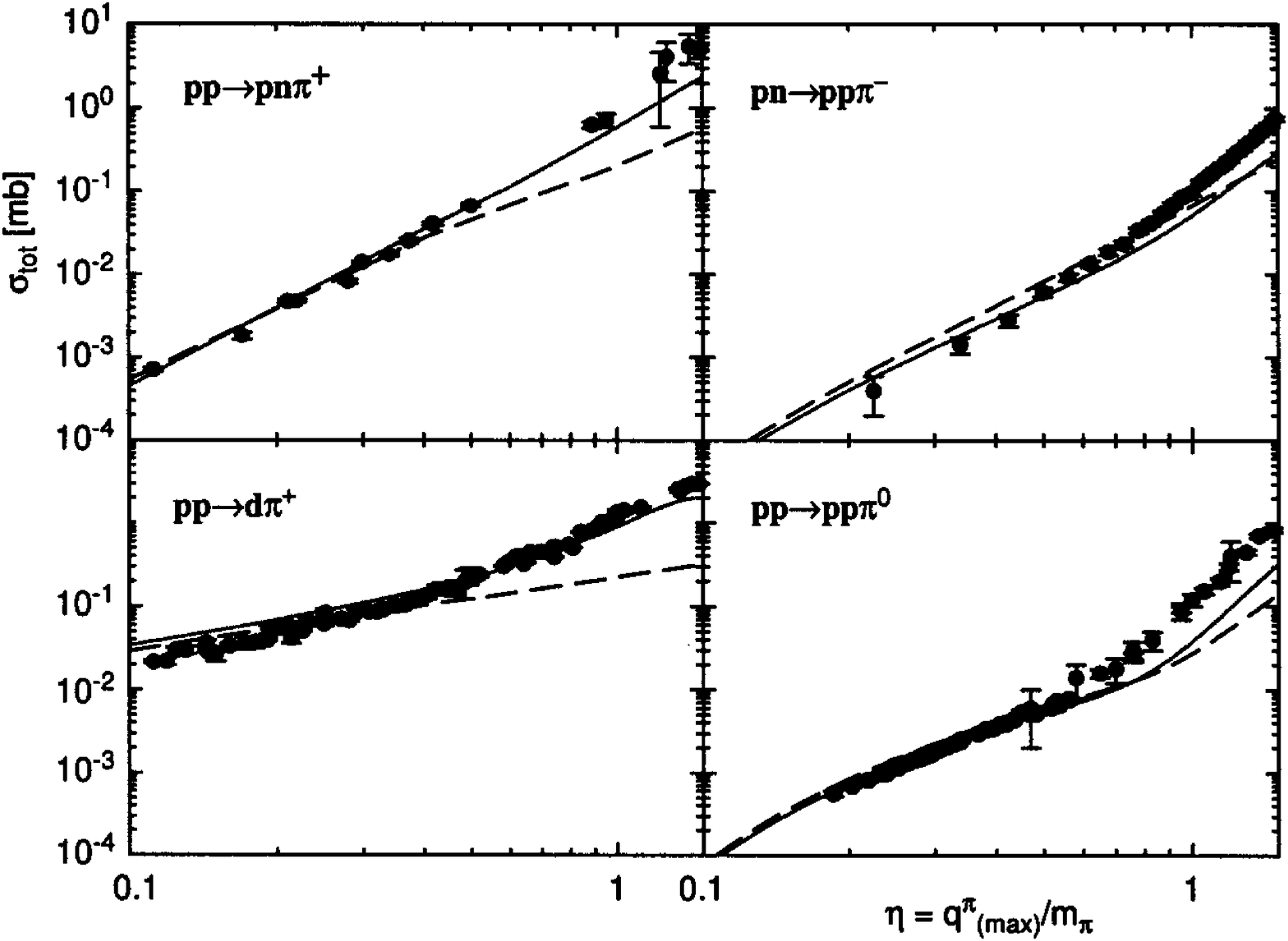}
  \captions[Comparison between the experimental values for the cross sections
   and the calculations of the J\"{u}lich phenomenological model
   for pion production reactions in $NN$ collisions]
   {Comparison of the J\"{u}lich phenomenological model of
  \cite{{Hanhart:1995ut},{Hanhart:1998za},{Hanhart:2000jj}}
  to the data. The solid lines show the result of the full model and the
  dashed lines show the results without the $\Delta$ contributions. \figtk{Hanhart:2003pg}.}
  \label{crossHanhart}
\end{center}
\end{figure}

The model describes qualitatively the data, as shown in
\fig{crossHanhart}. The most striking differences appear for
double polarisation observables in the neutral pion channel. As a
general pattern the amplitudes seem to be of the right order of
magnitude, but show a wrong interference pattern.
For charged pion production most of observables are described
satisfactorily. In contrast to the neutral channel, the charged
pion production was found to be completely commanded by two
transitions, namely, $^{3}P_{1} \rightarrow ^{3}S_{1}s$, which is
dominated by the isovector pion re-scattering, and $^{1}D_{2}
\rightarrow ^{3}S_{1}p$, which governs the cross section
especially in the regime of the $\Delta$ resonance.
%
%
\section{Chiral perturbation theory \label{ChiPertTh}}

\sprg
In the late 90's there was the hope that $\chi$PT might resolve
the true ratio of re-scattering and short-range contributions in
pion production.
It came as a big surprise, however, that the first results for the
reaction $pp \rightarrow pp \pi^{0}$
\cite{{Park:1995ku},{Cohen:1995cc}} showed that the $\chi$PT $\pi
N$ scattering amplitude interfered destructively with the direct
contribution, making the discrepancy with the data even more
severe, and thus suggesting a significant role for heavy meson
exchanges in $\pi^{0}$
production\cite{Hanhart:1997jd,Dmitrasinovic:1999cu}. In addition,
the same isoscalar re-scattering amplitude also worsened the
discrepancy in the $\pi^{+}$ channel\cite{daRocha:1999dm}.

In low-energy pion physics, the constrains to an effective field
theory ($\chi$PT) come from chiral symmetry, since it forces not
only the mass of the pion to be low, but also the interactions to
be weak: the pion needs to be free of interactions in the chiral
limit for vanishing momenta. The first success of $\left(
\chi\text{Pt}\right)  $ was the application to meson-meson
scattering\cite{Bijnens:1997vq}. Treating baryons as heavy allowed
straightforward extension of the scheme to
meson-baryon\cite{Bernard:1995dp} as well as baryon-baryon
systems\cite{{Ordonez:1993tn},{Ordonez:1995rz},{Epelbaum:1998ka},{Epelbaum:1999dj}}.
The $\Delta$ isobar could also be included consistently in the
effective field theory\cite{Hemmert:1996xg}.
%
It was shown recently\cite{{Hanhart:2000gp},{Hanhart:2002bu}} that
when the new scale induced by the initial momentum
$p\sim\sqrt{m_{\pi}M}$\footnote{Since $\sqrt{Mm_{\pi}}$ is smaller
than the characteristic mass scale of QCD $\left( M_{QCD} \sim 1
\ugev \right)$, at least in the chiral limit, the contribution of
other states (the Roper, the $\rho$ meson, etc) can be buried in
short-range interactions.} for meson production in nucleon-nucleon
collisions is taken properly into account, the series indeed
converges.

The starting point for the derivation of the amplitude is an
appropriated Lagrangian density, constructed to be consistent with
the symmetries of QCD and ordered according to a particular
counting scheme. The leading order Lagrangian
is\cite{{Hanhart:2003pg},{daRocha:1999dm},{Bernard:1995dp}}
%
%
\begin{eqnarray}
\mathcal{L}^{\left(  0\right)  }  &
=&\frac{1}{2}\partial_{\mu}\boldsymbol{\pi
}\partial^{\mu}\boldsymbol{\pi}-\frac{1}{2}m_{\pi}^{2}\boldsymbol{\pi}%
^{2}+\frac{1}{f_{\pi}^{2}}\left[  \left(
\boldsymbol{\pi}\cdot\partial_{\mu }\boldsymbol{\pi}\right)
^{2}-\frac{1}{4}m_{\pi}^{2}\left(  \boldsymbol{\pi
}^{2}\right)  ^{2}\right] \label{LX0Hanhart} \\
& & \hspace{-0.4cm}+N^{\dagger}\left[
i\partial_{0}-\frac{1}{4f_{\pi}^{2}}\boldsymbol{\tau}\cdot\left(
\boldsymbol{\pi}\times\boldsymbol{\dot{\pi}}\right)  +\frac{g_{A}}{2f_{\pi}%
}\boldsymbol{\tau}\cdot\vec{\sigma}\cdot\left(
\vec{\nabla}\boldsymbol{\pi}+\frac
{1}{2f_{\pi}^{2}}\boldsymbol{\pi}\left(
\boldsymbol{\pi}\cdot\vec{\nabla
}\boldsymbol{\pi}\right)  \right)  \right]  N \nonumber \\
& & \hspace{-0.4cm}+\Psi_{\Delta}^{\dagger}\left(
i\partial_{0}-\Delta\right) \Psi_{\Delta
}+\frac{h_{A}}{2f_{\pi}}\left[  N^{\dagger}\left(  \boldsymbol{T}%
\boldsymbol{\cdot}\vec{S}\boldsymbol{\cdot}\vec{\nabla}\boldsymbol{\pi
}\right)  \Psi_{\Delta}+h.c.\right]  +... \nonumber
\end{eqnarray}
and the next-to-leading order Lagrangian is
\begin{eqnarray}
\mathcal{L}^{\left(  1\right)  }  &  =&\frac{1}{2M}\left[  N^{\dagger}%
\vec{\nabla}^{2}N+\Psi_{\Delta}^{\dagger}\vec{\nabla}^{2}\Psi_{\Delta}\right]
+\frac{1}{8Mf_{\pi}^{2}}\left[
iN^{\dagger}\boldsymbol{\tau}\boldsymbol{\cdot}\left(
\boldsymbol{\pi}\times\vec{\nabla}\boldsymbol{\pi}\right)
\boldsymbol{\cdot
}\vec{\nabla}N+h.c.\right] \label{LX1Hanhart} \\
&& \hspace{-0.4cm} +\frac{1}{f_{\pi}^{2}}N^{\dagger}\left[  \left(
c_{2}+c_{3}-\frac {g_{A}^{2}}{8M}\right)
\boldsymbol{\dot{\pi}}^{2}-c_{3}\left(  \vec{\nabla
}\boldsymbol{\pi}\right)  ^{2}-2c_{1}m_{\pi}^{2}\boldsymbol{\pi}^{2}\right. \nonumber \\
&& \hspace{-0.4cm} \left.  -\frac{1}{2}\left(
c_{4}+\frac{1}{4M}\right)\epsilon_{ijk}\epsilon
_{abc}\sigma_{k}\boldsymbol{\tau}_{c}\partial_{i}\boldsymbol{\pi}_{a}\partial
_{j}\boldsymbol{\pi}_{b}  \right]
N+\frac{\delta}{2}N^{\dagger}\left[
\tau_{3}-\frac{1}{2f_{\pi}^{2}}\pi_{3}\boldsymbol{\pi}\cdot\boldsymbol{\boldsymbol{\tau}
}\right]  N \nonumber \\
&& \hspace{-0.4cm} -\frac{g_{A}}{4Mf_{\pi}}\left[  iN^{\dagger}\boldsymbol{\tau}\boldsymbol{\cdot\dot{\pi}%
}\vec{\sigma}\boldsymbol{\cdot}\vec{\nabla}N+h.c.\right]  -\frac{h_{A}%
}{2Mf_{\pi}}\left[  iN^{\dagger}\boldsymbol{T}\boldsymbol{\cdot\dot{\pi}}%
\vec{S}\boldsymbol{\cdot}\vec{\nabla}\Psi_{\Delta}+h.c.\right] \nonumber \\
&&  \hspace{-0.4cm} -\frac{d_{1}}{f_{\pi}}N^{\dagger}\left(
\boldsymbol{\tau}\cdot\vec{\sigma}\cdot
\vec{\nabla}\boldsymbol{\pi}\right)  NN^{\dagger}N-\frac{d_{2}}{2f_{\pi}%
}\epsilon_{ijk}\epsilon_{abc}\partial_{i}\boldsymbol{\pi}_{a}N^{\dagger}%
\sigma_{j}\boldsymbol{\tau}_{b}NN^{\dagger}\sigma_{k}\boldsymbol{\tau}_{c}N+...
\nonumber
\end{eqnarray}
where $f_{\pi}$ is the pion decay constant in the chiral limit,
$g_{A}$ is the axial-vector coupling of the nucleon and $h_{A}$ is
the $\Delta N\pi$ coupling. The the nucleon, pion and $\Delta$
field are $N$, $\pi$ and $\Psi_{\Delta}$, respectively.
The isobar-nucleon mass difference is $\Delta$, and the quark mass
difference contribution to the neutron-proton mass difference is
$\delta$.
The $1/2 \rightarrow 3/2$ spin transition matrix operator is
$\vec{S}$ and the $1/2 \rightarrow 3/2$ isospin transition matrix
operator is $\boldsymbol{T}$. The are normalised such that
\begin{equation}
S_{i}S_{j}^{\dagger}=\frac{1}{3}\left(  2\delta_{ij}-i\epsilon
_{ijk}\sigma_{k}\right) \hspace{2.5cm}
T_{i}T_{j}^{\dagger}=\frac{1}{3}\left(  2\delta_{ij}-i\epsilon
_{ijk}\tau_{k}\right).
\end{equation}

\sp

\begin{figure}
\begin{center}
  \includegraphics[width=.48\textwidth,keepaspectratio]{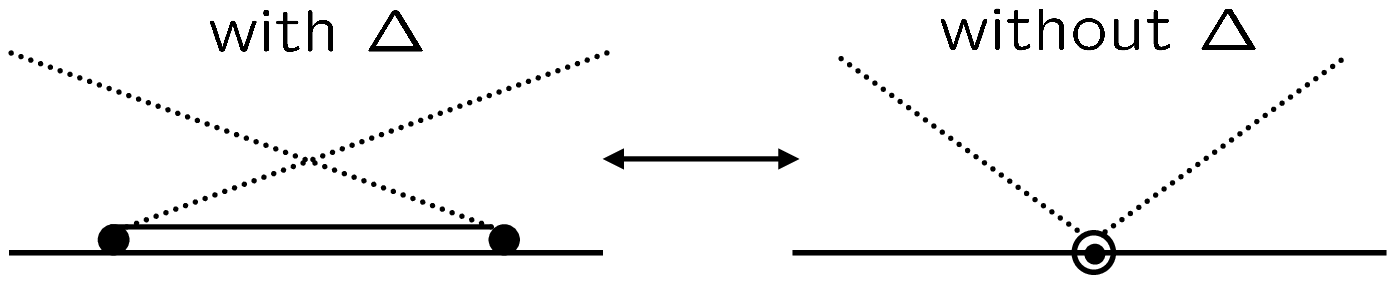}
  \captions[Illustration of resonance saturation]{Illustration of resonance
  saturation.
  \figtk{Hanhart:2003pg}.}
  \label{cDeltaHanhart}
\end{center}
\end{figure}

The constants $c_{i}$, not constrained by chiral symmetry, depend
on the details of QCD dynamics. They are at present unknown
functions of the fundamental QCD parameters, and can be extracted
from a fit to elastic $\pi N$ scattering\footnote{In the standard
case of processes involving momenta of order $m_{\pi}$, the
predictive power is not lost, because at any given order in the
power counting only a finite number of unknown parameters appear.
After these unknown parameters (\textit{low energy constants}) are
fitted to a finite set of data, all else can be predicted at that
order\cite{daRocha:1999dm}. }. The corresponding values are in
\tb{coeffci}.
In a theory without explicit $\Delta$'s, their effect is absorbed
in the low energy constants (see \fig{cDeltaHanhart}), which is
called resonance saturation (infinitely heavy or static limit).
Thus, in the case of a theory which considers explicit $\Delta$'s,
the $\Delta$ contribution\footnote{Note that there is some
sizeable uncertainty in the $\Delta$
contribution\cite{Bernard:1995dp}. The empirical values of the low
energy constants $c_{1}$, $c_{2}$, $c_{3}$, $c_{4}$ can be
understood from resonance exchange. In particular, assuming that
$c_{1}$ is saturated completely by scalar meson exchange (which is
in agreement with indications from the $NN$ force), the values for
$c_{2}$-$c_{4}$ can be understood from a combination of $\Delta$,
$\rho$ and scalar meson exchange\cite{Bernard:1996gq}:
$c_{2}^{\Delta}=-c_{3}^{\Delta}=2c_{4}^{\Delta}=-2.54.....-3.10\ugev^{-1}.
$} needs to be subtracted from the values given in the first two
columns of \tb{coeffci}.
\begin{table}
\begin{center}
\begin{tabular}
[c]{crrrr}\hline\hline $i\hspace{0.45cm}$ & $c_{i}^{\text{tree}}$
& $c_{i}^{\text{loop}}$ &
$c_{i}^{\text{tree}}\left(  \not \Delta\right)  $ & $c_{i}^{\text{loop}%
}\left(  \not \Delta\right)  $\\\hline
$1\hspace{0.45cm}$ & $-0.64$ & $-0.93$ & $-0.64$ & $-0.93$\\
$2\hspace{0.45cm}$ & $1.78$ & $3.34$ & $0.92$ & $0.64$\\
$3\hspace{0.45cm}$ & $-3.90$ & $-5.29$ & $-1.20$ & $-2.59$\\
$4\hspace{0.45cm}$ & $2.25$ & $3.63$ & $0.90$ &
$2.28$\\\hline\hline
\end{tabular}
\end{center}
\captions[$\chi$PT low-energy constants $c_{i}$ ] {Low-energy
constants $c_{i}$ for $\mathcal{L}^{\left(0 \right)}$ and
$\mathcal{L}^{\left(1 \right)}$ of \eq{LX0Hanhart}-\eq{LX1Hanhart}
(in $\ugev^{-1}$). The first two columns are from
\rf{Bernard:1995gx} and \rf{Bernard:1996gq}. The last two columns
are obtained subtracting a chosen $\Delta$ contribution of
$c_{2}^{\Delta}=-c_{3}^{\Delta}=2c_{4}^{\Delta}=2.7 \ugev^{-1}$
\cite{Hanhart:2003pg}.\\
}\label{coeffci}
\end{table}

\subsection{Power counting for the impulse term}
\sprg Within the framework of $\chi$PT, only pion exchange is
considered. Also, the impulse term (\fig{histdiagrams} (a)) is
included in the class of irreducible diagrams defined in
Weinberg's sense. A sub-diagram is considered reducible in
Weinberg's sense if it includes a small energy denominator of the
order of $\sim m_{\pi}^{2}/M$ and irreducible
otherwise\cite{Cohen:1995cc}.
In the following, we present the power counting for the impulse
term.

\begin{figure}
\begin{center}
  \includegraphics[width=.78\textwidth,keepaspectratio]{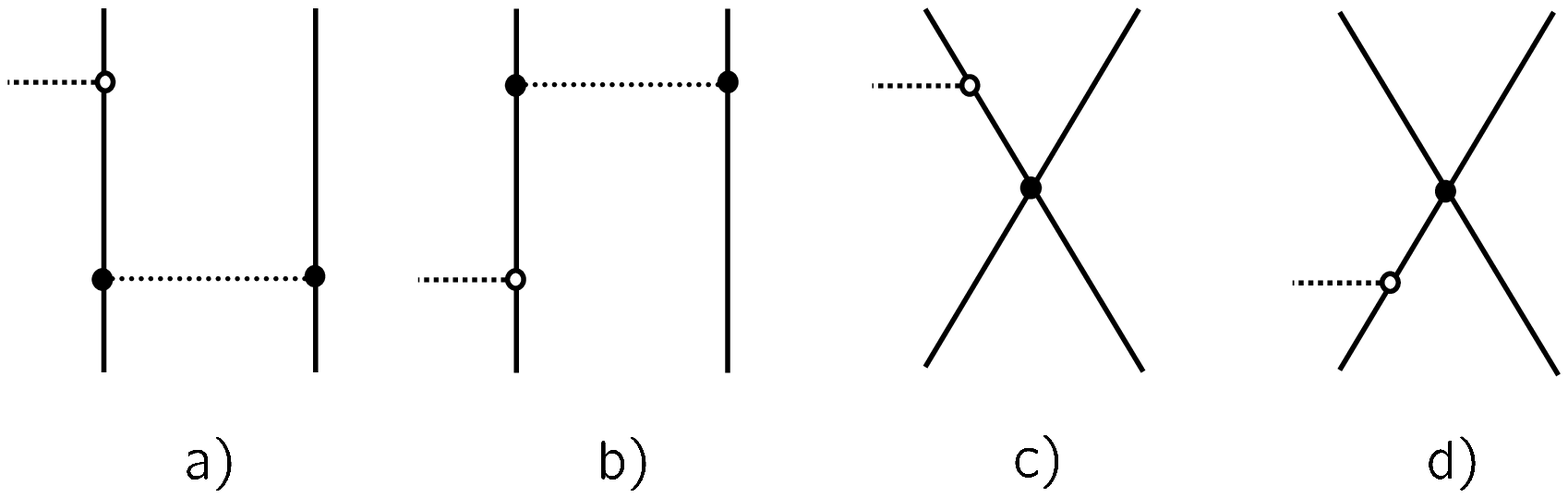}
  \captions[Diagrams contributing to the impulse term that are
   irreducible in the context
   of chiral power counting for the $pp \rightarrow pp \pi^{0}$ reaction]
   {Diagrams contributing to the impulse term that are irreducible in the context
   of chiral power counting for the $pp \rightarrow pp \pi^{0}$ reaction. The solid and
   dotted lines denote the nucleons and the pions, respectively. Leading(sub-leading)
   vertices are represented by solid(open) circles. They are originated by the
   three-momentum(energy) dependence of the vertex.}
  \label{impulseXpt}
\end{center}
\end{figure}

Since near threshold the pion carries an energy of the order of
the pion mass, at least one of the nucleon intermediate states,
before or after pion emission, must be off mass-shell by $\sim
m_{\pi}$. The transition to an off-mass-shell state is induced by
a relatively high-momentum $\left(\sim \sqrt{m_{\pi} M} \right)$
meson exchange mechanism. Therefore, the irreducible sub-diagrams
of \fig{histdiagrams} (a) are in lowest order to be drawn as in
\fig{impulseXpt}.
%

\noindent The corresponding power counting is done as follows:
\begin{itemize}
\item[---] Close to threshold, the $\vec{\sigma} \cdot \vec{q}$
term of the $\pi NN$ vertex is suppressed, and the pion-nucleon
interaction proceeds via the Galilean term $\left[N^{\dagger}\boldsymbol{\tau}\boldsymbol{\cdot\dot{\pi}%
}\vec{\sigma}\boldsymbol{\cdot}\vec{\nabla}N \right]$ of
\eq{LX1Hanhart}. This yields a factor of
\begin{equation}
\frac{1}{f_{\pi}}\frac{pE_{\pi}}{M} \sim \frac{\sqrt{M
m_{\pi}}m_{\pi}}{f_{\pi}M} = \frac{m_{\pi}^{3/2}}{f_{\pi}M^{1/2}}.
\end{equation}
\item[---] Since the nucleon-nucleon interaction originates from
virtual (static) pion exchange, as in \fig{impulseXpt} (a) and
(b), each three-momentum dependent $\pi NN$ vertex contributes
with a factor of $\sqrt{m_{\pi} M}/f_{\pi}$ and the pion
propagator contributes with a factor of $\left(m_{\pi} M
\right)^{-1}$, thus giving the overall factor of $f_{\pi}^{-2}$
from the nucleon-nucleon interaction. If the nucleon-nucleon
interaction arises from exchange of a heavier meson, as in
\fig{impulseXpt} (c) and (b), the overall factor is also
$\left(\frac{g_{NNh}}{m_{h}} \right)^{2} \sim
\frac{1}{f_{\pi^{2}}}$.
\item[---] Near threshold, the contribution of the
non-relativistic two-nucleon propagator is $
\left(E_{\mathrm{intermediate}}-E_{\mathrm{initial}} \right)^{-1}
\sim m_{\pi}^{-1}$.
\item[---] Finally, the overall contribution of the impulse term
is then
\begin{equation}
\frac{m_{\pi}^{3/2}}{f_{\pi}M^{1/2}} \times
\frac{1}{f_{\pi}^{2}}\frac{1}{m_{\pi}}=\frac{1}{f_{\pi}^{3}}\sqrt{\frac{m_{\pi}}{M}}.
\end{equation}
\end{itemize}
In conclusion, the impulse term is not irreducible in the
Weinberg's sense. Within $\chi$PT the impulse term is calculated
as given by the (distorted) diagrams in \fig{impulseXpt}.

%
\subsection{Why $\pi^0$ is problematic}

\sprg
%
For neutral pion production there is no meson exchange operator at
leading order and the nucleonic operator gets suppressed by the
poor overlap of the initial and final state wave functions (an
effect not captured by the power counting) and interferes
destructively with the direct production of the $\Delta$. Thus the
first significant contributions appear at NNLO. As there is a
large number of diagrams at NNLO, the different short-range
mechanisms found in the literature and discussed before are of
similar importance and capable of removing the discrepancy between
the Koltun and Reitan result and the data. Charged pion production
is expected to be significantly better under control, since there
is a meson exchange current at leading order and there are
non-vanishing loop contributions\cite{daRocha:1999dm}.
\subsubsection{The importance of the pion loops}
\sprg The most prominent diagram for neutral pion production close
to threshold is the pion re-scattering via the isoscalar
$T$-matrix that, for the kinematics given, is dominated by
one-sigma exchange\cite{Hanhart:2003pg}. Within the effective
field theory the isoscalar potential is built up perturbatively.
The leading piece of the one-sigma exchange gets cancelled by
other loops that cannot be interpreted as a re-scattering diagram
and therefore are not included in the phenomenological approaches.
This is an indication that in order to improve the
phenomenological approaches, at least in case of neutral pion
production, pion loops should be considered as well.
\subsection{Charged pion production in $\chi$PT}

\begin{figure}
\begin{center}
  \includegraphics[width=0.99\textwidth,keepaspectratio]{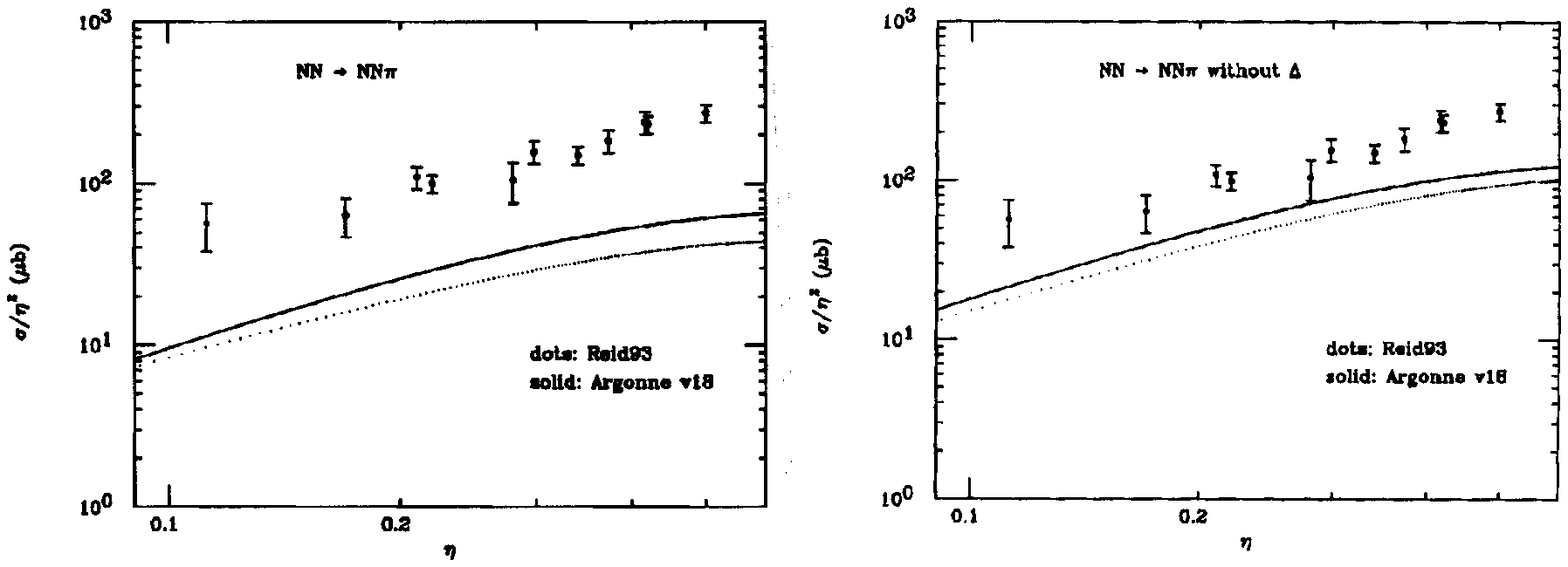}
  \captions[Effect of the $\Delta$ contribution on the cross section
   for $pp \rightarrow p n \pi^{+}$ within $\chi$PT]
  {Reduced cross section for $pp \rightarrow p n \pi^{+}$
  with (left panel) and without (right panel) the $\Delta$ contribution.
  The dotted and solid lines correspond, respectively to the Reid 93 and
  Argonne 18 $NN$ potentials. \figtk{daRocha:1999dm}.}
  \label{crossRocha}
\end{center}
\end{figure}

\sprg
The main calculations on $pp \rightarrow pn \pi^{+}$ in the
framework of $\chi$PT are those of \rf{daRocha:1999dm}, which also
included the mechanism proposed in
\rfs{Lee:1993xh,Horowitz:1993hk},  where the short-range
interaction is supposed to be originated from $Z$-diagrams
mediated by $\sigma$ and $\omega$ changes. The Coulomb interaction
is disregarded in the $p n \pi^{+}$ final state.
The channels considered were $^{3}P_{1} \rightarrow
\left(^{3}S_{1} \right)p$ and $^{3}P_{0} \rightarrow
\left(^{1}S_{0} \right)s$. For the $^{3}S_{1}$ final state, the
Weinberg-Tomozawa term contribution is found to be the largest.
Most of the other contributions were much smaller and tended to
cancel each other to some extent. The exception was the $\Delta$
contribution which had a significant destructive interference with
the Weinberg-Tomozawa term.
Although the theory produces the correct shape for the $\eta$
dependence, it fails in magnitude by a factor of $\sim 5$ (see
\fig{crossRocha}).

\rf{daRocha:1999dm} suggested that the $\Delta$ contribution may
be overestimated by a factor of $50 \%$ or more. Actually, the
cross sections with and without the $\Delta$ contribution
(respectively, left and right panel of \fig{crossRocha}), differed
by almost a factor of $2$. This could arise from the uncertainties
on the $\pi N \Delta$ coupling constant, or from the neglecting of
the $\Delta$-$N$ mass difference in energies. The kinetic energy
of the $\Delta$ was also neglected, account of which would further
decrease its amplitude.
%
%
\subsection{The $V_{\mathrm{low-k}}$ approach}

\sprg
%
As there are no $\chi$PT $NN$ potentials available yet for pion
production calculations, in practices one usually uses a hybrid
(not much consistent) $\chi$PT approach, in which the transition
operators are derived from $\chi$PT but the nuclear wave functions
are generated from high-precision phenomenological $NN$
potentials.
However, a conceptual problem underlying these hybrid $\chi$PT
calculations is that, whereas the transition operators are derived
assuming that relevant momenta are sufficient small compared with
the chiral scale $\chi \sim 1 \ugev$, the wave functions generated
by a phenomenological potential can in principle contain momenta
of any magnitude\cite{Kim:2005kd}.

A systematic method based on the renormalisation group approach
was recently developed\cite{Bogner:2003wn} to construct from a
phenomenological ``bare" $NN$ potential an effective $NN$
potential, $V_{\mathrm{low-k}}$, by integrating out momentum
components above a specified cutoff scale $\Lambda$.
For a $\Lambda=2.1 \ufm^{-1}$, it was found\cite{Bogner:2003wn}
that the low momentum behaviour of the $NN$ wave functions
calculated from $V_{\mathrm{low-k}}$ is essentially model
independent.
%

The very recent first calculation\cite{Kim:2005kd} for $pp
\rightarrow pp\pi^{0}$ using the $V_{\mathrm{low-k}}$ approach led
to a cross section closer to the experimental data than the one
calculated with ``bare" potentials.
%

\section{Energy prescription for the exchanged pion}

\begin{figure}
\begin{center}
  \includegraphics[width=.60\textwidth,height=0.33\textheight]{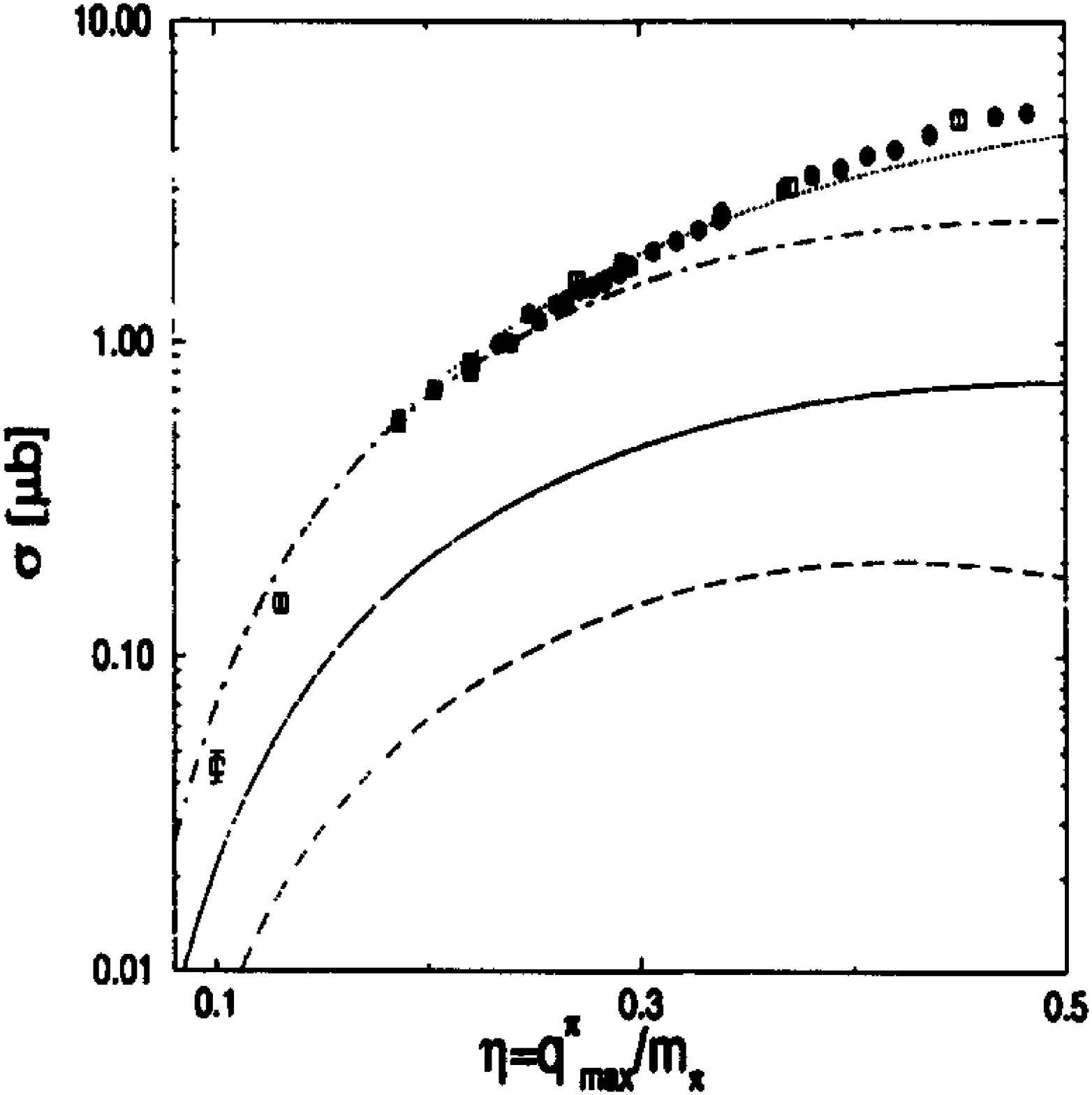}
  \captions[Effects of approximations for the energy of the pion in
  the re-scattering vertex within the J\"{u}lich phenomenological model]
  {Cross section for $pp \rightarrow pp \pi^{0}$.
  The solid curve is the full calculation of the re-scattering and direct
  production diagrams whereas the dashed line corresponds to the direct production
  diagram only. The dashed-dotted curve is the full result scaled by a factor
  of $\sim 4$. The dotted curve is the full calculation with the
  approximate treatment of the pion production vertex of \rf{Hernandez:1995kj}
  (frozen kinematics). \figtk{Hanhart:1995ut}.}
  \label{EnHanhart}
\end{center}
\end{figure}

\begin{figure}
\begin{center}
  \includegraphics[width=.58\textwidth,keepaspectratio]{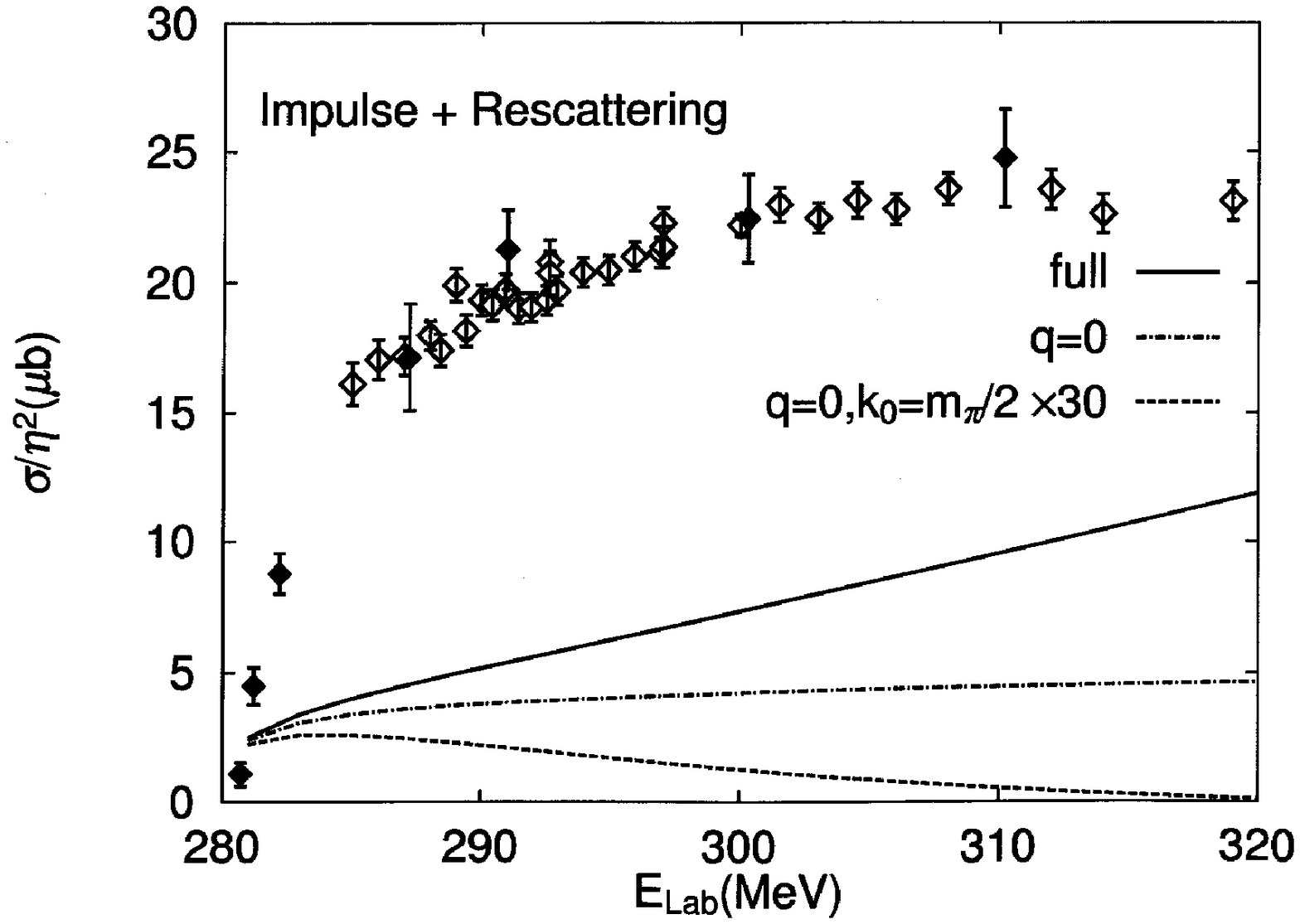}
  \captions[Effects of approximations for the energy of the pion in
  the re-scattering vertex within $\chi$PT]{Cross section for $pp \rightarrow pp \pi^{0}$,
  including both impulse and re-scattering amplitudes. The solid
  line is the calculation assuming four-momentum conservation and
  that the intermediate particles are on-mass-shell. The dotted and dashed-dotted
  lines correspond to two different fixed-threshold prescriptions for the
  energy of the exchanged pion. \figtk{Sato:1997ps}.}
  \label{EnSato}
\end{center}
\end{figure}

\sprg
In most of the calculations on pion production, several
approximations to the pion production vertex and the kinematics
have been tacitly assumed from the very first investigation by
Koltun and Reitan\cite{Koltun:1966pr}. However, these
prescriptions were found to have a significant effect both in the
magnitude and energy dependence of the cross
section\cite{Pena:1999hq,{Hanhart:1995ut},{Sato:1997ps}}, as it is
illustrated by \fig{EnHanhart} and by \fig{EnSato}.

\fig{EnHanhart} refers to the calculations\cite{Hanhart:1995ut}
for $pp \rightarrow pp \pi^{0}$ based on the J\"{u}lich model
presented in \sec{Phenapproach}.
%
%
It differs from the calculation of \rf{Hernandez:1995kj} since the
momentum of the exchanged pion is not fixed to the value
corresponding to the particular kinematic situation at the pion
production threshold when there are no distortions in the initial
and final $pp$ states.
%
%
%

The full treatment of the pion production vertex (solid line of
\fig{EnHanhart}) reduced the $\pi^{0}$ production rate by a factor
of $2$. Therefore, the work of \rf{Hanhart:1995ut} concluded that
the enhancement of the re-scattering amplitude due to offshellness
falls a bit too short to explain the scale of the $pp \rightarrow
pp\pi^{0}$ cross section.

In \fig{EnSato} we show a calculation for the $pp \rightarrow pp
\pi^{0}$ reaction within $\chi$PT, which also aimed to investigate
the effect of the simplifying assumptions on the energy-momentum
flow in the re-scattering diagram\cite{Sato:1997ps}. For this
case, the discrepancy between the approximations was found to be
of a factor of $3$.
%
%

A clarification of these formal issues is thus necessary before
one can draw conclusions about the physics of the pion production
processes. More specifically, one needs to obtain a
three-dimensional formulation from the more general Feynman
procedure. This will be the subject of the next Chapter, and was
the starting point of this thesis work.

%% file: Chapter3.tex
\setcounter{minitocdepth}{2}
\chapter{From Field Theory to DWBA} \label{FTtoDWBA}

\minitoc {\bfseries Abstract:} DWBA calculations, which are
currently applied to pion production, are based on a
quantum-mechanical three-dimensional formulation for the initial-
and final-$NN$ distortion, which is not obtained from the
four-dimensional field-theoretical Feynman diagrams. As a
consequence, the energy of the exchanged pion has been treated
approximately in the calculations performed so far. This Chapter
will discuss the validity of the DWBA approach through the link
with the time-ordered perturbation theory diagrams which result
from the decomposition of the corresponding Feynman diagram.

\newpage
All the meson production mechanisms described in the previous
Chapter are derived from relativistic Feynman diagrams.
Nevertheless, within the framework of DWBA, the evaluation of the
corresponding matrix elements for the cross section proceeds
through non-relativistic initial and final nucleonic wave
functions. Therefore the calculations apply a three-dimensional
formulation in the loop integrals for the nucleonic distortion,
which is not obtained from the underlying field-theoretical
Feynman diagrams. Namely, the energy of the exchanged pion in the
re-scattering operator (both in the $\pi N$ amplitude and in the
exchanged pion propagator) has been treated approximately and
under different prescriptions in calculations performed till now.
In other words, the usual approach, as in
\rfs{{Park:1995ku},{vanKolck:1996dp},{Pena:1999hq},{Cohen:1995cc},{daRocha:1999dm},{Gedalin:1999ht}},
is to use an educated ``guess" for the energy-dependence of the
virtual pion-nucleon interaction and then to use a Klein-Gordon
propagator for the exchanged pion propagator. In particular, most
of the calculations assume a typical threshold kinematics
situation even when the exchanged pion is expected to be largely
off-shell.

A theoretical control of the energy for the loop integration
embedding the non-relativis\-tic reduction of the Feynman
$\pi$-exchange diagram in the non-relativistic nucleon-nucleon
wave functions became then an important issue, as pointed out in
\rfs{Hanhart:2000wf,Motzke:2002fn}.

In this Chapter, following the work of \rf{Malafaia:2003wx}, we
deal with the isoscalar re-scattering term for the $pp \rightarrow
p p \pi^{0}$ reaction (near threshold). Although for $\pi^{0}$
production, as mentioned in Chapter~\ref{StateArt}, the
re-scattering mechanism is indeed small, the amount of its
interference with the (also small) impulse term
depends quantitatively on the calculation method. We note,
furthermore, that the pion isoscalar re-scattering term, which is
energy dependent, increases away from threshold, and that for the
charged pion production reactions the isovector term is important,
and also depends on the exchanged pion energy. Thus the knowledge
gained from the application discussed here to the $pp \rightarrow
pp\pi^{0}$ reaction near threshold  is useful for other
applications.

Specifically, this Chapter will focus on the investigation of
\begin{itemize}
\item[i)] the validity of the traditionally employed DWBA
approximation. We will use as a reference the result obtained from
the decomposition of the Feynman diagram into TOPT diagrams, and
realise how the last ones link naturally to an appropriate
quantum-mechanical DWBA matrix element. This study for a realistic
$\pi NN $ coupling was not done before;
\item[ii)] the (numerical) importance of the three-body
logarithmic singularities of the exact propagator of the exchanged
pion, which are not present when the usual approximations in DWBA
are considered; this study was not done before.
\end{itemize}
The work of \rf{Malafaia:2003wx} generalised the work of
\rfs{Hanhart:2000wf,Motzke:2002fn} which considered a solvable toy
model for scalar particles and interactions and treated the
nucleons as distinguishable and therefore pion emission to proceed
only from one nucleon. It is described by the following Lagrangian
\begin{eqnarray}
\mathcal{L}  &  = & \sum_{i=1,2}N_{i}^{\dagger}\left(
i\partial_{0}+\frac
{\nabla^{2}}{2M}\right)  N_{i}+\frac{1}{2}\left[  \left(  \partial_{\mu}%
\pi\right)  ^{2}-m_{\pi}^{2}\pi^{2}\right]  +\frac{1}{2}\left[
\left(
\partial_{\mu}\sigma\right)  ^{2}-m_{\sigma}^{2}\sigma^{2}\right]
+\nonumber\\
& &  \frac{g_{\pi}}{f_{\pi}}N_{2}^{\dagger}N_{2}\pi+g_{\sigma}\sum_{i=1,2}%
N_{i}^{\dagger}N_{i}\sigma+\frac{c}{f_{\pi}^{2}}\sum_{i=1,2}N_{i}^{\dagger
}N_{i}\left(  \partial_{0}\pi\right)  ^{2}
\label{lagrangianHanhart}
\end{eqnarray}
where $N_{i}$ and $\pi$ are the nucleon and pion field,
respectively. \eq{lagrangianHanhart} further assumes a Yukawa
coupling for the pion-nucleon coupling, $\frac {g_{\pi}}{f_{\pi}}$
and describes the pion re-scattering through a $\pi N$ seagull
vertex inspired by the chiral $\pi N$ interaction Lagrangian of
\eq{LX1Hanhart}, $\frac {c}{f_{\pi}^{2}}Q_{0}^{\prime}Q_{0}$,
where $c$ is an arbitrary constant. In this toy model, the nuclear
interactions are described through $\sigma$-exchange, which also
couples to the nucleons via a Yukawa coupling, $g_{\sigma}$, and
the nucleons are treated non-relativistically (in particular,
contributions from nucleon negative-energy states are not
considered).

There are some nonrealistic features of this model that also
motivated our work and the one of \rf{Malafaia:2003wx}. First of
all, the model does not satisfy chiral symmetry, which would have
required a derivative coupling of the pion to the nucleon instead
of the simpler Yukawa coupling. Furthermore, in this toy model the
nucleons, interacting via $\sigma$ exchange, have a stronger
overlap than in a more realistic model, since it does not include
short-range repulsive nucleon-nucleon interactions that keep the
nucleons apart. Note also that near threshold kinematics, the
scalar particle is produced in a $s$-wave state, as is the final
nucleon-nucleon pair (and the initial nucleon -nucleon pair too,
by angular momentum conservation requirements). This is not the
case for the production of a real pion, since it is a
pseudo-scalar particle and, therefore, near threshold the
production of $s$-wave pions calls for a initial $P$-wave nucleon
state.

Our calculation employs a physical model for nucleons and pions
and investigates how much of the features found in
\rfs{Hanhart:2000wf,Motzke:2002fn} survive in a more realistic
calculation which uses a pseudo-vector coupling for the $\pi N N$
vertex, the $\chi$PT $\pi N$ amplitude\cite{Park:1995ku} and the
Bonn B potential\cite{Machleidt:1987hj} for the nucleon-nucleon
interaction.
%
%
\section{Extraction of the effective production operator \label{Extrprop}}
%
\subsection{Final-state interaction diagram \label{Extrpropfsi}}
\subsubsection{The amplitude}
\begin{figure}
\begin{center}
  \includegraphics[width=1.02\textwidth,keepaspectratio]{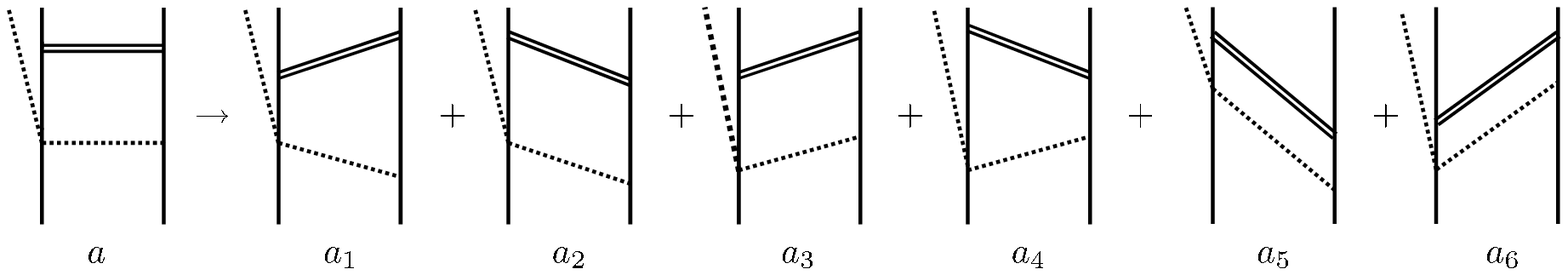}
  \captions[Decomposition of the Feynman diagram in terms of six
  time-ordered diagrams for the final-state interaction]
  {Decomposition of the Feynman diagram in terms of six
  time-ordered diagrams for the final-state interaction. The
  pion(sigma) field is represented by a dashed(solid double) line.
  The nucleons are represented by solid lines. The DWBA amplitude
  may be identified to the first four time-ordered diagrams ($a_1$
  to $a_4$). The last two diagrams ($a_5$ to $a_6$) are usually
  called stretched boxes.} \label{toptfsi}
\end{center}
\end{figure}

The Feynman diagram for the reaction $pp \rightarrow pp \pi^0$
where the $NN$ final-state interaction (FSI) proceeds through
sigma exchange is represented in part $a$ of \fig{toptfsi}.
After the nucleon negative-energy states are neglected, it
corresponds to the amplitude
\begin{eqnarray}
\mathcal{M}_{FSI} = && ig_{\sigma}^{2}  \int
\frac{d^{4}Q^{\prime}}{\left(2\pi\right)^{4}} f\left(
Q_{0}^{\prime}\right)  \frac{1}{Q_{1}^{0}
-\omega_{1}+i\varepsilon}\frac{1}{Q_{2}^{0}-\omega_{2}+i\varepsilon}\frac
{1}{K_{0}^{2}-\omega_{\sigma}^{2}+i\varepsilon} \label{mtoyfsi} \\
&& \frac{1}{Q_{0}^{\prime2}-\omega_{\pi}^{2}+i\varepsilon}
\nonumber
\end{eqnarray}
where the exchanged pion has four-momentum
$Q'=\left(Q^{\prime}_0,\vec{q}^{\,\prime}\right)$. The
four-momenta of the intermediate nucleons are $Q_{1}$ and $Q_{2}$
and the four-momentum of the exchanged $\sigma$ is $K$. The
on-mass-shell energies for the intermediate nucleons, exchanged
pion and sigma mesons are $\omega_{1,2}$, $\omega_{\pi}$,
$\omega_{\sigma}$, respectively. In \apx{Apkindiagrams} details on
these functions are given. All quantities are referred to the
three-body centre-of-mass frame of the $\pi N N$ final state.

In \eq{mtoyfsi} $f\left(  Q_{0}^{\prime}\right) $ is a short-hand
notation for both the $\mathcal{M}_{\pi NN}$ and
$\mathcal{M}_{\pi\pi NN}$. We note that $f$ in fact may depend on
the three-momenta and energies of both the exchanged and produced
pion, and also on the nucleon spin. However, for simplicity, only
the dependence on the energy of the exchanged pion,
$Q^{\prime}_{0}$, is emphasised, because this is the important
variable for the main considerations of these Chapter. For the toy
model described by the lagrangian of \eq{lagrangianHanhart}, it is
\begin{equation}
f\left(  Q_{0}^{\prime}\right)  =\left(
\frac{g_{\pi}}{f_{\pi}}\right) \times\left(
\frac{c}{f_{\pi}^{2}}Q_{0}^{\prime}E_{\pi}\right)
\text{,}\label{vtoy}
\end{equation}
but for the more realistic case of \rf{Malafaia:2003wx} which uses
a pseudo-vector coupling for the $V_{\pi NN}$ vertex and the
$\chi$PT $\pi N$ amplitude\cite{Park:1995ku}, it reads
\begin{equation}
f\left(  Q_{0}^{\prime}\right)  =\left(  \frac{g_{A}}{f_{\pi}}\vec{\sigma}%
^{2}\cdot\vec{q}^{\,\prime}\right)
\times\frac{m_{\pi}^{2}}{f_{\pi}^{2}}\left[
2c_{1}+\left(  c_{2}+c_{3}-\frac{g_{A}^{2}}{8M}\right)  \frac{E_{\pi}%
Q_{0}^{\prime}}{m_{\pi}^{2}}+c_{3}\frac{\vec{q}^{\prime}\cdot\vec{q}_{\pi}%
}{m_{\pi}^{2}}\right]  \label{vxpt}
\end{equation}
In \eq{vtoy} and \eq{vxpt}, $E_{\pi}$$\left(\vec{q}_{\pi} \right)$
is the energy (three-momentum) of the produced pion.  As mentioned
before, $f_{\pi}$ is the pion decay constant and $g_{A}$ is the
nucleon axial vector coupling constant, which is related to the
pseudovector coupling
constant $f_{\pi NN}$ by the Goldberger-Treiman relation $\frac{g_{A}}%
{2f_{\pi}}=\frac{f_{\pi NN}}{m_{\pi}}$.
The low-energy constants $c_{i}^{\prime}$s are defined on
\tb{coeffci}.

Using four-momentum conservation, the amplitude of \eq{mtoyfsi}
can be written as
\begin{eqnarray}
\mathcal{M}_{FSI} & =
&-ig_{\sigma}^{2}\int\frac{d^{4}Q^{\prime}}{\left(
2\pi\right)  ^{4}}f\left(  Q_{0}^{\prime}\right)  \frac{1}{E_{1}+Q_{0}%
^{\prime}-E_{\pi}-\omega_{1}+i\varepsilon}\frac{1}{Q_{0}^{\prime}-E_{2}%
+\omega_{2}-i\varepsilon}\label{mtoyfsiencons}\\
&&
\frac{1}{F_{2}-E_{2}+Q_{0}^{\prime}-\omega_{\sigma}+i\varepsilon}\frac
{1}{F_{2}-E_{2}+Q_{0}^{\prime}+\omega_{\sigma}-i\varepsilon}\frac{1}%
{Q_{0}^{\prime}-\omega_{\pi}+i\varepsilon}\frac{1}{Q_{0}^{\prime}+\omega_{\pi
}-i\varepsilon}\nonumber
\end{eqnarray}
where the functions $E_{1,2}$$\left(F_{1,2} \right)$ stand for the
energy of the initial(final) nucleons.

\begin{figure}
\begin{center}
  \includegraphics[width=.52\textwidth,keepaspectratio]{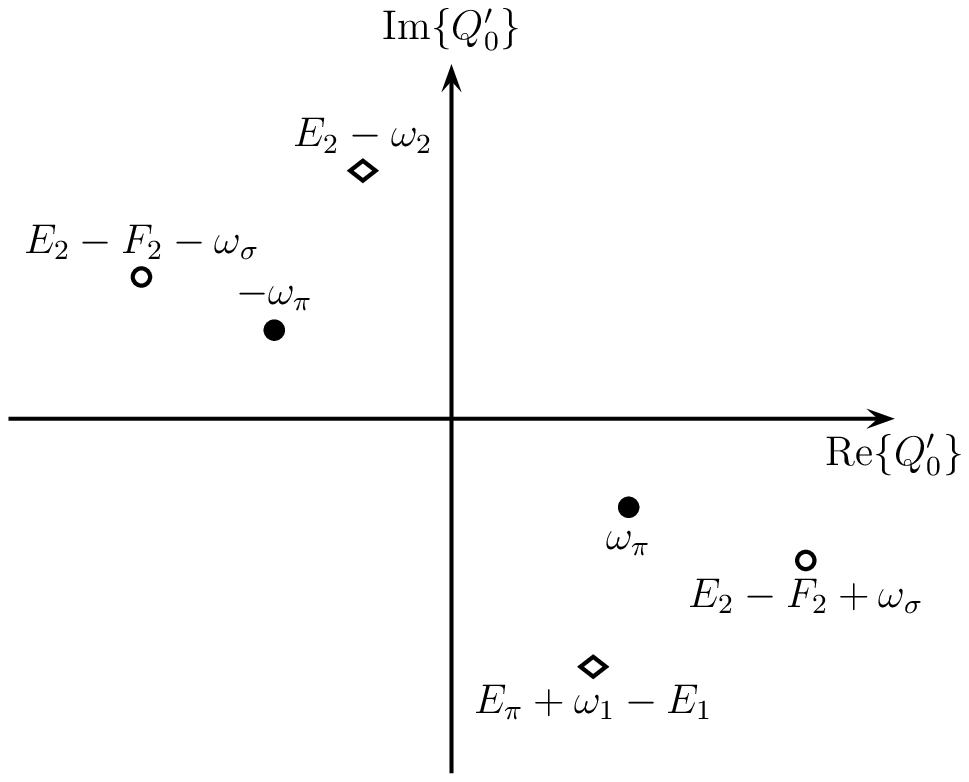}
  \captions[Schematic position of the poles in the energy
  $Q^{\prime}_{0}$ of the exchanged pion for the FSI amplitude]
  {Schematic position (scaled for a  kinematics close to pion
  production threshold) of the poles in $Q^{\prime}_0$ of the amplitude
  of \eq{mtoyfsiencons}. The diamonds,
  open circles and bullets represent, the nucleon
  poles, the sigma poles and the pion poles, respectively.}
  \label{polestoyfsi}
\end{center}
\end{figure}

The integrand in \eq{mtoyfsiencons} has three poles in the upper
half-plan and three poles in lower half-plan. They are
schematically reproduced on \fig{polestoyfsi}.

To integrate \eq{mtoyfsiencons} over the energy variable
$Q^{\prime}_{0}$ we can close the contour on one of the
half-planes and pick each of the three poles enclosed. However, to
get a more straightforward connection with the DWBA formalism
through time-ordered perturbation theory, it is better to perform
a partial fraction decomposition to isolate the pion poles before
integrating over $Q_{0}^{\prime}$. All details of this
decomposition are given in \apx{ApTopt}.

\begin{figure}[h!]
\begin{center}
  \includegraphics[width=.92\textwidth,keepaspectratio]{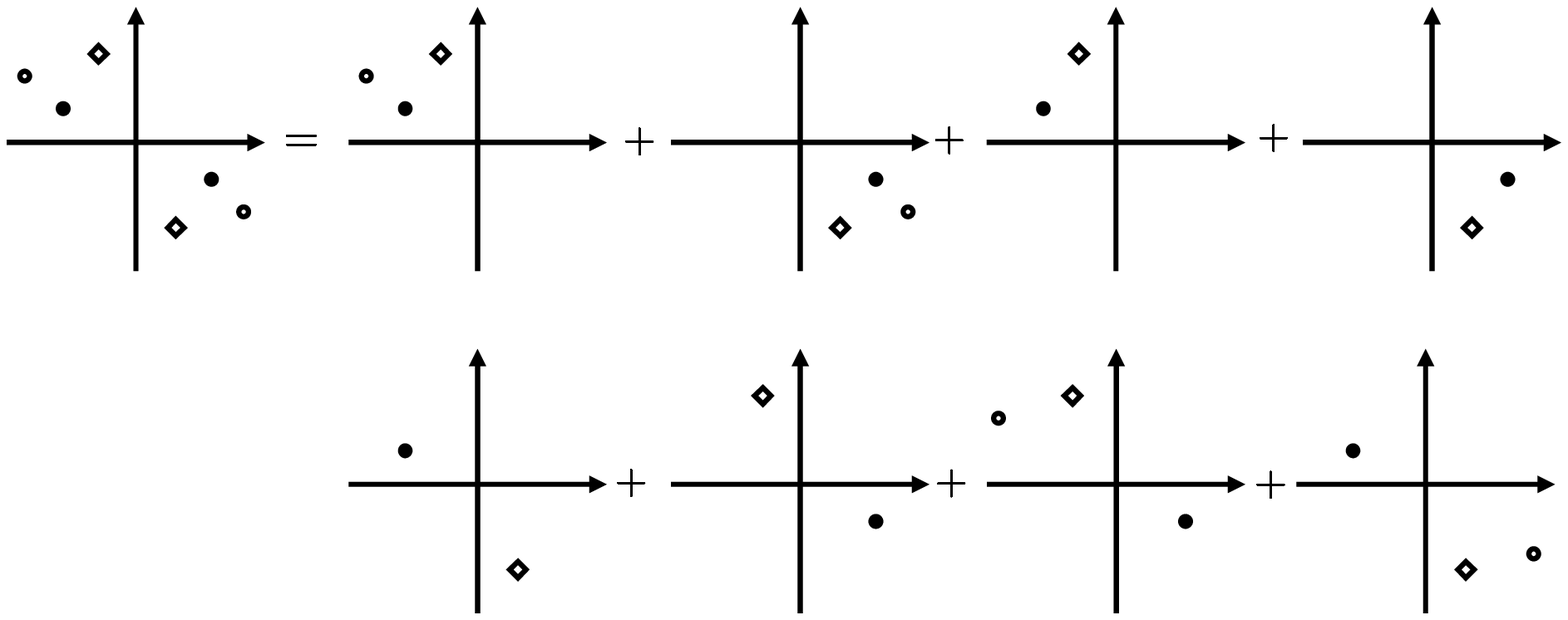}
  \captions[Schematic position of the poles in the energy of the
  exchanged pion resulting from the partial fraction decomposition of the FSI amplitude]
  {Schematic position of the poles in $Q^{\prime}_0$ resulting
  from the partial fraction decomposition of the amplitude
  of \eq{mtoyfsiencons}. The meaning of the symbols is the same of \fig{polestoyfsi}.
  }
  \label{polestoptfsi}
\end{center}
\end{figure}

As a result, \eq{mtoyfsiencons} is expressed as a sum of eight
types of terms, with each one having at the maximum three poles,
as represented in \fig{polestoptfsi}. From this figure one
realises that only $4$ terms do not vanish after the
$Q^{\prime}_{0}$ integration: they correspond to the location of
the poles as represented in the last $4$ diagrams of
\fig{polestoptfsi}.

After the $Q^{\prime}_0$ integration, one obtains for
\eq{mtoyfsiencons}:
\begin{eqnarray}
&&\mathcal{M}^{FSI}_{TOPT}=-g_{\sigma}^{2}\int \frac{d^{3}q^{\prime }}{\left( 2\pi \right) ^{3}}\frac{1%
}{4\omega _{\sigma }\omega _{\pi }}\times  \label{atoptfsi} \\
&&\left[ \frac{f\left( \omega _{\pi }\right) }{\left(
E_{tot}-E_{\pi }-\omega _{1}-\omega _{2}\right) \left(
E_{tot}-E_1-\omega _{2}-\omega _{\pi
}\right) \left( E_{tot}-F_1-E_\pi-\omega _{2}-\omega _{\sigma }\right) }\right.  \nonumber \\
&&\left. +\frac{f\left( \omega _{\pi }\right) }{\left(
E_{tot}-E_{\pi }-\omega _{1}-\omega _{2}\right) \left(
E_{tot}-E_1-\omega _{2}-\omega _{\pi
}\right) \left( E_{tot}-F_2-E_{\pi }-\omega _{1}-\omega _{\sigma }\right) }%
\right.  \nonumber \\
&&\left. +\frac{f\left( -\omega _{\pi }\right) }{\left(
E_{tot}-E_{\pi }-\omega _{1}-\omega _{2}\right) \left(
E_{tot}-E_{2}-E_{\pi }-\omega _{1}-\omega
_{\pi }\right) \left( E_{tot}-F_1-E_\pi-\omega _{2}-\omega _{\sigma }\right) }\right.  \nonumber \\
&&\left. +\frac{f\left( -\omega _{\pi }\right) }{\left(
E_{tot}-E_{\pi }-\omega _{1}-\omega _{2}\right) \left(
E_{tot}-E_2-E_{\pi }-\omega _{1}-\omega _{\pi }\right) \left(
E_{tot}-F_2-E_{\pi }-\omega _{1}-\omega _{\sigma
}\right) }\right.  \nonumber \\
&&\left. +\frac{f\left( \omega _{\pi }\right) }{\left(
E_{tot}-E_1-\omega _{2}-\omega _{\pi }\right) \left(
E_{tot}-F_2-E_{\pi }-\omega _{1}-\omega _{\sigma }\right) \left(
E_{tot}-E_1-F_{2}-\omega _{\pi }-\omega _{\sigma }\right)
}\right.  \nonumber \\
&&\left. +\frac{f\left( -\omega _{\pi }\right) }{\left(
E_{tot}-E_2-E_{\pi }-\omega _{1}-\omega _{\pi }\right) \left(
E_{tot}-F_1-E_\pi-\omega _{2}-\omega _{\sigma }\right) \left(
E_{tot}-E_2-F_1-E_\pi-\omega _{\pi }-\omega _{\sigma }\right)
}\right] \nonumber
\end{eqnarray}
with $E_{tot}=2 E=F_1+F_2+E_\pi$ and $E \equiv E_1=E_2$.

This equation evidences that there are six contributions to the
amplitude. These six terms, originated by the four
propagators of the loop, can be interpreted as time-ordered
diagrams. They are represented by diagrams $a_1$ to $a_6$ in
\fig{toptfsi}. This interpretation justifies the extra subscript
label $TOPT$ for the $\mathcal{M}_{FSI}$ amplitude in
\eq{atoptfsi}.

Obviously, the result of direct integration is the same of the
result obtained doing a partial fraction decomposition before
integrating. For the $f\left( Q_{0}^{\prime}\right)$ functions
considered, since they have a simple\footnote{The linear
dependence in $ Q_{0}^{\prime}$ of the functions $f\left(
Q_{0}^{\prime}\right)$ guarantees that the integral over the curve
of radius $R$ which closes the contour vanishes when $R
\rightarrow \infty$.} (linear) dependence in $Q_{0}^{\prime}$, the
integrals in which the two or three poles are all in the same
half-plan vanish (first four diagrams of \fig{polestoptfsi}), and
we end up with only six terms, corresponding to four distributions
of the poles (four last terms of \fig{polestoptfsi}).

Because the partial fraction decomposition of the propagators in
\eq{mtoyfsiencons} was done prior to the integration over the
variable $Q'_0$, we have the following outcome which is
independent of the choice of the contour of this integration: the
only terms from the decomposition which do contribute to the
integral correspond to the ones with only one pole, which happens
to be the $Q^{\prime}_0=\omega_\pi=\sqrt{m_\pi^2+\vec{q}^{\,\prime
2}}$ or the
$Q^{\prime}_0=-\omega_\pi=-\sqrt{m_\pi^2+\vec{q}^{\,\prime 2}}$
pion poles. The other terms, with nucleon poles and/or sigma
poles, together or not with pion poles, have all these poles
located on the same half-plane and consequently their contribution
vanish (upper part of \fig{polestoptfsi}).

We stress at this point that this method for the energy
integration implies effectively that the $\pi N$ re-scattering
amplitude is evaluated only for on-mass-shell pion energies. In
this way, off-shell extrapolations which are not yet solidly
constrained are avoided, which is an advantage. Other methods may
need the contribution of the off-shell amplitude in the integrand
with the form shown in \eq{mtoyfsiencons}. But the net result is
the same, provided that all the contributions from all (nucleon,
sigma and pion) propagator poles are considered. So far,
calculations\cite{{Park:1995ku},{vanKolck:1996dp},{Pena:1999hq},{Cohen:1995cc},{daRocha:1999dm},{Gedalin:1999ht}}
did not consider the pion propagator poles, since they approximate
that propagator by a form free of any singularity. Consequently,
they exhibit a strong dependence on the $\pi N$ amplitude at
off-mass-shell energies of the incoming pion.

\subsubsection{The effective pion propagator for FSI distortion}
\begin{figure}
\begin{center}
  \includegraphics[width=.98\textwidth,keepaspectratio]{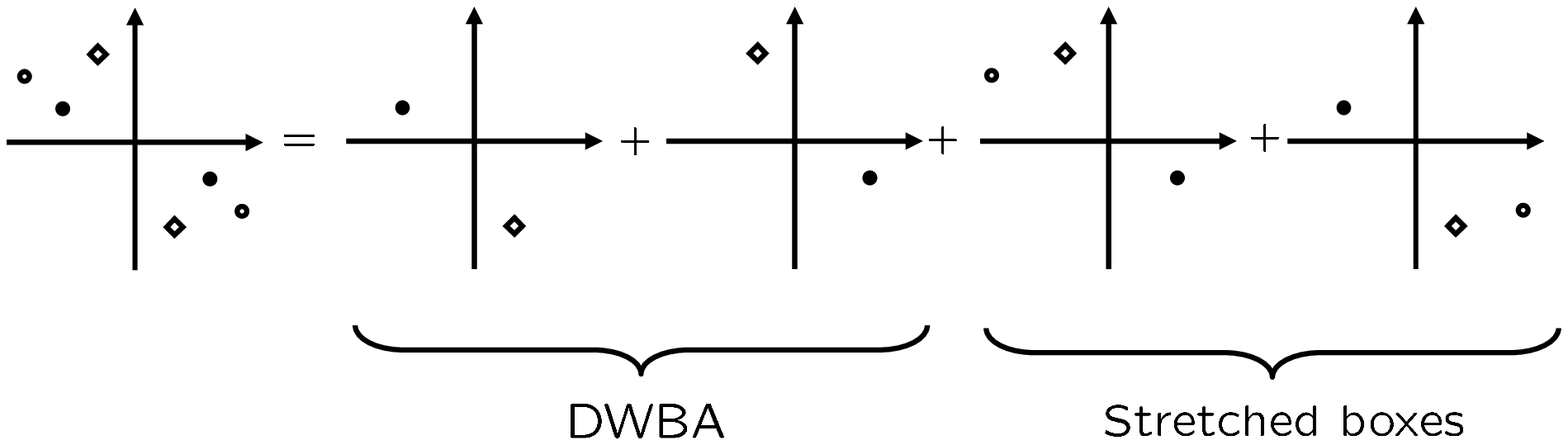}
  \captions[Schematic position of the poles in $Q^{\prime}_{0}$
  for the DWBA and stretched boxes terms (FSI case)]
  {Schematic position of the poles in $Q^{\prime}_0$ resulting
  from the partial fraction decomposition of the amplitude
  of \eq{mtoyfsiencons} for the DWBA and stretched boxes terms.
  The meaning of the symbols is the same of \fig{polestoyfsi}.
  }
  \label{polesdwbafsi}
\end{center}
\end{figure}

From the six terms in \eq{atoptfsi}, the first four terms
(corresponding to diagrams $a_1$ to $a_4$ of \fig{toptfsi}) have
the special feature that any cut through the intermediate state
intersects only nucleon legs. Thus, they may be identified to the
traditional DWBA amplitude for the final-state distortion. In
contrast, in the last two diagrams $a_5$ and $a_6$ of
\fig{toptfsi}, any cut through the intermediate state cuts not
only the nucleon legs, but also the two exchanged particles in
flight simultaneously. They are called the stretched
boxes\cite{Hanhart:2000wf}. The corresponding pole distribution is
illustrated in \fig{polesdwbafsi}.

Because of the identification of diagrams $a_1$ to $a_4$ with
DWBA, we may collect the four first terms of \eq{atoptfsi} and
obtain what we may call the {\it reference} expression for the
DWBA amplitude:
\begin{equation}
\mathcal{M}_{DWBA}^{FSI}=\frac{1}{2} \int \frac{d^3 q'}{\left(2
\pi \right)^3 } \left[ \tilde{f} \left( \omega_{\pi} \right)
G_{\pi}
 \right]
\frac{1}{\left(E_1+E_2-E_{\pi}-\omega_1-\omega_2 \right)}
T_{NN}^{FSI}. \label{dwbafsi}
\end{equation}
Here $T$ stands for the transition-matrix of the final-state
interaction. We have used both sigma exchange, (with
$m_{\sigma}=550\umev$) as in \rf{Hanhart:2000wf}, which makes
\eq{dwbafsi} coincide exactly with \eq{atoptfsi}, and also the
T-matrix calculated from the Bonn B
potential\cite{Machleidt:1987hj}. For the $\sigma$-exchange case,
one gets
\begin{equation}
V_{\sigma}^{DWBA}=\frac{g_{\sigma}^{2}}{2 \omega_\sigma}
\left[\frac{1}{\left( E_{tot}-F_1-E_\pi-\omega_2-\omega_\sigma
\right)}+ \frac{1}{\left(E_{tot}-F_2-E_\pi-\omega_1-\omega_\sigma
\right)} \right]. \label{fvsigmafsi}
\end{equation}
 The details of the $T$-matrix calculation are in \apx{ApTm}. In the derivation of the
integrand of \eq{dwbafsi} the propagators for the two nucleons in
the intermediate state fused into only one overall propagator with
the non-relativistic form,
\begin{equation}
G_{NN}=\frac{1}{\left(E_1+E_2-E_{\pi}-\omega_1-\omega_2 \right)},
\end{equation}
The function $\tilde{f} \left(\omega_\pi \right)$ includes the
contribution of the two pion poles $\omega_\pi$ and $-\omega_\pi$
corresponding to two different time-ordered diagrams,
\begin{equation}
\tilde{f} = \frac{f \left(\omega_\pi \right)
\left(E_1-E_\pi-\omega_1-\omega_\pi \right)+f \left(- \omega_\pi
\right) \left( E_2 - \omega_2-\omega_\pi \right)}{\omega_\pi}
\label{ftildefsi}
\end{equation}
where the function $f \left( \omega_\pi \right)$ is the product of
the $\pi N$ amplitude with the $\pi N N$ vertex. The
multiplicative kinematic factors $f_{1}^{f}  =
\frac{E_1-E_\pi-\omega_1-\omega_\pi}{\omega_\pi}$ and $f_{2}^{f} =
\frac{E_2-\omega_1-\omega_2}{\omega_\pi}$
%
may be treated as form factors (see \fig{ffactorvtoy}).

In the derivation of \eq{dwbafsi} from  \eq{atoptfsi} the function
$G_{\pi}$ for the pion propagator turns to be exactly
\begin{equation}
G_{\pi}=\frac{1}{\left[\frac{\omega_1-\omega_2}{2}+\frac{E_{\pi}}{2}
\right]^2- \left[\left(
E_{tot}-E-\frac{E_{\pi}}{2}\right)-\frac{\omega_1+\omega_2}{2}-\omega_{\pi}
\right]^2}, \label{propagator}
\end{equation}
which gives the form of the effective pion propagator appropriate
for a DWBA final-state calculation, and can also be written as
\begin{equation}
G_{\pi}=\frac{1}{\left[{\omega_1+E_{\pi}-E_{1}+\omega_\pi}\right]\left[{E_{2}-\omega_2-\omega_\pi}\right]}.
\label{gpiefffsi}
\end{equation}
The same approach will next be applied to the ISI case.
\subsection{Initial-state interaction diagram \label{Extrpropisi}}
\begin{figure}
\begin{center}
\includegraphics[width=1.02\textwidth,keepaspectratio]{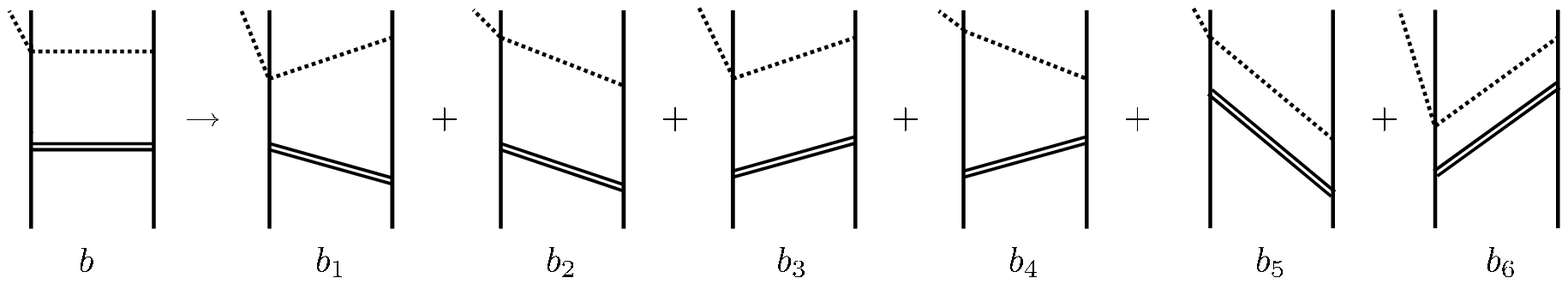}
  \captions[Decomposition of the Feynman diagram in terms of six
   time-ordered diagrams for the initial-state interaction]
   {Decomposition of the Feynman diagram in terms of six time-ordered
   diagrams for the initial-state interaction. The pion(sigma) field
   is represented by a dashed(solid double) line. The nucleons are
   represented by solid lines. The DWBA amplitude corresponds to the
   first four time-ordered diagrams ($b_1$ to $b_4$) and the
   stretched boxes to the last two ($b_5$ to $b_6$).} \label{toptisi}
\end{center}
\end{figure}
\subsubsection{The amplitude}
The corresponding  Feynman diagram ($b$ in \fig{toptisi}), for the
$NN$ initial-state interaction (ISI) when it proceeds through
sigma exchange, after the nucleon negative-energy states are
neglected, generates the amplitude:
\begin{eqnarray}
\mathcal{M}^{ISI} &=&  -ig_{\sigma}^{2}\int \frac{d^4q'}{\left( 2
\pi \right)^4}f \left(Q'_0 \right)
\frac{1}{Q'_0+F_2-\omega_{2}+i\varepsilon} \frac{1}{F_1+E_{\pi}-Q'_0-\omega_{1}+i\varepsilon}  \label{feyisi} \\
&& \frac{1}{Q'_0-E_2+F_2-\omega_{\sigma}+i \varepsilon}
\frac{1}{Q'_0-E_2+F_2+\omega_{\sigma}-i
\varepsilon}\frac{1}{Q'_{0}-\omega_{\pi}+i
\varepsilon}\frac{1}{Q'_{0}+\omega_{\pi}-i \varepsilon} \nonumber
,
\end{eqnarray}
where we have used a notation analogous to one used for the
final-state amplitude of \eq{mtoyfsi}.

As before, in order to perform the integration over the exchanged
pion energy $Q^{\prime}_0$, a partial fraction decomposition to
isolate the poles of the pion propagator was done. By closing the
contour such that only the residues of the
$Q'_0=\pm\omega_\pi=\pm\sqrt{q^{'2}+m_\pi^2}$ poles contribute, in
an entirely similar way to the FSI case, one obtains the
amplitude:
\begin{eqnarray}
&\mathcal{M}^{ISI}_{TOPT}&=-g_{\sigma}^{2}\int \frac{d^{3}q^{\prime }}{\left( 2\pi \right) ^{3}}\frac{1%
}{4\omega _{\sigma }\omega _{\pi }}\times  \label{atoptisi}\\
&&\left[ \frac{-f\left( -\omega _{\pi }\right) }{\left(
E_{tot}-\omega _{1}-\omega _{2}\right) \left( E_{tot}-E_1-\omega
_{2}-\omega _{\sigma }\right)
\left( E_{tot}-F_1-E_\pi-\omega _{2}-\omega _{\pi }\right) }\right.  \nonumber \\
&&\left. +\frac{-f\left( \omega _{\pi }\right) }{\left(
E_{tot}-\omega _{1}-\omega _{2}\right) \left( E_{tot}-E_1-\omega
_{2}-\omega _{\sigma }\right)
\left( E_{tot }-F_{2}-\omega _{1}-\omega _{\pi }\right) }\right.  \nonumber \\
&&\left. +\frac{-f\left( -\omega _{\pi }\right) }{\left(
E_{tot}-\omega _{1}-\omega _{2}\right) \left( E_{tot}-E_2-\omega
_{1}-\omega _{\sigma }\right)
\left( E_{tot}-F_1-E_\pi-\omega _{2}-\omega _{\pi }\right) }\right.   \nonumber \\
&&\left. +\frac{-f\left( \omega _{\pi }\right) }{\left(
E_{tot}-\omega _{1}-\omega _{2}\right) \left( E_{tot}-E_2-\omega
_{1}-\omega _{\sigma }\right)
\left( E_{tot}-F_{2}-\omega _{1}-\omega _{\pi }\right) }\right.  \nonumber \\
&&\left. +\frac{-f\left( \omega _{\pi }\right) }{\left(
E_{tot}-E_1-\omega _{2}-\omega _{\sigma }\right) \left(
E_{tot}-E_1-F_{2}-\omega _{\pi }-\omega
_{\sigma }\right) \left( E_{tot }-F_{2}-\omega _{1}-\omega _{\pi }\right) }%
\right.   \nonumber \\
&& \hspace{-1.5cm} \left. +\frac{-f\left( -\omega _{\pi }\right)
}{\left( E_{tot}-E_2-\omega _{1}-\omega _{\sigma }\right) \left(
E_{tot}-F_1-E_\pi-\omega _{2}-\omega _{\pi }\right) \left(
E_{tot}-E_{2}-E_{\pi }-F_{1}-\omega _{\pi }-\omega _{\sigma
}\right) }\right] \nonumber
\end{eqnarray}
The six terms in \eq{atoptisi} are interpreted as contributions
from time-ordered diagrams represented in \fig{toptisi}, $b_1$ to
$b_6$. This interpretation justifies the extra subscript label
$TOPT$ for the $\mathcal{M}^{ISI}$ amplitude. Although
\eq{atoptfsi} and \eq{atoptisi} are formally alike, for the
initial state an extra pole is present (besides that from the
nucleons propagator), since it is energetically allowed for the
exchanged pion to be on-mass-shell. All the singularities are
decisive for the real and imaginary parts of \eq{atoptfsi} and
\eq{atoptisi}.
\subsubsection{The effective pion propagator for the ISI distortion}
\begin{figure}[t!]
\begin{center}
\includegraphics[width=.78\textwidth,keepaspectratio]{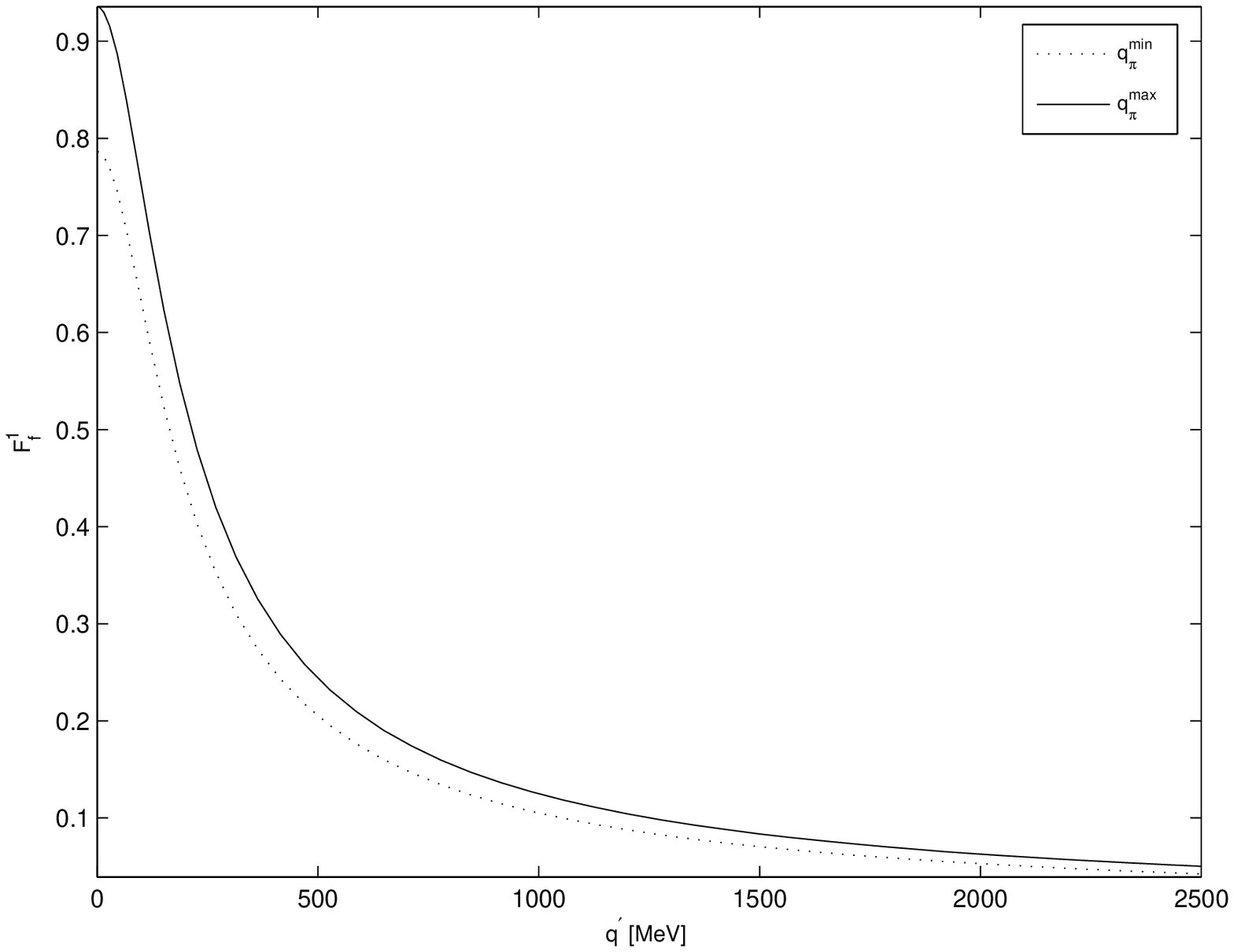}
  \captions[Form-factor like behaviour of the kinematic factors for
   the simple case when $f \left( \omega_{\pi} \right)=-f
   \left(-\omega_{\pi} \right)$, as a function of the exchanged pion
   momentum]
   {Form-factor like behaviour of the kinematic multiplicative factors
   for the simple case when $f \left( \omega_{\pi} \right)=-f
   \left(-\omega_{\pi} \right)$, as a function of the exchanged pion
   momentum. The dotted line corresponds to threshold and the solid
   line to the maximum momentum of the emitted pion for $T_{lab}= 440
   \umev$. } \label{ffactorvtoy}
\end{center}
\end{figure}

Analogously to the final-state case, the first four terms in
\eq{atoptisi} (diagrams $b_1$ to $b_4$ of \fig{toptisi}, where any
cut of the intermediate state intersects only nucleon lines) are
identified with the DWBA amplitude for the initial-state
distortion, and the last two (diagrams $b_5$ and $b_6$ of
\fig{toptisi}, with two exchanged particles in flight in any cut
of the intermediate state) to the stretched boxes. In other words,
the decomposition obtained in \eq{atoptisi} allows to write the
exact or reference expression for the DWBA amplitude for the
initial-state distortion, by collecting the four first terms,
which have intermediate states without exchanged particles in
flight, i.e.,
\begin{equation}
\mathcal{M}_{DWBA}^{ISI}=-\frac{1}{2} \int \frac{d^3 q'}{\left(2
\pi \right)^3 } \left[ \tilde{f} \left( \omega_{\pi} \right)
G_{\pi} \right] \frac{1}{\left(E_1+E_2-\omega_1-\omega_2 \right)}
T_{NN}^{ISI}, \label{dwbaisi}
\end{equation}
where
\begin{equation}
\tilde{f} \left(\omega_\pi \right) = \frac{f \left(\omega_\pi
\right)\left(E_\pi+F_1-\omega_1-\omega_\pi \right) +f
\left(-\omega_\pi \right) \left( F_2-\omega_2-\omega_\pi \right)
}{\omega_\pi}. \label{f0tilisi}
\end{equation}
Again, $f \left(\omega_\pi \right)$ stands for the product of the
$\pi N$ amplitude with the $\pi N N$ vertex. As before, the
multiplicative factors $f_{1}^{i} =
\frac{E_\pi+F_1-\omega_1-\omega_\pi }{\omega_{\pi}}$ and
$f_{2}^{i} = \frac{F_2-\omega_2-\omega_\pi  }{\omega_{\pi}}$
%
%
have a form-factor like behaviour (see \fig{ffactorvtoy}), cutting
the high momentum tail.

In \eq{dwbaisi} we made also the replacement of the exchanged
particle potential by the $NN$ scattering transition-matrix. The
two-nucleon propagator
\begin{equation}
G_{NN}=\frac{1}{\left(E_1+E_2-\omega_1-\omega_2 \right)}
\end{equation}
is the non-relativistic global $NN$ propagator. As we did for the
final-state interaction, we extracted from \eq{dwbaisi} the
effective pion propagator to be included in the initial state
distortion in a DWBA-type calculation. It reads,
\begin{equation}
G_{\pi}=\frac{1}{\left[\frac{\omega_2-\omega_1}{2}+\frac{F_1-F_2}{2}+\frac{E_{\pi}}{2}
\right]^2- \left[\left(
\frac{E_{\pi}}{2}-\frac{\omega_1+\omega_2}{2}+\frac{F_1+F_2}{2}\right)-\omega_{\pi}
\right]^2},
\end{equation}
or,
\begin{equation}
G_{\pi}=-\frac{1}{\left(F_{2}-\omega_{2}-\omega_{\pi} \right)
\left(F_{1}+E_{\pi}-\omega_{1}-\omega_{\pi} \right)}
\label{gpieffisi}.
\end{equation}
For the $\sigma$-exchange case, one gets
\begin{equation}
V_{\sigma}^{DWBA}=\frac{g_{\sigma}^{2}}{2 \omega_\sigma}
\left[\frac{1}{\left( E_{tot}-E_{1}-\omega_2-\omega_\sigma
\right)}+ \frac{1}{\left(E_{tot}-E_{2}-\omega_1-\omega_\sigma
\right)} \right]. \label{fvsigmaisi}
\end{equation}
\newpage

\section{Stretched Boxes vs. DWBA}
\sprg
In this section, the $NN$ and $(NN) \pi$ channels considered in
all calculations refer to the transition $^3P_0 \rightarrow
\left(^1S_0 \right)s$, which is the dominant one for $pp
\rightarrow pp\pi^{0}$ (see Chapter~\ref{Chargedandneutral}). The
details of the partial wave decomposition analytical formulae are
given in \apx{ApPWAmp}.

\begin{figure}
\begin{center}
\includegraphics[width=.78\textwidth,keepaspectratio]{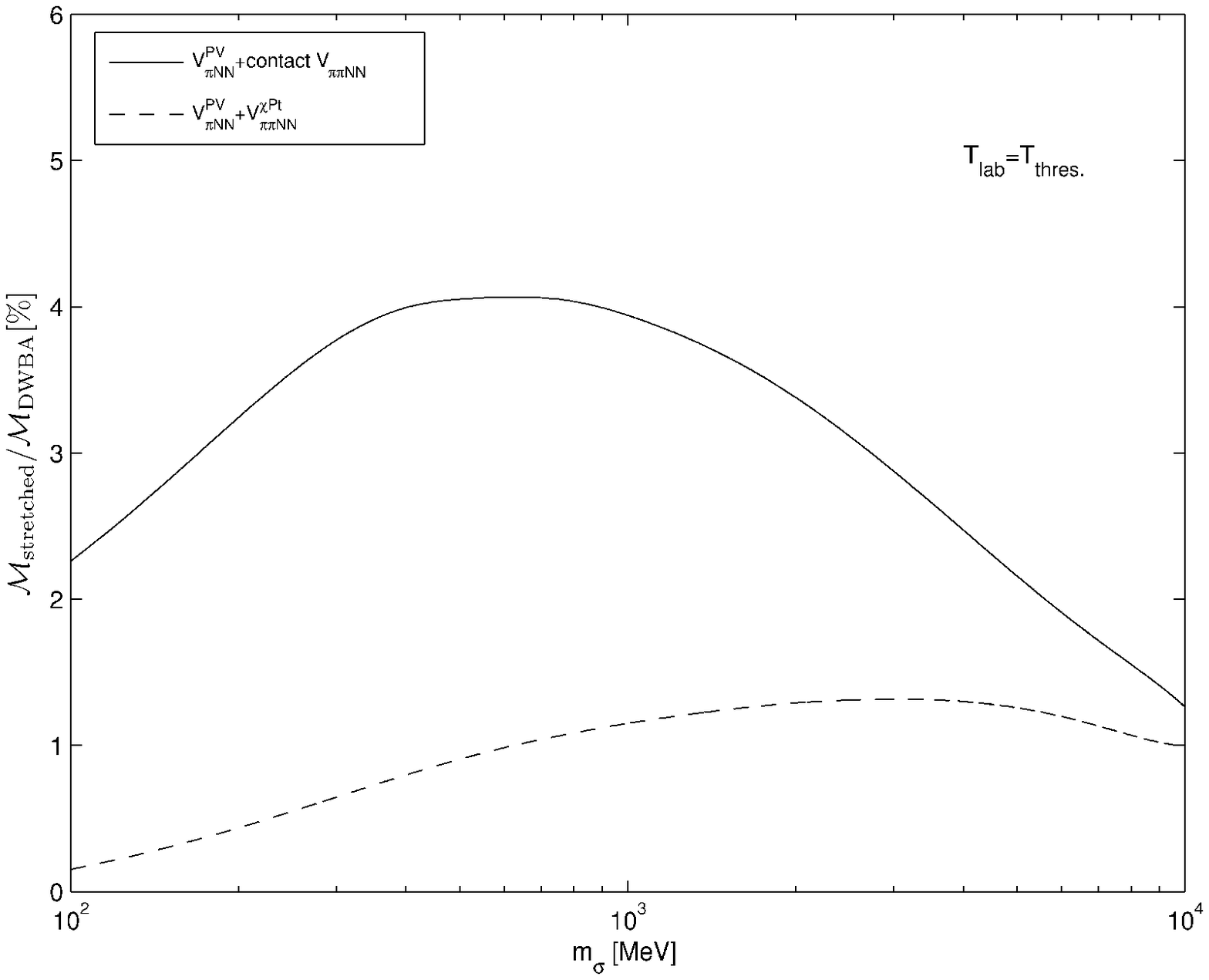}
  \captions[Importance of the stretched boxes compared to the DWBA
   amplitude for FSI]
   {Importance of the stretched boxes relatively to the
   DWBA FSI amplitude $\mathcal{M}^{FSI}_{DWBA}$ of \eq{dwbafsi}, as a function of
   the mass of the scalar particle for the final $NN$ interaction.
   The $\pi N $ amplitude is contact re-scattering vertex as in \eq{vtoy} (solid
   line) and the $\chi$PT amplitude as in \eq{vxpt} (dashed line). The energy is taken at the
   pion production threshold. Absolute values of the amplitudes are
   considered.} \label{msbvsdwbafsi}
\end{center}
\end{figure}

The DWBA amplitude was found to be clearly dominant over the
stretched boxes in the realistic model considered, as the dashed
line in \fig{msbvsdwbafsi} documents for the FSI case.

It is interesting to compare  this result with the one of
\rf{Hanhart:2000wf}. As shown in \fig{msbvsdwbafsi}, the stretched
boxes amplitude is less than $1 \% $, of the total amplitude and
therefore is about 6 times more suppressed than in the dynamics of
the toy model used in that reference. Replacing the $\pi N$
amplitude from $\chi$PT by a simple contact amplitude, the
stretched boxes amplitude becomes slightly more important, but
they still do not exceed $4 \% $ of the DWBA amplitude (solid line
in \fig{msbvsdwbafsi}).

\begin{figure}[t!]
\begin{center}
\includegraphics[width=.99\textwidth,height=.24\textheight]{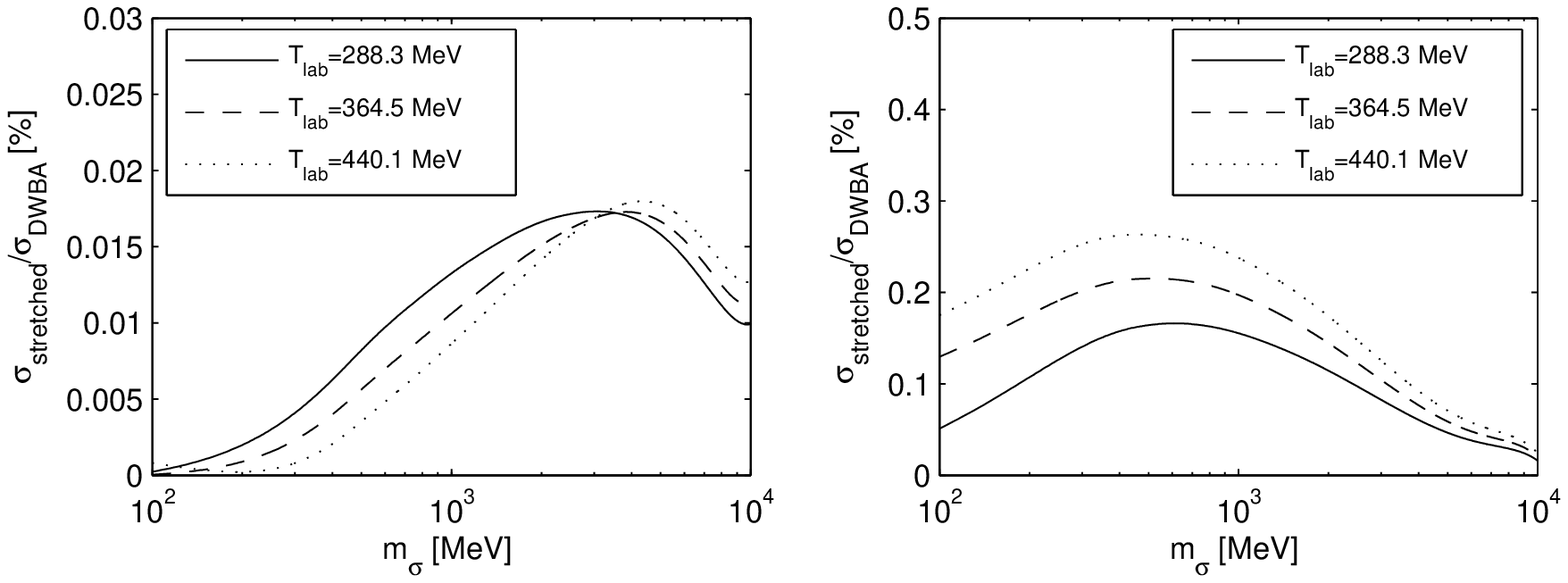}
  \captions[Importance of the stretched boxes compared to the DWBA
   cross section for FSI] {Importance of the stretched boxes
   compared to the DWBA (FSI) for the cross section as a function of
   the mass of the scalar particle for the final $NN$ interaction.
   The $\pi N$ amplitude is the $\chi$PT amplitude (left) and a
   contact re-scattering vertex (right).} \label{ssbvsdwbafsi}
\end{center}
\end{figure}

In terms of the cross section, the weight of the stretched boxes
relatively to DWBA is even smaller, of the order of $0.02 \%$ at
most. This is seen in \fig{ssbvsdwbafsi} where we compare, for
three different values of the laboratory energy, the cross section
obtained with only the DWBA contribution ($\sigma_{DWBA}$), with
the one obtained with only the stretched boxes terms
($\sigma_{stretched}$). In both cases considered, $V_{\pi N
N}^{PV} +V_{\pi \pi N N}^{\chi Pt} $ and $V_{\pi N N}^{PV}$ +
contact (left and right panel in \fig{ssbvsdwbafsi}, respectively)
the cross section with the DWBA terms is clearly dominant, and in
a more pronounced way when compared to the less realistic case of
\rf{Hanhart:2000wf}, not only at threshold but even for higher
energies as $440 \umev$.

However, we note that the ratios and energy dependence of the
amplitudes and of the cross sections are significantly influenced
by the $\pi N$ amplitude used in the calculation. The stretched
boxes are seen to be amplified by the contact $\pi N$
re-scattering amplitude, due to an interplay between the $\pi N$
and the $NN$ amplitudes. Relatively to the more realistic $\chi
$Pt amplitude, the contact $\pi N$ amplitude gives a larger weight
to the low-momentum transfer. The realistic $\pi N$ amplitude
satisfies chiral symmetry. This implies cutting small momenta and
giving more weight to the region of large momentum transfer, which
however in turn is cut by the nucleonic interactions. The
difference between the two $\pi N$ amplitudes is clearly seen by
comparing the behaviour of each curve on the left panel of
\fig{ssbvsdwbafsi} with the corresponding curves on the right
panel, for small values of the $NN$ interaction cut-off.

\begin{figure}[t!]
\begin{center}
\includegraphics[width=.78\textwidth,keepaspectratio]{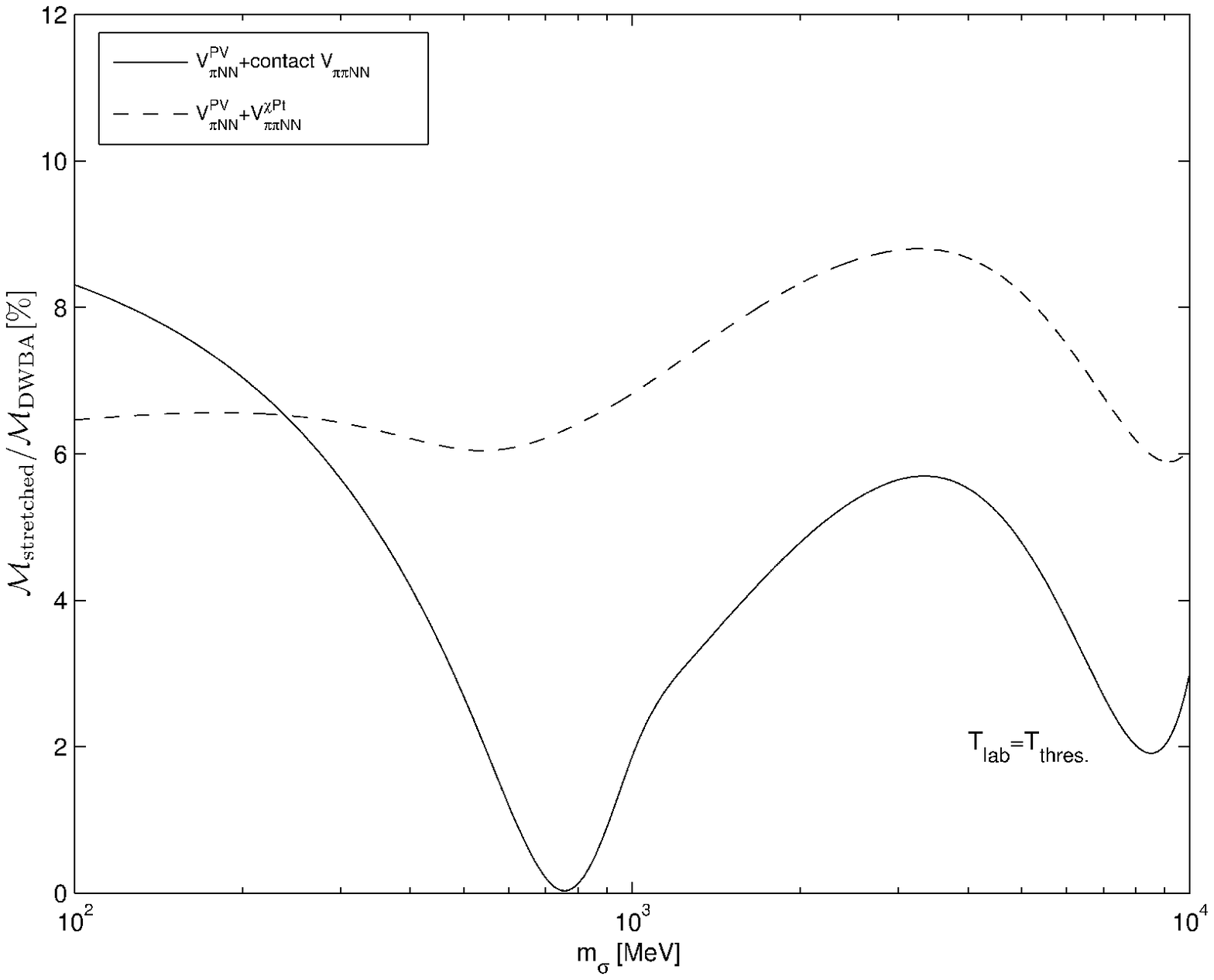}
  \captions[Importance of the stretched boxes compared to the DWBA
   amplitude for ISI]{The same as \fig{msbvsdwbafsi} but for the DWBA
   ISI amplitude $\mathcal{M}^{ISI}_{DWBA}$ of \eq{dwbaisi}. } \label{msbvsdwbaisi}
\end{center}
\end{figure}

In the case of the initial-state amplitude, the stretched boxes
amplitude is also much smaller than the DWBA amplitude for the two
cases shown in \fig{msbvsdwbaisi}.

\begin{figure}[h!]
\centering
\includegraphics[width=0.99\textwidth,keepaspectratio]{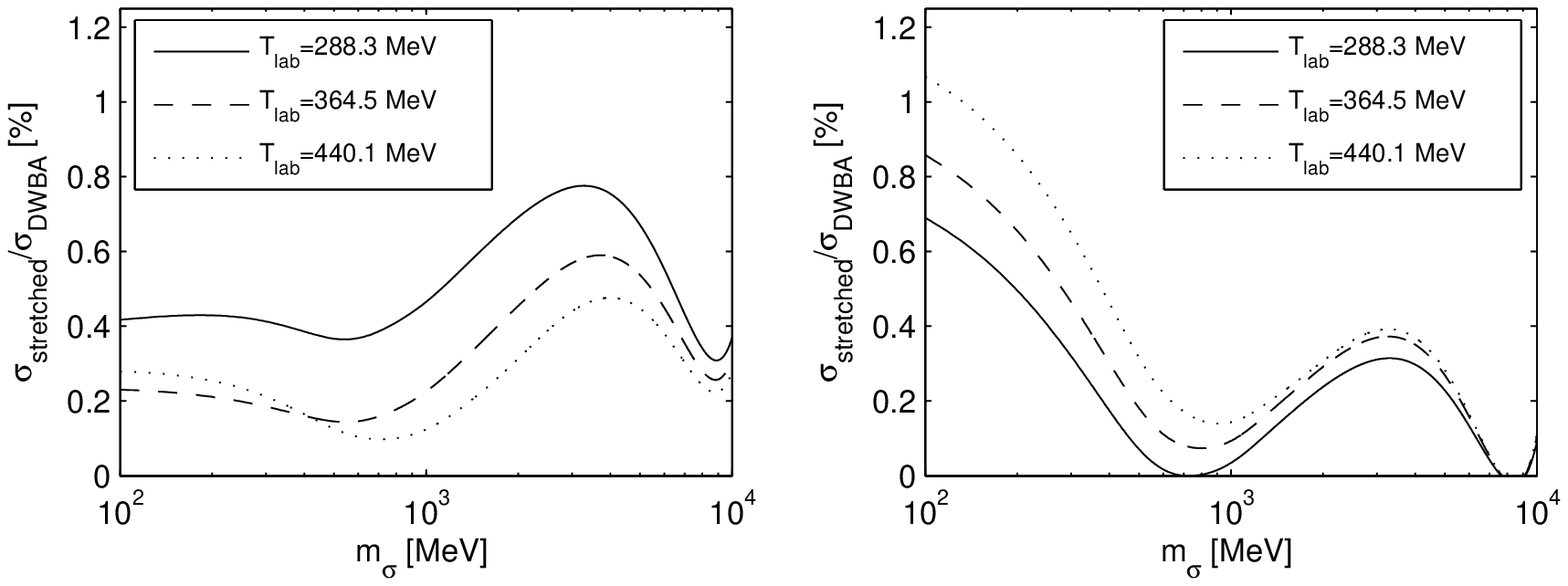}
  \captions[Importance of the stretched boxes compared to the DWBA
   cross section for ISI]{The same as \fig{ssbvsdwbafsi} but for the ISI case.}
\label{ssbvsdwbaisi}
\end{figure}

The cross sections obtained with only the stretched boxes terms
were found to be less than $1.2\% $ of the DWBA cross sections,
even for laboratory energies as high as $T_{lab} \sim 440\umev$
(see \fig{ssbvsdwbaisi}).

The results presented in \fig{msbvsdwbafsi} to \fig{ssbvsdwbaisi}
justify the DWBA treatment for pion production.
\vspace{3cm}

%

%
%

\newpage

\section{The logarithmic singularity in the pion propagator (ISI) \label{SecSing}}
\sprg
From the evaluation of the effective pion propagator for the ISI
case of \eq{gpieffisi}, at threshold,
\begin{equation}
G_{\pi}^{th.}= \frac{1}{\left(\frac{m_{\pi}}{2} \right)^{2}-\left(
\frac{m_{\pi}}{2}-\omega_{\pi}\right)^{2}}=\frac{1}{\omega_{\pi}
\left(m_{\pi}-\omega_{\pi}\right)},
\end{equation}
it is straightforward to conclude that $G_{\pi}$ has a pole, since
it is energetically allowed for the exchanged pion to be
on-mass-shell $\left(\omega_{\pi}=m_{\pi} \right)$. The treatment
of the pole leads to a moving singularity which was not considered
in the traditional DWBA calculations of
\rfs{{Park:1995ku},{vanKolck:1996dp},{Pena:1999hq},{Cohen:1995cc},{daRocha:1999dm},{Gedalin:1999ht}}.
 It requires a careful numerical treatment which was included in our calculations.

This is done by rewriting $G_{\pi}$ of \eq{gpieffisi} as
\begin{equation}
G_{\pi}=\frac{1}{\alpha_{1}-\alpha_{2}}
\left(\frac{1}{\alpha_{1}-\omega_{\pi}}-\frac{1}{\alpha_{2}-\omega_{\pi}}
\right) \label{gpialpha},
\end{equation}
with
\begin{eqnarray}
\alpha_{1} & = & F_{2}-\omega_{2} \\
\alpha_{2} & = & F_{1}+E_{\pi}-\omega_{1}.
\end{eqnarray}
With the definitions $F_{1,2}$, $\omega_{1,2}$ and $\omega_\pi$
introduced in \sec{Extrprop} (see \apx{Apkindiagrams}), then
\eq{gpialpha} reads
\begin{equation}
G_{\pi }=\frac{1}{\alpha_{1}-\alpha_{2}}\left[ \frac{\alpha
_{1}+\omega _{\pi }}{q_{\pi }I_{2}\left( x_{2}\right)
}\frac{1}{y_{1}-x_{1}}- \frac{\alpha _{2}+\omega _{\pi }}{q_{\pi
}I_{2}\left( x_{2}\right) }\frac{1}{y_{2}-x_{1}}\right].
\end{equation}
where $I_{2}\left( x_{2}\right) = \left| \overrightarrow{q_u}-
\overrightarrow{q_{k}} \right| =\sqrt{q_u^{2}-2 q_u
q_{k}x_{2}+q_{k}^{2}}$.  The angles involved in the partial wave
decomposition performed for the amplitudes are
$x_{1}=\measuredangle \left(
\overrightarrow{q_u}-\overrightarrow{q_{k}},\overrightarrow{q_{\pi
}}\right)$ and $x_{2}=\measuredangle
\left(\overrightarrow{q_u},\overrightarrow{q_{k}}\right)$ (see
\apx{ApPWAmp}).
Here, $y_{i}$,  roots of the denominator of $G_\pi$, are given by:
\begin{equation}
y_{i}=\frac{\alpha _{i}^{2}-\beta ^{2}}{q_{\pi }I_{2}\left( x_{2}\right) }%
\qquad \text{with }\beta ^{2}\equiv m_{\pi }^{2}+\frac{q_{\pi }^{2}}{4}%
+I_{2}\left( x_{2}\right) ^{2}\text{ and }i=1,2
\end{equation}

Once $G_{\pi }$ is written in this form, for the calculation of
the partial wave decomposition, we have to deal with its poles at
$x_1=y_i$. We used the subtraction technique (see \apx{ApImpI})
exposing the three-body logarithmic singularity\footnote{ An
alternative treatment may be found in \rf{Motzke:2002fn}.}:
\begin{eqnarray}
\int_{-1}^{1}\frac{f\left( x_{1},x_{2}\right) }{y_{i}-x_{1}+i
\varepsilon}P_{L}\left( x_{1}\right) dx_{1}
&=&\mathbf{PV}\int_{-1}^{1}\frac{f\left( x_{1},x_{2}\right)
-f\left( y_{i},x_{2}\right) }{y_{i}-x_{1}}P_{L}\left(
x_{1}\right) +2f\left( y_{i},x_{2}\right) Q_{L}\left( y_{i}\right) - \nonumber \\
&&-i\pi P_{L}\left( y_{i}\right) f\left( y_{i},x_{2}\right)
\end{eqnarray}
where $P_{L}$ and are the Legendre polynomials of order $L$ and
$Q_L$ are the Legendre functions of the second kind of order
$L$\cite{Arfken:1995bk}. These last functions exhibit logarithmic
singularities which are given by the condition $y_i=\pm 1$,
which are determined analytically. Then, the integral over the
moving logarithmic singularities is handled by a variable mesh.
When the number of singularities (always between zero and two) is
$n_s$, the interval of integration is divided into $2n_{s}+1$
regions including as breaking points those given by the found
singularities. In each one of the regions we considered a Gaussian
mesh.
%
%
%
\section{Conclusions}

\sprg
In this Chapter we have generalised the work done in
\rfs{Hanhart:2000wf,Motzke:2002fn} on the study of the DWBA
approaches for the irreducible pion re-scattering mechanism of the
reaction $pp \rightarrow pp \pi^0$.  We considered a physical
model for nucleons and pions, combining a pseudo-vector coupling
for the $\pi N N$ vertex and the $\chi$PT $\pi N$ re-scattering
amplitude.
We treated the initial and final state distortion exactly, taking
into account in the first case the three-body moving singularities
in the pion propagator.

Both for the final- and initial-state interaction, our results
show that the DWBA formalism is quite adequate at threshold and
even at higher energies, since this part of the full amplitude is
clearly seen to be dominant over the stretched boxes.
This is independent of the model for the $\pi N$
re-scattering amplitude. Nevertheless, relatively to less
realistic models, a  chirally constrained amplitude reinforces
even more the relative importance of the DWBA amplitude in the
total amplitude.

The amplitudes given by Eqs.(\ref{dwbafsi}) and (\ref{dwbaisi})
obtained as non-relativistic reductions of the corresponding
Feynman diagrams for the final- and initial-state interactions,
can now be employed to study the commonly used prescriptions for
the energy of the exchanged pion, both in the propagator and at
the re-scattering vertex. This energy, not fixed in a
non-relativistic quantum-mechanical calculation, was fixed by TOPT
to be $\omega_{\pi}$, its on-mass-shell energy.

%
Since in the time-ordered diagrams energy is not conserved at
individual vertices, each of the re-scattering diagrams for the
initial- and final-state distortion defines a different off-energy
shell extension of the pion re-scattering amplitude, as calculated
in \sec{Extrpropfsi} and \sec{Extrpropisi}.
As a consequence, the two operators for FSI and ISI are different,
demanding the evaluation of two different matrix elements of such
operators between the quantum-mechanical $NN$ wave functions.
In the next Chapter, an alternative approach will be presented,
where this problem does not arise.

%% file: Chapter4.tex
\setcounter{minitocdepth}{2}
\chapter{The S-matrix approach for quantum mechanical calculations} \label{Smatrixapp}

\minitoc {\bfseries Abstract:} The pion re-scattering process is
certainly part of the pion production mechanism, but its
importance relative to other contributions varies considerably
depending on the approximations made in evaluation of the
effective operators. In this Chapter, we will re-examine the
nature and extent  of this uncertainty. The pion re-scattering
operator originated from the S-matrix construction, previously
applied only below pion production threshold, is seen here to
approximate well the exact DWBA result consistent with
time-ordered perturbation theory.

\newpage

%
%
\section{The pion re-scattering operator in the S-matrix technique}
%
\sprg
To derive the quantum mechanical effective operators for pion
production, and other effective nuclear operators in general, one
starts from the relativistic (effective) Lagrangian written in
terms of hadronic fields. The interactions, mediated by meson
exchanges before and after the production reaction takes place,
are included in the effective $NN$ (and nucleon-meson)
interaction, while from the irreducible parts connected to the
reaction mechanism (e.g., pion production) one obtains effective
operators, whose expectation values are to be evaluated between
the initial and final nucleonic wave functions. One aims to get
such effective operators consistent with the realistic description
of the $NN$ interaction, which can then be used in studies of the
corresponding reactions not only in the simplest (one or
two-nucleon) systems, but preferably also in heavier nuclei.

The covariant techniques based on the Bethe-Salpeter equation or
its quasipotential re-arrangements are these days practically
manageable only below meson production threshold. However, above
the threshold the dressings of the single hadron propagators and
interaction vertices via the meson loops have to be included
explicitly. For this reason the construction of the production
operator is so far realised mostly in the Hamiltonian
quantum-mechanical framework (usually non-relativistic, or, with
leading relativistic effects included perturbatively by means of
the decomposition in powers of $p/m$, where $p$ is the typical
hadronic momentum and $m$ is the nucleon mass).

The derivation of the nuclear effective operators below the meson
production threshold within the Hamiltonian framework -- leading
to hermitian and energy independent $NN$ and $3N$ potentials, and
conserved electromagnetic and partially conserved weak current
operators -- can be done in many different ways (see discussion in
\rfs{{Chemtob:1971pu},{Riska:1984vs},{Adam:1989ha},{Adam:1994pc}}
and references therein). At the non-relativistic order the results
are determined uniquely. As for the leading order relativistic
contributions, they were shown to be unitarily
equivalent\cite{Adam:1994pc}. The unitary freedom allows to choose
the $NN$ potentials to be static (in the CM frame of the
two-nucleon system) and identify them with the successful static
semi-phenomenological potentials.

Also, above the threshold energy the static limit is commonly
employed, since more elaborate descriptions which include the
mesonic retardation and loop effects are technically considerably
more complex
\cite{{Elster:1988zu},{Elster:1988pp},{Schwamb:1999qd}},
especially for systems of more than two nucleons. Both the static
approaches and the ones including ``retardation'' typically
consider contributions of several one-meson exchanges, and the
potentials are fitted to describe the data. It is therefore
difficult to assess how well do they approximate the covariant
amplitudes (corresponding to the same values of physical masses
and coupling constants) which are so far outside the scope of
existing calculational schemes, but which we believe do provide in
principle the consistent description of the considered reactions.

Thus, the ultimately exact approach to the description of pion
production (and in particular of the pion re-scattering
contribution) would be either the covariant Bethe-Salpeter or
quasipotential frameworks (extended above the pion threshold) or
the quantum mechanical coupled-channel technique including
retardation. In these approaches one has to treat consistently the
non-hermitian energy-dependent $NN$ interaction (fitted to the
data also above pion production threshold) and consider besides
the effects of renormalisation of vertices, masses and wave
functions via meson loops.

In this Chapter (following
\rfs{{Hanhart:2000wf},{Motzke:2002fn},{Malafaia:2003wx},{Malafaia:2004cu}})
we rather numerically estimate the range of the predictions from
several frequently used simplifying approximations, and compare
them to the result obtained from the reduction of the
corresponding covariant Feynman diagrams for the pion
re-scattering operator. This reduction coincides with the
time-ordered perturbation theory\cite{Malafaia:2003wx}. To this
end we deal with retardation effects in the exchanged pion
propagator, i.e., its energy dependence, as well as with the
energy dependence of the $\pi$N scattering amplitude in the
vertex, from which the produced pion is emitted.
\subsection{Factorisation of the effective re-scattering operator}

\begin{figure}
\begin{center}
\includegraphics[width=.58\textwidth,keepaspectratio]{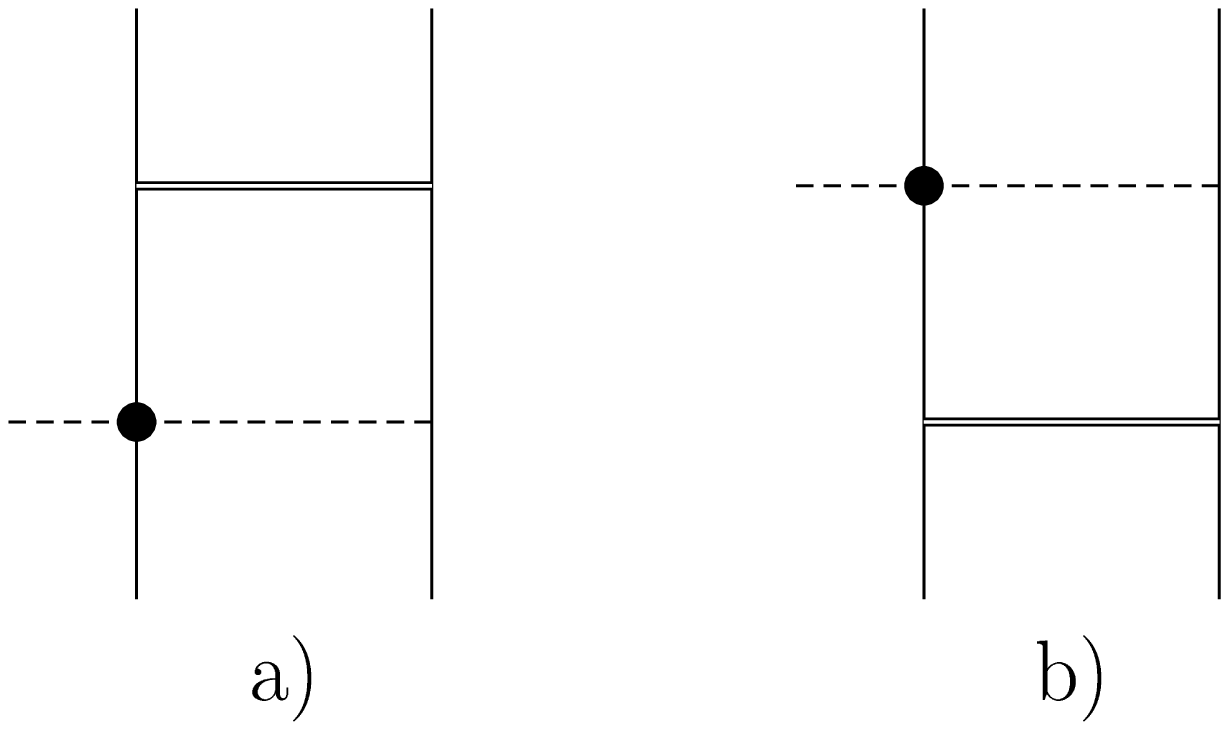}
  \captions[Feynman diagrams for pion re-scattering (FSI and ISI cases)]{Feynman
   diagrams for pion re-scattering. The pion field is represented by
   a dashed line, the $NN$ interaction by solid double line and the
   nucleons by solid lines.} \label{diagrams1}
\end{center}
\end{figure}

\sprg
In the previous Chapter,  we made the connection to the usual
representation of the pion re-scattering operator for
non-relativistic calculations, following the work of
\rf{Malafaia:2003wx}. We started from the covariant two-nucleon
Feynman amplitudes including final and initial state interaction
(FSI and ISI, respectively), shown in Fig.~\ref{diagrams1}a) and
\ref{diagrams1}b)).

To obtain the effective re-scattering operator, the negative
energy contributions in the nucleon propagators (to be included in
complete calculations) were neglected. By integrating subsequently
over the energy of the exchanged pion the resulting Feynman
amplitudes were transformed into those following from time-ordered
perturbation theory.

\begin{figure}
\begin{center}
\includegraphics[width=.58\textwidth,keepaspectratio]{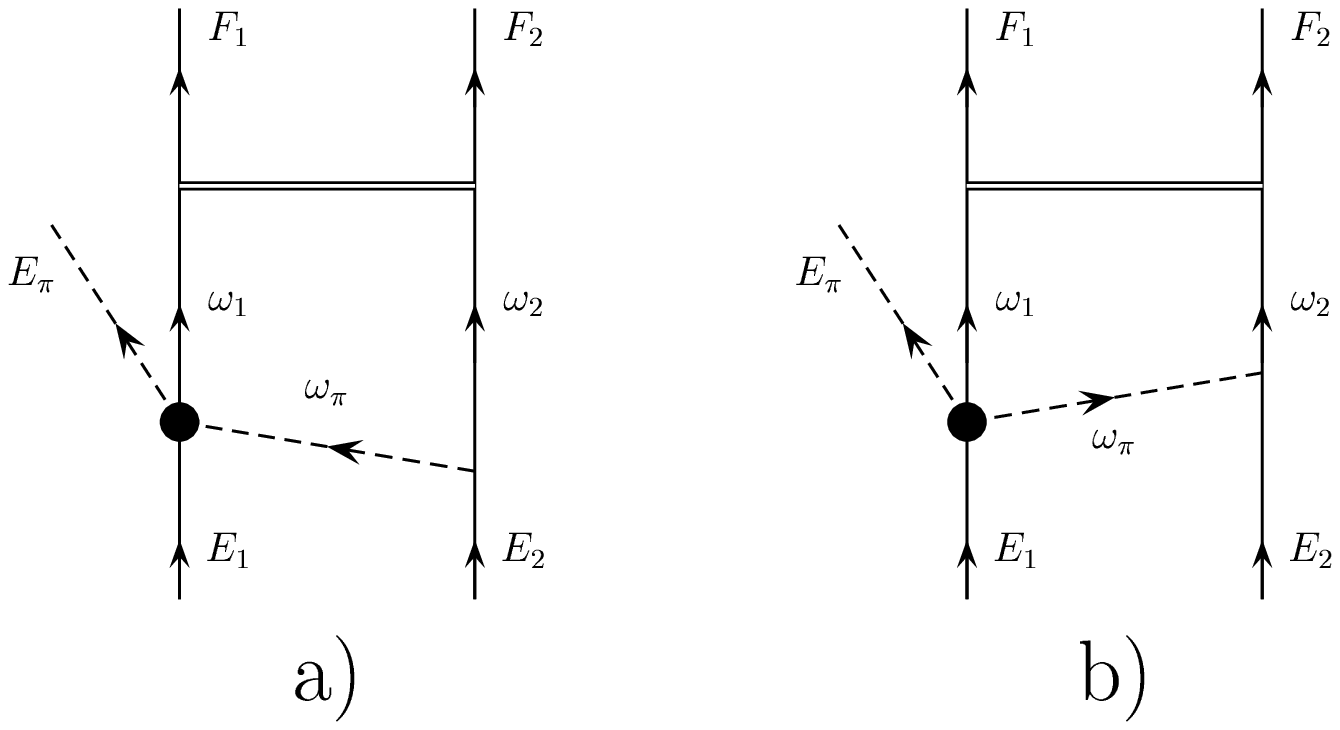}
  \captions[The two time-ordered diagrams which are connected with
   DWBA for the FSI case]{The two time-ordered diagrams for FSI
   considered here. The additional stretched box diagrams are
   neglected.} \label{diagrams2}
\end{center}
\end{figure}

The irreducible ``stretched box diagrams'' (i.e., those with more
than one meson in flight in the intermediate states) give a very
small contribution and can be therefore also neglected (as we
showed in the previous Chapter in
\fig{msbvsdwbafsi}-\fig{ssbvsdwbaisi}). Thus, the full covariant
amplitude is in the lowest order Born approximation well
approximated  by the product of the $NN$ potential and the
effective pion re-scattering operator, which can be extracted from
these time-ordered diagrams (\fig{diagrams2}).
Summarising, what was obtained in \chp{FTtoDWBA}, for the final
state distortion amplitude, was
\begin{eqnarray}
{\mathcal M}_{FSI}&=& \int \frac{d^3 q'}{(2\pi)^3}\,
 V_\sigma\, \frac{1}{F_1+F_2- \omega_1- \omega_2 + i\varepsilon}\,
 \hat{O}_{rs}   \label{afsi}\\
  \hat{O}_{rs}  &=& -\frac{1}{2\omega_\pi}
 \left[ \frac{f(\omega_\pi)}{E_2-\omega_2-\omega_\pi}+
 \frac{f(-\omega_\pi)}{E_1-\omega_1-E_\pi-\omega_\pi} \right]
= \frac{1}{2}\, \tilde{f}\, G_\pi . \label{mfsi}
\end{eqnarray}
The effective pion re-scattering operator was factorised into an
effective pion re-scattering  vertex $\tilde{f}$ and an effective
pion propagator $G_\pi$.  For the diagram with FSI
(\fig{diagrams2}) this factorisation reads (\eq{fvsigmafsi},
\eq{ftildefsi} and \eq{gpiefffsi}):
\begin{eqnarray}
  \tilde{f} &=& \frac{1}{\omega_\pi}\left[
  (E_1- \omega_1- E_\pi- \omega_\pi) f(\omega_\pi)+
  (E_2- \omega_2- \omega_\pi) f(-\omega_\pi)\right]  \label{vtil}\\
  G_\pi &=& - \frac{1}{(E_1- \omega_1- E_\pi- \omega_\pi)(E_2- \omega_2-
  \omega_\pi)} \label{gpi}\\
  V_\sigma&=& \frac{g_{\sigma}^{2}}{2\omega_\sigma} \left[
\frac{1}{F_2-\omega_2-\omega_\sigma}+
\frac{1}{F_1-\omega_1-\omega_\sigma} \right] . \label{vsigfsi}
\end{eqnarray}
As before, $\vec{q}^{\, \prime}$ is the momentum of the exchanged
pion, $\omega_\pi=\sqrt{ m_\pi^2+ \vec{q}^{\, \prime 2}}$ is its
on-mass-shell energy and $f(\omega_\pi)$ is the product of the
$\pi N$ amplitude with the $\pi NN$ vertex. The variables $E_i,
\omega_i, F_i$ are the on-mass-shell energies of the i-th nucleon
in the initial, intermediate and final state, respectively,
$E_\pi$ is the energy of the produced pion, $E_1+E_2= F_1+ F_2+
E_\pi$.

Again, we note that $f\left(\omega_\pi\right)$ in fact depends on
the three-momenta and energies of both (the exchanged and the
produced) pions and on the nucleon spin. We indicate explicitly
only the dependence on the exchanged pion energy, as the only one
important for the considerations below.

The inclusion of some pieces of the integrand of \eq{afsi} into
the propagator $G_\pi$ and of others in the modified vertex
$\tilde f$ is somewhat arbitrary. The appearance of the unusual
effective propagator $G_\pi$ and the effective vertex $\tilde{f}$
is the result of combining {\em two} time-ordered diagrams with
different energy dependence  into a {\em single} effective
operator.

The $NN$ interaction is in \fig{diagrams2} and  in \eq{afsi}
simulated by a simple $\sigma$-exchange potential. Though not
realistic, this interaction suffices for model studies of
approximations employed in derivations of the effective pion
re-scattering operator, as done in references
\rfs{{Hanhart:2000wf},{Motzke:2002fn}}. Since some results do
depend on the behaviour of the $NN$ scattering wave function, in
particular in the region of higher relative momenta, we perform
our calculations (as in \rf{Malafaia:2003wx}) also with $V_\sigma$
replaced by a full $NN$ T-matrix, generated from the realistic
Bonn B potential\cite{Machleidt:1987hj}.

We note that the meson poles are not neglected in the integration
over the energy $Q'_0$ of the exchanged pion, which generates
\eq{afsi} (see \sec{Extrprop}). A result similar to \eq{afsi} can
also be obtained for the amplitude with the initial state
interaction (ISI).
\begin{eqnarray}
{\mathcal M}_{ISI}&=& -\int \frac{d^3 q'}{(2\pi)^3}\,
 V_\sigma\, \frac{1}{E_1+E_2- \omega_1- \omega_2 + i\varepsilon}\,
 \hat{O}_{rs}   \label{aisi}\\
  \hat{O}_{rs}  &=& -\frac{1}{2\omega_\pi}
 \left[ \frac{f(\omega_\pi)}{E_{\pi}+F_{1}-\omega_{1}-\omega_\pi}+
 \frac{f(-\omega_\pi)}{F_{2}-\omega_2-\omega_\pi} \right]
= \frac{1}{2}\, \tilde{f}\, G_\pi  \label{misi} \\
  \tilde{f} &=& \frac{1}{\omega_\pi}\left[
  (F_2- \omega_2-  \omega_\pi) f(\omega_\pi)+
  (E_\pi+F_{1}- \omega_1- \omega_\pi) f(-\omega_\pi)\right]  \label{vtilisi}\\
  G_\pi &=& - \frac{1}{(E_\pi+F_{1}- \omega_1- \omega_\pi)
  (F_2- \omega_2-  \omega_\pi)} \label{gpiisi}\\
  V_\sigma&=& \frac{g_{\sigma}^{2}}{2\omega_\sigma} \left[
\frac{1}{E_2-\omega_2-\omega_\sigma}+
\frac{1}{E_1-\omega_1-\omega_\sigma} \right] . \label{vsigisi}
\end{eqnarray}
The two amplitudes differ however in the contribution from the
pion poles to the remaining integration over the three-momentum.
For the amplitude with FSI there are no such poles. However, for
the ISI case there are values of the exchanged pion three-momentum
for which the propagator $G_\pi$  has poles. These poles have been
considered in all our numerical calculations for the cross section
(see \sec{SecSing}). As we will see in this Chapter, they are one
of the main reasons for deviations between several approximations
and the reference results calculated from
Eqs.~(\ref{afsi}-\ref{vsigfsi}) and
 Eqs.~(\ref{aisi}-\ref{vsigisi}) .

It is worth mentioning that although the FSI and ISI diagrams
graphically separate the $NN$ interaction and the pion
re-scattering part (when the stretched boxes are neglected), they
do not define a {\em single} effective operator (as a function of
nucleon three-momenta and the energy of emitted pion). Since in
these time-ordered diagrams energy is not conserved at individual
vertices, each of these diagrams defines a different off-energy
shell extension of the pion re-scattering amplitude. This is an
unpleasant feature, since one would have to make an analogous
construction for diagrams with both FSI and ISI. Moreover, one
would have to repeat the whole analysis for systems of more than
two nucleons. Only after the on-shell approximation is made
consistently (in the next subsection), the pion re-scattering
parts of FSI and ISI diagrams coincide and one can identify them
with a single effective re-scattering operator.
\subsection{The S-matrix technique}

\sprg
The S-matrix technique is a simple prescription to derive
effective nuclear operators from the corresponding covariant
Feynman diagrams
\cite{{Chemtob:1971pu},{Riska:1984vs},{Adam:1989ha},{Adam:1994pc}}.
For electromagnetic operators and also for $NN$ and $3N$
potentials at energies below the first nucleonic inelasticities,
the S-matrix approach analytically reproduces the results of more
laborious constructions, based on time-ordered or non-relativistic
diagram techniques. We will prove here that the same is
numerically true for energies above pion production threshold.

This approach is well defined and understood below the meson
production threshold. As a simple tool it was  employed also above
the threshold in \rfs{{Lee:1993xh},{Pena:1999hq}}, but only to
derive the Z-diagram operators.
The two-nucleon effective operators are by definition identified
with the diagrams describing the irreducible mechanism of the
corresponding reaction. The only exception are the nucleon Born
diagrams from which the iteration of the one-nucleon operator has
to be subtracted. The operators of the nuclear electromagnetic and
weak currents, as well as the pion absorption operators and
nuclear potentials, are obtained by a straightforward
non-relativistic reduction of the corresponding Feynman diagrams,
in which the intermediate particles are off-mass-shell but energy
is conserved at each vertex: therefore the derived effective
operators are also defined on-energy-shell. The nuclear currents
and other transition operators are defined to be consistent with a
hermitian energy independent $NN$ potential, which has the usual
one boson exchange form employed in realistic models of the $NN$
interaction, and can be used also in systems of more than two
nucleons.

\subsubsection{The $\sigma$-exchange potential}
For the $\sigma$-exchange potential the S-matrix technique in the
lowest order of non-relativistic reduction yields
\begin{equation}
V_\sigma^{S}=  g_{\sigma}^{2}\frac{1}{\Delta^2-\left(m_\sigma^2+
\vec{q}_\sigma^{\, 2}\right) } , \label{vsigos}
\end{equation}
where $\Delta=\Delta_1=\Delta_2$, $\Delta_1= \epsilon^{\prime}_1-
\epsilon_1$ and $\Delta_2= \epsilon_2-\epsilon^{\prime}_2$, with
$\epsilon^{\prime}_i$ and $\epsilon_i$ being the on-mass-shell
energies of the i-th nucleon after and before the meson exchange,
respectively. As pointed out above, this defines the potential
only on-energy-shell (where actually $\Delta=0$).
However, the Lippmann-Schwinger equation and even the first order
Born approximation to the wave function require the potential
off-energy-shell.

The S-matrix prescription takes the symmetric combination
\begin{equation}
\Delta=\frac{\epsilon^{\prime}_1-\epsilon_1}{2}+
\frac{\epsilon_{2}-\epsilon^{\prime}_2}{2}=\frac{\Delta_{1}+\Delta_{2}}{2}.
\label{Smatprescpot}
\end{equation}
%
%
Most realistic $NN$ potentials, namely those fitted to the data
below pion threshold, in particular the Bonn
B\cite{Machleidt:1987hj} potential used in this work, are
energy-independent and static in the nucleon CM frame and are
therefore consistent with this construction.
The extended S-matrix approach
\cite{{Chemtob:1971pu},{Riska:1984vs},{Adam:1989ha},{Adam:1994pc}}
given by \eq{vsigos} defines the most general off-energy-shell
continuation of $V_\sigma$ as a class of unitarily equivalent
potentials parameterised by the ``retardation parameter'' $\nu$.

\begin{figure}
\begin{center}
\includegraphics[width=.88\textwidth,keepaspectratio]{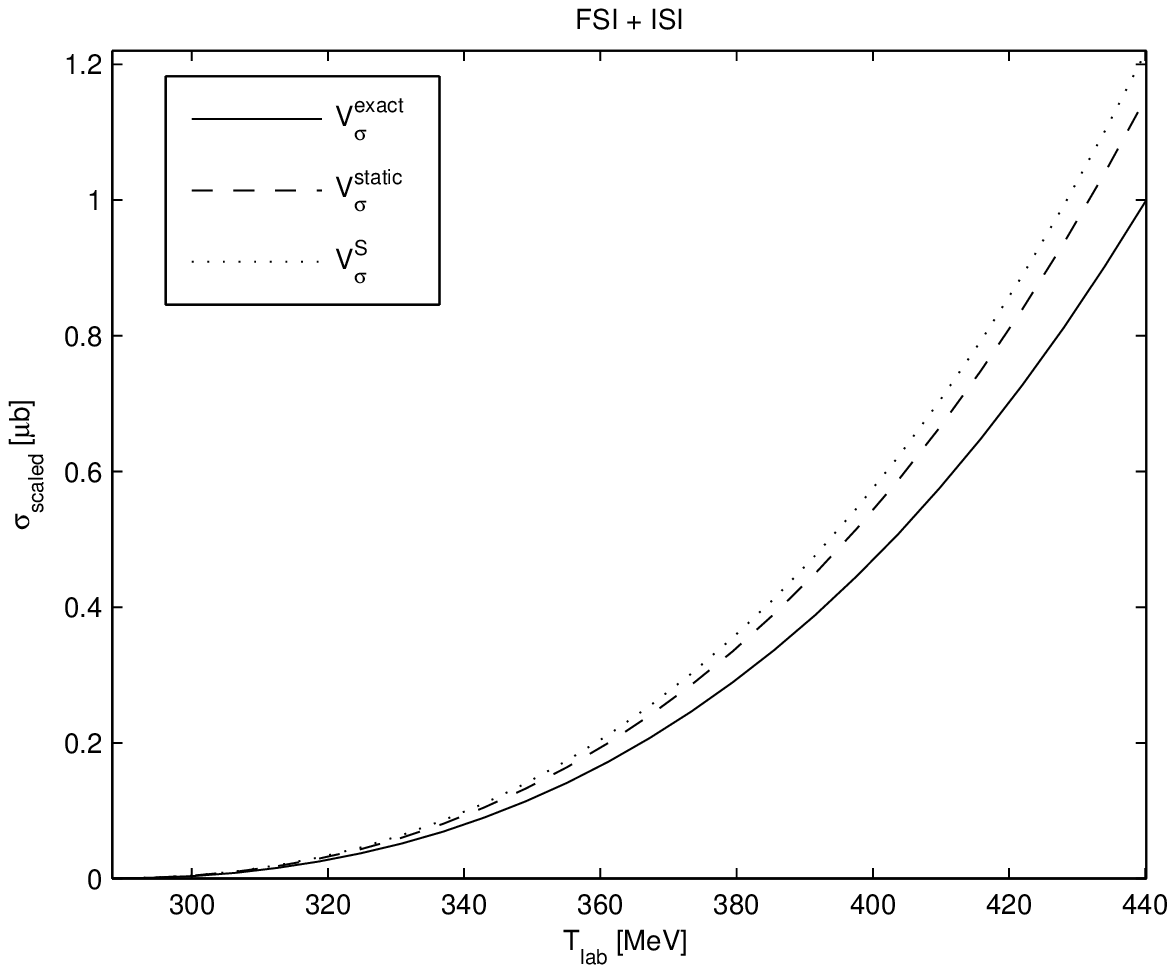}
  \captions[Comparison of the cross section for $pp \rightarrow pp \pi^{0}$
   calculated with the static and on-shell
   approximations for the $\sigma$-exchange potential to the
   reference result coming from TOPT]{Comparison of the cross section for
   $pp \rightarrow pp \pi^{0}$ calculated with the static
   and on-shell approximations for the $\sigma$-exchange potential
   (dashed and dotted lines, respectively) to the reference result
   of \eq{afsi} and \eq{aisi} (solid line), as a
   function of the laboratory energy $T_{lab}$. The
   values of the cross sections are scaled by requiring a reference
   result of $1$ for $T_{lab}=440 \umev$.  \label{vsigmaboth}}
\end{center}
\end{figure}

We show in \fig{vsigmaboth} the comparison of the cross section
for $pp \rightarrow pp \pi^{0}$ with the approximation of
\eq{vsigos} and the static approximation ($\Delta=0$) for the
potential, to the ``exact" reference result of \eq{afsi} and
\eq{aisi}.
From \fig{vsigmaboth}, one concludes that the effect of the
approximations for $V_{\sigma}$ is small till $Q \approx
0.1m_{\pi}$. As expected, the deviations of the described
approximations (dashed and dotted line, respectively) from the
reference result (solid line) increase with energy, with a maximum
deviation of about $20\%$.
\subsubsection{The pion re-scattering operator}
\begin{figure}
\begin{center}
\includegraphics[width=.58\textwidth,keepaspectratio]{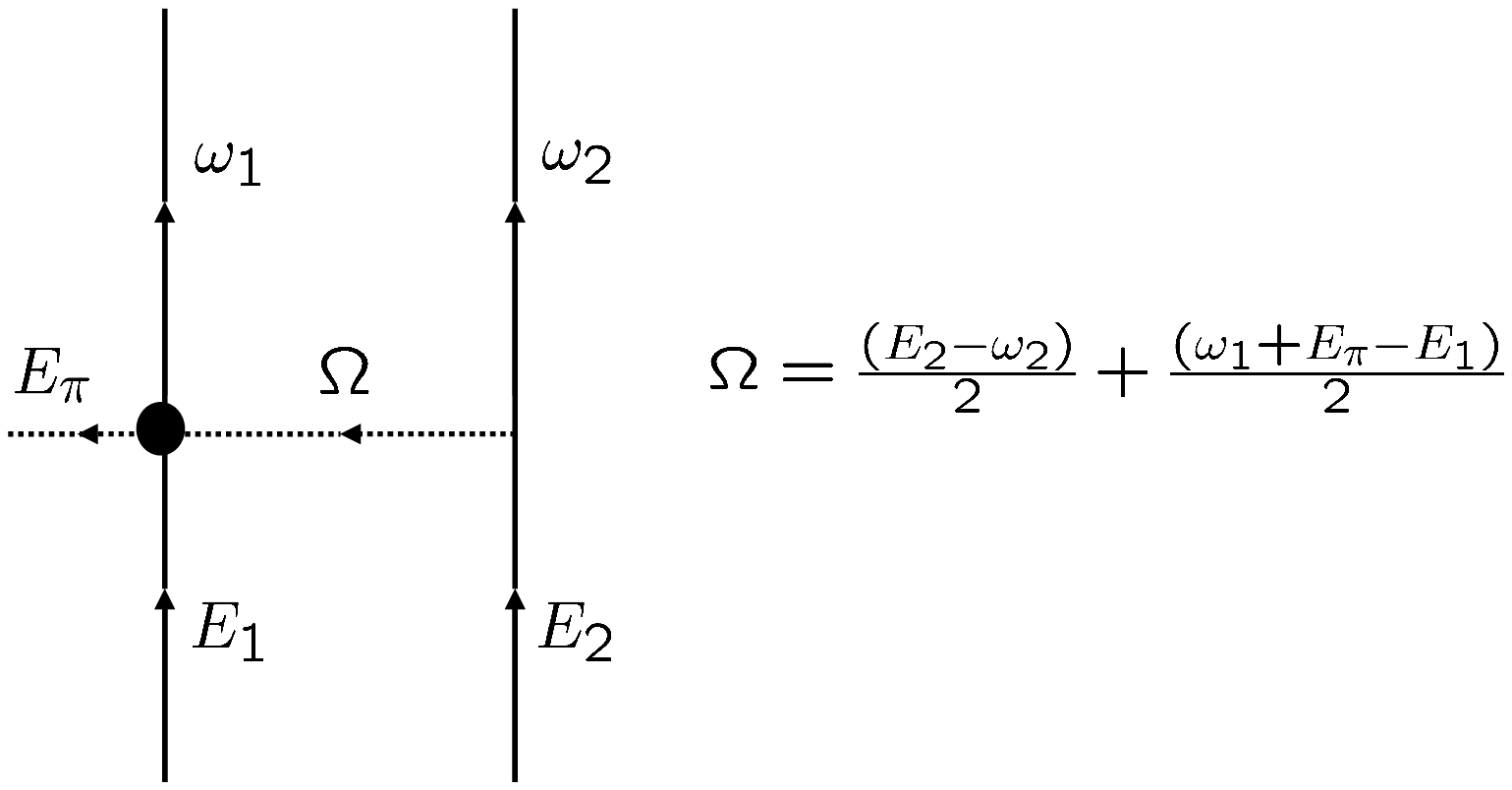}
 \captions{Illustration of the S-matrix prescription.
\label{EconvSm}}
\end{center}
\end{figure}

\sprg For the pion re-scattering diagram in the FSI case
(\eq{mfsi}), according to the S-matrix approach one assumes
energy conservation at each vertex. Then the exchanged pion is no
longer on-mass-shell and we have to replace $f(\omega_\pi)
\rightarrow f(E_2-\omega_2)$ and $f(-\omega_\pi) \rightarrow
f(E_2-\omega_2)$: in the first time-ordered diagram the virtual
pion is entering the re-scattering vertex and in the second one it
is emitted from this vertex (as defined on \fig{diagrams2}).
Therefore, one obtains
\begin{equation}
\hat{O}_{rs} \rightarrow \hat{O}_{rs}^{S} = \frac{f
\left(E_{2}-\omega_{2} \right)}{\left(E_{2}-\omega_{2}
\right)^{2}-\omega_{\pi}^{2}}, \label{mfsion}
\end{equation}
which is of the S-matrix form\cite{Malafaia:2004cu}. The nucleon
energies before and after pion emission are $E_{2}$ and
$\omega_{2}$, respectively (see \fig{EconvSm}).

For three-momentum values corresponding to the on-mass-shell
conditions of the nucleons and the emitted pion, the
characterisation (\ref{mfsion}) of the S-matrix technique
suffices. However, whenever the operators are used in convolution
integrals for the nucleon states distortion, the momentum variable
is free. The on-mass shell condition has then to be reinforced
explicitly by demanding the conservation of energy
$\omega_\pi=E_2-\omega_2=-\left(E_1-\omega_1-E_\pi \right)$. This
is implemented by setting additionally
\begin{equation}
\Omega=\frac{\left(E_2-\omega_2\right)}{2}+\frac{\left(\omega_1+E_\pi-E_{1}
\right)}{2}, \label{Smatpresc}
\end{equation}
as it is illustrated on \fig{EconvSm}.
The general form of the S-matrix
prescription\cite{Malafaia:2004cu} for the pion re-scattering
operator becomes henceforth
\begin{equation}
 \hat{O}_{rs}^{S}=
 \frac{f(\Omega)}{\Omega^2-\left( m_\pi^2+ \vec{q}^{\, \prime 2}\right)
 }.
\label{msmat}
\end{equation}

Note that the S-matrix approach defines a \textit{single}
effective operator (\ref{msmat}) both for the FSI and ISI
diagrams. In \eq{msmat}, the energy of the exchanged pion is
determined by applying consistently energy conservation in
\textit{both} vertices of each diagram and guaranteeing that
intermediate particles are on-mass shell. The prescription
(\ref{Smatpresc}) is also consistent with the most realistic $NN$
potentials (see \eq{Smatprescpot}).

\section{Energy prescriptions for the exchanged pion}
%
\sprg
The \textit{on-shell approximation} (also labelled the ``$E-E'$''
approximation\cite{Hanhart:2000wf}) as introduced in
\rfs{{Sato:1997ps},{Hanhart:2000wf},{Motzke:2002fn}} actually
coincides with the approximation defined by \eq{mfsion}.
The re-scattering operator of \eq{mfsion} can indeed be obtained
directly from \eq{vtil} and \eq{gpi} by the substitutions
following from the on-energy-shell prescription and the energy
conservation in individual vertices
$\Omega=E_2-\omega_2=-\left(E_1-\omega_1-E_\pi \right)$ as
follows:
\begin{equation}
 \hat{O}_{rs} \rightarrow -\frac{1}{2 \left(E_2-\omega_2 \right)}
 \frac{-2 \left(E_2-\omega_2 \right) \times  f\left(E_2-\omega_2 \right) +
 0 \times f \left(\omega_2-E_2 \right)}{\left(E_2-\omega_2-\omega_\pi \right)
 \left( \omega_2 -E_2-\omega_\pi \right)}=\hat{O}_{rs}^{on} \, .
\label{vfullon}
\end{equation}
We note that this form of $\hat{O}_{rs}$ only coincides with
\eq{msmat} for on-mass shell nucleons, which is not the case in
intermediate states, where the momentum integration variable does
not in general satisfy the condition of \eq{Smatpresc}.

%
In the work of \rf{Malafaia:2003wx} the extra kinematic factors in
\eq{vtil} multiplying the function $f\left(\omega_\pi \right)$
were interpreted as form factors and kept unaltered, i.e., the
substitution $f(\omega_\pi) \rightarrow
f(E_2-\omega_2),f(-\omega_\pi) \rightarrow f(E_2-\omega_2)$ was
made only in $G_\pi$ and $f(\omega_\pi)$ of \eq{mfsi}, not in the
kinematic factors included in the function $\tilde{f}$.

In the following we will analyse in detail the S-matrix approach
as well as the effects of other approximations frequently used for
the energy of the exchanged pion, namely the {\em on-shell}, {\em
fixed threshold-kinematics} and {\em static} approximations.
The fixed kinematics approximation, besides considering
on-mass-shell particles and energy conservation at the vertices,
further assumes that the energy of the emitted pion is $m_{\pi}$,
its threshold value ($ \Omega \rightarrow m_{\pi}/2$ in
\eq{msmat}). On the other hand, the static approximation assumes
that no energy is exchanged ($ \Omega \rightarrow 0$ in
\eq{msmat}).


\tb{sumupapxopi} lists the different approximations considered
here and in other works for the full re-scattering operator. The
corresponding approximations for the pion propagator $G_{\pi}$ and
the re-scattering vertex $\tilde{f}$ are in \tb{sumupapxgpi} and
\tb{sumupapxfpi}, respectively.

\begin{table}
\begin{center}
\begin{tabular}
[c]{l|cc}\hline\hline
$\hat{O}_{rs}$ & FSI & ISI\\
&  & \\\hline
&  & \\
TOPT & $-\frac{1}{2\omega_{\pi}}\left[  \frac{f\left(
\omega_{\pi}\right)
}{E_{2}-\omega_{2}-\omega_{\pi}}+\frac{f\left(  -\omega_{\pi}\right)  }%
{E_{1}-\omega_{1}-E_{\pi}-\omega_{\pi}}\right]  $ & $-\frac{1}{2\omega_{\pi}%
}\left[  \frac{f\left(  \omega_{\pi}\right)  }{F_{2}-\omega_{2}-\omega_{\pi}%
}+\frac{f\left(  -\omega_{\pi}\right)
}{F_{1}+E_{\pi}-\omega_{1}-\omega
_{\pi}}\right]  $\\
&  & \\
on-shell & $\frac{f\left(  E_{2}-\omega_{2}\right)  }{\left(  E_{2}%
-\omega_{2}\right)  ^{2}-\omega_{\pi}^{2}}$ & $\frac{f\left(
\omega
_{2}-F_{2}\right)  }{\left(  \omega_{2}-F_{2}\right)  ^{2}-\omega_{\pi}^{2}}%
$\\
&  & \\
fixed-kin. & $\frac{f\left(  \frac{m_{\pi}}{2}\right)  }{\left(
\frac
{m_{\pi}}{2}\right)  ^{2}-\omega_{\pi}^{2}}$ & $\frac{f\left(  \frac{m_{\pi}%
}{2}\right)  }{\left(  \frac{m_{\pi}}{2}\right)  ^{2}-\omega_{\pi}^{2}}$\\
&  & \\
static & $-\frac{f\left(  0\right)  }{\omega_{\pi}^{2}}$ &
$-\frac{f\left(
0\right)  }{\omega_{\pi}^{2}}$\\
&  & \\
S-matrix & $\dfrac{f\left(  \frac{\left(  E_{2}-\omega_{2}\right)
+\left( \omega_{1}+E_{\pi}-E_{1}\right)  }{2}\right) }{\left[
\frac{\left(
E_{2}-\omega_{2}\right)  +\left(  \omega_{1}+E_{\pi}-E_{1}\right)  }%
{2}\right]  ^{2}-\omega_{\pi}^{2}}$ & $\dfrac{f \left(\frac{
\left(\omega_{2}-F_{2}\right)+\left(
F_{1}+E_{\pi}-\omega_{1}\right)  }%
{2}\right)  }{\left[  \frac{\left( \omega_{1}-E_{\pi}-F_{1}\right)
+\left(
F_{2}-\omega_{2}\right)  }{2}\right]  ^{2}-\omega_{\pi}^{2}}$\\
&  & \\\hline\hline
\end{tabular}
  \captions[Prescriptions frequently used for the energy of the
   exchanged pion in the full re-scattering operator]
   {Energy prescriptions for the exchanged
   pion in the full re-scattering operator. The first line is the
   reference result coming from TOPT of \eq{mfsi} and
   \eq{misi}. The second, third and fourth lines are for the
   frequently used prescriptions for the energy of the exchanged pion
   (on-shell\cite{{Pena:1999hq},{Sato:1997ps},{Hanhart:2000wf},{Motzke:2002fn}},
   fixed-kinematics\cite{{Park:1995ku},{Adam:1997pe},{vanKolck:1996dp},{Cohen:1995cc}}
   and static approximation,
   respectively). The last line presents the explicit expressions for
   the full production operator in the S-matrix approach of
   \eq{msmat}.\\
   Note that the S-matrix approach defines a single effective
   operator: although they may look different because the labels for the energies
   in the FSI and ISI cases are different,
   the operators in the last line have the same dependence on the
   on-shell energies.
   \label{sumupapxopi}}
\end{center}
\end{table}
\begin{table}
\begin{center}
\begin{tabular}
[c]{l|cc}\hline\hline
$G_{\pi}$ & FSI & ISI\\
&  & \\\hline
&  & \\
TOPT & $\dfrac{1}{\left(
\omega_{1}+E_{\pi}-E_{1}+\omega_{\pi}\right) \left(
E_{2}-\omega_{2}-\omega_{\pi}\right)  }$ & $\dfrac{1}{\left(
\omega_{1}-E_{\pi}-F_{1}+\omega_{\pi}\right)  \left(  F_{2}-\omega_{2}%
-\omega_{\pi}\right)  }$\\
&  & \\
on-shell & $\dfrac{1}{\left(  E_{2}-\omega_{2}\right)
^{2}-\omega_{\pi}^{2}}$
& $\frac{1}{\left(  \omega_{2}-F_{2}\right)  ^{2}-\omega_{\pi}^{2}}$\\
&  & \\
fixed-kin. & $\dfrac{1}{\left(  \frac{m_{\pi}}{2}\right)
^{2}-\omega_{\pi
}^{2}}$ & $\dfrac{1}{\left(  \frac{m_{\pi}}{2}\right)  ^{2}-\omega_{\pi}^{2}}%
$\\
&  & \\
static & $-\dfrac{1}{\omega_{\pi}^{2}}$ & $-\dfrac{1}{\omega_{\pi}^{2}}$\\
&  & \\
S-matrix & $\dfrac{1}{\left[  \frac{\left( E_{2}-\omega_{2}\right)
+\left( \omega_{1}+E_{\pi}+E_{1}\right)  }{2}\right]
^{2}-\omega_{\pi}^{2}}$ & $\dfrac{1}{\left[  \frac{\left(
\omega_{1}-E_{\pi}-F_{1}\right)  +\left(
F_{2}-\omega_{2}\right)  }{2}\right]  ^{2}-\omega_{\pi}^{2}}$\\
&  & \\\hline\hline
\end{tabular}
  \captions[Prescriptions frequently used for the energy of the
   exchanged pion in the pion propagator]
   {Frequently used prescriptions
   (on-shell\cite{{Pena:1999hq},{Sato:1997ps},{Hanhart:2000wf},{Motzke:2002fn}},
    fixed-kinematics\cite{{Park:1995ku},{Adam:1997pe},{vanKolck:1996dp},{Cohen:1995cc}}
     and static approximation) for the
   energy of the exchanged pion in the pion propagator. The first
   line is the reference result coming from TOPT of \eq{gpi}
   and \eq{gpiisi}. The last line presents the explicit
   expressions for $G_{\pi}$ within the S-matrix approach.
\label{sumupapxgpi}}
\end{center}
\end{table}
\begin{table}
\begin{center}
\begin{tabular}
[c]{l|cc}\hline\hline
$\tilde{f}\left(  \omega_{\pi}\right)  $ & FSI & ISI\\
&  & \\\hline
&  & \\
TOPT & $f\left(  \omega_{\pi}\right)  f_{1}^{f}+f\left(
-\omega_{\pi}\right) f_{2}^{f}$ & $f\left(  \omega_{\pi}\right)
f_{1}^{i}+f\left(  -\omega_{\pi
}\right)f_{2}^{i}  $\\
&  & \\
on-shell & $f\left(  E_{2}-\omega_{2}\right)  f_{1}^{f}+f\left(
\omega
_{2}-E_{2}\right)  f_{2}^{f}$ & $f\left(  \omega_{2}-F_{2}\right)  f_{1}%
^{i}+f\left(  F_{2}-\omega_{2}\right)  f_{2}^{i}$\\
&  & \\
fixed-kin. & $f\left(  \frac{m_{\pi}}{2}\right)  f_{1}^{f}+f\left(
-\frac{m_{\pi}}{2}\right)  f_{2}^{f}$ & $f\left(
\frac{m_{\pi}}{2}\right)
f_{1}^{i}+f\left(  -\frac{m_{\pi}}{2}\right)  f_{2}^{i}$\\
&  & \\
static & $f\left(  0\right)  f_{1}^{f}+f\left(  0\right)
f_{2}^{f}$ &
$f\left(  0\right)  f_{1}^{i}+f\left(  0\right)  f_{2}^{i}$\\
&  & \\\hline\hline
\end{tabular}%
  \captions[Prescriptions frequently used for the energy of the
   exchanged pion in the pion re-scattering vertex alone]
   {The same of \tb{sumupapxgpi} but for the re-scattering vertex $\tilde{f}$ alone. The
   first line is the reference result of \eq{vtil} and
   \eq{vtilisi}. The functions $f^{f}_{k}$ and $f^{i}_{k}$
    $\left(k=1,2 \right)$ refer to the multiplicative
    factors mentioned in \sec{Extrprop}.   \label{sumupapxfpi}}
\end{center}
\end{table}
\newpage

As in \chp{FTtoDWBA}, for numerical calculations we considered the
$NN \rightarrow(NN)\pi$ transition in partial waves $^3P_0
\rightarrow (^1S_0)s$ for the $pp \rightarrow pp\pi^{0}$ reaction.
The amplitudes and cross sections are evaluated both with the
simple interaction $V_\sigma$ of \eq{vsigfsi} and \eq{vsigisi} and
with the Bonn B potential\cite{Machleidt:1987hj}.

We tested the S-matrix prescription for the re-scattering operator
(\ref{msmat}) and the approximations discussed in the previous
section (see \tb{sumupapxopi}). We include also the results for
the on-shell, fixed threshold-kinematics, static and the S-matrix
approximations for the effective pion propagator $G_\pi$, as
listed in \tb{sumupapxgpi}.

\begin{figure}
\begin{center}
\includegraphics[width=.98\textwidth,keepaspectratio]{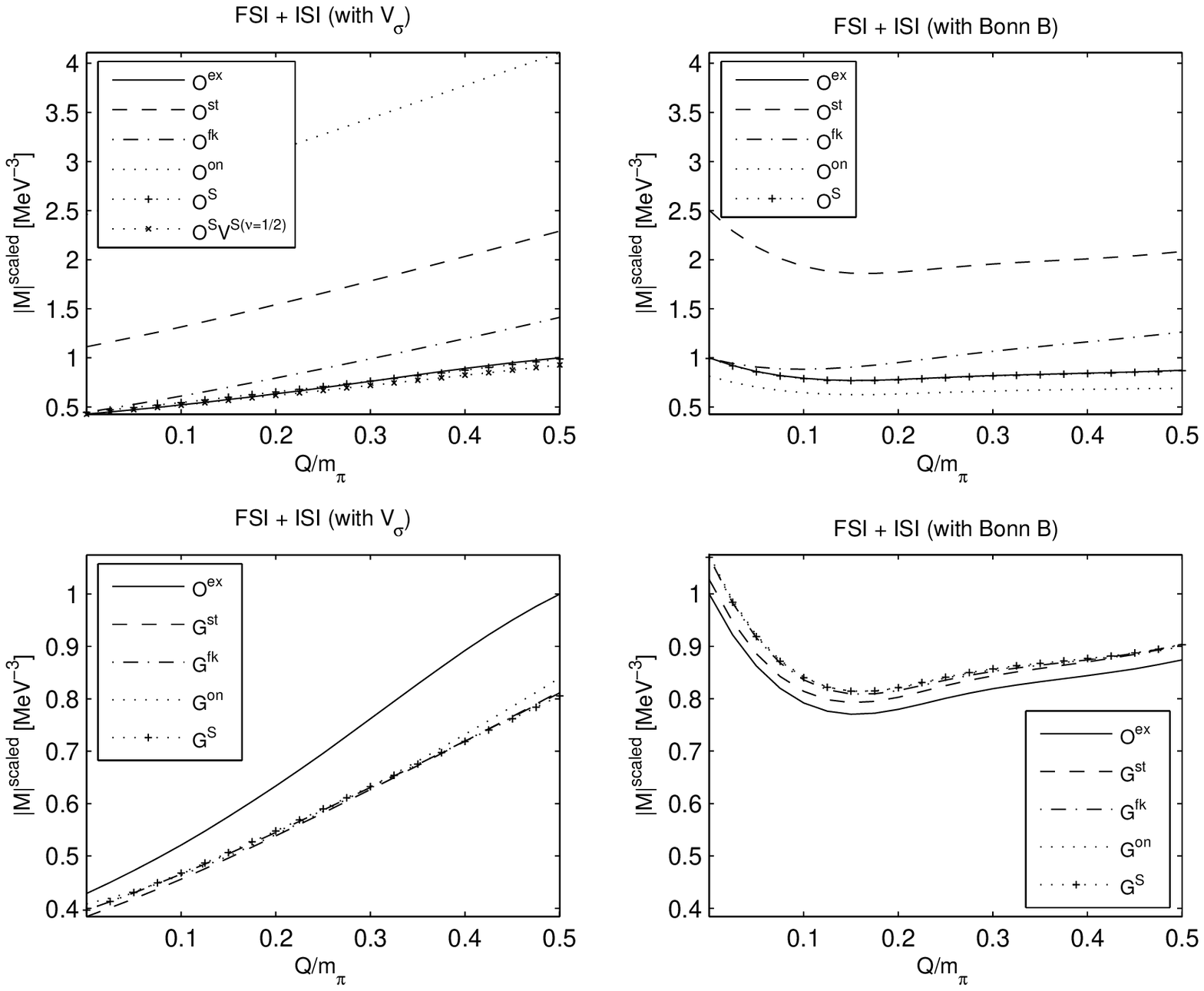}
  \captions[Absolute values of the FSI + ISI amplitude as a function
   of the excess energy $Q$ for different energy prescriptions in the
   full re-scattering operator and in the pion propagator]
   {Absolute values of the FSI + ISI
   amplitude as a function of the excess energy $Q=2E-2M-E_\pi$ (in
   units of $m_\pi$). The right(left) panels correspond to the
   amplitudes with $\sigma$-exchange(Bonn B potential) for the $NN$
   interaction. The amplitudes are taken at the maximum pion momentum
   $q_\pi^{max}$, determined by $Q$. The upper panels correspond to
   approximations for the whole operator $ \hat{O}_{rs}$; the lower
   panels to approximations for the pion propagator $G_\pi$ only. The
   solid line $(O^{ex})$ is the reference calculation. The dashed,
   dashed-dotted, dotted lines and dotted-line with +'s
   correspond to the static, fixed threshold-kinematics and on-shell
   approximations and the S-matrix approach, respectively. The
   corresponding operators are $O^{st}$, $O^{fk}$, $O^{on}$ and
   $O^{S}$, and $G^{st}$, $G^{fk}$, $G^{on}$ and $G^{S}$. The dotted
   line with x's corresponds to both the pion production operator
   and $\sigma$-exchange potential in the S-matrix approach. All
   amplitudes were normalised by a factor defined by the maximum
   value of the reference result. } \label{zmboth}
\end{center}
\end{figure}

\begin{figure}
\begin{center}
\includegraphics[width=.98\textwidth,keepaspectratio]{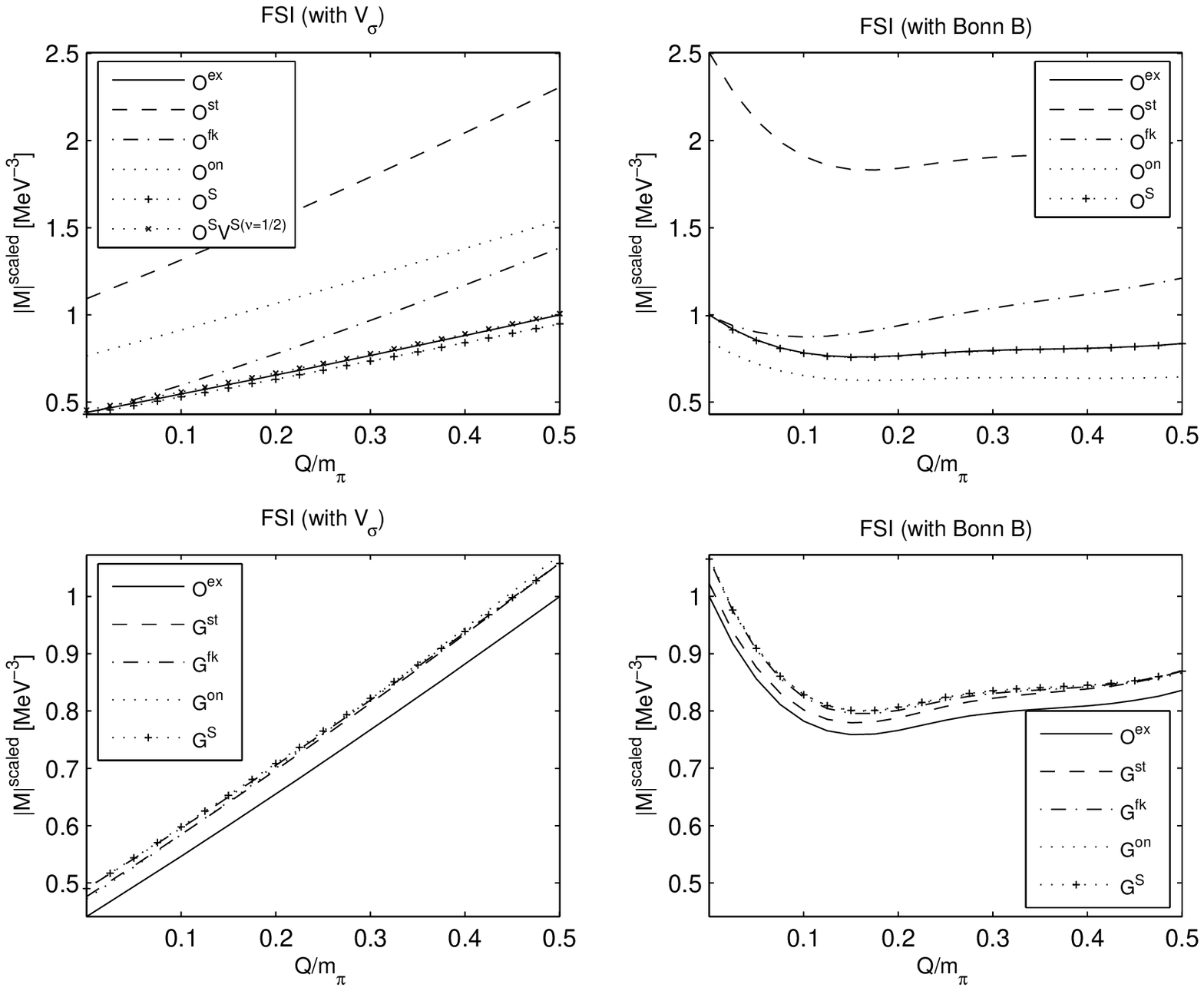}
  \captions[Absolute values of the FSI amplitude as a function of
   the excess energy $Q$ for different energy prescriptions in the
   full re-scattering operator and in the pion propagator]
   {The same of \fig{zmboth} but for the
   FSI amplitude. } \label{zmfsi}
\end{center}
\end{figure}

\begin{figure}
\begin{center}
\includegraphics[width=.98\textwidth,keepaspectratio]{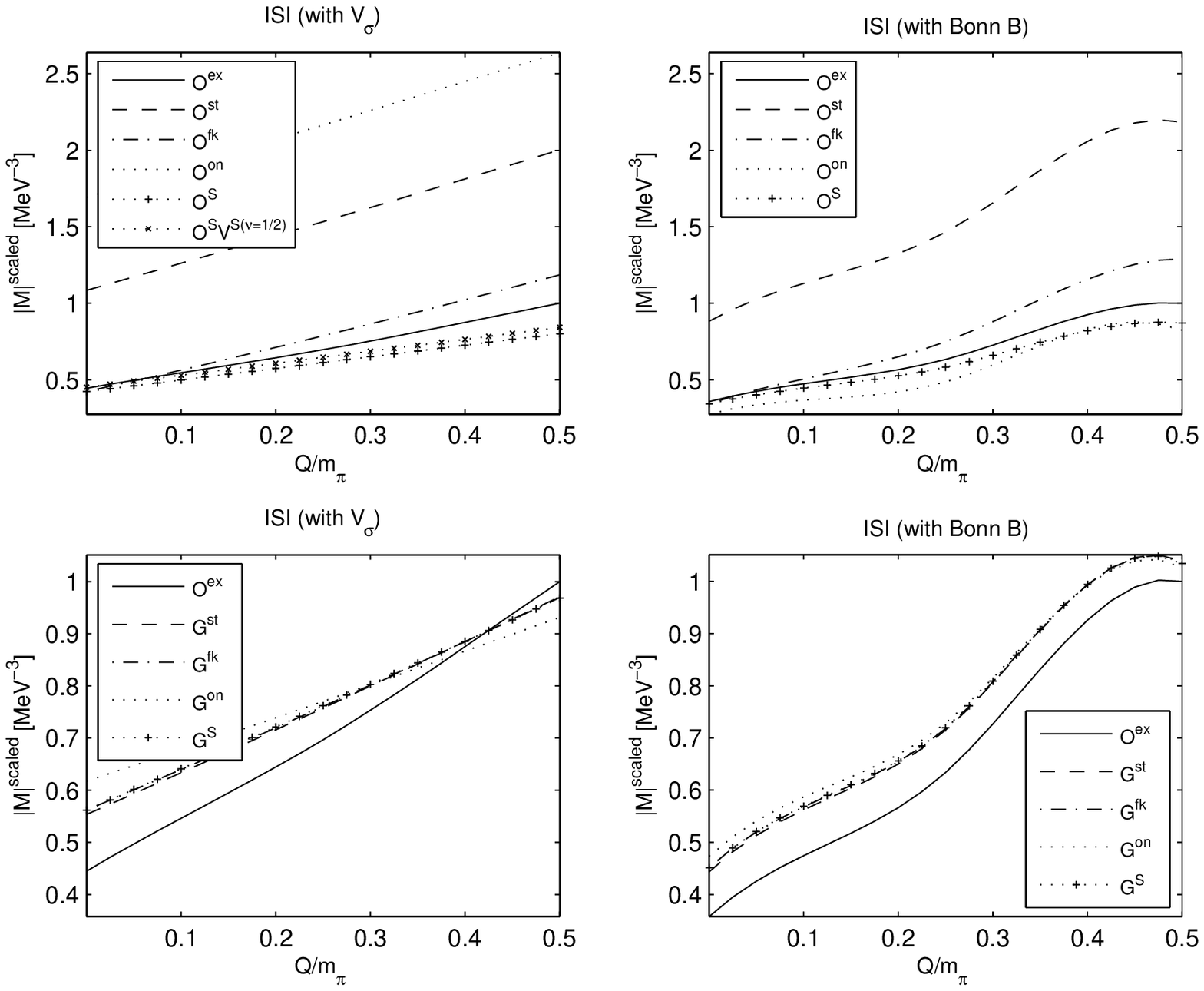}
  \captions[Absolute values of the ISI amplitude as a function of
   the excess energy $Q$ for different energy prescriptions in the
   full re-scattering operator and in the pion propagator]
   {The same of \fig{zmboth} but for the
   ISI amplitude. } \label{zmisi}
\end{center}
\end{figure}

In \fig{zmboth} we show that the amplitudes with the S-matrix
operator $O^S$ (dotted line with +'s on the upper panels) are the
closest to the reference result (solid line). Using the same
approach both for the operator and for $V_\sigma$ increases
slightly the gap from the reference result (line with x's versus
solid line on the upper left panel). The fixed
threshold-kinematics version of $ \hat{O}_{rs}$, denoted as
$\hat{O}^{fk}$, works well for small values of the excess energy
$Q$, but starts to deviate rapidly with increasing $Q$
(dashed-dotted line in the upper panels of \fig{zmboth}). The
on-shell approximation for the operator (dotted-line in the upper
panels), although deviates largely from the reference result when
the $NN$ interaction is described through $\sigma$-exchange, is
close to the reference result for the Bonn B potential (right
panel)\footnote{A more detailed analysis of these differences can
be found in \sec{ssexpg}.}. The static approximation for the
re-scattering operator ($\hat{O}^{st}$) overestimates
significantly the amplitude (dashed versus solid lines in the
upper panels). The same conclusions hold individually for the FSI
and ISI amplitudes (\fig{zmfsi} and \fig{zmisi}, respectively).

\begin{figure}
\begin{center}
\includegraphics[width=.98\textwidth,keepaspectratio]{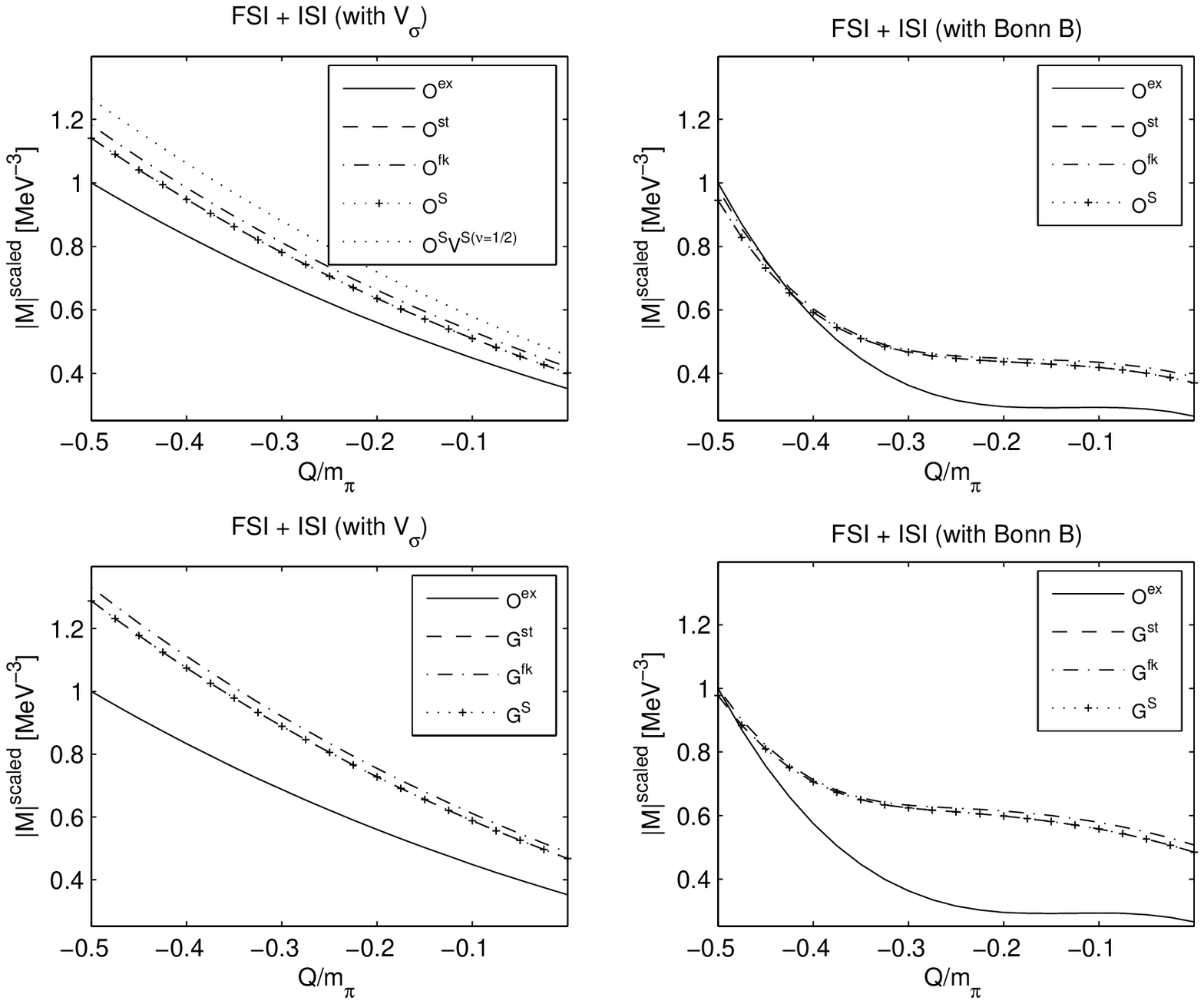}
  \captions[Absolute values of the FSI + ISI amplitude,
    below pion production threshold, as a function
   of the symmetric excess energy $Q$ for different energy
   prescriptions in the full re-scattering operator and in the pion propagator]
    {Absolute values of the FSI + ISI amplitude
   as a function of the symmetric of excess energy $Q$
   (in units of $m_\pi$), below pion production threshold. The
   meaning of the lines is the same of \fig{zmboth}. }
\label{zmbelowboth}
\end{center}
\end{figure}

The investigation of how much of these features survive below pion
production threshold is on \fig{zmbelowboth}. The fixed kinematics
approximation (dashed-dotted line) is now found to disagree the
most with the reference result. The deviation of the static
approximation (dashed line) from the reference result (solid line)
is much smaller when compared to the situation above threshold of
\fig{zmboth}. This justifies the traditional approximation of
static exchange below threshold. As before, the several
approximations for the pion propagator yield results very close to
each other (lower panels of \fig{zmbelowboth}). Also, the S-matrix
approach (dotted line with +'s in the upper panels) describes
quite satisfactorily the reference amplitude.
\clearpage
\subsection{Expansion of the effective pion propagator \label{ssexpg}}

\sprg
The lower panels on \fig{zmboth} show that all considered
approximations taken only for the effective pion propagator do not
differ much from each other, as already found on
\rf{Malafaia:2003wx}. It means that the choices for the energy of
the exchanged pion in the effective propagator alone are not very
decisive (solid line versus dotted, dashed, dashed-dotted and
dotted line with x's). We notice however that for
$\sigma$-exchange there is a considerable deviation
of all these approximations from the reference result.

In order to understand this we considered the expansion of the
effective pion propagator $G_\pi$ in Eq.~$\left(\ref{gpi}\right)$
in terms of an ``off-mass-shell" dimensionless parameter $y$:
\begin{equation}
y=-\frac{2 E-E_\pi-\omega_1-\omega_2}{\omega_1-\omega_2+E_\pi} \,
,
\end{equation}
which measures the deviation of the total energy from the energy
of the intermediate state with all three particles
on-mass-shell\cite{Malafaia:2004cu}.
As before, $E=E_{1}=E_{2}$.
This Taylor series expansion can give an insight on the small
effect of retardation effects in the propagator, and it reads
\begin{equation}
{G_\pi}=\underbrace{\frac{1}{\left(\frac{E_\pi+\omega_1-\omega_2}{2}\right)^2-
\omega_\pi^2}}_{G^{(1)}_{Tay}}
\left[1+\frac{\left(-2E+E_\pi+\omega_1+\omega_2\right)}
{\left(\frac{E_\pi+\omega_1-\omega_2}{2}\right)^2-\omega_\pi^2}
+...\right] \label{gexpfsi}
\end{equation}
where $G^{(1)}_{Tay}$ has the form of the usual Klein-Gordon
propagator.

We notice here that in the case of the ISI amplitude, the
representation of the pion propagator $G_{\pi}$ by its Taylor
series, the first term of which is $G^{(1)}_{Tay}$, fails due to
the presence of a pole in the propagator.

\begin{figure}[t!]
\begin{center}
\includegraphics[width=.98\textwidth,keepaspectratio]{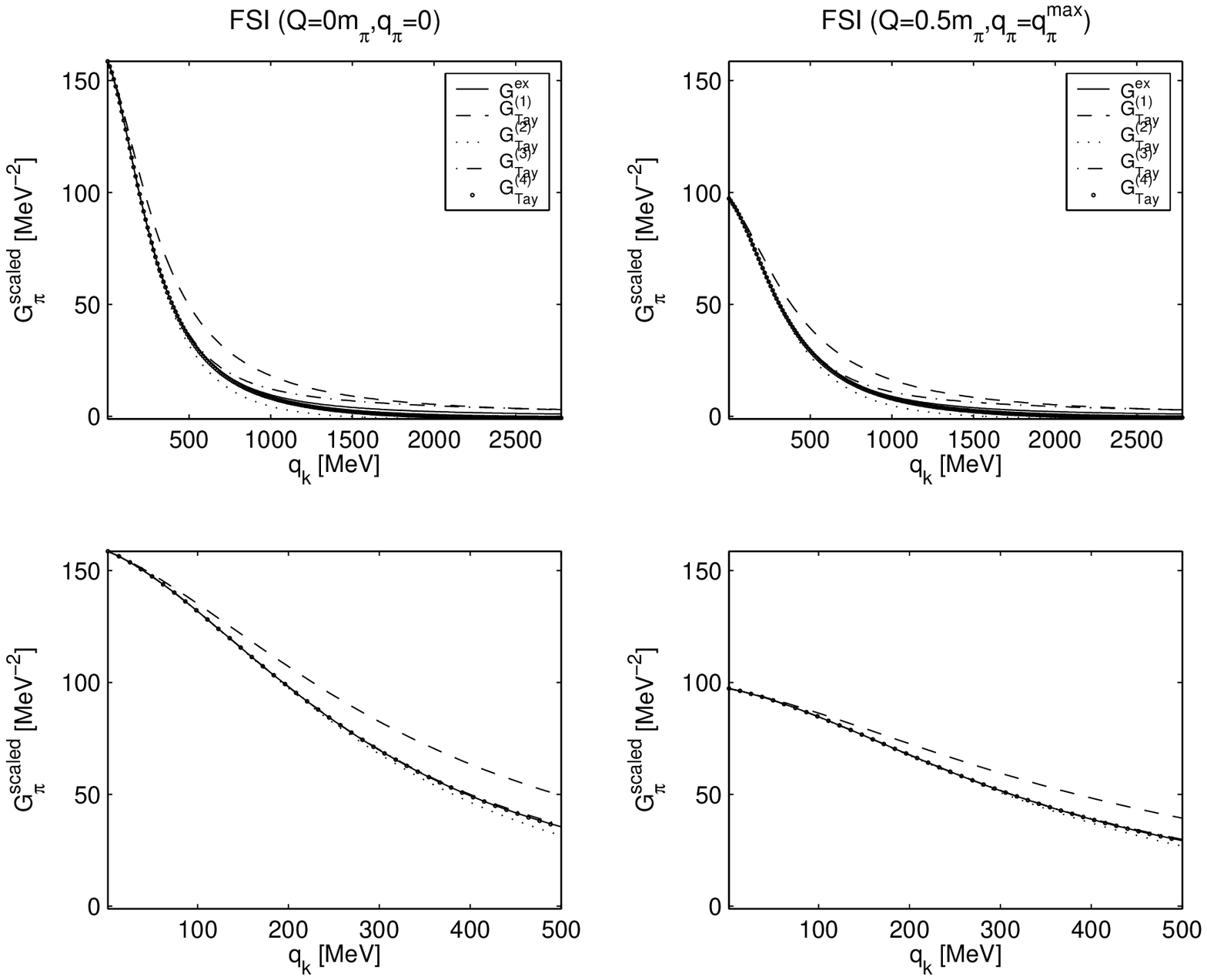}
  \captions[Convergence of the Taylor expansion of the pion
   propagator $G_{\pi}$ in the FSI amplitude, as a function of the
   two-nucleon relative momentum]
   {Convergence of the Taylor expansion (\ref{gexpfsi})
   of the pion propagator $G_{\pi}$ in the FSI amplitude
   as a function of the two-nucleon relative
   momentum. The solid line is the full effective propagator of
   \eq{gpi}. The dashed, dotted and dashed-dotted lines correspond to
   the first three terms of the Taylor expansion. The
   bullets represent the fourth term. Left panel: at threshold; right
   panel: above threshold at maximum pion momentum for an excess
   energy $Q=0.5m_\pi$. The bottom panels zoom into the region of
   low relative momentum ($q_{k}<500 \umev$).} \label{gexpansion}
\end{center}
\end{figure}

Figure~\ref{gexpansion} compares the first four terms
$G_{Tay}^{(i)} (i=1,...,4)$ of this expansion (dashed, dotted and
dashed-dotted line and bullets, respectively) with the full
effective propagator in \eq{gpi} (solid line), as a function of
the two-nucleon relative momentum $q_k$, for two different values
of the excess energy $Q$. The convergence of the series demands at
least 4 terms. Besides, as expected, this convergence is
momentum-dependent.

\begin{figure}[t!]
\begin{center}
\includegraphics[width=.98\textwidth,keepaspectratio]{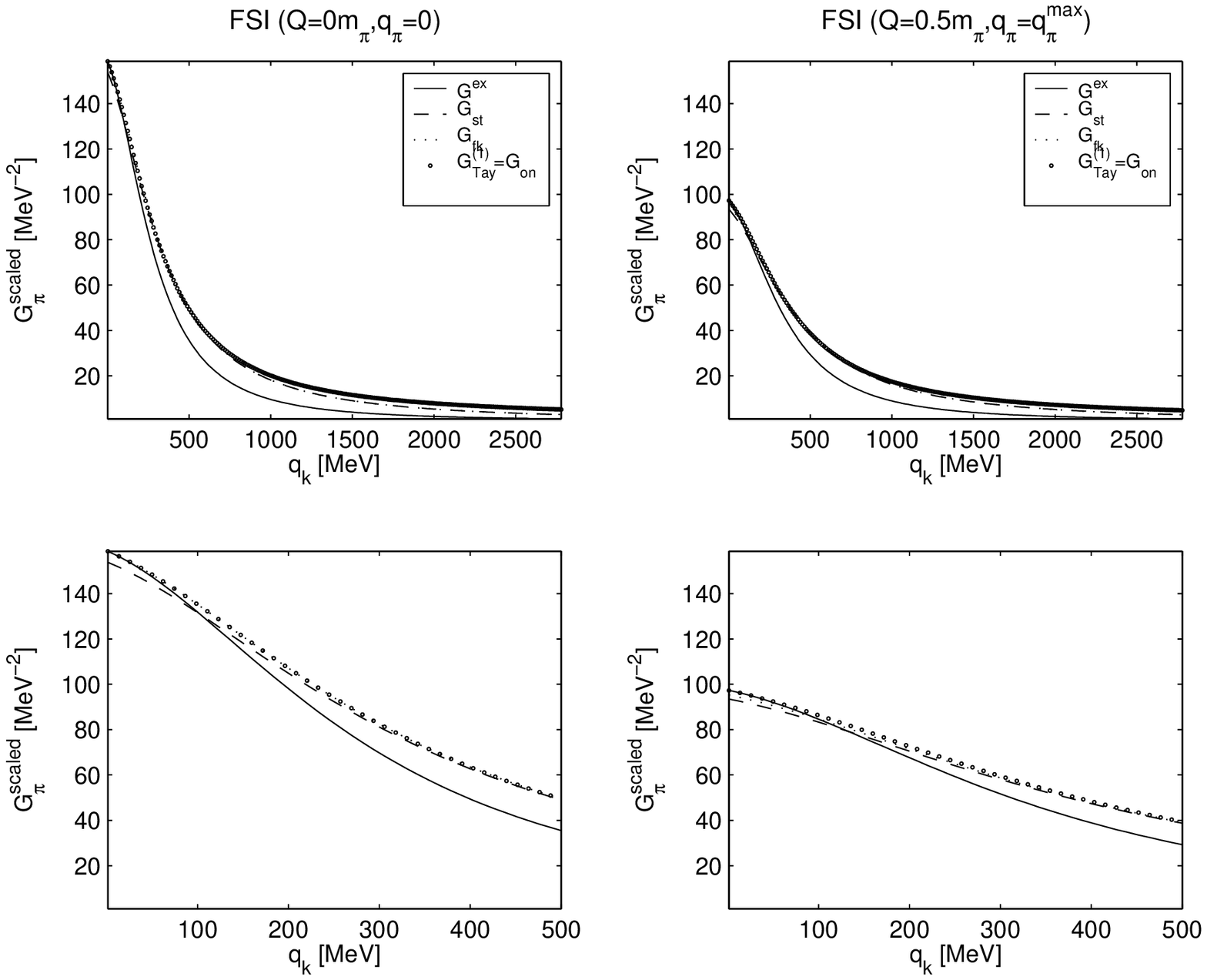}
  \captions[Comparison of the approximations for $G_{\pi}$ to
   the first term of the
   Taylor series (FSI case)]{Approximations for $G_{\pi}$ in the FSI
   diagram and the first term of the Taylor series. Left and right
   panels have the same meaning as on \fig{gexpansion}.}
\label{gapprox}
\end{center}
\end{figure}

We have also compared the first term of this expansion with the
already considered on-shell, fixed threshold-kinematics and static
approximations for the pion propagator. These results are shown in
\fig{gapprox}. We realise that all these approximations are very
near to the 1st order term of the Taylor series. The corrections
arising from higher order terms in the expansion are negligible
only for low momentum transfer, more precisely in the range $q_k<
100\umev$.

The deviations of $G^{st}$, $G^{fk}$ and $G^{on}$ from the
effective propagator $G_\pi$ given by  \eq{gpi} cannot explain the
relatively large deviations obtained on the bottom-left panel of
\fig{zmboth} between considered approximations and the reference
result. These deviations follow from the ISI contribution.

\begin{figure}
\begin{center}
\includegraphics[width=.64\textwidth,keepaspectratio]{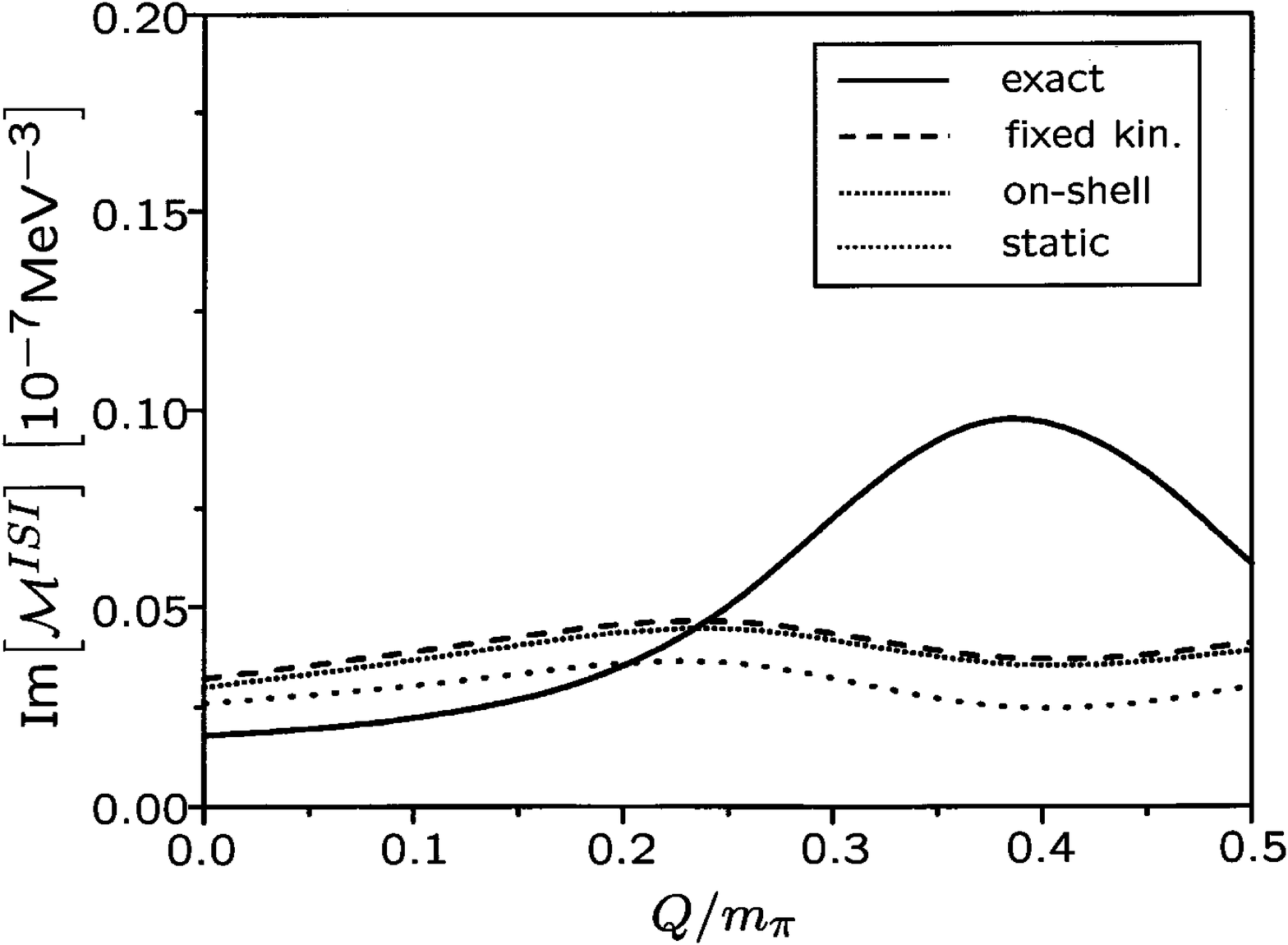}
  \captions[Imaginary part of the ISI amplitude $\mathcal{M}_{DWBA}$
   as a function of the excess energy $Q$]{Imaginary part
   of the ISI amplitude $\mathcal{M}_{DWBA}$ of \eq{aisi} as a
   function of the excess energy. The exact result
   (full line) and {\it fixed kinematics} (dashed line), {\it
   on-shell} (short-dashed line) and {\it static} (dotted line)
   approximations for the pion propagator are shown.} \label{imagisi}
\end{center}
\end{figure}

When we compare the results for the final state interaction (lower
panels of \fig{zmfsi}) with the results obtained for the initial
state interaction (lower panels of \fig{zmisi}), we recognise
however that the deviation between the approximate results and the
exact one is much more significant for the ISI amplitude. The
on-shell $G^{on}$ approximation (and also the fixed kinematics
approximation) gives a larger amplitude. Its effect is much more
pronounced than it is for the final-state distortion. This
difference can be understood since the initial state interaction
induces high off-shell energies in the intermediate nucleons. This
enlarges the gap between the $G^{on}$ or $G^{fk}$ and  the
$G^{exact}$ calculations, compared to what happens for the FSI
case, where the nucleons emit the outcoming pion before their
interaction happens. Also, what is behind the approximations
$G^{on}$, $G^{fk}$ or $G^{st}$ being worse representations of the
exact amplitude for the ISI case than for the FSI case,  is the
physics related to the logarithmic singularities of the exact pion
propagator. These singularities are present only for the ISI
amplitude, but are absent in the case of the approximate forms for
that propagator. Their effect is most important for the imaginary
part of the amplitude, which is shown in \fig{imagisi}.

Also, the weight of the ISI term depends on the $NN$ interaction
employed. It is comparable to the FSI term for $V_\sigma$ (for
which the deviations are large, as seen on the bottom-left panel
of \fig{zmboth}), but it is less important for the full Bonn B
potential (and therefore the corresponding deviations on the
bottom-right panel of \fig{zmboth} are indeed much smaller).

\begin{figure}[t!]
\begin{center}
\includegraphics[width=.98\textwidth,keepaspectratio]{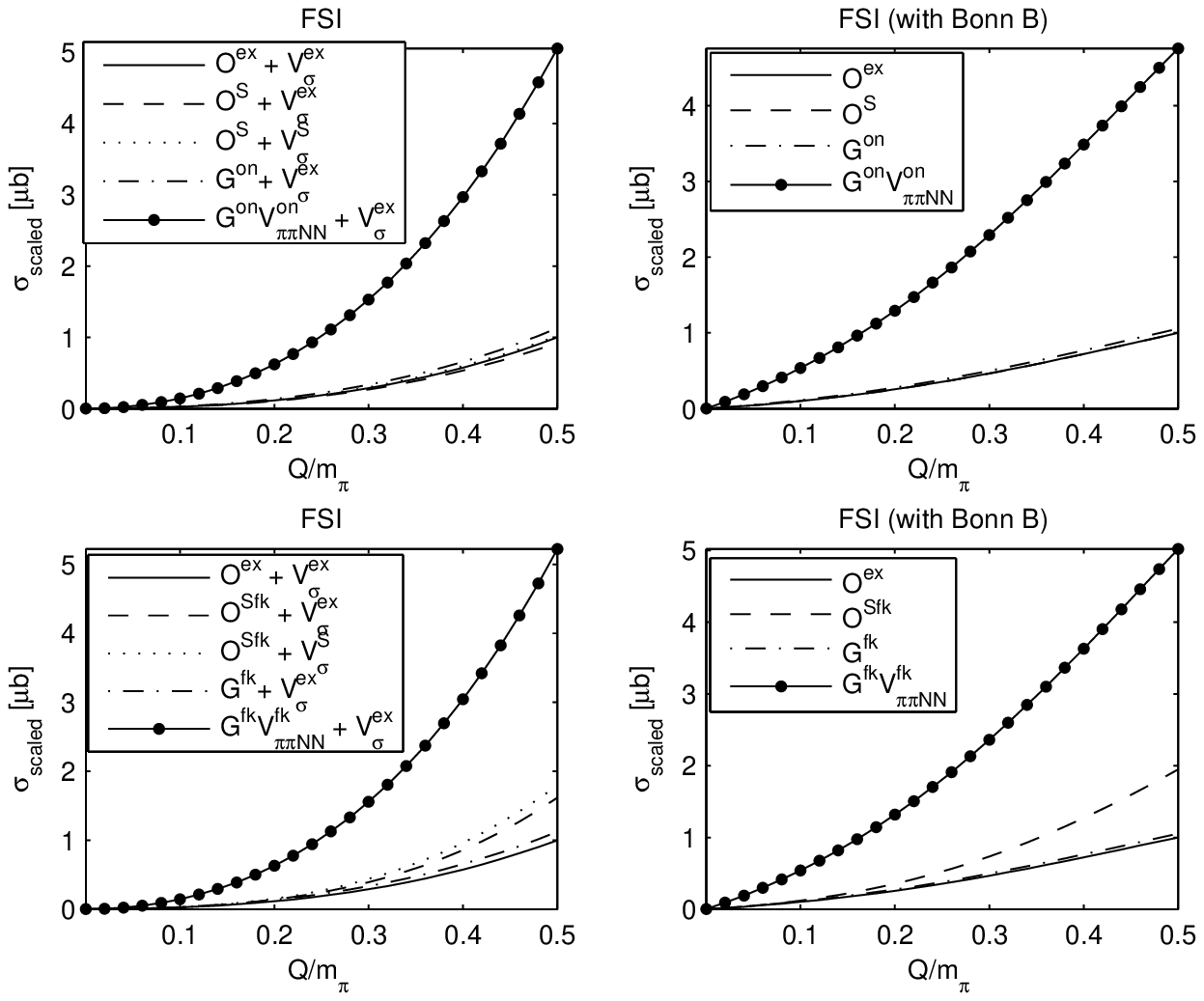}
  \captions[Effects of the approximations for the re-scattering
   operator $\hat{O}_{rs}$ and for the effective pion propagator
   $G_\pi$ as a function of the excess energy $Q$ (FSI only)]{Effects
   of the approximations for the re-scattering operator
   $\hat{O}_{rs}$ and for the effective pion propagator $G_\pi$ as a
   function of the excess energy $Q$. The cross section curves shown
   correspond to the FSI amplitude alone. The upper panels correspond
   to the re-scattering operator $\hat{O}_{rs}$ given by Eq.
   $\left(\ref{msmat} \right)$ and the lower panels to fixed
   threshold-kinematics approximation. The solid line is the
   reference calculation $\left(\ref{afsi} \right)$. The dashed line
   is the S-matrix calculation for the re-scattering operator
   $\hat{O}_{rs}$ given by Eq. $\left(\ref{msmat} \right)$ (upper panels)
   and the fixed threshold-kinematics approximation for
   $\left(\ref{msmat}\right)$ (lower panels). The dotted line
   corresponds to take the S-matrix approximation not only for
   $\hat{O}_{rs}$, but also for the $\sigma$-exchange interaction
   (\ref{vsigos}). The dashed-doted line corresponds to the on-shell
   (upper panels) and  fixed threshold-kinematics (lower panels)
   prescriptions only for $G_\pi$. The solid lines with bullets refer
   to the energy prescriptions taken for the propagator $G_\pi$ and
   for $f(\omega_\pi)$ in \eq{vtil}, as in \cite{Malafaia:2003wx},
   but not for extra kinematic factors in $\tilde{f}$. All the
   cross sections were normalised with a factor defined by the
   maximum value of the reference result. \label{fsi}}
\end{center}
\end{figure}

All the findings for the amplitudes manifest themselves also in
the results for the cross section. We show in \fig{fsi} the
effects of the considered approximations on the cross section,
first taking only the FSI contribution. On the left panel the
amplitude includes $V_\sigma$ for the $NN$ interaction, and on the
right panel the Bonn B T-matrix is used. The curves compare the
reference result (solid line in all panels) with the S-matrix
results (upper panels) and  their fixed threshold-kinematics
version (lower panels). The S-matrix approach (dashed  line) is
the closest to the reference result (upper panels of \fig{fsi}).

For the case of the $NN$ interaction described by $V_\sigma$ we
also show the result following from the S-matrix prescription
applied to the $NN$ interaction (dotted line on left panels in
\fig{fsi}). For the fixed threshold-kinematics versions (bottom
panel) the deviations from the reference result increase
pronouncedly with the excess energy $Q$, as expected. The
approximations for the energy of the exchanged pion taken in the
pion propagator $G_\pi$ and in the re-scattering vertex
$f(\omega_\pi)$, but not in the kinematic factors of \eq{vtil}
(see \tb{sumupapxfpi}), overestimate the cross section by a factor
of 5 (solid line with bullets).

\begin{figure}[t!]
\begin{center}
\includegraphics[width=.98\textwidth,keepaspectratio]{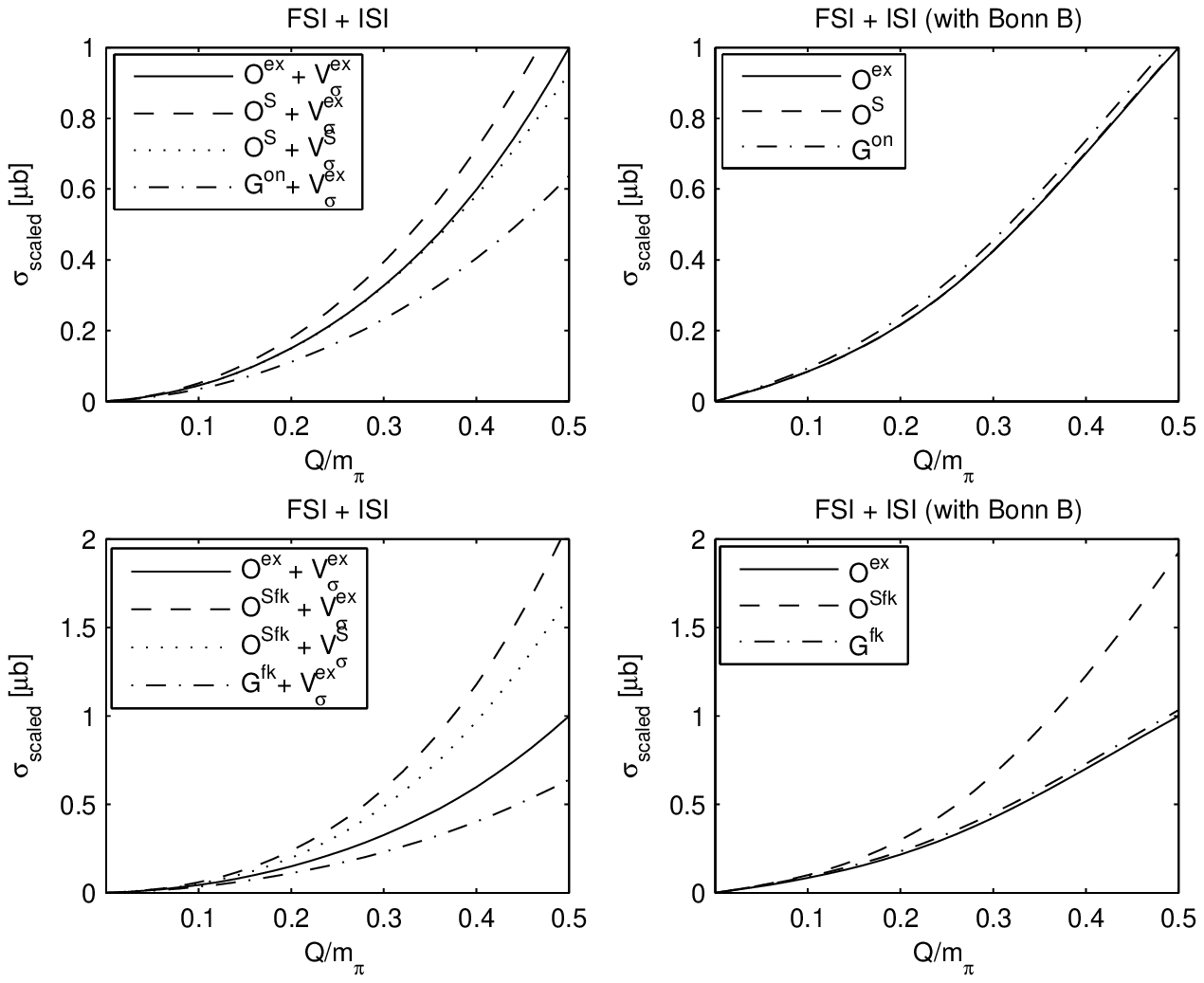}
  \captions[Effects of the approximations for the re-scattering
   operator $\hat{O}_{rs}$ and for the effective pion propagator
   $G_\pi$ as a function of the excess energy $Q$ (FSI + ISI)]{The
   same of Fig. \ref{fsi},  but for the total (FSI+ISI) cross-section
   and considering only the approximations for $\hat{O}$ and
   $G_\pi$.} \label{bothzoom}
\end{center}
\end{figure}

Finally, we present on \fig{bothzoom} the comparison between the
approximated total cross sections with both FSI and ISI included.
The approximation dictated by the S-matrix approach (dashed and
dotted lines on the upper panels of \fig{bothzoom}) is clearly
seen as the best one. For the Bonn potential calculation, it
practically coincides with the reference result. As shown in the
previous section, this procedure amounts to extend the on-shell
approximation, used in \rf{Malafaia:2003wx} for $G_\pi$ and
$f(\omega_\pi)$ alone, also to the multiplicative kinematic
factors showing up in the operator $\tilde{f}$ (see
Eqs.~(\ref{mfsi}-\ref{gpi})).

To conclude we notice, moreover, that for the realistic $NN$
interaction, the difference between the S-matrix approach (upper
right panel on \fig{bothzoom}) and its fixed threshold-kinematics
version (lower right panel of the same figure) is not very
important near threshold, provided the excess energy does not
exceed $\approx 30 \umev$ ($Q/m_\pi \sim 0.2$).
%
%
\section{Conclusions}
%
\sprg
This Chapter discussed the choices for the exchanged pion energy
which are unavoidable in the three-dimensional non-relativistic
formalism underlying DWBA. The main conclusions are:

1) The usual approximations to the effective pion propagator
\cite{{Park:1995ku},{vanKolck:1996dp},{Cohen:1995cc},{Sato:1997ps},{Malafaia:2003wx}},
obtained from a quantum-mechanical reduction of the Feynman
diagram describing the pion re-scattering process, are rather
close to the first order term of a Taylor series in a parameter
measuring off-mass-shell effects in the intermediate states. The
series converges rapidly for the FSI amplitude near threshold. As
a consequence, retardation effects are not decisive in the pion
re-scattering mechanism near the threshold energy for pion
production.

2) As for the pion energy in the $\pi N$  re-scattering amplitude,
the on-shell approach when used only in $f(\omega_\pi)$
overestimates significantly  the reference result.

3) Nevertheless, and this is the key point of this Chapter, this
deviation is dramatically reduced if the approximation coming from
the S-matrix approach is used consistently in the whole effective
operator\cite{Malafaia:2004cu}. This procedure amounts to extend
the on-shell approximation used in \rf{Malafaia:2003wx} for
$G_\pi$ and $f (\omega_\pi)$ to the full operator $\hat{O}_{rs}$,
including kinematic factors which differently weight the two
dominant time-ordered diagrams. The amplitudes and cross sections
obtained with the S-matrix effective operator are very close to
those obtained with the time-ordered one in the considered
kinematic region. Nevertheless, the static approximation works
well below threshold.

\sp

The re-scattering operator for the neutral pion production in the
isoscalar $\pi$N channel indeed seems to be relatively
unimportant: its enhancement reported in previous papers followed
from inconsistent or too crude (static or fixed
threshold-kinematics) treatment of the energy dependence of the
effective operator.  Our findings\cite{Malafaia:2004cu} explain
why the calculation of \rf{{Sato:1997ps}}, where the on-shell
approximation is used, artificially enhances the contribution of
the isoscalar re-scattering term.
On the other hand, our results indicate that the fixed kinematics
choice done in
\rfs{{Park:1995ku},{Adam:1997pe},{vanKolck:1996dp},{Cohen:1995cc}}
for the different production operators considered, is valid in the
restricted kinematic region where $Q \lesssim 0.2 m_{\pi}$.

The re-scattering mechanism is filtered differently by other
spin/isospin channels in pion production reactions. For charged
pion production reactions the general irreducible re-scattering
operator comprises also the dominant isovector Weinberg-Tomozawa
term of the $\pi$N amplitude, and its importance is therefore
enhanced. Next Chapter investigates these reactions within the
S-matrix approach.

%% file: Chapter5.tex
\setcounter{minitocdepth}{2}
\chapter{Charged and neutral pion production reactions} \label{Chargedandneutral}
\minitoc
{\bfseries Abstract:} The S-matrix technique, shown to reproduce
the results of time-ordered perturbation theory for the
re-scattering mechanism in $\pi^{0}$ production, is applied both
to charged and neutral pion production reactions. The
contributions from the direct-production, pion re-scattering,
Z-diagrams and explicit $\Delta$-isobar excitation terms are
considered. High angular momentum partial wave channels are
included and the convergence of the partial wave series is
investigated.
\\

\newpage
%
%
\section{Pion production operators\label{prodoperators}}
%
\sprg
\chp{Smatrixapp}, following the work of \rf{Malafaia:2004cu}, has
shown that the S-matrix approach reproduces the energy dependent
re-scattering operator for neutral pion production derived from
time-ordered diagram techniques.  This conclusion provides the
motivation to apply this approach further to total cross sections
for not only neutral, but also charged pion production. As
discussed before, the S-matrix technique has the advantage to
define a single effective operator, contrarily to what happens
when deriving the quantum-mechanical production operator in the
framework of non-relativistic, time-ordered perturbation theory.

Another usual problem of pion production calculations concerns the
need of a realistic nucleon-nucleon potential valid for the
energies above the pion production threshold, which necessarily
are needed for the initial nucleonic states. In this Chapter,
we apply for the first time an $NN$ interaction including two
boson exchange potentials which describes well not only $NN$
phase-shifts but also inelasticities.
\subsection{The mechanisms and their operators}
\begin{figure}
\begin{center}
\includegraphics[width=.88\textwidth,keepaspectratio]{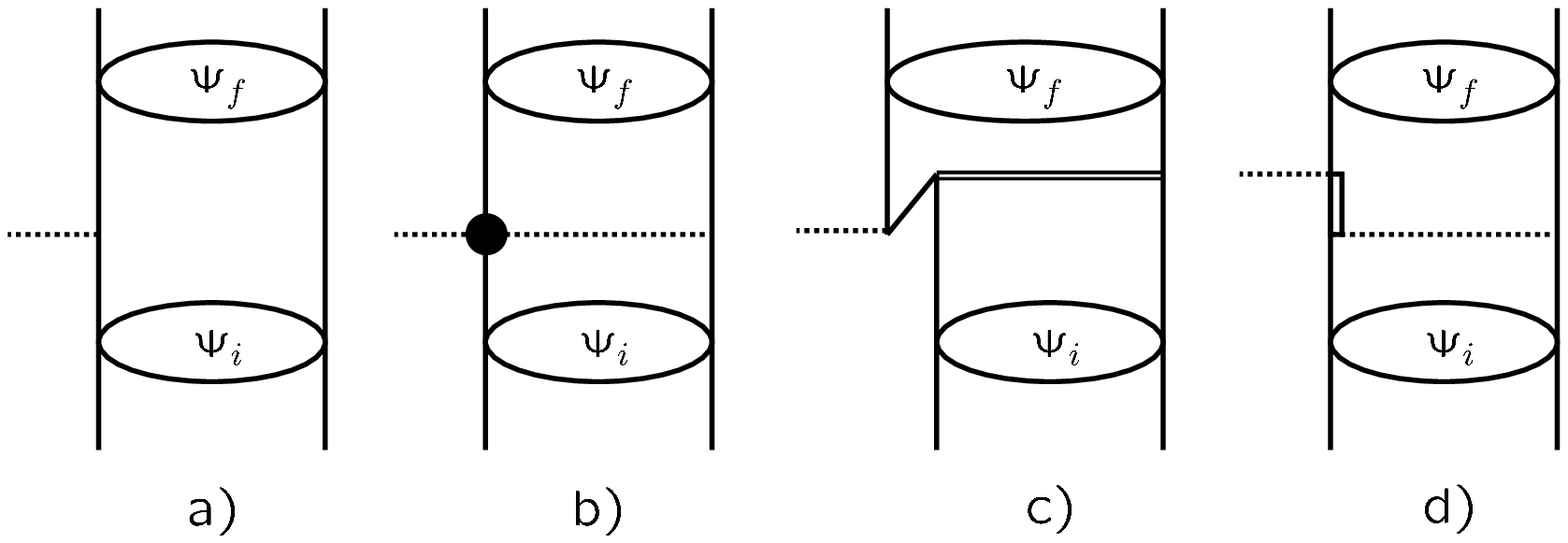}
  \captions[Mechanisms considered for charged and neutral pion
   production: direct production, re-scattering, Z-diagrams and
   $\Delta$-isobar contribution]
   {Mechanisms considered for charged and neutral pion production. a)
   direct production, b) re-scattering, c) Z-diagrams and d)
   $\Delta$-isobar contribution. $T$ is calculated from realistic
   $NN$ interactions.} \label{mechanisms}
\end{center}
\end{figure}

\sprg
We considered the contributions of the impulse, re-scattering,
$\Delta$-isobar mechanisms and Z-diagrams as represented in
\fig{mechanisms}\footnote{Our calculations do not include
contributions to the diagrams on \fig{mechanisms} involving the
two $T$-matrices. We note however that all the contributions with
at least one $T$-matrix are considered, together with the
undistorted operator.}.

The $\pi N$ amplitudes are those of chiral perturbation theory
from \rf{daRocha:1999dm}, resulting from the leading-order
Lagrangian of \eq{LX0Hanhart} and the next-to-leading order
Lagrangian of \eq{LX1Hanhart}.

From \eq{LX0Hanhart} and \eq{LX1Hanhart}, the single-nucleon
emission vertex is
\begin{equation}
\mathcal{M}^{\pi NN}_{a}=\frac{f_{\pi
NN}}{m_{\pi}}\tau_{a}^{\left( 1\right) }\left[
-\left( \vec{\sigma}^{\left( 1\right) }\cdot\vec{q}_{\pi}\right) +\frac{%
E_{\pi}}{2M}\vec{\sigma}^{\left( 1\right) }\cdot\left( \vec{p}_{1}+\vec{q}%
_{1}\right) \right] \label{opimpulse}
\end{equation}
where $\vec{p}_1$ and $\vec{q}_1$ are the nucleon momentum before
and after pion emission, $\vec{q}_\pi$ is the momentum of the
emitted pion and $a$ is an isospin index. The second term of
\eq{opimpulse} accounts for the nucleon recoil effect.
For the re-scattering $\pi N$-$\pi N$ vertex (with no isospin
flip), one has
\begin{eqnarray}
\mathcal{M}^{\pi\pi NN}_{resc} &
=-i\frac{m_{\pi}^{2}}{f_{\pi}^{2}}\frac{g_{A}}{f_{\pi
}}\left[ 2c_{1}-\left( c_{2}+c_{3}-\frac{g_{A}^{2}}{8M}\right) \frac {%
Q^{\prime}_{0}E_{\pi}}{m_{\pi}^{2}}+c_{3}\frac{\vec{q}_{\pi}\cdot\vec{q}^{\,\prime}}{m_{\pi}^{2}}%
\right] \left[ \tau_{a}^{\left( 2\right) }\vec{\sigma}^{\left(
2\right)
}\cdot\vec{q}^{\,\prime}\right] \label{oprescattering}\\
&
-i\frac{1}{f_{\pi}^{2}}\frac{g_{A}}{f_{\pi}}\frac{\delta_{q}}{8}\left[
\tau_{3}^{\left( 2\right) }\tau_{a}^{\left( 1\right) }\vec{\sigma
}^{\left( 2\right) }\cdot\vec{q}^{\,\prime}+\delta_{3a}\left(
\tau^{\left( 1\right)
}\cdot\tau^{\left( 2\right) }\right) \left( \sigma^{\left( 2\right) }\cdot%
\vec{q}^{\,\prime}\right) \right] \nonumber
\end{eqnarray}
and for the Weinberg-Tomozawa term,
\begin{equation}
\mathcal{M}^{\pi\pi
NN}_{WT}=\frac{g_{A}}{2f_{\pi}}\frac{1}{4f_{\pi}^{2}}\epsilon
_{abc}\tau_{b}^{\left( 1\right) }\tau_{c}^{\left( 2\right) }\left[
Q^{\prime}_{0}+ E_{\pi}\right] \vec{\sigma}^{\left( 1\right) }\cdot%
\vec{q}^{\,\prime},
\end{equation}
where $Q^{\prime}_{0}$ is the zeroth component of the exchanged
pion four-momentum and $\epsilon _{abc}$ the Levi-Civita tensor
(as before, $a$, $b$ and $c$ are isospin indices).

The $\Delta$-isobar contribution was calculated from
\eq{LX0Hanhart} and reads
\begin{equation}
\mathcal{M}^{\pi N \Delta} = i\frac{g_{A}}{2f_{\pi }}
\left(\frac{h_{A}}{2 f_{\pi}} \right)^{2} \frac{4}{9} \left[ 2
\tau_{a}^{\left( 2 \right)}+\tau_{b}^{\left( 1 \right)}
\tau_{c}^{\left(2 \right)}\varepsilon^{abc} \right]\left(
\vec{q}_{\pi} \cdot \vec{q}^{\,\prime} \right)
\vec{\sigma}^{\left( 2 \right)} \cdot \vec{q}^{\prime}
\frac{1}{\Delta-E_{\pi}}\frac{1}{\Omega^{2}-\omega_{\pi}^{2}}\label{opdelta},
\end{equation}
where $\Omega$ is given by \eq{Smatpresc}. Since the
$\Delta$-resonance is here included explicitly, its contribution
needs to be subtracted from the values of the low energy constants
$c_{i}$'s of \eq{oprescattering}. This was done taking the heavy
baryon limit of \eq{opdelta}, giving the contribution
$c_{3}^{\Delta} \approx -\frac{h_{A}^{2}}{18 m_{\pi}} =
-2.78\ugev^{-1}$ which was subtracted from the $-5.29 \ugev^{-1}$
value of the  $c_{3}$ constant appearing in the
$\left(\vec{q}_{\pi}\cdot\vec{q}^{\,\prime} \right)$ term of
\eq{oprescattering}.

%

The heavy-meson mediated $Z$-diagrams contributions given in
\rf{Lee:1993xh} and used in \rf{Pena:1999hq} for the $\pi^{0}$
production calculation, were also included:
\begin{equation}
\mathcal{M}_{a}^{Z\_\sigma}=-i\frac{f_{\pi NN}}{m_{\pi}}\frac{%
V_{S}^{+}\left( k\right)
}{M}\frac{E_{\pi}}{2M}\vec{\sigma}^{\left( 1\right) }\cdot\left(
\vec{p}_{1}+\vec{q}_{1}\right) \tau_{a}^{\left( 1\right) }
\end{equation}
for $\sigma$-exchange, and
\begin{equation}
\mathcal{M}_{a}^{Z\_\omega}=+i\frac{f_{\pi NN}}{m_{\pi}}\frac{E_{\pi}}{2M}%
\frac{V_{V}^{+}\left( k\right) }{M}\tau_{a}^{\left( 1\right) }\vec{\sigma}%
^{\left( 1\right) }\cdot\left[ \left( \vec{p}_{2}+\vec{q}_{2}\right) +i\vec{%
\sigma}^{\left( 2\right) }\times\vec{k}\right],
\end{equation}
for $\omega$ exchange, where $V_{S}^{+}$ and $V_{V}^{+}$ are
respectively, the scalar component and isospin independent vector
component of the nucleon-nucleon interaction. The i-th nucleon
final momentum is $\vec{q}_i$ and $\vec{k}=\vec{q}_2-\vec{p}_2$.
The kinematics conventions for the diagrams may be found in
\apx{Apkindiagrams}. Details for the partial wave decomposition of
the amplitudes are given in \apx{ApPWAmp}.

\subsection{Nucleon-nucleon potentials at intermediate energies}

\sprg
Since the energies considered for the nucleonic initial state
interaction are necessarily above the pion production threshold,
we took a recently developed $NN$ interaction fitted to the
nucleon-nucleon scattering
data\cite{{Elster:1988zu},{Schwick:2004},{Schwick:2005cc}}. This
nucleon-nucleon interaction was developed by the Ohio group as an
extension of the Bonn family\cite{Machleidt:1987hj} of realistic
nucleon-nucleon potentials well into the intermediate and high
energies region.
%

The free $NN$ T-matrices are obtained from a meson-exchange model
including $\Delta$-isobar and N*(1440)/N*(1535) degrees of freedom
to account for pion-production above the pion
threshold\cite{{Elster:1988zu},{Schwick:2004},{Schwick:2005cc}}.
To properly generate inelasticities, the model incorporates the
particle data book nucleon resonances as intermediate excitations
within two meson exchanges loops.
As an extension of the family of the Bonn potentials, it describes
well the $NN$ phase-shifts and inelasticities up to $1$~GeV.

The model is based on the solution of the relativistic Thomson
equation. This equation was favoured to time-ordered field theory
since it gives a good description of the nucleon inelasticities,
without the numerical complexities from the one-pion exchange cut
singularities. In time-ordered perturbation theory the last ones
are formally unavoidable, but actually, the main contribution to
the nucleon singularities arise from the nucleon resonances, and
the effect from the one-pion exchange cut is indeed quite small.

\begin{figure}
\begin{center}
  \includegraphics[width=.78\textwidth,keepaspectratio]{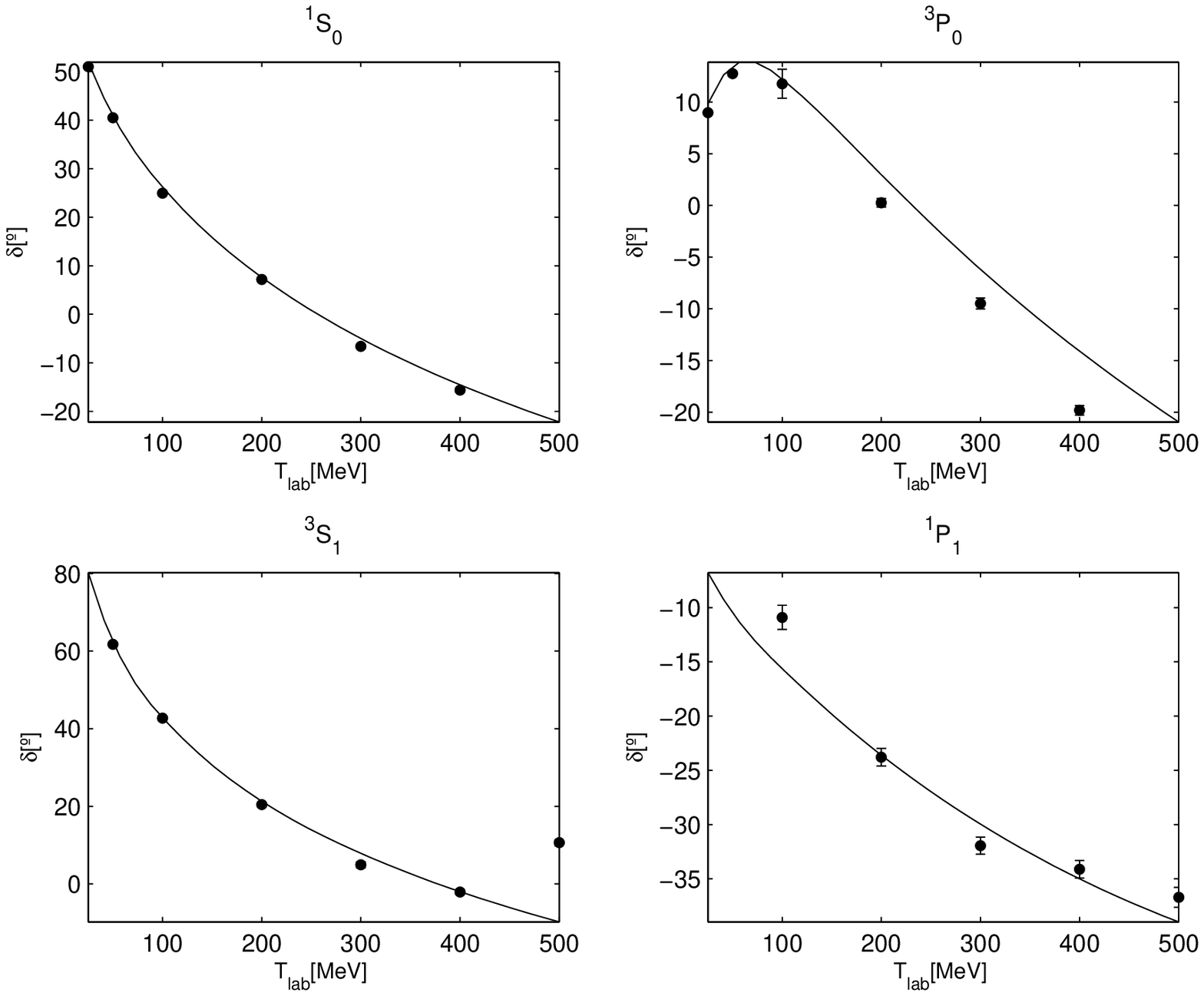}
  \captions[Phase shifts calculated with the Ohio group $NN$ potential]
  {Phase shifts calculated with the Ohio group $NN$ potential.}
  \label{ohiophsf}
\end{center}
\end{figure}

The Thomson equation for $NN$ scattering was solved in order to
obtain the fit of the $NN$ interaction to both phase-shifts and
inelasticities.
\tb{potparam} lists the parameters employed by the Ohio group $NN$
potential and consistently used in all the calculations in this
work. \fig{ohiophsf} shows the quality of the phase shifts in the
relevant energy region for the initial and final nucleonic states.
\begin{table}[h!]
\begin{center}
\begin{tabular}{cccc}
\hline\hline Meson & $m$ $\left[ \text{MeV}\right] $ &
$\frac{g^{2}}{4\pi }$ & $\Lambda $ $\left[ \text{MeV}\right] $ \\
\hline
$NN\alpha $-vertices &  &  &  \\
$\pi $ & $138.03$ & $13.8$ & $1519$ \\
$\eta $ & $547.30$ & $5.81$ & $830$ \\
$\omega $ & $782.60$ & $23.1$ & $1382$ \\
$\rho $ & $769.00$ & $1.09$ & $1275$ \\
$a_{0}$ & $983.00$ & $4.75$ & $1004$ \\
$\sigma \left( T=1\right) $ & $497$ & $5.502$ & $1807$ \\
$\sigma \left( T=0\right) $ & $482$ & $3.513$ & $1990$ \\
$\eta ^{\prime }$ & $958$ & $1.62$ & $1433$ \\
$f_{0}\left( 980\right) $ & $980$ & $2.50$ & $1274$ \\ \hline
$N\Delta \alpha $-vertices &  & $\frac{f^{2}}{4\pi }$ &  \\
$\pi $ & $138.03$ & $0.224$ & $640$ \\
$\rho $ & $769.00$ & $20.45$ & $1508$ \\ \hline\hline
\end{tabular}
\captions[$NN$-meson and $N \Delta$-meson vertices of the Ohio
group $NN$ model]{$NN$-meson and $N \Delta$-meson vertices of the
Ohio group $NN$ model. The meson mass is $m$, the coupling
constants are $g$ and $f$, and $\Lambda$ is the form-factor
cutoff. All of the corresponding form factors are of the monopole
type $\left(
\frac{\Lambda^{2}-m^{2}}{\Lambda^{2}+k^{2}}\right)$\cite{Schwick:2004}.
\label{potparam}}
\end{center}
\end{table}
\section{Calculational details}
%
\subsection{Three-body kinematics}

\sprg
The phase-space of the reaction $NN \rightarrow B_{1}B_{2}x$ is
five-dimensional: the three particles in the final state introduce
$3 \times 3=9$ degrees of freedom, but the four-momentum
conservation reduces this number to $5$. For energies close to
threshold, the final state can be treated non-relativistically.
The natural coordinate system is therefore given by the Jacobi
coordinates\cite{Glocke:1983} where first the relative momentum of
one pair of particles is constructed, and then the momentum of the
third particle is calculated relatively to the two-body system.
Obviously, there are three equivalent sets of variables possible.

For reactions of the type $NN \rightarrow B_{1} B_{2} x$ it is
common to work with the relative momentum of the two-nucleon
system and to treat the meson $x$ separately, as the spectator
particle.
This choice is theoretically the most convenient, since one is
working already with the relative momentum of the dominant final
state interaction.
Close to threshold, for the final-state distortion (but not for
the initial-state distortion), the nucleons can be treated as
non-relativistic, and thus for any given relative energy of the
outgoing two-nucleon system,
\begin{equation}
\varepsilon=\frac{q_{k}^{2}}{M}.
\end{equation}
The modulus of the meson momentum $\left| \vec{q}_{x} \right|$ is
fixed by energy conservation. The phase-space is then
\begin{equation}
\xi = \left\{\varepsilon, \Omega_{k},\Omega_{\pi} \right\}
\end{equation}
where $\Omega_{k}$ and $\Omega_{\pi}$ are the solid angle for the
relative $NN$ momentum and for the pion momentum, respectively
(see \apx{Apcrosssection}).
initial state will be denoted by $\vec{p}$. Explicit expressions
for the vectors appearing are given in \apx{Apkinematics}.
As long as the initial state is unpolarised, the system has
azimuthal rotation symmetry, reducing the number of degrees of
freedom from $5$ to $4$.
%
%

\subsection{The role of the different isospin channels}
\sprg
Meson production reactions in nucleon-nucleon collisions give also
the possibility to manipulate another internal degree of freedom:
the isospin. Nucleons are isospin $\frac{1}{2}$ particles in a
$\mathcal{SU} \left( 2 \right)$ doublet,
\begin{equation}
\left\vert p\right\rangle \equiv\left\vert t=\frac{1}{2},t_{3}=+\frac{1}%
{2}\right\rangle \text{ \ and \ }\left\vert n\right\rangle
\equiv\left\vert t=\frac{1}{2},t_{3}=-\frac{1}{2}\right\rangle,
\end{equation}
where $\left\vert p\right\rangle$ and $\left\vert n\right\rangle$
represent the proton and neutron wave functions, respectively, and
$t_{3}$ is the third component of the isospin operator
$\mathbf{t}$.

A two-nucleon pair can be either in an isotriplet state (total
isospin $T=1$ with the three possible projections of the total
isospin $T_{3}=+1$ for a $pp$ state, $T_{3}=0$ for a $pn$ state
and $T_{3}=-1$ for a $nn$ state) or in an isosinglet state ($T=0$
with $T_{3}=0$ for a $pn$ state).

Even in an isospin conserving situation, various transitions are
possible, depending on whether an isovector or an isoscalar
particle is produced. In particular, pion production in $NN$
collisions can be expressed in terms of only three independent
reactions with cross sections denoted by $\sigma_{10}$,
$\sigma_{01}$ and $\sigma_{11}$ (here, the reference of
\rf{Rosenfeld:1954pr} is adopted, where $T_{i}$ and $T_{f}$ in
$\sigma_{T_{i}T_{f}}$ are the $NN$ pair initial and final total
isospin, respectively). Isospin conservation forbids the existence
of $\sigma_{00}$. In total, there are seven relations for $NN
\rightarrow NN \pi$ reactions\cite{Machner:1999ky,Watson:1951pr}:
\begin{eqnarray}
\sigma\left(  pp\rightarrow pp\pi^{0}\right)    & =&\sigma_{11}\\
\sigma\left(  pp\rightarrow pn\pi^{+}\right)    &
=&\sigma_{10}  +\sigma_{11}\\
\sigma\left(  np\rightarrow np\pi^{0}\right)    &
=&\frac{1}{2}\left[
\sigma_{10} +\sigma_{01}\right]  \\
\sigma\left(  np\rightarrow pp\pi^{-}\right)    &
=&\frac{1}{2}\left[
\sigma_{11}+\sigma_{01}\right]  \\
\sigma\left(  np\rightarrow nn\pi^{+}\right)    &
=&\frac{1}{2}\left[
\sigma_{11}+\sigma_{01}\right]  \\
\sigma\left(  nn\rightarrow np\pi^{-}\right)    &
=&\sigma_{10} +\sigma_{11}\\
\sigma\left(  nn\rightarrow nn\pi^{0}\right)    & =&\sigma_{11}%
\end{eqnarray}
Note that none of the reactions involves more than two of the
fundamental cross sections and that the last two reactions are
related with the first two by charge symmetry. The relative
strength of the different transition amplitudes can thus provide
significant information on the production operator.  At medium
energies, the cross sections $\sigma_{11}$ and $\sigma_{10}$ are
dominated by the excitation of the intermediate $\Delta_{33}$
resonance and are well measured even close to threshold. In
contrast, the isoscalar cross section $\sigma_{01}$, which has to
be extracted from pion production data in neutron-proton and
proton-proton collisions is not well known. Actually, the
determination of $\sigma_{01}$ includes a principal model
dependence, as
\begin{equation}
\sigma_{np \rightarrow pp \pi^{-} }=\frac{1}{2}\left(
\sigma_{11}+\sigma_{01} \right)
\end{equation}
only holds in the case of exact isospin invariance. However, due
to the different particle masses entering in the reactions $np
\rightarrow pp \pi^{-}$, $np \rightarrow nn \pi^{+}$ and $pp
\rightarrow pp \pi^{0}$, isospin invariance is only an approximate
symmetry, and thus the comparison of the cross sections can not be
performed at the same beam energy, as mentioned
before\cite{Daum:2001yh}.

Meson production data close to threshold in $np$ collisions suffer
from the fact that only relative cross section measurements have
been performed, as a consequence of the (poor) quality of the
neutron beam. Unfortunately, the isospin cross section
$\sigma_{01}$ can only be accessed via these $np$ interactions. It
is obtained from the reaction cross section for $np \rightarrow pp
\pi^{-}$ by subtracting the $\sigma_{11}$ part, which results in
larger uncertainties than in other isospin cross sections.
\subsection{Partial wave analysis}

\sprg
Some of the complexity inherent to three-body final state
reactions can be dealt using partial wave analysis. For energies
close to threshold, since the total kinetic energy in the final
state is small, the reactions are dominated by transitions to
states with low orbital angular momentum values. Therefore, an
expansion of the outgoing spherical waves will involve mainly
low-order spherical harmonics, and thus the observables should
have a relatively simple angular dependence.

The amplitude of the outgoing wave
in spin state $\sigma _{1}\sigma _{2}\sigma _{x}$ obtained when
the nucleons in the
initial state have spin projections $\sigma _{a}$ and $\sigma _{b}$ \cite%
{{Knutson:1999dy},{Meyer:2001gj}} is given by
\begin{align}
M_{\sigma _{a}\sigma _{b}}^{\sigma _{1}\sigma _{2}\sigma _{x}}&
=iK_{i}\sum_{\alpha _{i}\alpha _{f}\alpha _{m}}\sqrt{\frac{2L_{i}+1}{2J+1}}%
\left\langle s_{a}s_{b}\sigma _{a}\sigma _{b}\left\vert
S_{i}\sigma _{i}\right. \right\rangle \left\langle
L_{i}S_{i}0\sigma _{i}\left\vert
JM\right. \right\rangle   \label{ms123} \\
& \left\langle s_{1}s_{2}\sigma _{1}\sigma _{2}\left\vert
S_{f}\sigma _{f}\right. \right\rangle \left\langle
S_{f}L_{if}\sigma _{f}\lambda _{f}\left\vert j_{f}m_{f}\right.
\right\rangle \left\langle s_{x}l_{x}\sigma
_{x}\lambda _{x}\left\vert j_{x}m_{x}\right. \right\rangle   \notag \\
& \left\langle t_{a}t_{b}\nu _{a}\nu _{b}\left\vert T_{i}\nu
_{i}\right. \right\rangle \left\langle t_{1}t_{2}\nu _{1}\nu
_{2}\left\vert T_{f}\nu _{f}\right. \right\rangle \left\langle
T_{f}t_{x}\nu _{f}\nu _{x}\left\vert
T_{i}\nu _{i}\right. \right\rangle   \notag \\
& \left\langle j_{f}j_{x}m_{f}m_{x}\left\vert JM\right.
\right\rangle U_{\alpha _{i}\alpha _{f}}Y_{\lambda
_{f}}^{L_{f}}\left( \hat{q}_{k}\right) Y_{\lambda
_{x}}^{l_{x}}\left( \hat{q}\right)   \notag
\end{align}%
where $L_{i}$, $S_{i}$ and $J$ are respectively, the two initial
nucleon orbital angular momentum, spin and total angular momentum,
and $U_{\alpha _{i}\alpha _{f}}$ is a matrix element depending on
the quantum numbers in both initial
state $\alpha _{i}=\left\{ L_{i}\text{, }S_{i}\text{, }J\text{ and }%
T_{i}\right\} $ and final state $\alpha _{f}=\left\{ S_{f}\text{, }L_{f}%
\text{, }l_{x}\text{, }j_{f}\text{, }j_{x}\text{ and
}T_{f}\right\} $.
This matrix element is also a function of the energy sharing parameter $%
\varepsilon $.

The individual nucleon spin quantum numbers are $s_{a}$ and $s_{b}
$, with the corresponding magnetic quantum numbers $\sigma _{a}$
and $\sigma _{b}$.
In the final state the nucleon spin quantum numbers are
represented by $s_{1}$ and $s_{2}$ with spin projections $\sigma
_{1}$ and $\sigma _{2}$, and the $NN$ total spin quantum number is
$S_{f}$.
The $NN$ relative orbital angular momentum is $L_{f}$ and the $NN$
total angular momentum is $j_{f}$.

For the produced meson, the spin is represented by quantum numbers
$s_{x}$ and $\sigma _{x}$, the angular momentum relative to the
$NN$ centre of mass is $l_{x}$ and the vector sum of $s_{x}$ and
$l_{x}$ is $j_{x}$.
The isospin quantum numbers are represented by $t$ and $\nu $.
$K_{i}$ is a kinematic constant arising from phase-space (as
defined in \eq{Kidef}). The sum in $\left( \ref{ms123}\right) $
extends over the initial and final state quantum numbers $\alpha
_{1}$ and $\alpha _{2}$, and
also the magnetic quantum numbers $\alpha _{m}=\left\{ \sigma _{i}\text{, }%
\sigma _{f}\text{, }M\text{, }m_{f}\text{, }m_{x}\text{, }\lambda _{f}\text{%
, }\lambda _{x}\text{, }\nu _{i}\text{ and }\nu _{f}\right\} $.
These notations are summarised in Table~\ref{notqt}.
\begin{table}[tbp]
\begin{center}
\begin{tabular}{lccccc}
\hline\hline
& \vspace{-0.4cm} &  &  &  &  \\
& \vspace{-0.4cm} $N$ & Tot$_{NN_{\text{initial}}}$ & $N$ & Tot$_{NN_{\text{%
final}}}$ & $\pi $ \\
&  &  &  &  &  \\ \hline
& \vspace{-0.4cm} &  &  &  &  \\
\textbf{Spin} & $s_{a}$ and $s_{b}$ & $S_{i}$ & $s_{1}$ and
$s_{2}$ & $S_{f}$
& $s_{x}$ \\
\qquad projection & $\sigma _{a}$ and $\sigma _{b}$ & $\sigma _{i}$ & $%
\sigma _{1}$ and $\sigma _{2}$ & $\sigma _{f}$ & $\sigma _{x}$ \\
& \vspace{-0.4cm} &  &  &  &  \\
\textbf{Orbital mom.} &  & $L_{i}$ &  & $L_{f}$ & $l_{x}$ \\
\qquad projection &  &  &  & $\lambda _{f}$ & $\lambda _{x}$ \\
& \vspace{-0.4cm} &  &  &  &  \\
\textbf{Total ang. mom.} &  & $J$ &  & $j_{f}$ & $j_{x}$ \\
\qquad projection &  & $M$ &  & $m_{f}$ & $m_{x}$ \\
& \vspace{-0.4cm} &  &  &  &  \\
\textbf{Isospin} & $t_{a}$ and $t_{b}$ & $T_{i}$ & $t_{1}$ and $t_{2}$ & $%
T_{f}$ & $t_{x}$ \\
\qquad projection & $\nu _{a}$ and $\nu _{b}$ & $\nu _{i}$ & $\nu _{1}$ and $%
\nu _{2}$ & $\nu _{f}$ & $\nu _{x}$ \\
& \vspace{-0.4cm} &  &  &  &  \\ \hline\hline
\end{tabular}%
\end{center}
\captions{Summary of the notations for the nucleon and pion spin,
orbital momentum, total angular momentum and isospin.}
\label{notqt}
\end{table}

Note that as in the two particle case, the complexity of the
angular pattern is determined by the final state orbital angular
momentum values $L_{f}$ and
$l_{x}$. Since $s_{x}=\sigma _{x}=0\Rightarrow j_{x}=l_{x}$ and $%
m_{x}=\lambda _{x}$, $\left( \ref{ms123}\right) $ simplifies to%
\begin{align}
M_{\sigma _{a}\sigma _{b}}^{\sigma _{1}\sigma _{2}\sigma _{x}}&
=iK_{i}I_{\pi }I_{s}\sum_{\alpha _{1}\alpha _{2}\alpha _{m}}\sqrt{\frac{2L_{i}+1}{%
2J+1}}\left\langle s_{a}s_{b}\sigma _{a}\sigma _{b}\left\vert
S_{i}\sigma _{i}\right. \right\rangle \left\langle
L_{i}S_{i}0\sigma _{i}\left\vert
JM\right. \right\rangle   \label{ms123I} \\
& \left\langle s_{1}s_{2}\sigma _{1}\sigma _{2}\left\vert
S_{f}\sigma _{f}\right. \right\rangle \left\langle
S_{f}L_{f}\sigma _{f}\lambda _{f}\left\vert j_{f}m_{f}\right.
\right\rangle \left\langle
j_{f}l_{x}m_{f}\lambda _{x}\left\vert JM\right. \right\rangle   \notag \\
& U_{\alpha _{1}\alpha _{2}}Y_{\lambda _{f}}^{L_{f}}\left( \widehat{q_{k}}%
\right) Y_{\lambda _{x}}^{l_{x}}\left( \hat{q}\right)   \notag
\end{align}%
with%
\begin{equation}
I_{\pi }=\left\langle t_{a}t_{b}\nu _{a}\nu _{b}\left\vert
T_{i}\nu _{i}\right. \right\rangle \left\langle t_{1}t_{2}\nu
_{1}\nu _{2}\left\vert T_{f}\nu _{f}\right. \right\rangle
\left\langle T_{f}t_{x}\nu _{f}\nu _{x}\left\vert T_{i}\nu
_{i}\right. \right\rangle   \label{Ipidef}
\end{equation}
If one follows the obvious procedure of including only initial and
final angular momentum states that are consistent with the
anti-symmetrisation requirements, then the resulting formula will
satisfy all the necessary symmetries for identical particles in
the initial and final states. The only requirement  is that when
the initial state has identical particles, an extra factor of $2$
must be included in the cross section to reproduce the appropriate
incident flux normalisation\cite{Knutson:1999dy}\footnote{The
corresponding requirement for the particles in the final state is
included in \eq{ms123I} through the definition of $K_{i}$
(\eq{Kidef}).}. This extra factor of $2$ is already included in
\eq{ms123I} via the definition $\left( \ref{Isdef}\right) $ of
$I_{s }$.
\begin{equation}
I_{s}=\left\{
\begin{array}{ccc}
\sqrt{2} &  & \text{for }\pi ^{0} \\
\sqrt{2} &  & \text{for }\pi ^{+} \\
1&  & \text{for }\pi ^{-}%
\end{array}%
\right. \text{production}  \label{Isdef}
\end{equation}

\subsubsection{Expansion of the cross section}

The simplest observable is the unpolarised cross section, which
reads\cite{{Knutson:1999dy},{Meyer:2001gj}}:
\begin{equation}
\sigma _{0}=\frac{1}{\left( 2s_{a}+1\right) \left( 2s_{b}+1\right) }\mathrm{%
Tr}\left[MM^{\dag }\right].  \label{unpsigtr}
\end{equation}
To obtain a formula for the partial-wave expansion of the cross
section which shows explicitly the various allowed angular
dependencies, one substitutes Eq.~(\ref{ms123I}) in
Eq.~(\ref{unpsigtr}) and then combine the spherical harmonics and
carry out an angular momentum reduction. More
precisely, using $\left( \ref{cleborth1}\right) $, $\left( \ref{racsumcleb}%
\right) $, $\left( \ref{9jsumcleb}\right) $ and $\left( \ref{prodY}\right) $%
, one obtains,
\begin{equation}
\sigma _{0}=\frac{\left\vert K_{i}\right\vert ^{2}}{\left(
2s_{a}+1\right) \left( 2s_{b}+1\right) }\left( \frac{1}{4\pi
}\right) \sum_{L_{p}L_{q}L}\sum_{\alpha \alpha ^{\prime
}}C_{L_{p}L_{q}L}^{\alpha
\alpha ^{\prime }}U_{\alpha }U_{\alpha ^{\prime }}^{\ast }\mathcal{Y}%
_{L_{p}L_{q}}^{L0}\left( \hat{q}_{k},\hat{q}\right)
\label{expcross}
\end{equation}%
with the coefficients\cite{{Knutson:1999dy},{Meyer:2001gj}}

\begin{align}
C_{L_{p}L_{q}L}^{\alpha \alpha ^{\prime }}& =\left(I_{\pi }I_{s}
\right)^{2}\left( -1\right) ^{L_{f}+l_{x}^{\prime }+L_{i}^{\prime
}}\delta _{S_{i}S_{i}^{\prime }}\delta
_{S_{f}S_{f}^{\prime }}\sqrt{2L_{i}+1}\sqrt{2L_{i}^{\prime }+1}\sqrt{2L_{f}+1%
}\sqrt{2L_{f}^{\prime }+1}  \label{calpha} \\
& \sqrt{2l_{x}+1}\sqrt{2l_{x}^{\prime }+1}\sqrt{2j_{f}+1}\sqrt{%
2j_{f}^{\prime }+1}\sqrt{2J+1}\sqrt{2J^{\prime }+1}  \notag \\
& \left\langle L_{f}L_{f}^{\prime }00|L_{p}0\right\rangle
\left\langle l_{x}l_{x}^{\prime }00|L_{q}0\right\rangle
\left\langle L_{i}L_{i}^{\prime
}00|L0\right\rangle   \notag \\
& W\left( j_{f}S_{f}L_{p}L_{f}^{\prime };L_{f}j_{f}^{\prime
}\right) W\left( L_{i}S_{i}LJ^{\prime };JL_{i}^{\prime }\right)
\left\{
\begin{array}{ccc}
j_{f} & l_{x} & J \\
j_{f}^{\prime } & l_{x}^{\prime } & J^{\prime } \\
L_{p} & L_{q} & L,%
\end{array}%
\right\} ,  \notag
\end{align}%
and
\begin{equation}
\left(I_{\pi }I_{s} \right)^{2}=\left\{
\begin{array}{ccc}
1 &  & \text{for }\pi ^{0} \\
\left[ \left\langle \frac{1}{2}\frac{1}{2}\nu _{1}\nu
_{2}\left\vert T_{f}0\right. \right\rangle \left\langle
T_{f}101\left\vert 11\right.
\right\rangle \right] ^{2} &  & \text{for }\pi ^{+} \\
\frac{1}{2}\left[ \left\langle \frac{1}{2}\frac{1}{2}\nu _{a}\nu
_{b}\left\vert T_{i}0\right. \right\rangle \left\langle
111-1\left\vert
T_{i}0\right. \right\rangle \right] ^{2} &  & \text{for }\pi ^{-}%
\end{array}%
\right. \text{production}  \label{Ipi}
\end{equation}
In \eq{expcross} and \eq{calpha}, $\alpha =\left\{
L_{i},S_{i},J,S_{f},L_{f},l_{x},j_{f},\sigma _{i},\sigma
_{f},M,m_{f},\lambda _{f},\lambda _{x}\right\} $ is a shorthand
for summation over all the indices. The values for $\left(I_{\pi
}I_{s} \right)^{2}$ are listed in \tb{Ipivalues}.
\begin{table}
\begin{center}
\begin{tabular}
[c]{cccccc}
\hline\hline
& \vspace{-0.6cm}  &  &  &  & \\
$\left(I_{\pi
}I_{s} \right)^{2}$ &  \hspace{0.2cm}$\pi^{0}$ \hspace{0.2cm} &  & $\pi^{+}$ &  & $\pi^{-}$\\
& \vspace{-0.6cm}  &  &  &  & \\\hline
& \vspace{-0.4cm}  &  &  &  & \\
$ \vspace{0.4cm} T=0$ & $-$ &  & $1$ &  & $\dfrac{1}{6}$\\
$ T=1$ & $1$ &  & $\dfrac{1}{2}$  &  & $\dfrac{1}{4}$\\
& \vspace{-0.4cm} &  &  &  & \\\hline\hline
\end{tabular}
\captions{$\left(I_{\pi }I_{s} \right)^{2}$ values for the several
pion production reactions considered.} \label{Ipivalues}
\end{center}
\end{table}

\sp

The various cross section terms in $\left( \ref{expcross}%
\right) $ arise from the interference between the two matrix
elements depending on the quantum numbers $\alpha $ and $\alpha
^{\prime }$. Moreover, from \eq{calpha} one sees that two partial
waves may interfere only if the $NN$ spin quantum numbers match in
both the initial and final states. Since the $\delta
_{S_{f}S_{f}^{\prime }}$ is also present in the general formula
for the polarisation observables, one can divide the partial waves
in groups, where interference is only permitted within each group.
For instance, for $\pi ^{0}$ production, these first two groups
are the $S_{f}=0$ group and the $S_{f}=1$ group, respectively
$\left\{ Ss,Sd,Ds\right\} $ and $\left\{ Ps,Pp\right\} $. Since
$Ss$ is dominant at energies close to threshold, one concludes
that the $Sd$ and $Ds$ terms may be more important at low energies
that one might otherwise have thought. Note also that the
expansion of the
cross section involves only angular functions $\mathcal{Y}%
_{L_{p}L_{q}}^{L\Lambda }$ with $\Lambda =0$. This results from
the angular momentum algebra and is a consequence of the fact that
the cross section must be invariant under rotations about the
$z$-axis.

\subsection{Selection rules for $NN \rightarrow NN x$ \label{selectionrules}}

\sprg
The selection rules for $NN\rightarrow NNx$ reactions are based on
the symmetries of strong interaction that imply the conservation
of parity, total angular momentum and isospin.

For the two-nucleon system, its parity (in other words, the
behaviour of its wave function under a reflection of the
coordinate system through the origin), determined solely by the
relative angular momentum of the two particles, is $\left(-1
\right)^{L}$.
Also, the symmetry of the spin and isospin wave function is
$\left( -1\right) ^{S+1}$ and $\left( -1\right) ^{T+1}$.
Then, because the nucleons are fermions,
\begin{equation}
\left(-1
\right)=\left(-1\right)^{S_{i}+T_{i}+L_{i}}=\left(-1\right)^{S_{f}+T_{f}+L_{f}}
\label{Parinitial}
\end{equation}
where $L_{i}\left(L_{f}\right)$, $S_{i}\left(S_{f}\right)$ and
$T_{i}\left(T_{f}\right)$ denote the angular momentum, total spin
and total isospin of the initial(final) two-nucleon system.
%
In addition, for a reaction of the type $NN\rightarrow NNx$, one
finds from parity conservation\cite{Hanhart:2003pg}
\begin{equation}
\left( -1\right) ^{L_{i}}=\pi _{x}\left( -1\right) ^{\left(
L_{f}+l_{x}\right) }  \label{Parcons}
\end{equation}
where $\pi _{x}$ and $l_{x}$ denote, respectively, the intrinsic
parity and the angular momentum (with respect to the outgoing
two-nucleon system) of the particle $x$.
The two criteria $\left( \ref{Parinitial}%
\right) $ and $\left( \ref{Parcons}\right)$ can now be combined to give%
\begin{equation}
\left( -1\right) ^{\left( \Delta S+\Delta T\right) }=\pi
_{x}\left( -1\right) ^{l_{x}}  \label{deltaP}
\end{equation}%
where $\Delta S\left( \Delta T\right) $ is the change in total
(iso)spin when going from the initial to the final $NN$ system.
We can now analyse in detail the channels contributing to the
different charge reactions:


\begin{itemize}
\item $pp\rightarrow pp\pi^{0}$

The $pp\rightarrow pp\pi^{0}$ reaction corresponds to a
$\left(T_{i}^{NN}=1\right) \rightarrow \left(T_{f}^{NN}=1\right)$
transition. The possible partial waves for $pp\rightarrow
pp\pi^{0}$ result from applying \eq{Parcons} and \eq{deltaP}. They
are listed in \tb{pwpi0o} up to $J=3$, and restricted to pion $s$-
and $p$-wave states relatively to the final $NN$ system. The
channels are grouped in sets of equal $J$. Within each set, the
different channels were ordered according to the total angular
momentum $j_{f}$ of the final $NN$ pair.
\end{itemize}
\begin{table}[h!]
\begin{center}
\begin{tabular}{c|c|ccc|cccc}
\hline\hline $\left( NN\right) _{i}$ & $\left( NN\right)
_{f}\text{ }l_{x}$ & $S_{i}$ &
$L_{i}$ & $J$ & $S_{f}$ & $L_{f}$ & $j_{f}$ & $l_{x}$ \\
\hline $^{3}P_{0}$ & $\left( ^{1}S_{0}\right) s$ & $1$ & $1$ & $0$
& $0$ & $0$ & $0$
& $0$ \\
$^{1}S_{0}$ & $\left( ^{3}P_{0}\right) s$ & $0$ & $0$ & $0$ & $1$
& $1$ & $0$
& $0$ \\
$^{3}P_{0}$ & $\left( ^{3}P_{1}\right) p$ & $1$ & $1$ & $0$ & $1$
& $1$ & $1$
& $1$ \\
$^{3}P_{1}$ & $\left( ^{3}P_{0}\right) p$ & $1$ & $1$ & $1$ & $1$
& $1$ & $0$
& $1$ \\
$^{3}P_{1}$ & $\left( ^{3}P_{2}\right) p$ & $1$ & $1$ & $1$ & $1$
& $1$ & $2$
& $1$ \\
$^{3}P_{1}$ & $\left( ^{3}P_{1}\right) p$ & $1$ & $1$ & $1$ & $1$
& $1$ & $1$
& $1$ \\
$^{3}P_{2}$ & $\left( ^{3}P_{1}\right) p$ & $1$ & $1$ & $2$ & $1$
& $1$ & $1$
& $1$ \\
$^{3}F_{2}$ & $\left( ^{3}P_{1}\right) p$ & $1$ & $3$ & $2$ & $1$
& $1$ & $1$
& $1$ \\
$^{1}D_{2}$ & $\left( ^{3}P_{2}\right) s$ & $0$ & $2$ & $2$ & $1$
& $1$ & $2$
& $0$ \\
$^{3}P_{2}$ & $\left( ^{3}P_{2}\right) p$ & $1$ & $1$ & $2$ & $1$
& $1$ & $2$
& $1$ \\
$^{3}F_{2}$ & $\left( ^{3}P_{2}\right) p$ & $1$ & $3$ & $2$ & $1$
& $1$ & $2$
& $1$ \\
$^{3}F_{3}$ & $\left( ^{3}P_{2}\right) p$ & $1$ & $3$ & $3$ & $1$
& $1$ & $2$ & $1$ \\ \hline\hline
\end{tabular}%
\end{center}
\captions[The lowest partial waves for $\left( NN_{T=1}\right)
\rightarrow \left( NN_{T=1}\right) \pi $, ordered by increasing
values of $J$]{The lowest partial waves for $\left(
NN_{T=1}\right) \rightarrow \left( NN_{T=1}\right) \pi $, ordered
by increasing values of $J$.
The notation is $S_{i}(S_{f})$, $%
L_{i}(L_{f})$ and $J(j_{f})$ for the total spin, the orbital
momentum and the total angular momentum of the initial(final)
two-nucleon pair, respectively. The orbital angular momentum of
the pion relative to the final $NN$ pair is $l_{x}$. The different
channels are grouped in sets with the same total angular momentum
$J$. Within each set, the order of the channels is ruled by the
total angular momentum $j_{f}$ of the final $NN $ pair.}
\label{pwpi0o}
\end{table}

\begin{itemize}
\item $pp\rightarrow pn\pi^{+}$

The partial waves considered for $pp\rightarrow pn\pi^{+}$ are
both in \tb{pwpi0o} and \tb{pwpipo}, since $pn$ can be either in a
isospin isotriplet state $T^{NN}=1$ (\tb{pwpi0o}) or in an
isosinglet state $T^{NN}=0$ (\tb{pwpipo}). The selection rules for
this last case follow from that $\left( L_{i}+S_{i}\right) $ must
be even and $\left(L_{f}+S_{f}\right) $ must be odd, since
$T_{i}^{NN}=1$ and $T_{f}^{NN}=0$.
\end{itemize}
\begin{table}[h!]
\begin{center}
\begin{tabular}{c|c|ccc|cccc}
\hline\hline $\left( NN\right) _{i}$ & $\left( NN\right)
_{f}\text{ }l_{x}$ & $S_{i}$ &
$L_{i}$ & $J$ & $S_{f}$ & $L_{f}$ & $j_{f}$ & $l_{x}$ \\
\hline $^{1}S_{0}$ & $\left( ^{3}S_{1}\right) p$ & $0$ & $0$ & $0$
& $1$ & $0$ & $1$
& $1$ \\
$^{3}P_{0}$ & $\left( ^{1}P_{1}\right) p$ & $1$ & $1$ & $0$ & $0$
& $1$ & $1$
& $1$ \\
$^{3}P_{1}$ & $\left( ^{3}S_{1}\right) s$ & $1$ & $1$ & $1$ & $1$
& $0$ & $1$
& $0$ \\
$^{3}P_{1}$ & $\left( ^{1}P_{1}\right) p$ & $1$ & $1$ & $1$ & $0$
& $1$ & $1$
& $1$ \\
$^{1}D_{2}$ & $\left( ^{3}S_{1}\right) p$ & $0$ & $2$ & $2$ & $1$
& $0$ & $1$
& $1$ \\
$^{3}P_{2}$ & $\left( ^{1}P_{1}\right) p$ & $1$ & $1$ & $2$ & $0$
& $1$ & $1$
& $1$ \\
$^{3}F_{2}$ & $\left( ^{1}P_{1}\right) p$ & $1$ & $3$ & $2$ & $0$
& $1$ & $1$ & $1$ \\ \hline\hline
\end{tabular}%
\end{center}
\captions[The lowest partial waves for $\left( NN_{T=1}\right)
\rightarrow \left( NN_{T=0}\right) \pi $, ordered by increasing
values of $J$]{The same of \tb{pwpi0o} but $\left( NN_{T=1}\right)
\rightarrow \left( NN_{T=0}\right) \pi $.} \label{pwpipo}
\end{table}
%

%
%
\newpage

\begin{itemize}
\item $pn\rightarrow pp\pi^{-}$

The lowest partial waves for $pn\rightarrow pp\pi^{-}$ are in
\tb{pwpi0o} and \tb{pwpimo}, which correspond to the case when
$pn$ is in an isotriplet and isosinglet state, respectively.
\end{itemize}
\begin{table}[h!]
\begin{center}
\begin{tabular}{c|c|ccc|cccc}
\hline\hline $\left( NN\right) _{i}$ & $\left( NN\right)
_{f}\text{ }l_{x}$ & $S_{i}$ &
$L_{i}$ & $J$ & $S_{f}$ & $L_{f}$ & $j_{f}$ & $l_{x}$ \\
\hline $^{3}S_{1}$ & $\left( ^{1}S_{0}\right) p$ & $1$ & $0$ & $1$
& $0$ & $0$ & $0$
& $1$ \\
$^{3}D_{1}$ & $\left( ^{1}S_{0}\right) p$ & $1$ & $2$ & $1$ & $0$
& $0$ & $0$
& $1$ \\
$^{1}P_{1}$ & $\left( ^{3}P_{0}\right) p$ & $0$ & $1$ & $1$ & $1$
& $1$ & $0$
& $1$ \\
$^{3}S_{1}$ & $\left( ^{3}P_{1}\right) s$ & $1$ & $0$ & $1$ & $1$
& $1$ & $1$
& $0$ \\
$^{3}D_{1}$ & $\left( ^{3}P_{1}\right) s$ & $1$ & $2$ & $1$ & $1$
& $1$ & $1$
& $0$ \\
$^{1}P_{1}$ & $\left( ^{3}P_{1}\right) p$ & $0$ & $1$ & $1$ & $1$
& $1$ & $1$
& $1$ \\
$^{1}P_{1}$ & $\left( ^{3}P_{2}\right) p$ & $0$ & $1$ & $1$ & $1$
& $1$ & $2$
& $1$ \\
$^{3}D_{2}$ & $\left( ^{3}P_{2}\right) s$ & $1$ & $2$ & $2$ & $1$
& $1$ & $2$
& $0$ \\
$^{1}F_{3}$ & $\left( ^{3}P_{2}\right) p$ & $0$ & $3$ & $3$ & $1$
& $1$ & $2$ & $1$ \\ \hline\hline
\end{tabular}%
\end{center}
\captions[The lowest partial waves for $\left( NN_{T=0}\right)
\rightarrow \left( NN_{T=1}\right) \pi $, ordered by increasing
values of $J$]{The same of \tb{pwpi0o} but $\left( NN_{T=0}\right)
\rightarrow \left( NN_{T=1}\right) \pi $.} \label{pwpimo}
\end{table}

It is clear from \tb{pwpi0o} and \tb{pwpimo} that partial waves
with $Ss$ final states can only contribute to $\sigma_{11}$ and
partial waves with $Sp$ final states only to $\sigma_{01}$.

\sp

Reactions proceeding dominantly via the $\Delta$ resonance are
expected to be significantly larger than others where the
excitation of the $\Delta$ resonance is suppressed by selection
rules\cite{Machner:1999ky}. Since the $\Delta$ has isospin
$\frac{3}{2}$ and the nucleon has isospin $\frac{1}{2}$, the
$N\Delta$ system can couple to total isospin equal to $2$ or $1$.
On the other hand, the nucleon-nucleon system can couple to total
isospin equal to $1$ or $0$. Thus, the $N \Delta$ states do not
contribute to the nucleon-nucleon channels with isospin equal to
$0$ and only the $N \Delta$ channels with isospin equal to $1$
will couple to the nucleon-nucleon system\cite{Garcilazo:1990ws}.
\tb{cpndelta} lists the lowest partial wave states for the $N
\Delta$ system.
\begin{table}
\begin{center}
\begin{tabular}[c]{cccccc}
\hline\hline
& \vspace{-0.5cm} &  &  &  & \\
& $\left(  NN\right)  _{i}$ & $N\Delta$ & & $\left( NN\right)
_{f}$ &
$N\Delta$\\
& \vspace{-0.5cm} &  &  &  & \\\hline 
& \vspace{-0.5cm} &  &  &  & \\
$\sigma_{11}$ & $^{3}P_{0}$ & $^{3}P_{0}$ &  & $^{1}S_{0}$ & $^{5}D_{0}$\\
&  \vspace{-0.5cm}&  &  &  & \\
$\sigma_{10}$ & $^{3}P_{_{1}}$ & $^{3}P_{1}$ &  & $^{3}S_{1}$ & $-$\\
& $^{1}S_{0}$ & $^{5}D_{0}$ &  & $^{3}S_{1}$ & $-$\\
& $^{1}D_{2}$ & $^{5}S_{2}$ &  & $^{3}S_{1}$ & $-$\\
& \vspace{-0.5cm} &  &  &  & \\
$\sigma_{01}$ & $^{3}S_{1}$ & $-$ &  & $^{1}S_{0}$ & $^{5}D_{0}$\\
& $^{3}D_{1}$ & $-$ &  & $^{1}S_{0}$ & $^{5}D_{0}$\\
& \vspace{-0.5cm} &  &  &  & \\\hline\hline
\end{tabular}
\captions[The lowest $N \Delta$ partial wave states] {The lowest
$N\Delta$ partial wave
states\cite{{Machner:1999ky},{Garcilazo:1990ws}}.}
\label{cpndelta}
\end{center}
\end{table}
For instance, for $\sigma_{01}$, intervening in $\pi^{-}$
production, $\Delta$-excitation in the initial state is impossible
because a pion-nucleon state with isospin $\frac{3}{2}$ cannot
couple with a nucleon with isospin zero. A $\Delta$ excitation is
possible in the final state but then the intermediate $\Delta N$
system needs to be in a $L_{\Delta N}=2$ state (see
\tb{cpndelta}). The necessary rotational energy in the $\Delta N$
system can be estimated through\cite{Machner:1999ky}
\begin{equation}
\frac{L_{\Delta N} \left( L_{\Delta N}+1 \right)}{2 \mu r^{2}}
\end{equation}
with $\mu$ the reduced mass. Assuming a distance $r$ of $1 \ufm$
between the $\Delta$ and the nucleon, one obtains
\begin{eqnarray}
73 \umev \hspace{0.55cm}\text{for }L_{\Delta N}  & =&1\\
219 \umev \hspace{0.55cm}\text{for }L_{\Delta N}  & =&2,
\end{eqnarray}
which means that the excitation of the $\Delta$-resonance will be
strongly suppressed at threshold in this particular isospin
channel. The cross section $\sigma_{11}$ can proceed through the
$\Delta$ but here the $\Delta N$ system, like the initial $NN$
system, has at least an orbital angular momentum of $1$. For the
cross section $\sigma_{10}$, and consequently for the $\pi^{+}$
production reaction, however, there is one partial wave where the
$\Delta N$ system is in an $s$-wave state and therefore, the
$\Delta$ excitation contribution is expected to be important given
the weight of the $\left(^{1}D_{2}\right)_{NN} \rightarrow
\left(^{5}S_{2} \right)_{N\Delta} \rightarrow \left( ^{3}S_{1}
\right)p$ transition\cite{Machner:1999ky}.

\sp

So far, most of the investigations considered only the
lowest-partial waves in the outgoing channel with the pion in an
$s$-wave state relative to the nucleon pair. Such calculations
permit only conclusions on the absolute magnitude of the
production cross section near threshold. The inclusion of higher
partial waves, in the $NN$ as well as in the $\pi N$ sector,
allows predictions for differential cross sections and, in
particular, spin dependent observables. Therefore, it is possible
to examine whether the considered production mechanisms lead to
the proper onset of higher partial waves, as suggested by the
data. Results of the latter kind are especially interesting
because they reflect the spin-dependence of the production
processes and thus should be very useful in discriminating between
different mechanisms\cite{Hanhart:1998za}.

Very close to threshold, $\pi^0$ production is naturally dominated
by s-wave production. To go beyond threshold, we made convergence
studies of the partial wave series by including partial waves up
to a total angular momentum of the system $J=3$.
%
%
In our calculations, the coupled channels were properly taken into
account in the convergence studies.


\section{Convergence of the partial wave series}
%

\begin{figure}[t!]
\begin{center}
\includegraphics[width=.98\textwidth,keepaspectratio]{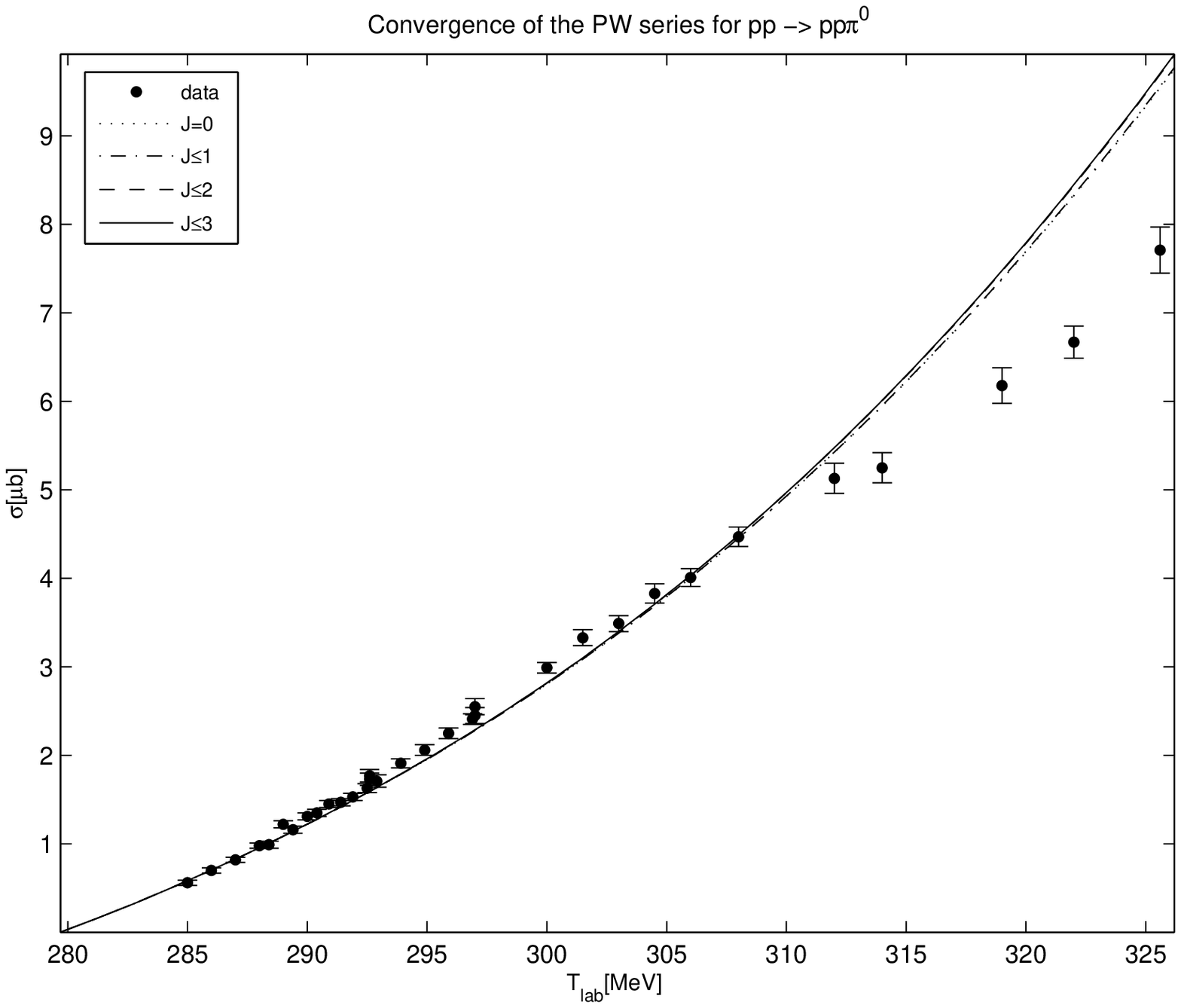}
  \captions[Convergence of the partial wave series for $pp
   \rightarrow pp \pi^{0}$]{Convergence of the partial wave series
   for $pp \rightarrow pp \pi^{0}$. The dotted, dashed-dotted, dashed
   and solid line correspond, respectively, to the cross sections
   with all the contributions
   up to $J=0$, $J=1$, $J=2$ and $J=3$, as listed in \tb{pwpi0o}. The
   nucleon-nucleon interaction is described through the Ohio $NN$ model. The data points are from \rf{Meyer:1992jt}.}
\label{pi0fcptb}
\end{center}
\end{figure}

\begin{figure}[h!]
\begin{center}
\includegraphics[width=.98\textwidth,keepaspectratio]{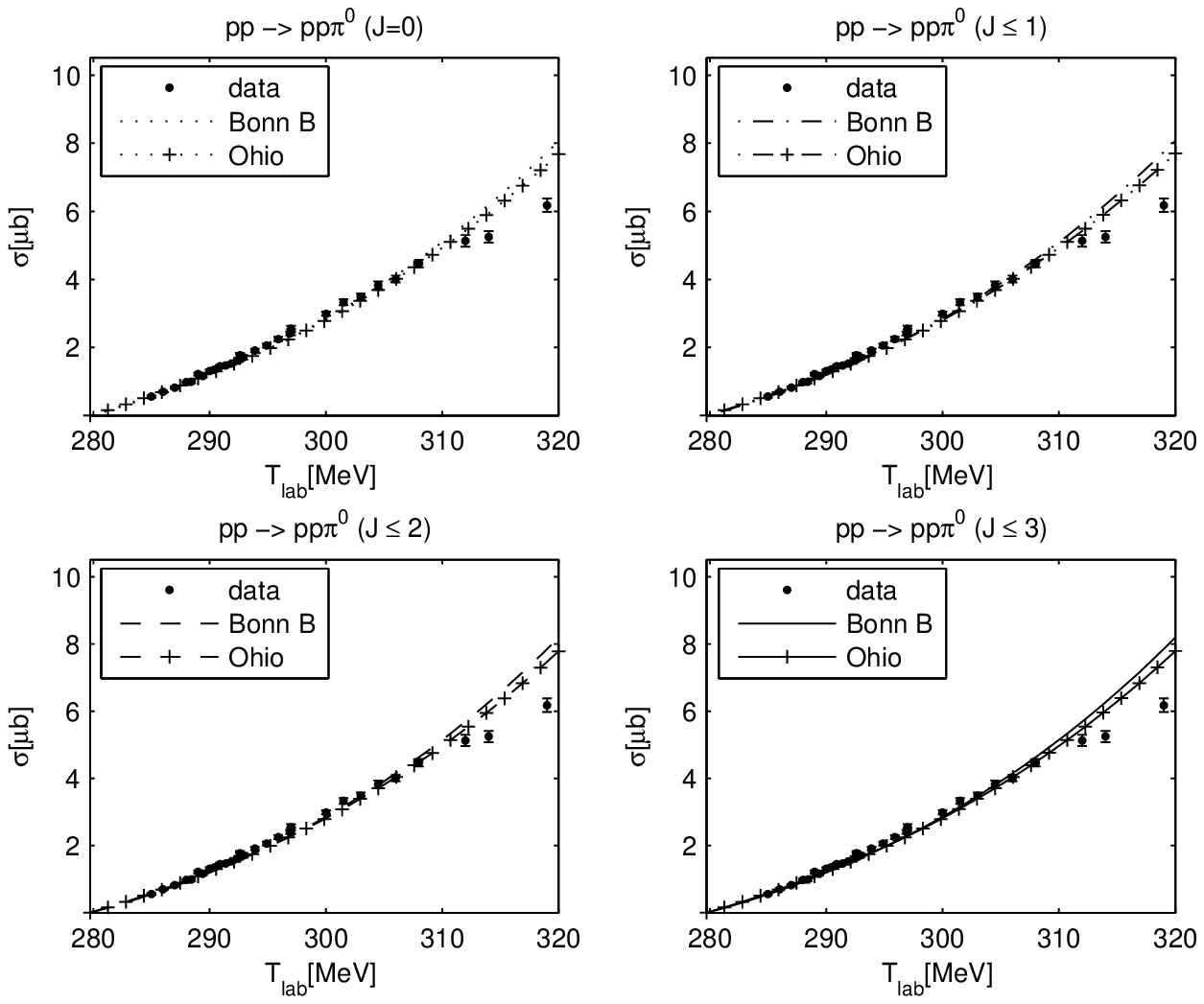}
  \captions[Effect of the nucleon-nucleon interaction in the cross
   section for $pp \rightarrow pp \pi^{0}$]{Effect of the
   nucleon-nucleon interaction in the cross section for $pp
   \rightarrow pp \pi^{0}$. The dotted, dashed-dotted, dashed and
   solid line correspond, respectively, to the cross section
   considering the Bonn B potential for the nucleon-nucleon
   interaction and all the contributions up to $J=0$, $J=1$, $J=2$
   and $J=3$. The corresponding lines with +'s consider the
   nucleon-nucleon interaction to be described by
   the Ohio $NN$ model. The data points are from \rf{Meyer:1992jt}.}
\label{pi0cmcp}
\end{center}
\end{figure}

\sprg
In \fig{pi0fcptb} we show the results for the convergence of the
partial wave series for the cross section of $\pi^{0}$ production
(with all the mechanisms discussed in \sec{prodoperators}
included).
As expected, the importance of the contributions from the channels
with higher $J's$ increases with the laboratory energy, $T_{lab}$.
Till $\sim 30\umev$ above threshold, the contributions from the
several sets of $J's$ are immaterial compared to the $J=0$ result,
which alone describes the data in that energy region.
The same conclusions hold when the $NN$ interaction is described
through the Bonn B potential\cite{Machleidt:1987hj}, due to a
smooth variation of the nucleon phase-shifts in that region. We
compare in \fig{pi0cmcp} the cross sections with both
interactions.

The cross sections with the Ohio $NN$ model for the
nucleon-nucleon interaction (dotted, dashed-dotted, dashed and
solid line with +'s) are slightly lower than the ones with the
Bonn B potential (dotted, dashed-dotted, dashed and solid lines).
Also, naturally, the deviations between the results obtained with
the two interactions increase with the energy, and also with $J$.
%

\begin{figure}[h!]
\begin{center}
\includegraphics[width=.98\textwidth,keepaspectratio]{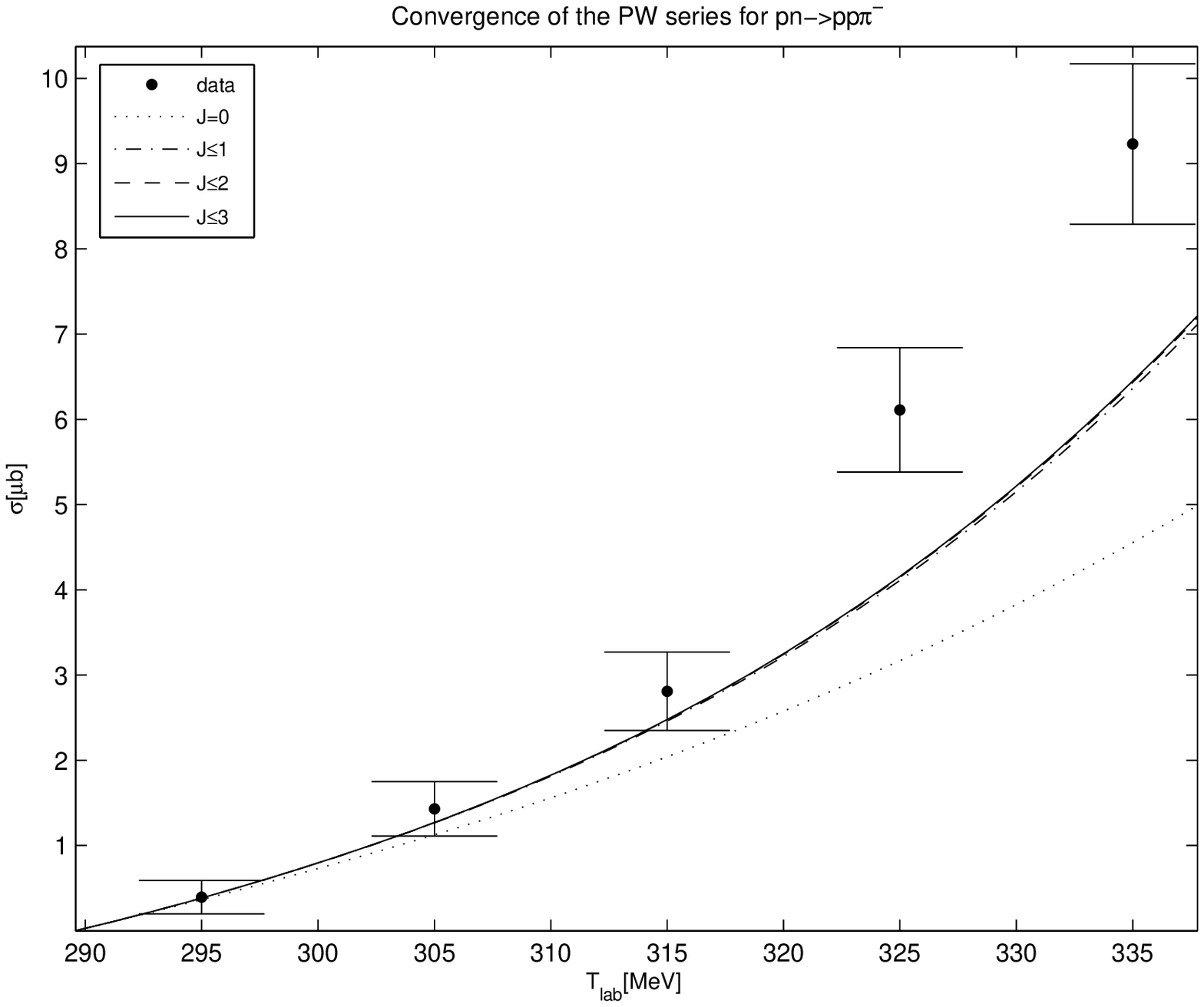}
  \captions[Convergence of the partial wave series for $pn
   \rightarrow pp \pi^{-}$]{Convergence of the partial wave series
   for $pn \rightarrow pp \pi^{-}$. The meaning of the lines is the
   same of \fig{pi0fcptb}. The data points are from
   \rf{Bachman:1995gn}.} \label{pimfcptb}
\end{center}
\end{figure}

\begin{figure}[t!]
\begin{center}
\includegraphics[width=.98\textwidth,keepaspectratio]{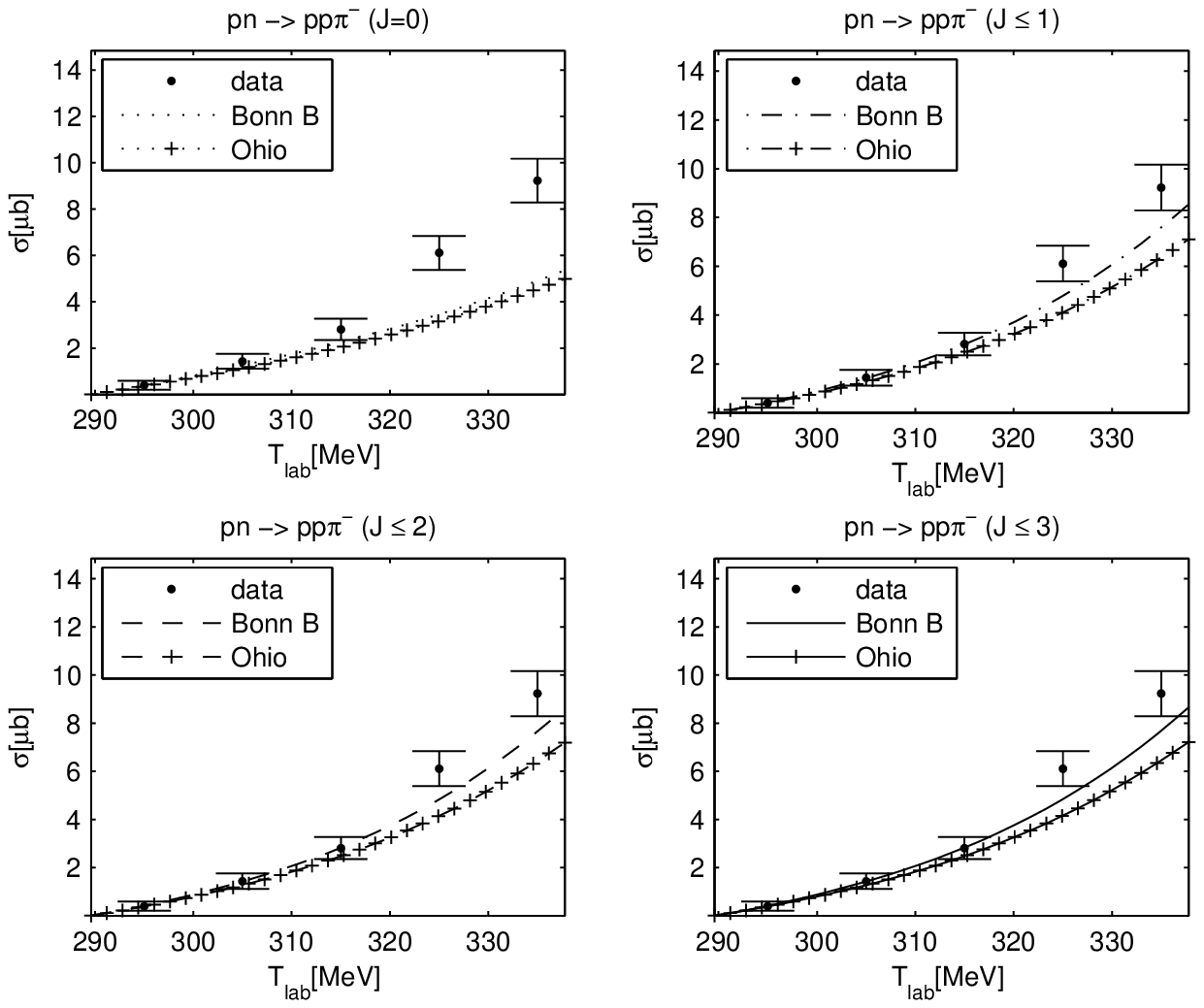}
  \captions[Effect of the nucleon-nucleon interaction on the cross
   section for $pn \rightarrow pp \pi^{-}$ ]{Cross section for $pn
   \rightarrow pp \pi^{-}$ for the Ohio $NN$ model and Bonn
   B potential for the $NN$ interaction. The meaning of the lines is
   the same of \fig{pi0cmcp}. The data points are from
   \rf{Bachman:1995gn}.} \label{pimcmcp}
\end{center}
\end{figure}

For $\pi^{-}$ production, the deviation between the cross section
with all the contributions up to $J=1$ (dashed-dotted line in
\fig{pimfcptb}), $J=2$ (dashed line), and up to $J=3$ (solid line)
is very small. However, unlikely to what happens for $\pi^{0}$,
the $J=0$ channels (dotted line) are not enough to describe the
data in the region of $30 \umev$ above threshold.
The same is found when the $NN$ interaction is described using the
Bonn B potential, as it is shown in \fig{pimcmcp}.

\begin{figure}[t!]
\begin{center}
\includegraphics[width=.98\textwidth,keepaspectratio]{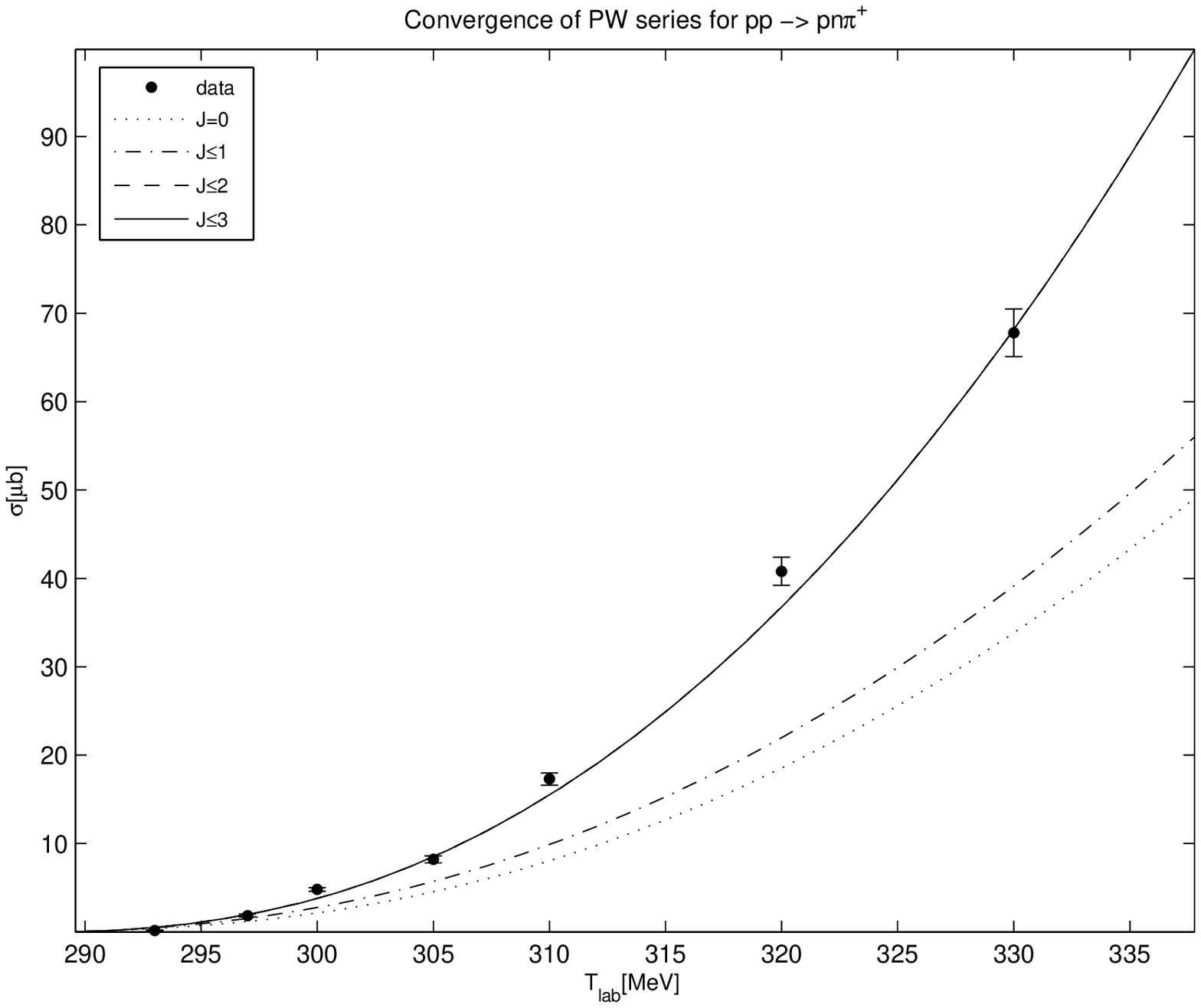}
  \captions[Convergence of the partial wave series for $pp
   \rightarrow pn \pi^{+}$]{Convergence of the partial wave series
   for $pp \rightarrow pn \pi^{+}$. The meaning of the lines is the
   same of \fig{pi0fcptb}. The data points are from
   \rf{Hardie:1997mg}.} \label{pipfcptb}
\end{center}
\end{figure}

We show in \fig{pipfcptb} the convergence of the partial wave
series for the $\pi^{+}$ production. For this reaction, although
the convergence is slower, the experimental data near threshold
are also described. As already found for $\pi^{-}$ production,
since the channels with $J=0$ (dotted line) are not enough to
describe the data, one needs to include at least the channels up
to $J=2$ (dashed line).

\begin{figure}[t!]
\begin{center}
\includegraphics[width=.98\textwidth,keepaspectratio]{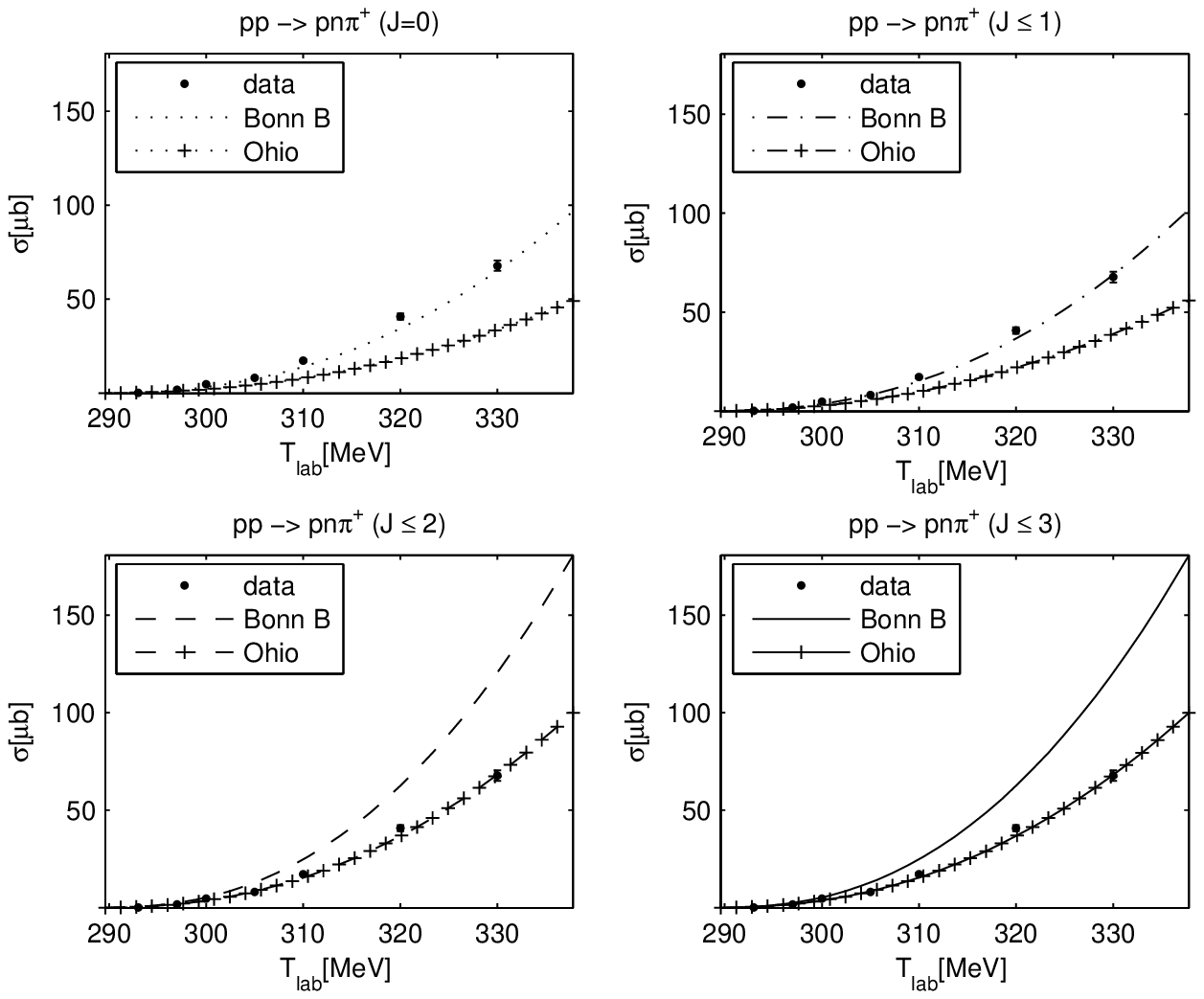}
  \captions[Effect of the nucleon-nucleon interaction on the cross
   section for $pp \rightarrow pn \pi^{+}$ ]{Cross section for $pp
   \rightarrow pn \pi^{+}$ for the Ohio $NN$ model and Bonn
   B potential for the $NN$ interaction.  The data points are from
   \rf{Hardie:1997mg}.} \label{pipcmcp}
\end{center}
\end{figure}

Moreover, the $NN$ potential used to describe the $NN$ final- and
initial-state interaction can be decisive for the case of
$\pi^{+}$ production. We compare in \fig{pipcmcp} the calculated
cross sections with the Bonn B and Ohio interactions, for all sets
of $J$'s considered. As expected, the deviation of both results
increase with $T_{lab}$ and with $J$, but for this reaction the
Bonn B potential fails in describing the data in a much wider
region than the Ohio $NN$ model (solid line vs. solid line with
+'s in the bottom-right panel of \fig{pipcmcp}).
%

\clearpage

%
\section{The role of the different production mechanisms}
%

\sprg
\begin{figure}[t!]
\begin{center}
\includegraphics[width=.98\textwidth,keepaspectratio]{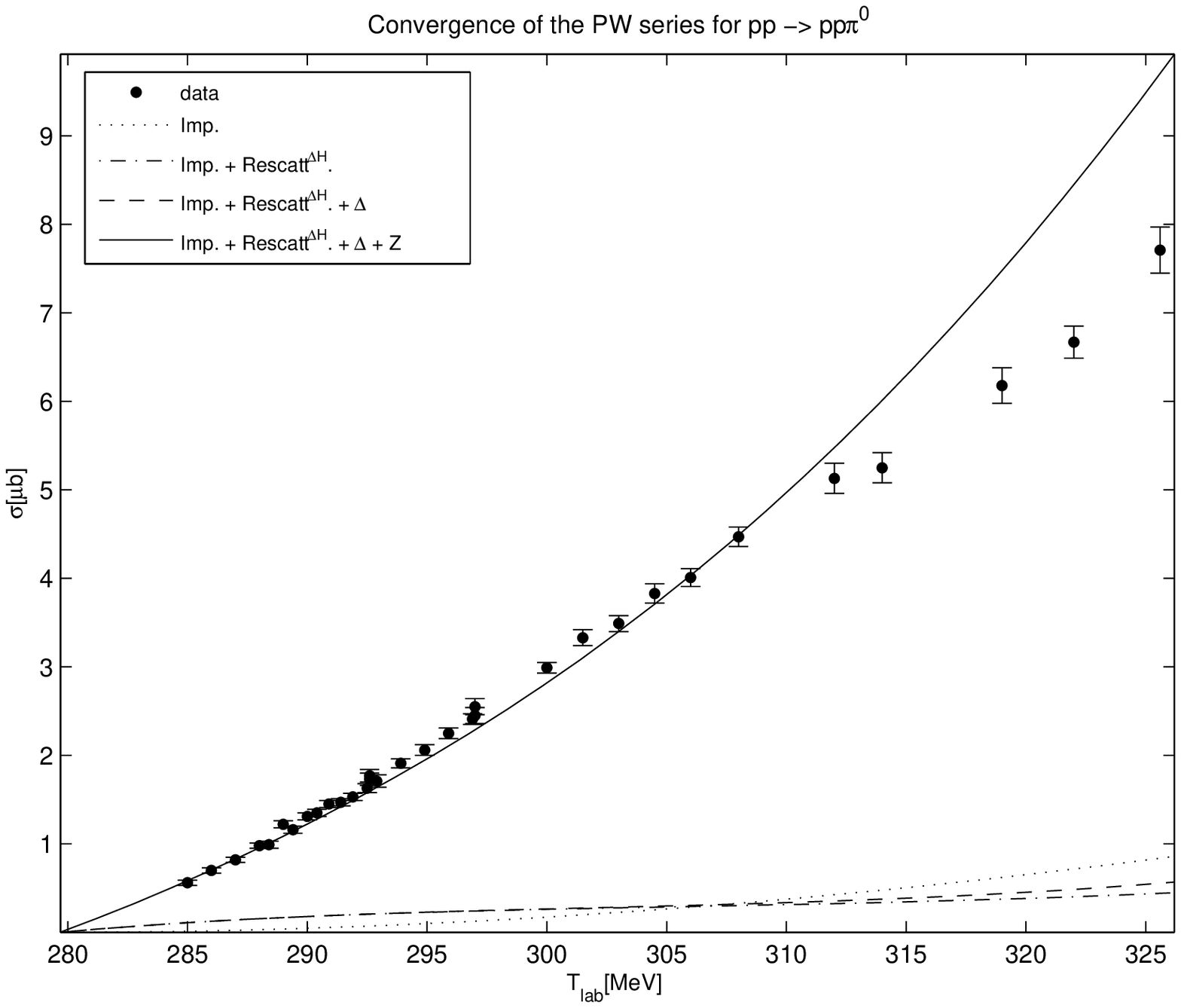}
  \captions[Comparison between the mechanisms contributing to
   $\pi^{0}$ production]{Comparison between the contributions for
   $\pi^{0}$ production of the mechanisms of \fig{mechanisms}. The
   dotted line corresponds to the direct-production mechanism alone
   (diagram a)) and the dashed-dotted line considers the additional
   contribution of the re-scattering mechanism. In the legend, $\Delta_{H}$
   refers to the subtraction of the infinitely heavy limit of
   \eq{opdelta} from the c$_{i}$ values of \eq{oprescattering}.
   The dashed line
   further includes the explicit $\Delta$-resonance excitation. The solid
   line represents all the contributions (diagrams a)-d)). The data
   points are from \rf{Meyer:1992jt}.} \label{pi0fcptbci}
\end{center}
\end{figure}

We compare in \fig{pi0fcptbci} the contributions of the pion
production mechanisms that are illustrated on \fig{mechanisms}.
The direct-production and re-scattering mechanisms alone
(dashed-dotted line), highly suppressed, are clearly not enough to
describe the $\pi^{0}$ production data. Also, the additional
mechanism considering the explicit excitation of a $\Delta$
(dashed line) was found to be small (this is natural, given the
discussion in \sec{selectionrules}). As already discussed on
\rf{Pena:1999hq}, the $Z$-diagrams (solid line) are decisive to
remove the discrepancy with the experimental data.

\begin{figure}
\begin{center}
\includegraphics[width=.98\textwidth,keepaspectratio]{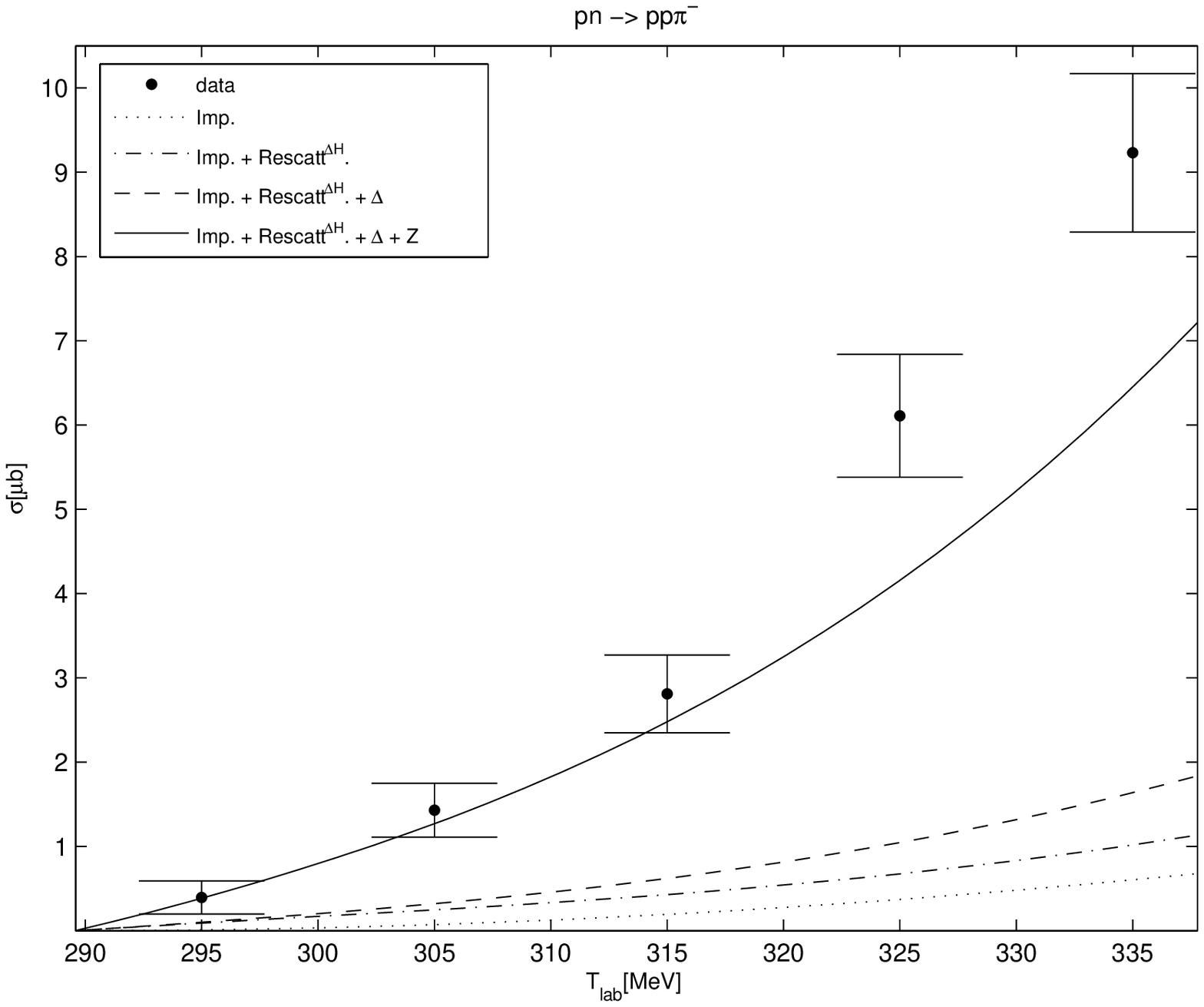}
  \captions[Comparison between the mechanisms contributing to
   $\pi^{-}$ production]{Comparison between the contributions for
   $\pi^{-}$ production of the mechanisms of \fig{mechanisms}. The
   meaning of the lines is the same of \fig{pi0fcptbci}. The
   re-scattering term now comprises also the Weinberg-Tomozawa
   mechanism. The data points are from \rf{Bachman:1995gn}.
\label{pimfcptbci}}
\end{center}
\end{figure}

For $\pi^{-}$ production, the re-scattering operator comprises
also the Weinberg-Tomozawa term, which does not contribute to
$\pi^{0}$ production, and its importance relative to the single
nucleon emission term is therefore enhanced (dotted line vs.
dashed-dotted line in \fig{pimfcptbci}). For this case, the
$\Delta$-isobar mechanism is also found to play an important role
(dashed line in \fig{pimfcptbci}).

\begin{figure}[t!]
\begin{center}
\includegraphics[width=.98\textwidth,keepaspectratio]{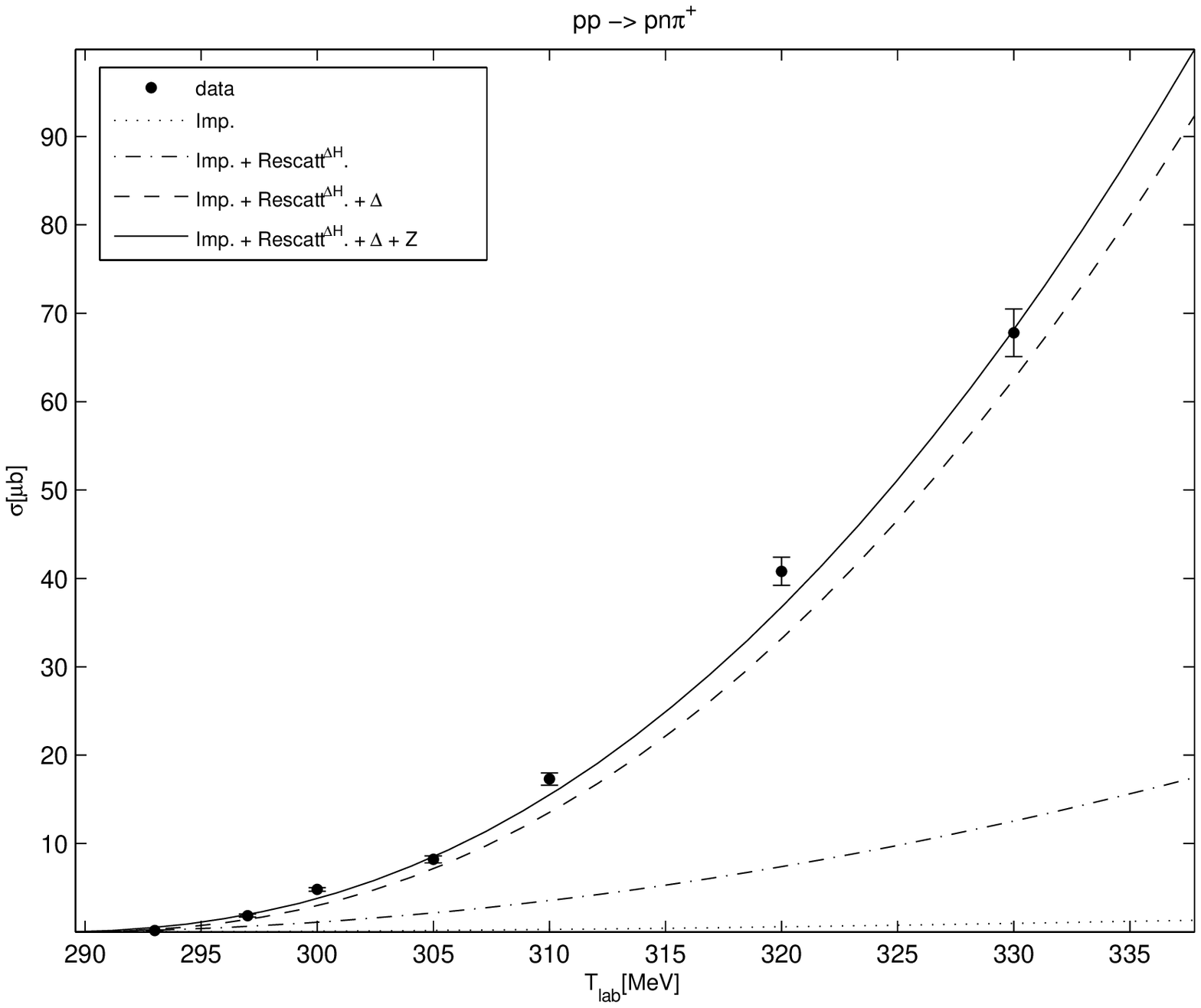}
  \captions[Comparison between the mechanisms contributing to
   $\pi^{+}$ production]{The same of \fig{pi0fcptbci} but for $pp
   \rightarrow pn \pi^{+}$. The data points are from
   \rf{Hardie:1997mg}.} \label{pipfcptbci}
\end{center}
\end{figure}

The relative importance of the several mechanisms of
\fig{mechanisms}, for $\pi^{+}$ production, is shown in
\fig{pipfcptbci}, where the re-scattering mechanism is found to
contribute significantly (dashed-dotted line).

\begin{figure}[h!]
\begin{center}
\includegraphics[width=.99\textwidth,keepaspectratio]{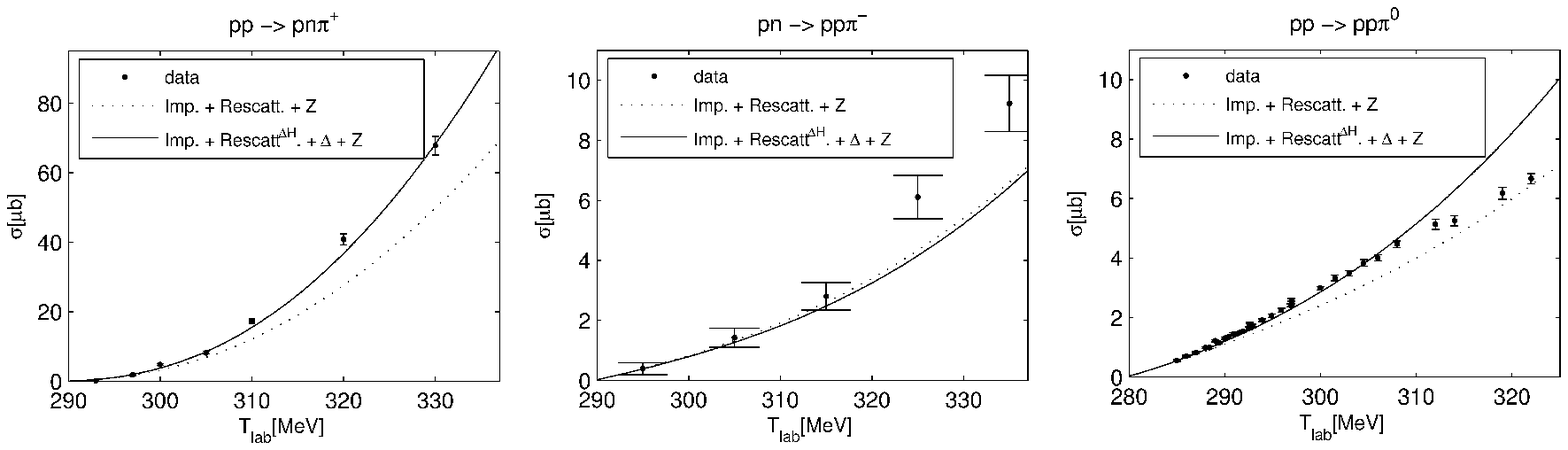}
  \captions[Effect on the cross sections of
   not considering the $\Delta$ as an explicit degree of
   freedom]{Effect on the cross section of not considering the
   $\Delta$ as an explicit degree of freedom.
   The solid line is the full calculation of \fig{pi0fcptbci}-\fig{pipfcptbci}
   considering the mechanisms of direct-production, re-scattering, $\Delta$-excitation
   and Z-diagrams (Imp. + Rescatt$^{\Delta_{H}}$. + $\Delta$ + Z). The dotted line corresponds the calculation
   with only the direct-production,
   re-scattering terms and Z diagrams (Imp. + Rescatt. + Z).} \label{svsd}
\end{center}
\end{figure}

Since the $\Delta$-resonance is included explicitly in all our
calculations, its contribution was subtracted from the values of
the low energy constants $c_{i}$'s of the $\pi N$ scattering
amplitude in \eq{oprescattering}, which was done taking the
infinitely heavy or static limit of \eq{opdelta} (this was
indicated by the superscript $\Delta_{H}$ in the legends of
\fig{pi0fcptbci}, \fig{pimfcptbci} and \fig{pipfcptbci}).
We show in \fig{svsd} the effect of not including explicitly the
mechanism of pion production through the $\Delta$-excitation. In
that figure, in contrast to previous ones, the re-scattering
amplitude (in the dotted line) is taken with the low energy
constant $c_{3}$ including the $\Delta$ contribution. The solid
lines are the full calculation for $\pi^{0}$, $\pi^{-}$ and
$\pi^{+}$. They coincide with the full lines in \fig{pi0fcptbci},
\fig{pimfcptbci} and \fig{pipfcptbci}, respectively.

%
For all the cases, the inclusion of the $\Delta$ improves the
description of the experimental data. As expected, its importance
increases with the energy. Also, both for $\pi^{0}$ and $\pi^{+}$
production, the inclusion of the $\Delta$ increases the cross
section. These general trends were already found for the
J\"{u}lich
model\cite{{Hanhart:1995ut},{Hanhart:1998za},{Hanhart:2000jj}},
presented in \sec{Phenapproach} (see \fig{crossHanhart}). However,
contrarily to this phenomenological model, in which all the short
range mechanisms are included through $\omega$-exchange (diagram
c) of Fig.~\ref{mechanisms}) and adjusted to reproduce the total
cross section of the reaction $pp \rightarrow pp \pi^{0}$ close to
threshold, in our calculations no adjustment is made, as the
parameters for the $Z$-diagrams are taken consistently from the
$NN$ interaction employed.

\section[Importance of the different orbital contributions]
{Importance of the different orbital contributions in the
three-body final states}

\begin{figure}[t!]
\begin{center}
\includegraphics[width=.98\textwidth,keepaspectratio]{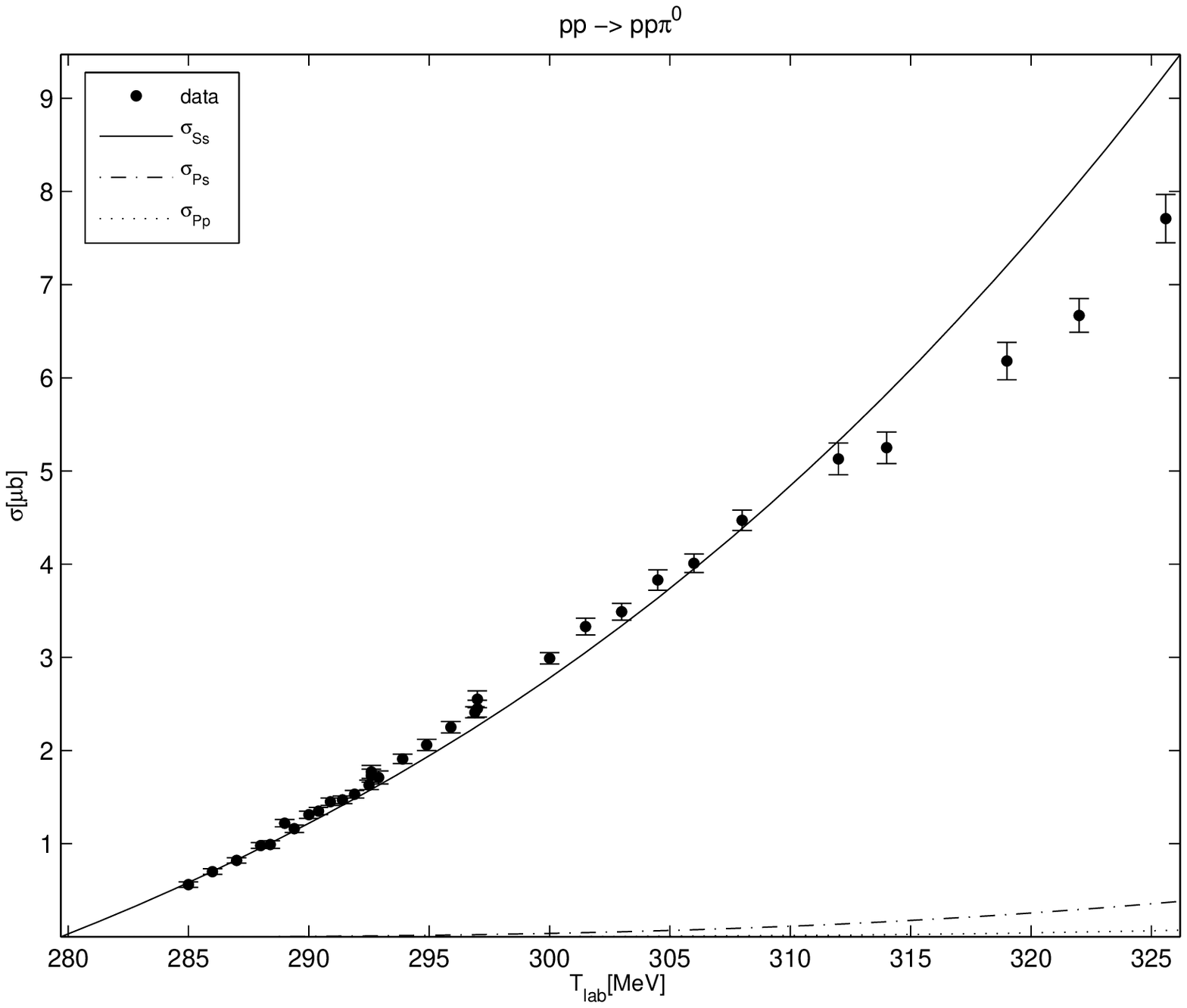}
  \captions[Relative importance of the $\left( NN \right) \pi$ final
   states in $pp \rightarrow pp \pi^{0}$]{Relative importance of the
   $\left( NN \right) \pi$ final states in $pp \rightarrow pp
   \pi^{0}$. The cross sections
   including only the $Ss$, $Ps$ and $Pp$ $\left(NN \right) \pi$
   final states are $\sigma_{Ss}$ (solid line), $\sigma_{Ps}$ (dashed-dotted
   line) and $\sigma_{Pp}$ (dotted line), respectively. The data points are from
\rf{Meyer:1992jt}.} \label{pi0fcptbl}
\end{center}
\end{figure}

\begin{figure}[t!]
\begin{center}
\includegraphics[width=.98\textwidth,keepaspectratio]{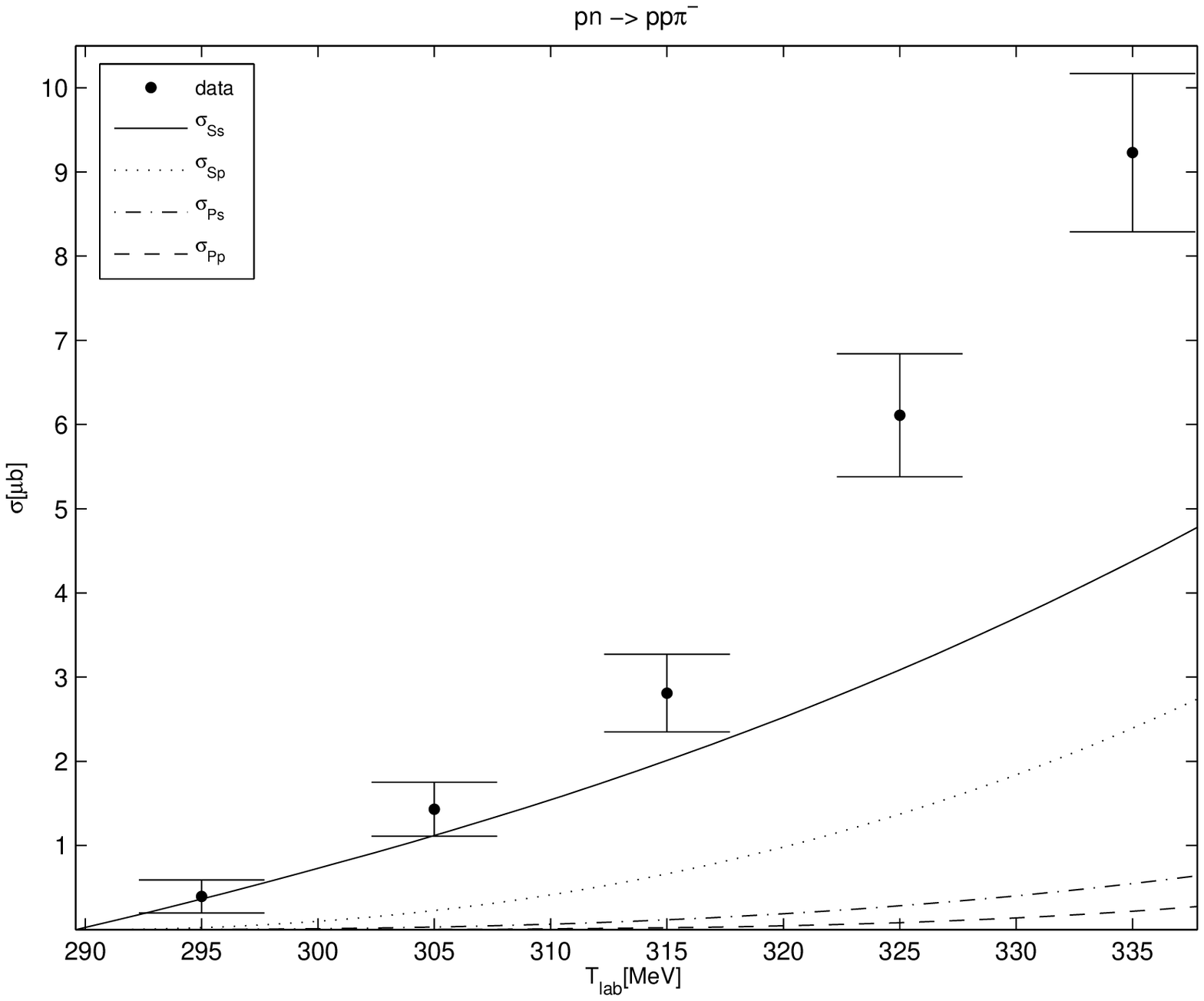}
  \captions[Relative importance of the $\left( NN \right) \pi$ final
   states in $pn \rightarrow pp \pi^{-}$]{Relative importance of the
   $\left( NN \right) \pi$ final states in $pn \rightarrow pp
   \pi^{-}$. The cross sections including only the $Ss$, $Sp$ ,$Ps$ and
   $Pp$ $\left(NN \right) \pi$ final states are $\sigma_{Ss}$ (solid line),
   $\sigma_{Sp}$ (dotted line), $\sigma_{Ps}$ (dashed-dotted line) and
   $\sigma_{Pp}$ (dashed line), respectively. The data
   points are from \rf{Bachman:1995gn}.} \label{pimfcptbl}
\end{center}
\end{figure}

\begin{figure}[t!]
\begin{center}
\includegraphics[width=.98\textwidth,keepaspectratio]{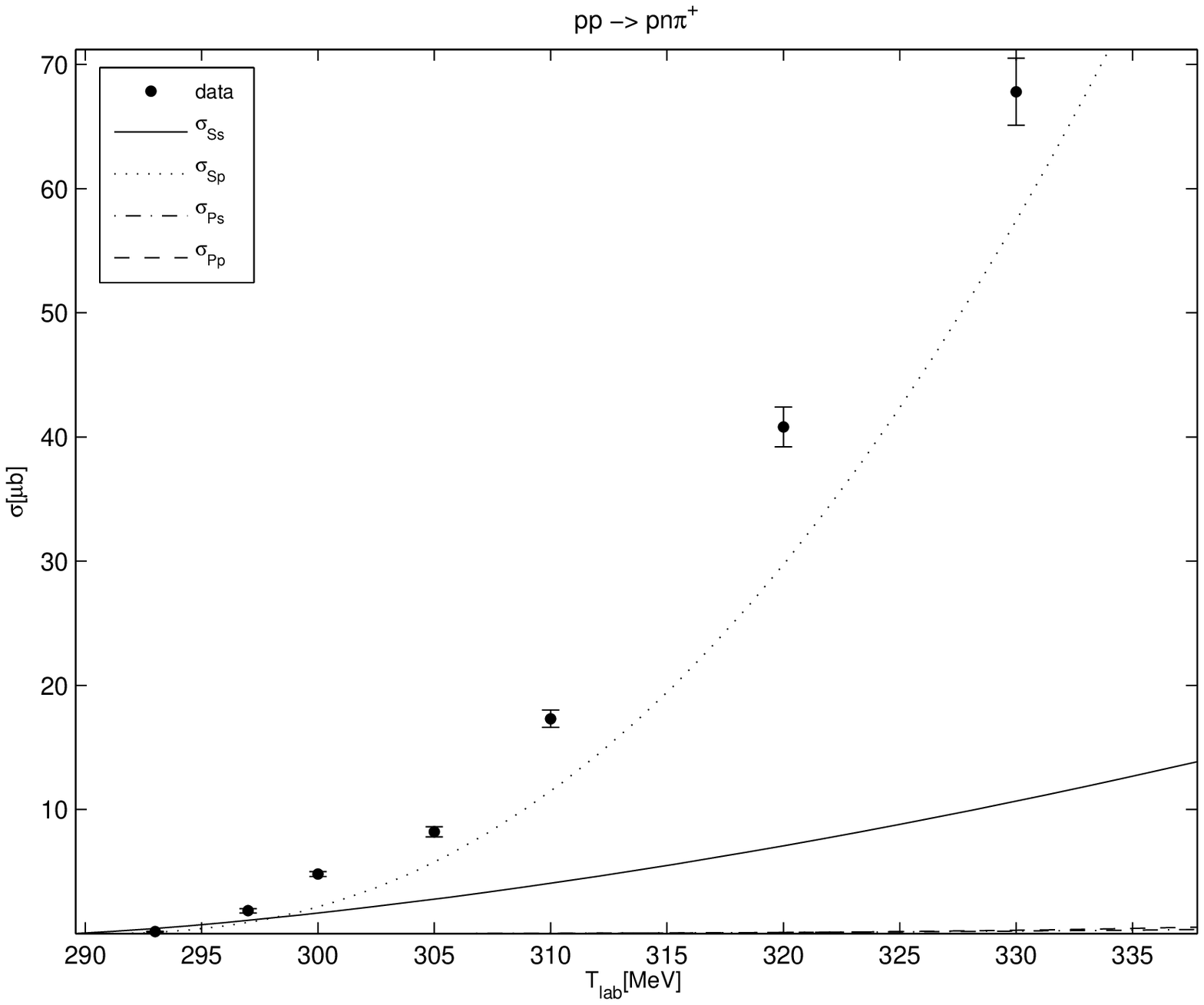}
  \captions[Relative importance of the $\left( NN \right) \pi$ final
  states in $pp \rightarrow pn \pi^{+}$]{Relative importance of the
  $\left( NN \right) \pi$ final states in $pp \rightarrow pn
  \pi^{+}$. The meaning of the lines is the same of \fig{pimfcptbl}.
  The data points are from
  \rf{Hardie:1997mg}.} \label{pipfcptbl}
\end{center}
\end{figure}

%
\sprg
For all the production reactions, the $Ss$ and $Sp$ $\left(NN
\right)\pi$ final states were found to be the dominant ones (solid
line in \fig{pi0fcptbl}, \fig{pimfcptbl} and \fig{pipfcptbl} and
dotted line in \fig{pimfcptbl} and \fig{pipfcptbl}, respectively).

In particular, for $\pi^{0}$ production, the cross section with
only $Ss$ final states (solid line in \fig{pi0fcptbl}) is enough
to describe the data in a wide energy range. For charged-pion
production reactions, the cross section results from an interplay
between the $Ss$ final states (which dominate for $\pi^{-}$
production) and the $Sp$ final states (which dominate for
$\pi^{+}$ production) (solid and dotted line in \fig{pimfcptbl}
and \fig{pipfcptbl}).
\clearpage
\section{Approximations for the energy of the exchanged pion}
%

\sprg
In this section we start by showing that for all charged pion
reaction channels the S-matrix approach for the effective
operators, which describe the physical meson production
mechanisms, is a very good approximation to the DWBA result
obtained directly from TOPT.

\begin{figure}[t!]
\begin{center}
\includegraphics[width=1.01\textwidth,keepaspectratio]{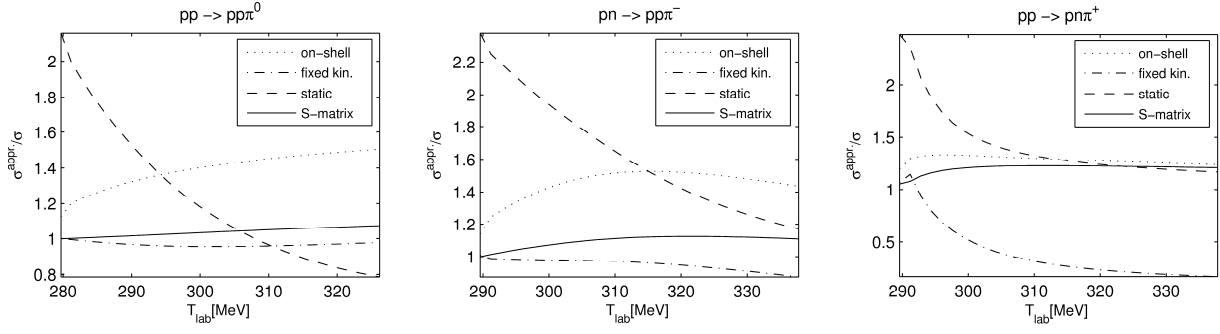}
  \caption[Comparison of the cross sections with different energy
  prescriptions for the production operator (on-shell, fixed kinematics, static and the
  S-matrix) to the TOPT result]
  {Ratio between the cross section with different energy prescriptions
   and the TOPT cross section.
   The dotted, dashed-dotted and dashed line corresponds on-shell approximation,
   fixed-kinematics and static approximation, respectively. The solid line is the
   calculation within the S-matrix approach. The Ohio interaction is used for the
   $NN$ distortions.} \label{pifcptbepr}
\end{center}
\end{figure}

We compare in \fig{pifcptbepr} the result from time-ordered
perturbation theory to the total cross sections of the several
charge production reactions obtained with the on-shell,
fixed-kinematics, static and S-matrix approximations.
The validity of these approximations has been investigated
previously\cite{Malafaia:2004cu} but only for the re-scattering
operator and $\pi^0$ production (see \chp{Smatrixapp}).

For all the production reactions, the deviation of the S-matrix
approach result from the TOPT cross section is much less than the
deviations obtained with the other approximations. The maximum
deviation, on the other hand, occurs for the static approximation
(as checked in \chp{Smatrixapp}, at variance, below threshold, as
expected, the static approximation performs better, giving a basis
for the traditional static interactions at low energies.)

\begin{figure}[h!]
\begin{center}
\includegraphics[width=.98\textwidth,keepaspectratio]{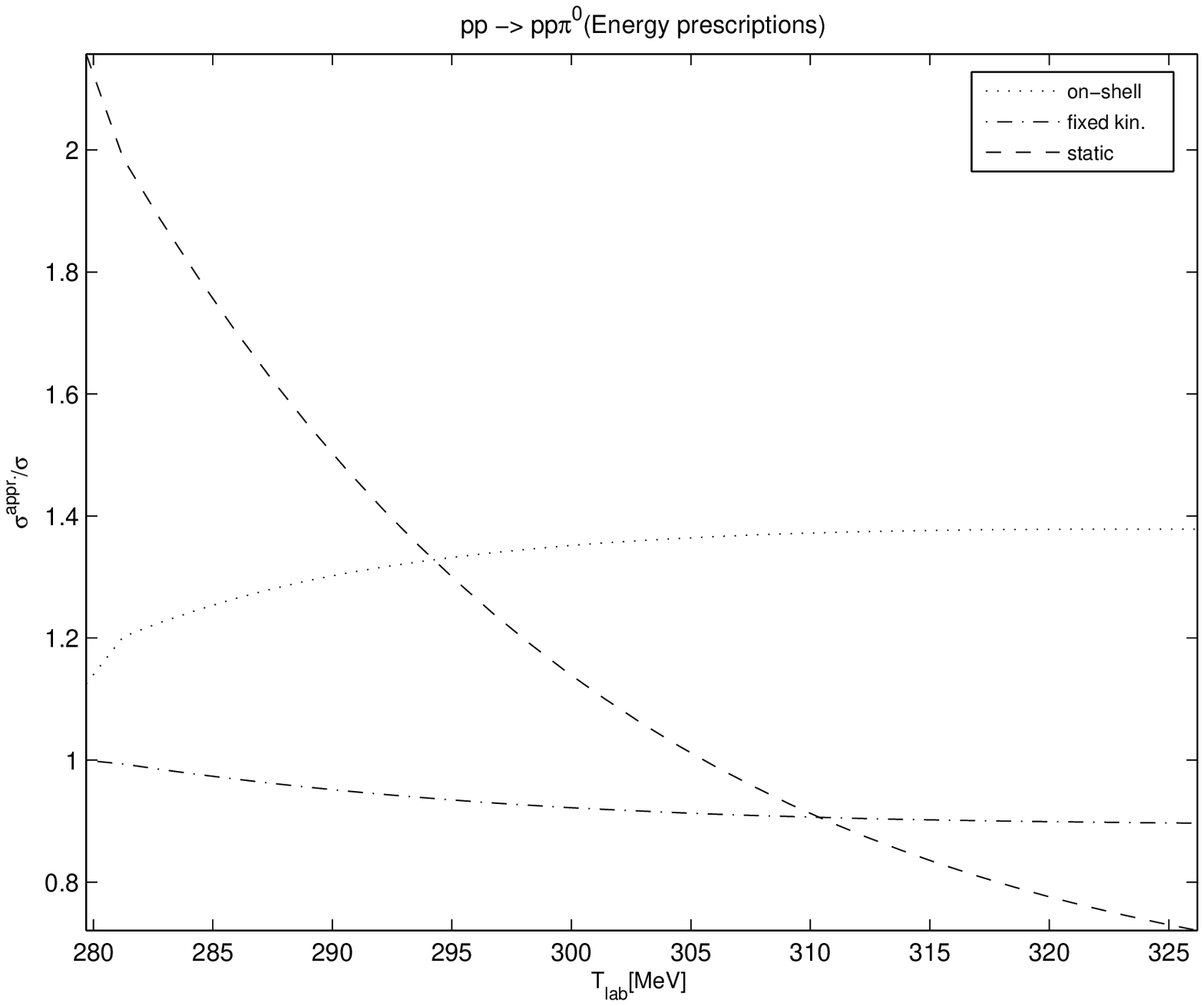}
  \captions[Effect of the approximations for the energy of the
   exchanged pion on the cross section for $\pi^{0}$ production]
   {Ration between the cross section for $pp \rightarrow pp\pi^{0}$
   with different energy prescriptions for the
   production operator and the S-matrix (coinciding with the ``exact" TOPT result,
   as shown in \fig{pifcptbepr})
   cross section.
   The dotted, dashed-dotted and dashed line corresponds on-shell approximation,
   fixed-kinematics and static approximation, respectively.} \label{pi0fcptbepr}
\end{center}
\end{figure}

\begin{figure}[h!]
\begin{center}
\includegraphics[width=.98\textwidth,keepaspectratio]{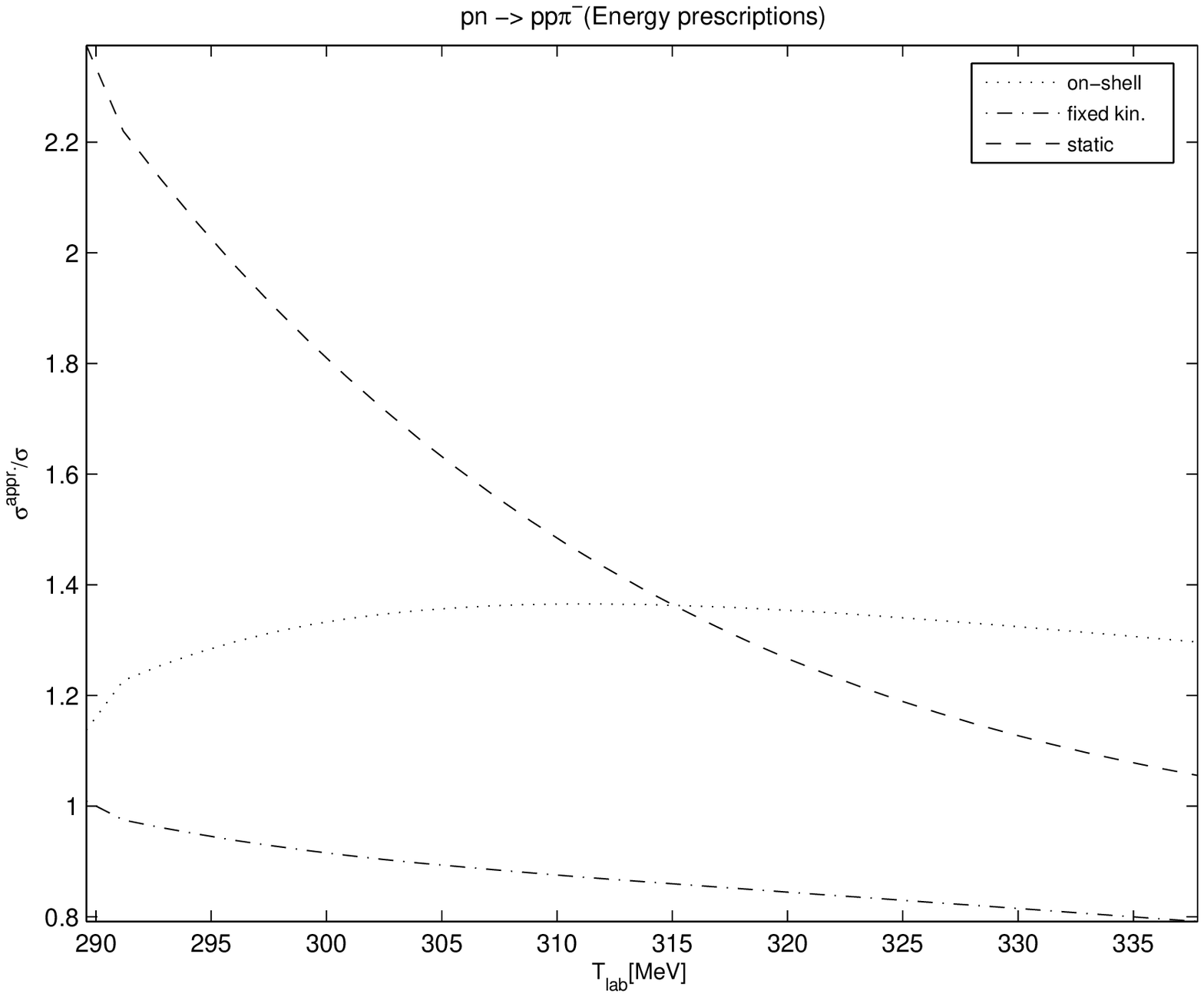}
  \captions[Effect of the approximations for the energy of the
   exchanged pion on the cross section for $\pi^{-}$ production]{The same of
   \fig{pi0fcptbepr} but for $pn \rightarrow pp \pi^{-}$.}
\label{pimfcptbepr}
\end{center}
\end{figure}

\begin{figure}[h!]
\begin{center}
\includegraphics[width=.98\textwidth,keepaspectratio]{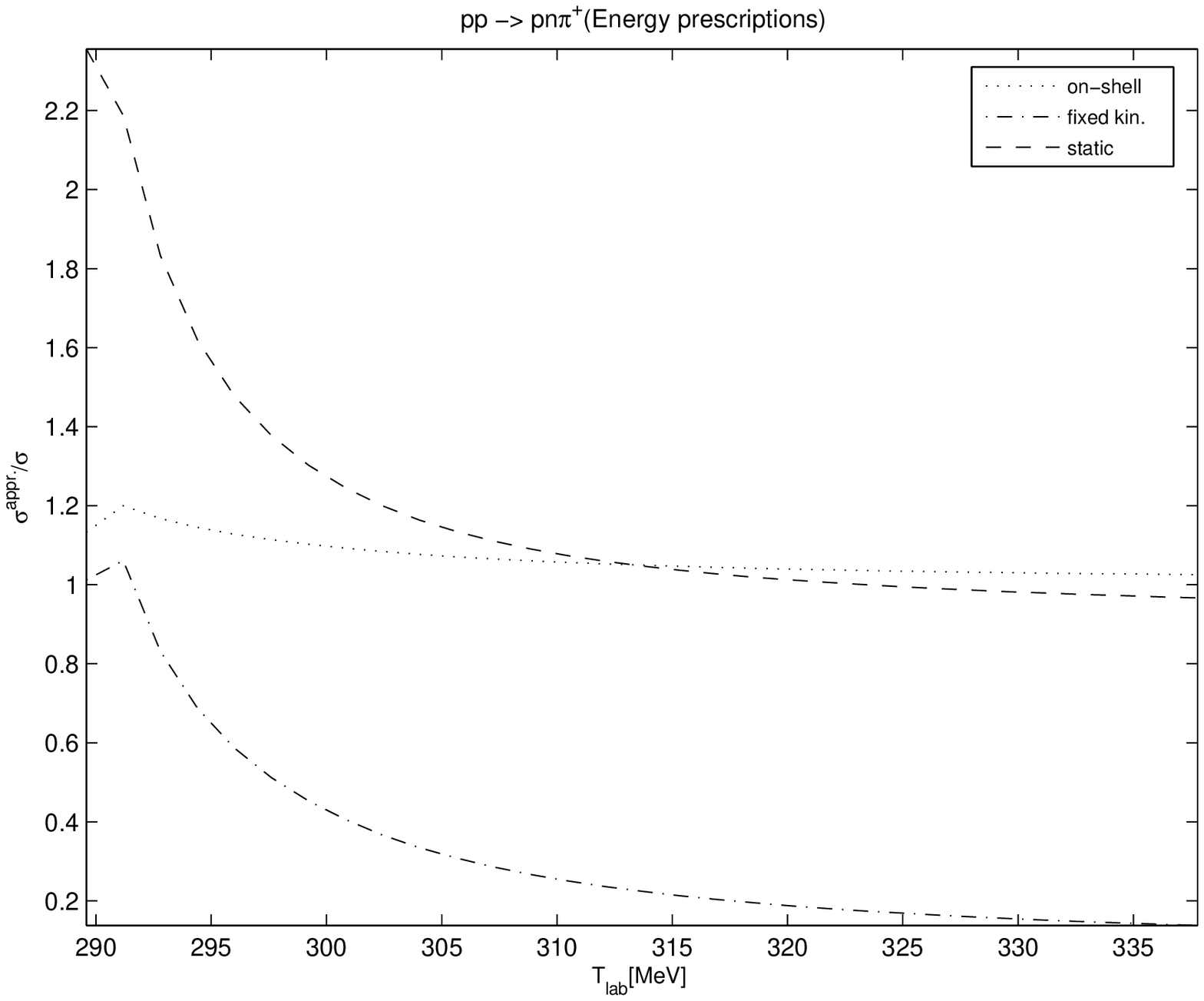}
\captions[Effect of the approximations for the energy of the
exchanged pion on the cross section for $\pi^{+}$ production]{The
same of \fig{pi0fcptbepr} but for $pp \rightarrow pn \pi^{+}$. }
\label{pipfcptbepr}
\end{center}
\end{figure}

We compare in \fig{pi0fcptbepr}, \fig{pimfcptbepr} and
\fig{pipfcptbepr} the effects of the on-shell, fixed-kinematics
and static approximations on the total cross section to the result
within the S-matrix approach. As expected, these results are in
agreement with those shown in \fig{pifcptbepr}.

\newpage

The deviation from the cross section calculated within the
S-matrix approach is maximum for the static approximation, which
considers the energy of the exchanged pion to be zero (dashed
line).
For energies close to threshold, the static approximation
overestimates the cross section by a factor larger than $2$.
The on-shell approximation (``$E-E'$" approximation, which
considers the energy of the exchanged pion to be the difference
between the nucleon on-shell energies before and after pion
emission) deviates also from the reference result, but it may be
off $20 \%$ for all the charge channels, even for energies close
to threshold.

For the $\pi^{0}$ and $\pi^{-}$ production cases
(\fig{pi0fcptbepr} and \fig{pimfcptbepr}, respectively), about the
same happens for the fixed kinematics approximation (energy of the
exchanged pion set to its threshold value, $m_{\pi}/2$) as seen in
the dashed-dotted lines.
As for ${\pi}^{+}$ production, represented in \fig{pipfcptbepr},
the cross section within the fixed-kinematics approximation is
underestimated by a factor of $1.5-2.5$ in the region near
threshold, and, for higher energies, by a factor of $\sim 5$.

\begin{figure}
\begin{center}
\includegraphics[width=.98\textwidth,keepaspectratio]{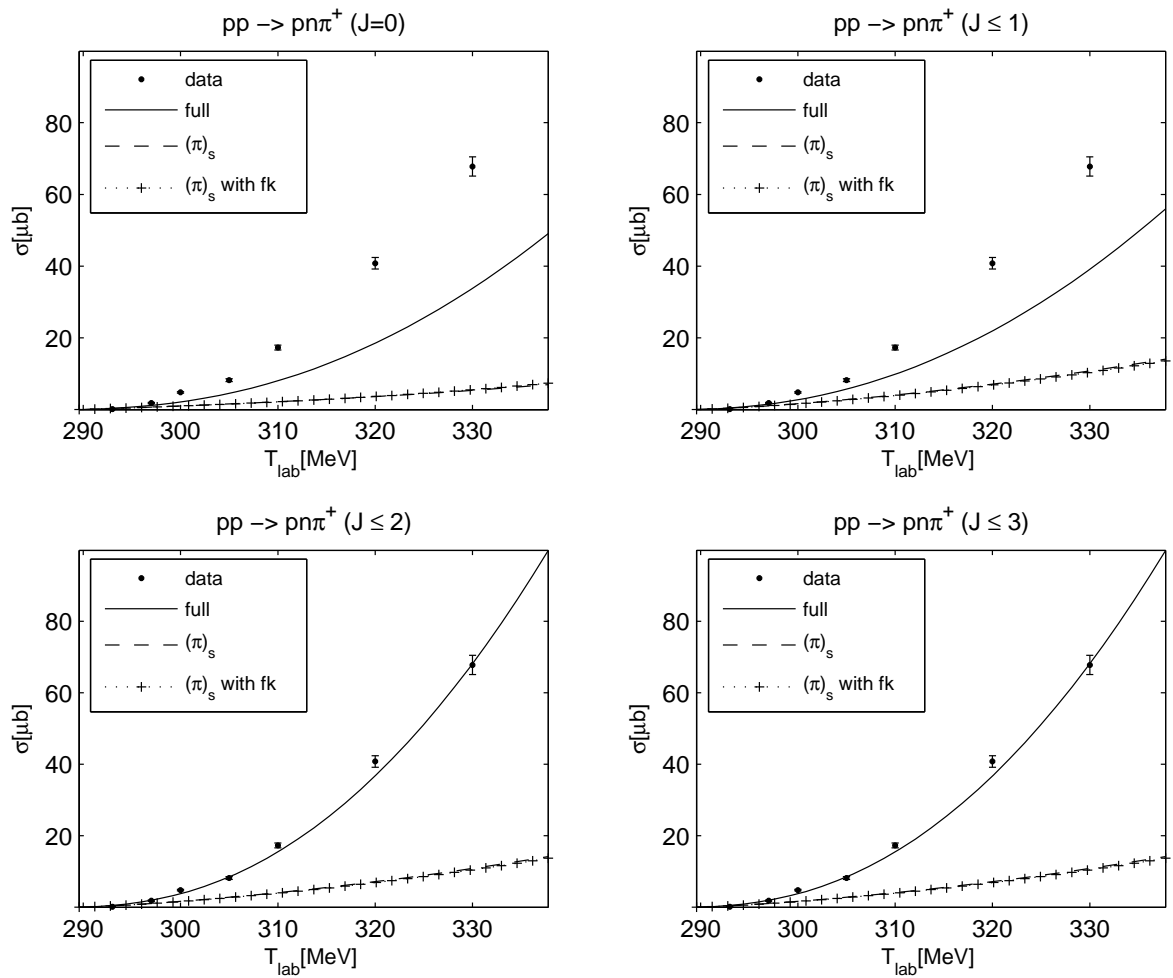}
  \caption[Dependence of the effect of the fixed kinematics approximation for the cross section
  for $pp \rightarrow pn \pi^{+}$ on the orbital $\pi\left(NN \right)$ angular momentum]
  {Dependence of the effect of the fixed kinematics approximation for the cross section
  for $pp \rightarrow pn \pi^{+}$ on the orbital $\pi\left(NN \right)$ angular momentum.
  The solid line is the full calculation.
  The dashed line is the cross section considering only $s$-wave contributions
  to the $\pi N$ amplitude. The dotted line with +'s further assumes the fixed kinematics
  approximation. The Ohio interaction is used for the
   $NN$ distortions.} \label{pipswave}
\end{center}
\end{figure}

Our results for $\pi^{+}$ production are consistent with
discrepancy with the experimental data reported in
\rf{daRocha:1999dm}, where the calculation also used the $\chi$PT
$\pi$N amplitude, but the fixed kinematics approximation was
assumed\footnote{Also, only two $NN$ channels were considered,
namely the $^{3}P_{1} \rightarrow ^{3}S_{1}$ and $^{3}P_{0}
\rightarrow ^{1}S_{0}$. The Argonne V18 and the Reid93 potentials
were employed for the $NN$ interaction.}, leading to a calculated
cross section which underestimated  the data by a factor of $5$ or
by a factor of $2$, depending on the $\Delta$ contribution being
included or not included\cite{daRocha:1999dm}. We verified these
reduction factors, when the same channels were introduced in our
calculation. We furthermore concluded that the deviations observed
for the $\pi^+$ production reaction are originated by the
$p$-waves contributions to the $\pi$N amplitude, as illustrated in
\fig{pipswave}.

In other words, the crucial role of the $\Delta$ in $\pi^{+}$
production enhances the importance of the $\pi N$ $p$-wave states
resulting in a slower convergence and in a higher sensitivity to
the fixed kinematics approximation.

\clearpage

\section{Conclusions}

\sprg
This Chapter investigated charged and neutral pion production
reactions within the S-matrix approach. The main conclusions are:

1) the S-matrix approach for the description of the pion
production operators reproduces numerically the DWBA (TOPT)
result\footnote{As this thesis was being concluded, an
investigation (arXiv:nucl-th/0511054) identified  within $\chi$PT
the irreducible contributions of loop diagrams to be included
consistently with the distorted $NN$ wave functions. That
reference provides the general recipe of calculating the energy
dependence of the vertices by using its corresponding on-shell
value, in agreement with the optimal prescription that we found
numerically in this work.}.
We note that the S-matrix approach simplifies tremendously the
numerical effort demanded in TOPT by the presence of logarithmic
singularities in the pion propagator for ISI,

2) the contributions from the impulse, re-scattering,
$\Delta$-isobar and $Z$-diagrams, within the S-matrix approach
successfully describe the cross sections not only for neutral but
also for charged pion production, near threshold,

3) the newly developed realistic Ohio $NN$ interaction tuned well
above pion production threshold was found to improve the
description of the experimental data. For $\pi^{+}$ production,
its effect is crucial to reproduce the data in the near-threshold
region,

4) the reason for previously reported\cite{daRocha:1999dm}
failures to describe the $pp \rightarrow pn \pi^{+}$ reaction
with the $\chi$PT $\pi$N amplitude (giving strengths for the cross
section from $1/5$ to $1/2$ of the experimental one) stems from
the approximation taken then for the energy of the exchanged pion,

5) convergence of the partial wave series for neutral pion
production, and for energies up to $30\umev$ above threshold, does
not require channels with total angular momentum higher than
$J=0$. However, for the charged $\pi^-$ and $\pi^+$ production
cases and at the same energies, channels with $J=1$ and $J=2$ at
least are needed, due to the importance of $p$ waves in the $\pi
N$ and $NN$ interactions, and to the admixtures of both $T=0$ and
$T=1$ $NN$ isospin states.

We add that a more detailed description of the experimental data
at higher energies may further require a complete coupled-channel
$N \Delta$ calculation. The S-matrix approach, shown here to be
successful near threshold, should naturally be tested also by the
more sensitive polarisation observables and, above all, in high
precision calculations, as for instance in charge symmetry
breaking studies.

%% file: Chapter6.tex
\setcounter{minitocdepth}{2}

\chapter{Conclusions} \label{Conclusions}


\sprg
Understanding pion production, in all its possible charge
channels, in nucleon-nucleon collisions near threshold has been a
challenge over the last decades. With the experimental data from
the new cooler rings, the field entered in a new domain of
precision. These unprecedented high quality data on unpolarised,
as well as on polarised observables, triggered a plethora of
theoretical investigations.

\sp

The physics of meson production in nucleon-nucleon collisions is
very rich. As the first strong inelasticity for the $NN$ system,
the phenomenology of the $NN \rightarrow NN \pi$ reaction is
closely linked to that of elastic $NN$ scattering. The fact that
low- and medium- energy strong interactions are controlled by
chiral symmetry led the hope that chiral effective theories could
be used to analyse these processes, and achieve a fundamental
understanding of the production processes. This provides the
opportunity to improve the phenomenological approaches via
matching the chiral expansion and inversely, to constrain the
chiral contact terms via resonance saturation. This early
excitement was quickly abated by the realisation that proper
evaluation within $\chi$PT involves surmounting several severe
difficulties, which are caused by the high-momentum transfer
nature of this process.

\sprg

The cross section for the reaction $pp \rightarrow pp \pi^{0}$
near threshold was especially elucidative, since the main pion
exchange contribution is there ruled out by isospin conservation.
It clearly showed that the transition amplitude results from a
delicate interference between the single-nucleon term, the pion
re-scattering term, and various additional and individually
important contributions from shorter range mechanisms
(``Z-diagrams" from heavy meson exchanges). To establish the
importance of the latter in an overall description of the
different charge channels of pion production, was the main
motivation for this work.

\sp

DWBA approaches apply a three-dimensional quantum-mechanical
formulation for the $NN$ distortion, which is not obtained from
the Feynman diagrams describing the production mechanisms. As a
consequence, the energy dependence of the pion production operator
has been treated approximately and under different prescriptions
in calculations performed so far. A clarification of these formal
issues was thus needed to enable sound conclusions about the
physics of the pion production processes, and it was the aim of
Chapters \ref{FTtoDWBA} and \ref{Smatrixapp}.

\sp

Moreover, advance in understanding pion production in $NN$
collisions must follow not from the exclusive study of $pp
\rightarrow pp\pi^{0}$, but also from the global understanding of
the effects that afflict all channels. Also, to pin down the
different production mechanisms, one needs to go beyond the lowest
partial waves. These were the main motivations for the work of
Chapters \ref{Smatrixapp} and \ref{Chargedandneutral}.

\sp

In short, this thesis aimed to address the problem of charged and
neutral pion production in nucleon-nucleon collisions. The main
conclusions are:
\begin{enumerate}
\item \textbf{From field theory to DWBA}

\mp The effective pion re-scattering operator was obtained
starting from the corresponding four-dimensional Feynman diagram.
By integrating over the energy of the exchanged pion, the
amplitudes were transformed into those following from time-ordered
perturbation theory and could be identified with six time-ordered
diagrams.

\mp From these diagrams, the ``stretched box diagrams'' (i.e.,
those with more than one meson in flight in the intermediate
states), were seen quantitatively to give a very small
contribution, and therefore were neglected. The remaining diagrams
were identified with the distorted-wave Born approximation (DWBA)
amplitude, allowing to extract an effective production operator.
The calculations considered a physical realistic model for pions
and nucleons and their interaction vertices.
\item \textbf{Energy prescriptions for the production operator}

\mp The DWBA amplitude originated by the decomposition of the
Feynman diagram into the corresponding time-ordered perturbation
theory (TOPT) terms, was used as a reference result to study the
commonly used approximations (\textit{on-shell},
\textit{fixed-kinematics} and \textit{static}) for the energy of
the exchanged pion. This energy, not fixed in the
quantum-mechanical three-dimensional, non-relativistic formalism,
was fixed as $\omega_{\pi}$, the on-shell energy of the exchanged
pion.

\mp Near threshold, the retardation effects in the pion
re-scattering mechanism were seen to be not very decisive. On the
other hand, the effect on the re-scattering vertex alone was found
to be significant, largely overestimating the cross section, as
unnaturally obtained in previous calculations with ad-hoc choices
for the pion energy.
\item \textbf{The S-matrix approach}

\mp When applying TOPT, each of the re-scattering diagrams for the
initial- and final-state distortion defines a different off-energy
shell extension of the pion re-scattering amplitude, since energy
is not conserved at individual vertices.
This imposes the evaluation of two different matrix elements
between the quantum-mechanical $NN$ wave functions, for such
operators.

\mp Within the S-matrix technique, however, after the on-shell
approximation is made consistently, the pion re-scattering parts
of FSI and ISI diagrams coincide and one can identify them with a
single effective re-scattering operator.

\mp The amplitudes and cross sections obtained with the S-matrix
effective operator were verified to be very close to those
calculated with the ``exact" TOPT result, up to $150\umev$ above
threshold.
\item \textbf{Charged and neutral pion production}

\mp If one considers the contributions from the single-nucleon
emission term, re-scattering, $\Delta$-isobar and $Z$-diagrams,
the S-matrix approach successfully describes the cross sections
for both
neutral and charged pion production, near threshold. \\
The S-matrix approach reproduces numerically the results from
TOPT.

\mp For neutral (charged) pion production, the summation of all
channels up to total angular momentum $J=0$ ($J=2$) is enough up
to $30\umev$.

\mp The $NN$ interaction was found to be decisive to reproduce the
data in the near-threshold region (in particular for $\pi^{+}$
production).
In this channel we found also that the correct treatment of the
exchanged pion energy in the re-scattering term beyond the fixed
kinematics approximation, is essential to remove the previously
reported too low cross section strength.

\mp In the overall, even for energies close to threshold, the
cross sections with the approximations for the energy of the
exchanged pion other than the S-matrix, deviate at least $20 \%$
from the full calculation, being larger for the $\pi^{+}$ channel.
\end{enumerate}

\sp

Our findings for the cross section are in agreement with those of
the J\"{u}lich model.  However, contrarily to this approach, for
which all the short range mechanisms are included only through
$\omega$-exchange and adjusted to reproduce the total cross
section of the reaction $pp \rightarrow pp \pi^{0}$ close to
threshold, in our calculations no adjustment is made, as the
parameters for the $Z$-diagrams are taken consistently from the
$NN$ interaction employed, providing a good description of all
charge channels.

\sp

There are several issues which would undoubtedly be a step further
in the investigations of this thesis. Since the importance of the
$\Delta$ contribution increases with energy, a detailed
description of the experimental data in regions of higher energy
may further require a complete coupled-channel $N \Delta$
calculation. Also, although the effect of other higher mass
nucleon resonances was found to be small, these mechanisms should
also be considered in future calculations.

\sp

Moreover, the S-matrix approach, which successfully describes the
near-threshold cross sections for neutral and charged pion
production, should be submitted to the most stringent test of
describing pion (and other mesons) polarisation observables.
Finally, since the S-matrix approach passed so well the test of
pion production, it is natural to extend the method to the $\eta$
meson sector.

\sp

In recent years, the improvement of experimental facilities,
detectors and accelerators permitted the acquisition of a great
deal of high-precision data, leading to the improvement of its
theoretical understanding (both in $\chi$PT and in other
phenomenological approaches), and are a promise of very exciting
times to come in the near future of meson production physics,
where for instance, $\eta$ and $\eta^{\prime}$ production still
pose interesting problems.

%% file: Appendix_A.tex
\clearpage\chapter{General remarks on kinematics
\label{Apkinematics}}

\begin{figure}[h!]
\begin{center}
  \includegraphics[width=.29\textwidth,keepaspectratio]{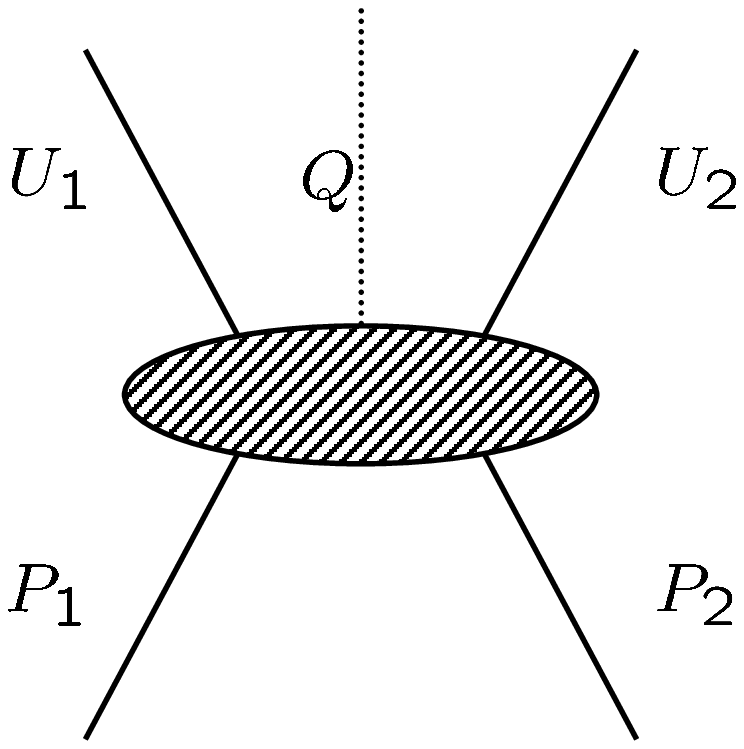}
  \captions[Kinematics: illustration of the choice of variables for $NN \rightarrow NN \pi$]
  {Illustration of the choice of variables for $NN \rightarrow NN \pi$. The solid lines
  and dashed lines represent the nucleon and pion, respectively. The initial(final)
  nucleon four-momenta are $P_{1}$ and
  $P_{2}$$\left(U_{1} \text{ and } U_{2}\right)$. The four-momentum of the produced pion is $Q$.}
  \label{kinconventions}
\end{center}
\end{figure}

\fig{kinconventions} illustrates the choice of variables adopted
in this work.
\section*{Final and initial momenta}
The initial nucleon four-momenta in the overall centre of mass
system are given by
\begin{eqnarray}
P_{1}&=&\left(E_{1},\vec{p} \right) \\
P_{2}&=&\left(E_{2},-\vec{p} \right),
\end{eqnarray}
with $E_{1,2}=\sqrt{M^{2}+p^{2}}\equiv E$.
%
For energies close to the pion production threshold, the
non-relativistic expressions for the energy of the final nucleons
can be used.
Thus, in the centre-of-mass of the final three-body system,
\begin{eqnarray}
U_{1}&=&\left(F_{1},\vec{q}_{u}-\frac{\vec{q}_{\pi}}{2} \right) \\
U_{2}&=&\left(F_{2},-\vec{q}_{u}-\frac{\vec{q}_{\pi}}{2} \right),
\end{eqnarray}
where $\vec{q}_{u}$ is the relative momentum of the final nucleons
and $\vec{q}_{\pi}$ is the centre-of-mass momentum of the produced
pion. From energy conservation it follows that
\begin{eqnarray}
2E& =       &F_{1}+F_{2}+E_{\pi} \\
  & \approx &2M + \frac{q_{u}^{2}}{2 \mu}+\frac{q_{\pi}^{2}}{4 M}+
  \sqrt{m_{\pi}^{2}+q_{\pi}^{2}}.
\end{eqnarray}
\section*{Laboratory energy and maximum pion momentum}
For the kinematics of the pion production reactions investigated
in this work, the nucleons were considered as having equal masses
($M$). The laboratory energy can then be determined from
\begin{equation}
T_{lab}=\frac{s-4M^{2}}{2M}, \label{ApTlab}
\end{equation}
with $ s \equiv \left(P_{1}+P_{2} \right)^{2}$.
The excess energy $Q$ is given by
\begin{equation}
Q=\sqrt{s}-\left(2M+m_{\pi} \right) \label{ApQ}.
\end{equation}
The maximum pion momentum, which occurs when $q_{u}=0$, can be
calculated (in units of $m_{\pi}$) using
\begin{equation}
\eta=\frac{1}{m_{\pi}}\left[\frac{1}{2 \sqrt{s}}\sqrt{\left(
s-m_{\pi}^{2}-4M^{2}\right)^{2}-16M^{2}m_{\pi}^{2}} \right].
\label{Apeta}
\end{equation}
Eqs.~(\ref{ApTlab}), (\ref{ApQ}) and (\ref{Apeta}) can be easily
generalised for the case when the nucleons have different
masses\cite{Moskal:2002jm}:
\begin{equation}
T_{lab}=\frac{s-M_{i1}^{2}-M_{i2}^{2}-2M_{i1}M_{i2}}{2 M_{i1}},
\end{equation}
\begin{equation}
Q=\sqrt{s}-\left(M_{f1}+M_{f2}+m_{\pi} \right),
\end{equation}
and
\begin{equation}
\eta=\frac{1}{m_{\pi}}\left[\frac{1}{2 \sqrt{s}}\sqrt{\left(
s-m_{\pi}^{2}-M^{2}_{f}\right)^{2}-4M^{2}_{f}m_{\pi}^{2}} \right].
\end{equation}
where the masses of the initial (final) nucleons are $M_{i1}$ and
$M_{i2}$ ($M_{f1}$ and $M_{f2}$), and $M_{f}=M_{f1}+M_{f2}$.

\sp

It is important to remark that the relativistic and non
relativistic calculations of the threshold laboratory energy
$T_{thr.}$ differ significantly,
\begin{eqnarray}
T_{thr.}^{Rel.} &  =& \frac{4M^{2}+4Mm_{\pi}+m_{\pi}^{2}-4M^{2}}{2M}\\
&  =&2m_{\pi}+\left(  \frac{m_{\pi}^{2}}{2M}\right)  \\
&  =&T_{thr.}^{NRel.}+\left(  \frac{m_{\pi}^{2}}{2M}\right),
\end{eqnarray}
since $\left(  \frac{m_{\pi}^{2}}{2M}\right) \approx 10\umev$.
This difference can have decisive effects on the results,
especially at threshold.

%% file: Appendix_B.tex
\clearpage\chapter{Kinematic definitions for the diagrams
\label{Apkindiagrams}}
%
%
\section*{Re-scattering diagram}
\begin{figure}[h!]
\begin{center}
  \includegraphics[width=.85\textwidth,keepaspectratio]{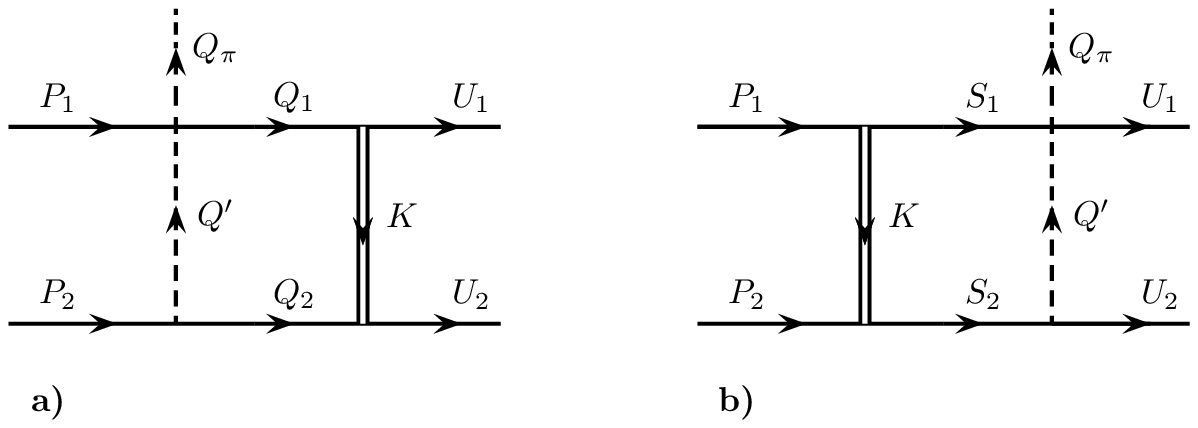}
  \captions[Kinematics for the re-scattering diagram]
  {Illustration of the choice of variables for the re-scattering diagram for
  a) FSI and b) ISI. The solid lines
  and dashed lines represent the nucleon and pion, respectively.
  The initial(final) nucleon four-momenta are $P_{1}$ and
  $P_{2}$$\left(U_{1} \text{ and } U_{2}\right)$.
  The four-momentum of the produced pion is $Q_{\pi}$.
  The four-momentum of the exchanged pion($\sigma$ meson) is $Q^{\prime}$$\left(K\right)$.
  The four-momenta of the intermediate nucleons for the FSI(ISI) case
  are $Q_{1} \text{ and } Q_{2}$$\left(S_{1} \text{ and } S_{2}\right)$.}
  \label{kinconvrescatt}
\end{center}
\end{figure}

%
The kinematic conventions adopted for the re-scattering diagram
are represented on \fig{kinconvrescatt}. The relations resulting
from four-momentum conservation are listed on \tb{Fourmomrescatt}.
\begin{table}
\begin{center}
\begin{tabular}[c]{cc}
\hline\hline \hspace{-0.85cm}FSI & ISI\\\hline
\multicolumn{1}{l}{$P_{1}+Q^{\prime}-Q_{\pi}-Q_{1}=0\hspace{0.85cm}$}
&
\multicolumn{1}{l}{$P_{1}-K-S_{1}=0$}\\
\multicolumn{1}{l}{$P_{2}-Q^{\prime}-Q_{2}=0$} & \multicolumn{1}{l}{$P_{2}%
+K-S_{2}=0$}\\
\multicolumn{1}{l}{$Q_{1}-K-U_{1}=0$} &
\multicolumn{1}{l}{$S_{1}+Q^{\prime
}-Q_{\pi}-U_{1}=0$}\\
\multicolumn{1}{l}{$Q_{2}+K-U_{2}=0$} &
\multicolumn{1}{l}{$S_{2}-Q^{\prime }-U_{2}=0$}\\\hline\hline
\end{tabular}
\end{center}
  \captions[Four-momentum conservation for the re-scattering
   diagram]{Four-momentum conservation for the re-scattering diagram.
\label{Fourmomrescatt}}
\end{table}
\newpage

For the final state $NN$ interaction (diagram a) of
\fig{kinconvrescatt}), the expressions for the on-shell energies
of the particles are:
\begin{eqnarray}
&&\omega_{\pi}   =\sqrt{m_{\pi}^{2}+\left\vert
-\vec{p}+\frac{\overrightarrow
{q_{\pi}}}{2}+\overrightarrow{q_{k}}\right\vert ^{2}}\text{\hspace
{1.17cm}exchanged }\pi \label{kinconvfi}\\
&& \omega_{2,1}   =\sqrt{M^{2}+\left\vert \overrightarrow{q_{k}}
\text{$\pm$}
\frac{\overrightarrow{q_{\pi}}}{2}\right\vert ^{2}}\hspace{2cm}%
\text{intermediate nucleons}\\
&& \omega_{\sigma}   =\sqrt{m_{\sigma}^{2}+\left\vert \overrightarrow{q_{k}%
}-\overrightarrow{q_{u}}\right\vert
^{2}}\text{\hspace{2.39cm}}\sigma\text{
exchanged}\\
&& F_{2,1}   =M+\frac{1}{2M}\left\vert \overrightarrow{q_{u}}
\text{$\pm$} \frac {\overrightarrow{q_{\pi}}}{2}\right\vert
^{2}\hspace{1.71cm}\text{final nucleons} \label{kinconvff}%
\end{eqnarray}
and, for the initial $NN$ interaction, the corresponding
expressions are
\begin{eqnarray}
&& \omega _{\pi }=\sqrt{m_{\pi }^{2}+\left| \overrightarrow{q_{u}}+\frac{%
\overrightarrow{q_{\pi }}}{2}-\overrightarrow{q_{k}}\right|
^{2}}\text{\hspace
{1.18cm}exchanged }\pi \label{kinconvii}\\
&& \omega _{1,2}=\sqrt{M^2+\left| \overrightarrow{q_k} \right|^2}\hspace{2.96cm}%
\text{intermediate nucleons}\\
&& \omega _{\sigma }=\sqrt{m_{\sigma }^{2}+\left| \overrightarrow{p}-%
\overrightarrow{q_{k}}\right|
^{2}}\text{\hspace{2.38cm}}\sigma\text{
exchanged}\\
&& F_{2,1}=M+\frac{1}{2M}\left| \overrightarrow{q_{u}}\text{$\pm$} \frac{\overrightarrow{%
q_{\pi }}}{2}\right| ^{2} \hspace{1.70cm}\text{final nucleons} \label{kinconvif}%
\end{eqnarray}
where $\vec{p}$ and $\overrightarrow{q_{\pi }}$ are the initial
nucleon three-momentum and the emitted pion three-momentum, respectively.
The relative three-momentum of the two intermediate(final)
nucleons is $\overrightarrow{%
q_{k}}\left( \overrightarrow{q_{u}}\right) $, and the energy of
the emitted pion is $E_{\pi }=\sqrt{m_{\pi }^{2}+\left|
\overrightarrow{q_{\pi }}\right| ^{2}}$.
The nucleon, the pion and the $\sigma$ mass are $M$, $m_{\pi}$ and
$m_{\sigma}$, respectively. All quantities are referred to the
three-body centre-of-mass frame of the $\pi NN$ final state.

\newpage
\section*{Direct-production diagram}
\begin{figure}[h!]
\begin{center}
  \includegraphics[width=.85\textwidth,keepaspectratio]{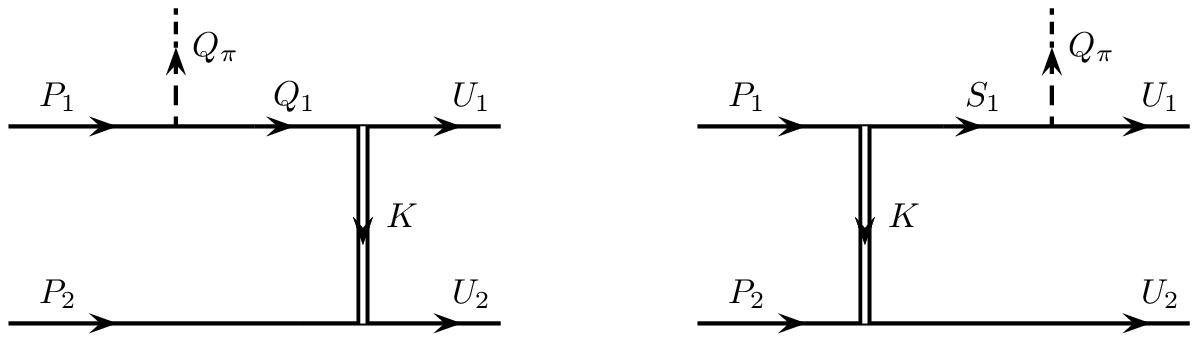}
  \captions[Kinematics for the direct-production diagram]
  {Illustration of the choice of variables for the direct-production diagram for
  a) FSI and b) ISI. The meaning of the variables is the same of \fig{kinconvrescatt}.}
  \label{kinconvimp}
\end{center}
\end{figure}

%
The kinematic conventions adopted for the direct-production
diagram are represented on \fig{kinconvimp}. The relations
resulting from four-momentum conservation are listed on
\tb{Fourmomimp}.
\begin{table}
\begin{center}
\begin{tabular}[c]{cc}
\hline\hline \hspace{-0.85cm}FSI & ISI\\\hline 
\multicolumn{1}{l}{$P_{1}-Q_{\pi}-Q_{1}=0\hspace{0.85cm}$} &
\multicolumn{1}{l}{$P_{1}-K-S_{1}=0$}\\
\multicolumn{1}{l}{$Q_{1}-K-U_{1}=0$} &
\multicolumn{1}{l}{$S_{1}-Q_{\pi}-U_{1}=0$}\\
\multicolumn{1}{l}{$P_{2}+K-U_{2}=0$} &
\multicolumn{1}{l}{$P_{2}+K-U_{2}=0$}\\\hline\hline
\end{tabular}
\end{center}
 \captions[Four-momentum conservation for the direct-production
 diagram]{Four-momentum conservation for the direct-production diagram.
\label{Fourmomimp}}
\end{table}

The expressions for the on-shell energies can be easily obtained
from \eq{kinconvfi}-\eq{kinconvff} and
\eq{kinconvii}-\eq{kinconvif}.

%% file: Appendix_C.tex
\clearpage

\chapter{Partial fraction decomposition and TOPT\label{ApTopt}}
For the loop diagrams considered in this work, we have in general
to evaluate integrals like
\begin{equation}
\mathcal{M}=\int\frac{dQ_{0}^{\prime}}{\left(  2\pi\right)
}f\left(
Q_{0}^{\prime}\right)  \times U_{1}L_{1}\times U_{2}L_{2}\times U_{3}%
L_{3}\label{mloopusls}%
\end{equation}
where $U_{i}$, $L_{i}$ are terms with the poles in the
upper-half plan and in the lower half-plan, respectively,
which can be generically written as:
\begin{equation}
U_{i}=\frac{1}{Q_{0}^{\prime}-u_{i}-i\varepsilon}\hspace{3cm}
L_{i}=\frac {1}{Q_{0}^{\prime}-l_{i}+i\varepsilon},
\end{equation}
and $f\left(  Q_{0}^{\prime}\right)  $ is a generic vertex
function which depends on the momentum of one of the exchanged
particles.

Direct integration of \eq{mloopusls} by closing the contour in one
of the half-plans and picking up the poles enclosed, give as a
result an expression where one has to evaluate the vertex $f$ for
three different energies (the pole contributions) of the exchanged
particle $Q_{0}^{\prime}$. For instance, if we close the contour
in the upper half-plan, $f$ must be calculated for
$Q_{0}^{\prime}=u_{1}$, $Q_{0}^{\prime}=u_{2}$ and
$Q_{0}^{\prime}=u_{3}$. We thus gain little insight on the
adequate choice to be done for $Q_{0}^{\prime}$ in the traditional
three-dimensional formalisms, where this energy is not fixed.

However, by performing a partial fraction decomposition of the
integrand before integrating in $Q_{0}^{\prime}$, provides us a
visible connection with time-ordered perturbation theory. The way
the partial fraction decomposition is done ensures that we have
only contributions from terms with one pole on the on-shell energy
of the particle interacting through $f\left( Q_{0}^{\prime}\right)
$ and the remaining one or two poles in the other half-plan.
Closing the contour in the half-plan where there is only one pole,
we can relate directly $Q_{0}^{\prime}$ at the vertex with the
on-shell energy of the considered particle. Of course, the final
result must not depend on the method adopted for integration.

\bigskip

Let $u_{3}$ and $l_{3}$ be the poles we want to isolate. We start
by writing the product $U_{3}L_{3}$ as a difference of two terms
\begin{equation}
U_{3}L_{3}=\alpha_{33}\left(  U_{3}-L_{3}\right)  \hspace{1cm}\text{with}%
\hspace{1cm}\alpha_{33}\equiv\frac{1}{u_{3}-l_{3}}\text{,}%
\end{equation}
and doing the same for the terms concerning the other exchanged
particle, $U_{2}L_{2}$:
\begin{equation}
U_{2}L_{2}=\alpha_{22}\left(  U_{2}-L_{2}\right)  \hspace{1cm}\text{with}%
\hspace{1cm}\alpha_{22}\equiv\frac{1}{u_{2}-l_{2}}%
\end{equation}
Then \eq{mloopusls} reads
\begin{eqnarray}
\mathcal{M} &  = &\int\frac{dQ_{0}^{\prime}}{\left(  2\pi\right)
}f\left(
Q_{0}^{\prime}\right)  \alpha_{22}\alpha_{33}U_{1}L_{1}U_{2}U_{3}%
 -\int\frac{dQ_{0}^{\prime}}{\left(  2\pi\right)  }f\left(
Q_{0}^{\prime
}\right)  \alpha_{22}\alpha_{33}U_{1}L_{1}U_{2}L_{3}\label{malfa12}\\
& & -\int\frac{dQ_{0}^{\prime}}{\left(  2\pi\right)  }f\left(
Q_{0}^{\prime }\right)  \alpha_{22}\alpha_{33}U_{1}L_{1}L_{2}U_{3}
+\int\frac{dQ_{0}^{\prime}}{\left(  2\pi\right)  }f\left(
Q_{0}^{\prime
}\right)  \alpha_{22}\alpha_{33}U_{1}L_{1}L_{2}L_{3}\label{malfa34}%
\end{eqnarray}
Now, in \lneq{malfa12}, we apart $L_{1}U_{2}$ and in
\lneq{malfa34}, we apart $U_{1}L_{2}$,
\begin{eqnarray}
\mathcal{M} &  =&\int\frac{dQ_{0}^{\prime}}{\left(  2\pi\right)
}f\left(
Q_{0}^{\prime}\right)  \alpha_{22}\alpha_{33}\alpha_{21}U_{1}U_{2}%
U_{3}\label{malfaalfa1}\\
&&  -\int\frac{dQ_{0}^{\prime}}{\left(  2\pi\right)  }f\left(
Q_{0}^{\prime
}\right)  \alpha_{22}\alpha_{33}\alpha_{21}U_{1}L_{1}U_{3}\label{malfaalfa2}\\
&&  -\int\frac{dQ_{0}^{\prime}}{\left(  2\pi\right)  }f\left(
Q_{0}^{\prime
}\right)  \alpha_{22}\alpha_{33}\alpha_{21}U_{1}U_{2}L_{3}\label{malfaalfa3}\\
&&  +\int\frac{dQ_{0}^{\prime}}{\left(  2\pi\right)  }f\left(
Q_{0}^{\prime
}\right)  \alpha_{22}\alpha_{33}\alpha_{21}U_{1}L_{1}L_{3}\label{malfaalfa4}\\
&&  -\int\frac{dQ_{0}^{\prime}}{\left(  2\pi\right)  }f\left(
Q_{0}^{\prime
}\right)  \alpha_{22}\alpha_{33}\alpha_{12}U_{1}L_{1}U_{3}\label{malfaalfa5}\\
&&  +\int\frac{dQ_{0}^{\prime}}{\left(  2\pi\right)  }f\left(
Q_{0}^{\prime
}\right)  \alpha_{22}\alpha_{33}\alpha_{12}L_{2}L_{1}U_{3}\label{malfaalfa6}\\
&&  +\int\frac{dQ_{0}^{\prime}}{\left(  2\pi\right)  }f\left(
Q_{0}^{\prime
}\right)  \alpha_{22}\alpha_{33}\alpha_{12}U_{1}L_{1}L_{3}\label{malfaalfa7}\\
&&  -\int\frac{dQ_{0}^{\prime}}{\left(  2\pi\right)  }f\left(
Q_{0}^{\prime
}\right)  \alpha_{22}\alpha_{33}\alpha_{12}L_{2}L_{1}L_{3}\label{malfaalfa8}%
\end{eqnarray}
and finally, in \lneq{malfaalfa2}, \lneq{malfaalfa4}, \lneq{malfaalfa5} and
\lneq{malfaalfa7}, we apart $U_{1}L_{1}$:
\begin{eqnarray}
\mathcal{M} &&  =  \int\frac{dQ_{0}^{\prime}}{\left(  2\pi\right)
}f\left( Q_{0}^{\prime}\right)
\alpha_{22}\alpha_{33}\alpha_{21}U_{1}U_{2}U_{3} \label{malfaaa1} \\
 &&  -\int\frac{dQ_{0}^{\prime}}{\left(  2\pi\right)
}f\left( Q_{0}^{\prime }\right)
\alpha_{22}\alpha_{33}\alpha_{21}\alpha_{11}U_{1}U_{3}+\int
\frac{dQ_{0}^{\prime}}{\left(  2\pi\right)  }f\left(
Q_{0}^{\prime}\right)
\alpha_{22}\alpha_{33}\alpha_{21}\alpha_{11}L_{1}U_{3}\label{malfaaa2}\\
&&  -\int\frac{dQ_{0}^{\prime}}{\left(  2\pi\right)  }f\left(
Q_{0}^{\prime
}\right)  \alpha_{22}\alpha_{33}\alpha_{21}U_{1}U_{2}L_{3}\label{malfaaa3}\\
&&  +\int\frac{dQ_{0}^{\prime}}{\left(  2\pi\right)  }f\left(
Q_{0}^{\prime }\right)
\alpha_{22}\alpha_{33}\alpha_{21}\alpha_{11}U_{1}L_{3}-\int
\frac{dQ_{0}^{\prime}}{\left(  2\pi\right)  }f\left(
Q_{0}^{\prime}\right)
\alpha_{22}\alpha_{33}\alpha_{21}\alpha_{11}L_{1}L_{3}\label{malfaaa4}
\bkarr
&&  -\int\frac{dQ_{0}^{\prime}}{\left(  2\pi\right)
}f\left( Q_{0}^{\prime }\right)
\alpha_{22}\alpha_{33}\alpha_{12}\alpha_{11}U_{1}U_{3}+\int
\frac{dQ_{0}^{\prime}}{\left(  2\pi\right)  }f\left(
Q_{0}^{\prime}\right)
\alpha_{22}\alpha_{33}\alpha_{12}\alpha_{11}L_{1}U_{3}\label{malfaaa5}\\
&&  +\int\frac{dQ_{0}^{\prime}}{\left(  2\pi\right)  }f\left(
Q_{0}^{\prime
}\right)  \alpha_{22}\alpha_{33}\alpha_{12}L_{1}L_{2}U_{3}\label{malfaaa6}\\
&&  +\int\frac{dQ_{0}^{\prime}}{\left(  2\pi\right)  }f\left(
Q_{0}^{\prime }\right)
\alpha_{22}\alpha_{33}\alpha_{12}\alpha_{11}U_{1}L_{3}-\int
\frac{dQ_{0}^{\prime}}{\left(  2\pi\right)  }f\left(
Q_{0}^{\prime}\right)
\alpha_{22}\alpha_{33}\alpha_{12}\alpha_{11}L_{1}L_{3}\label{malfaaa7}\\
&&  -\int\frac{dQ_{0}^{\prime}}{\left(  2\pi\right)  }f\left(
Q_{0}^{\prime
}\right)  \alpha_{22}\alpha_{33}\alpha_{12}L_{2}L_{1}L_{3} \label{malfaaa8}%
\end{eqnarray}
where, accordingly to our conventions, $\alpha_{ij}\equiv\frac{1}{u_{i}-l_{j}%
}$.

$\mathcal{M}$ is written as a sum of eight types of terms: terms
with three-poles in the upper-half plan \lneq{malfaaa1} or in the
lower half-plan \lneq{malfaaa8}; with two poles in the upper
half-plan (first term of \lneq{malfaaa2} and \lneq{malfaaa5}) or
in the lower half-plan (second term of \lneq{malfaaa4} and
\lneq{malfaaa7}); with one pole in each half-plan (second term of
\lneq{malfaaa2} and \lneq{malfaaa5} and first term of
\lneq{malfaaa4} and \lneq{malfaaa7}); and finally, terms with two
poles in the upper half-plan and one in the lower half-plan
\lneq{malfaaa3} or vice-versa \lneq{malfaaa6}. The corresponding
schematic is represented on \fig{polestoptfsi}.

Assuming that the integrals with all the poles in the same
half-plan vanish (which happens if  $f\left( Q_{0}^{\prime}\right)
$ is linear in $Q_{0}^{\prime}$), $\mathcal{M}$ is reduced to only
six terms:
\begin{eqnarray}
\mathcal{M} &  = &\int\frac{dQ_{0}^{\prime}}{\left(  2\pi\right)
}f\left(
Q_{0}^{\prime}\right)  \alpha_{22}\alpha_{33}\alpha_{21}\alpha_{11}L_{1}%
U_{3} +\int\frac{dQ_{0}^{\prime}}{\left(  2\pi\right)  }f\left(
Q_{0}^{\prime
}\right)  \alpha_{22}\alpha_{33}\alpha_{12}\alpha_{11}L_{1}U_{3}%
\label{dwbalu12}\\
&& \hspace{-0.4cm} +\int\frac{dQ_{0}^{\prime}}{\left(  2\pi\right)
}f\left( Q_{0}^{\prime
}\right)  \alpha_{22}\alpha_{33}\alpha_{21}\alpha_{11}U_{1}L_{3}%
 +\int\frac{dQ_{0}^{\prime}}{\left(  2\pi\right)  }f\left(
Q_{0}^{\prime
}\right)  \alpha_{22}\alpha_{33}\alpha_{12}\alpha_{11}U_{1}L_{3}%
\label{dwbalu34}\\
&& \hspace{-0.4cm} +\int\frac{dQ_{0}^{\prime}}{\left(  2\pi\right)
}f\left( Q_{0}^{\prime }\right)
\alpha_{22}\alpha_{33}\alpha_{12}L_{1}L_{2}U_{3}
-\int\frac{dQ_{0}^{\prime}}{\left(  2\pi\right)  }f\left(
Q_{0}^{\prime
}\right)  \alpha_{22}\alpha_{33}\alpha_{21}U_{1}U_{2}L_{3}\label{stretlu56}%
\end{eqnarray}
These terms can be directly identified with the ones resulting
from TOPT. More precisely, \lneq{dwbalu12}-\lneq{dwbalu34} can be
identified with the DWBA terms and \lneq{stretlu56} with the
stretched boxes, as is illustrated on \fig{polesdwbafsi} for the
FSI case.

The next section exemplifies this method for the FSI case.
\section*{Re-scattering diagram with final state interaction}
From \eq{mtoyfsiencons} we have
\begin{equation}
\left\{
\begin{array}
[c]{l}%
u_{1}=E_{2}-\omega_{2}+i\varepsilon\\
u_{2}=E_{2}-F_{2}-\omega_{\sigma}+i\varepsilon\\
u_{3}=-\omega_{\pi}+i\varepsilon
\end{array}
\right.  \qquad\text{and}\qquad\left\{
\begin{array}
[c]{l}%
l_{1}=E_{\pi}+\omega_{1}-E_{1}-i\varepsilon\\
l_{2}=E_{2}-F_{2}+\omega_{\sigma}-i\varepsilon\\
l_{3}=\omega_{\pi}-i\varepsilon
\end{array}
\right.
\end{equation}
and consequently,
\begin{eqnarray}
\frac{1}{\alpha_{11}} &  \equiv & u_{1}-l_{1}=E_{tot}-E_{\pi}-\omega_{1}%
-\omega_{2}+2 i \varepsilon\\
\frac{1}{\alpha_{22}} &  \equiv & u_{2}-l_{2}=-2\omega_{\sigma}+2 i \varepsilon\\
\frac{1}{\alpha_{33}} &  \equiv & u_{3}-l_{3}=-2\omega_{\pi}+2 i \varepsilon\\
\frac{1}{\alpha_{12}} &  \equiv &
u_{1}-l_{2}=F_{2}-\omega_{2}-\omega_{\sigma
}=E_{tot}-F_{1}-E_{\pi}-\omega_{2}-\omega_{\sigma}+2 i \varepsilon\\
\frac{1}{\alpha_{21}} &  \equiv &
u_{2}-l_{1}=E_{tot}-F_{2}-E_{\pi}-\omega
_{1}-\omega_{\sigma} +2 i \varepsilon%
\end{eqnarray}
with $E_{tot}=E_{1}+E_{2}=F_{1}+F_{2}+E_{\pi}$
. The residues of $U_{1}$, $U_{2}$, $L_{1}$ and $L_{2}$ are
\begin{eqnarray}
&&\text{Res}\left.  \bigskip U_{1}\right\vert
_{Q_{0}^{\prime}=l_{3}}
=\frac{1}{l_{3}-u_{1}}=-\frac{1}{E_{2}-\omega_{2}-\omega_{\pi}}=-\frac
{1}{E_{tot}-E_{1}-\omega_{2}-\omega_{\pi}} \hspace{3cm}\\
&&\text{Res}\left.  \bigskip U_{2}\right\vert
_{Q_{0}^{\prime}=l_{3}}
=\frac{1}{l_{3}-u_{2}}=-\frac{1}{E_{2}-F_{2}-\omega_{\sigma}-\omega_{\pi}%
}=-\frac{1}{E_{tot}-E_{1}-F_{2}-\omega_{\sigma}-\omega_{\pi}}%
\end{eqnarray}
and
\begin{eqnarray}
&&\text{Res}\left.  \bigskip L_{1}\right\vert
_{Q_{0}^{\prime}=u_{3}}
=\frac{1}{u_{3}-l_{1}}=\frac{1}{E_{1}-E_{\pi}-\omega_{1}-\omega_{\pi}}%
=\frac{1}{E_{tot}-E_{2}-E_{\pi}-\omega_{1}-\omega_{\pi}}\hspace{3cm}\\
&& \text{Res}\left.  \bigskip L_{2}\right\vert
_{Q_{0}^{\prime}=u_{3}}
=\frac{1}{u_{3}-l_{2}}=\frac{1}{F_{2}-E_{2}-\omega_{\sigma}-\omega_{\pi}%
}=\frac{1}{E_{tot}-F_{1}-E_{\pi}-E_{2}-\omega_{\sigma}-\omega_{\pi}}%
\end{eqnarray}

\sp

Therefore, the amplitude \eq{mtoyfsiencons} reads:
\begin{eqnarray}
&&\mathcal{M}^{FSI}_{TOPT}=-g_{\sigma}^{2}\int \frac{d^{3}q^{\prime }}{\left( 2\pi \right) ^{3}}\frac{1%
}{4\omega _{\sigma }\omega _{\pi }}\times  \label{aptoptfsi} \\
&&\left[ \frac{f\left( \omega _{\pi }\right) }{\left(
E_{tot}-E_{\pi }-\omega _{1}-\omega _{2}\right) \left(
E_{tot}-E_1-\omega _{2}-\omega _{\pi
}\right) \left( E_{tot}-F_1-E_\pi-\omega _{2}-\omega _{\sigma }\right) }\right.  \nonumber \\
&&\left. +\frac{f\left( \omega _{\pi }\right) }{\left(
E_{tot}-E_{\pi }-\omega _{1}-\omega _{2}\right) \left(
E_{tot}-E_1-\omega _{2}-\omega _{\pi
}\right) \left( E_{tot}-F_2-E_{\pi }-\omega _{1}-\omega _{\sigma }\right) }%
\right.  \nonumber \\
&&\left. +\frac{f\left( -\omega _{\pi }\right) }{\left(
E_{tot}-E_{\pi }-\omega _{1}-\omega _{2}\right) \left(
E_{tot}-E_{2}-E_{\pi }-\omega _{1}-\omega
_{\pi }\right) \left( E_{tot}-F_1-E_\pi-\omega _{2}-\omega _{\sigma }\right) }\right.  \nonumber \\
&&\left. +\frac{f\left( -\omega _{\pi }\right) }{\left(
E_{tot}-E_{\pi }-\omega _{1}-\omega _{2}\right) \left(
E_{tot}-E_2-E_{\pi }-\omega _{1}-\omega _{\pi }\right) \left(
E_{tot}-F_2-E_{\pi }-\omega _{1}-\omega _{\sigma
}\right) }\right.  \nonumber \\
&&\left. +\frac{f\left( \omega _{\pi }\right) }{\left(
E_{tot}-E_1-\omega _{2}-\omega _{\pi }\right) \left(
E_{tot}-F_2-E_{\pi }-\omega _{1}-\omega _{\sigma }\right) \left(
E_{tot}-E_1-F_{2}-\omega _{\pi }-\omega _{\sigma }\right)
}\right.  \nonumber \\
&&\left. +\frac{f\left( -\omega _{\pi }\right) }{\left(
E_{tot}-E_2-E_{\pi }-\omega _{1}-\omega _{\pi }\right) \left(
E_{tot}-F_1-E_\pi-\omega _{2}-\omega _{\sigma }\right) \left(
E_{tot}-E_2-F_1-E_\pi-\omega _{\pi }-\omega _{\sigma }\right)
}\right] \nonumber
\end{eqnarray}

%% file: Appendix_D.tex
\clearpage

\chapter{T-matrix equations and phase shifts \label{ApTm}}
%
%
\section*{$NN$ propagator}
The $NN$ propagator,
\begin{equation}
G_{NN}=\frac{1}{2\left( E_{0}-E\right) + i \varepsilon}
\end{equation}
is given for non-relativistic nucleons by
\begin{equation}
G_{NN}^{NR}=\frac{1}{2\left( \frac{k_{0}^{2}}{2M}-\frac{k^{2}}{2M}\right) + i \varepsilon}=\frac{%
M}{\left( k_{0}^{2}-k^{2}\right) + i \varepsilon} \label{gnrel}
\end{equation}
where $E=\frac{k^{2}}{2M}$. The nucleon-nucleon relative momentum
 (on-shell momentum) is $k$ $\left(k_{0}\right)$.
\eq{gnrel} formally coincides with the Blankenbecler-Sugar
propagator\cite{{Blankenbecler:1966sg},{Ramalho:2001pd}}
considered with minimal relativistic definitions of the
amplitudes. The relativistic nucleons can be described by the
Thompson propagator\cite{{Ramalho:2001pd},{Thompson:1970wt}},
\begin{equation}
G_{NN}^{R}=\frac{1}{2}\frac{E_{0}+E}{\left( E_{0}^{2}-E^{2}\right)+ i \varepsilon}=\frac{1}{2}\frac{E_{0}+E}{%
\left( k_{0}^{2}-k^{2}\right)+ i \varepsilon} \label{grel}
\end{equation}%
since $E=\sqrt{M^{2}+k^{2}}$. In a short-hand notation,
\begin{equation}
G_{NN}=\frac{\beta\left(k \right)}{\left(k_{0}^{2}-k^{2}\right) +
i \varepsilon} \label{gnnhand}
\end{equation}
with $\beta\left(k \right)=M$ for the propagator of \eq{gnrel} and
$\beta\left(k \right)=\frac{E+E_{0}}{2}$ for the propagator of
\eq{grel}.
%
%
\section*{Uncoupled channels}
\subsection*{Equations for $T$}
For uncoupled channels, the equation for $T$ is calculated from
the potential $V$ using
\begin{equation}
T\left( k,k_{0}\right) =V\left( k,k_{0}\right) +\int V\left(
k,k^{\prime }\right) G_{NN}\left( k^{\prime }\right) T\left(
k^{\prime },k_{0}\right) k^{\prime 2}dk^{\prime } \label{eqforT}
\end{equation}
where $k_{0}$ is the on-shell momentum. The $i\varepsilon$ term of
the denominator of $G_{NN}$ in \eq{eqforT} may be evaluated by
applying the Cauchy principal value theorem\cite{Marsden:1987bk}.
Numerically we used also the subtraction method with
regularisation (see \apx{ApImpI}) to calculate $T \left(k,k_0
\right)$:
\begin{eqnarray}
T\left( k,k_{0}\right) &=& V\left( k,k_{0}\right) + \mathcal{P}
\int \frac{g\left( k,k^{\prime }\right) -g\left(
k,k_{0}\right) }{k_{0}^{2}-k^{\prime 2}}k^{\prime 2}dk^{\prime }-i\pi \delta Res\left[ \frac{%
g\left( k,k^{\prime }\right) }{k_{0}^{2}-k^{\prime 2}}\right]
_{k^{\prime }=k_{0}} \label{eqforTreg}
\end{eqnarray}
Here, $g\left( k,k^{\prime }\right) = V\left( k,k^{\prime }\right)
\beta \left( k^{\prime }\right) T\left( k^{\prime },k_{0}\right)
k^{\prime 2}$ is the regular part of the integrand. Using
\eq{gnnhand}, \eq{eqforTreg} transforms into:
\begin{eqnarray}
T\left( k,k_{0}\right) = V\left( k,k_{0}\right) &+& \int
V\left(k,k^{\prime} \right) \beta\left(k^{\prime} \right) T
\left(k^{\prime},k_0 \right)\frac{k^{\prime 2}
}{k_{0}^{2}-k^{\prime 2}}dk^{\prime } \\
&-& \int V\left(k,k_0 \right) \beta\left(k_0 \right) T
\left(k_0,k_0 \right)\frac{k_0^{ 2} }{k_{0}^{2}-k^{\prime
2}}dk^{\prime } \\
&-& i\pi \frac{k_0}{2} V\left(k,k_0 \right) \beta \left(k_0
\right) T\left(k_0,k_0 \right)
\end{eqnarray}
The corresponding equation in the discrete form is
\begin{eqnarray}
T_{i0} = V_{i0}&+&\sum_{j\neq 0}\left[
V_{ij}T_{j0}k_{j}^{2}G_{j}w_{j}\right]
-V_{i0}k_{0}^{2} \beta\left(k_0\right) \left( \sum_{j\neq 0}\frac{1}{k_{0}^{2}-k_{j}^{2}}w_{j}\right) T_{00} \\
&+& i\alpha V_{i0}T_{00}
\end{eqnarray}
with $\alpha=-\frac{\pi Mk_{0}}{2}$ for the non-relativistic
 case (\eq{gnrel}), and $\alpha=-\pi
\frac{E_{0}k_{0}}{2}$ for the relativistic one (\eq{grel}).
Defining $M_{ij}=\delta _{ij}-\overline{M}_{ij}$, with
\begin{equation}
\overline{M}_{ij}=\left\{
\begin{array}{ccc}
V_{ij}G_{j}k_{j}^{2}w_{j} &  & j\neq 0 \\
V_{i0}k_{0}^{2}\beta \left(k_{0}\right) \left( \sum\limits_{j \neq 0} \frac{1}{k_{0}^{2}-k_{j}^{2}}%
w_{j}\right) +i\frac{\pi E_{0}k_{0}}{2}V_{i0} &  & j=0%
\end{array}%
\right. ,
\end{equation}
the $T$-matrix can be calculated from:
\begin{equation}
M_{ij}T_{j0}=V_{i0}\Rightarrow T_{j0}=M_{jk}^{-1}V_{k0}.
\end{equation}
\subsection*{Phase shifts}
\begin{equation}
T\left( k_{0}\right) =\frac{1}{\alpha }e^{i\delta }\sin \delta
\Rightarrow
tg\left(2\delta\right) =\frac{\alpha {\mathrm{Re}}\,{[T]}\left( k_{0}\right) }{\frac{1}{2}%
-\alpha {\mathrm{Im}}\,{[T]}\left( k_{0}\right) }
\end{equation}%

%
\section*{Coupled channels}
\subsection*{Equations for $T$}
For each value of $J\neq 0$ there are three possible values for
the orbital angular momentum, $L=J$ and the coupled $L=J\pm 1$.
The equations for the coupled case are
\begin{equation}
\left\{
\begin{array}{c}
T^{--}=V^{--}+V^{--}G_{0}T^{--}+V^{-+}G_{0}T^{+-} \\
T^{+-}=V^{+-}+V^{+-}G_{0}T^{--}+V^{++}G_{0}T^{+-} \\
T^{-+}=V^{-+}+V^{-+}G_{0}T^{++}+V^{--}G_{0}T^{-+} \\
T^{++}=V^{++}+V^{++}G_{0}T^{++}+V^{+-}G_{0}T^{-+}%
\end{array}%
\right.
\end{equation}%
which reads in the matrix form
\begin{equation}
\left(
\begin{array}{cc}
M_{ij}^{--} & -\overline{M}_{ij}^{-+} \\
-\overline{M}_{ij}^{+-} & M_{ij}^{++}%
\end{array}%
\right) \left(
\begin{array}{cc}
T_{j0}^{--} & T_{j0}^{-+} \\
T_{j0}^{+-} & T_{j0}^{++}%
\end{array}%
\right) =\left(
\begin{array}{cc}
V_{i0}^{--} & V_{i0}^{-+} \\
V_{i0}^{+-} & V_{i0}^{++}%
\end{array}%
\right)
\end{equation}
\subsection*{Phase-shifts}

\subsubsection*{Stapp, Ypsilantis and Metropolis parametrisation}
In the Stapp, Ypsilantis and Metropolis\cite{Stapp:1956mz}
parametrisation, the $S$-matrix is written as a product of the
following three matrices
\begin{eqnarray}
S &=&\left(
\begin{array}{cc}
e^{i\overline{\delta }_{-}} & 0 \\
0 & e^{i\overline{\delta }+}%
\end{array}%
\right) \left(
\begin{array}{cc}
\cos 2\overline{\varepsilon } & i\sin 2\overline{\varepsilon } \\
i\sin 2\overline{\varepsilon } & \cos 2\overline{\varepsilon }%
\end{array}%
\right) \left(
\begin{array}{cc}
e^{i\overline{\delta }_{-}} & 0 \\
0 & e^{i\overline{\delta }+}%
\end{array}%
\right) \\
&=&\left(
\begin{array}{cc}
\cos \left( 2\overline{\varepsilon }\right) e^{i2\overline{\delta
}_{-}} &
i\sin \left( 2\overline{\varepsilon }\right) e^{i\left( \overline{\delta }%
_{+}+\overline{\delta }_{-}\right) } \\
i\sin \left( 2\overline{\varepsilon }\right) e^{i\left( \overline{\delta }%
_{+}+\overline{\delta }_{-}\right) } & \cos \left( 2\overline{\varepsilon }%
\right) e^{i2\overline{\delta }+}%
\end{array}%
\right)
\end{eqnarray}
From
\begin{equation}
S^{-+}+S^{+-}=2i\sin \left( 2\overline{\varepsilon }\right)
e^{i\left( \overline{\delta }_{+}+\overline{\delta }_{-}\right) }
\end{equation}%
and%
\begin{equation}
S^{++}S^{--}=\cos ^{2}\left( 2\overline{\varepsilon }\right)
e^{i2\left(
\overline{\delta }_{+}+\overline{\delta }_{-}\right) }=\left[ \cos \left( 2%
\overline{\varepsilon }\right) e^{i\left( \overline{\delta }_{+}+\overline{%
\delta }_{-}\right) }\right] ^{2},
\end{equation}%
it follows that%
\begin{equation}
tg\left( 2\overline{\varepsilon }\right) =-\frac{i}{2}\frac{S^{-+}+S^{+-}}{%
\sqrt{S^{++}S^{--}}}
\end{equation}%
and thus%
\begin{equation}
\left( 2\overline{\varepsilon }\right) =-\frac{i}{2}\ln \left( \frac{1-\frac{%
i}{2}\frac{S^{-+}+S^{+-}}{\sqrt{S^{++}S^{--}}}}{1+\frac{i}{2}\frac{%
S^{-+}+S^{+-}}{\sqrt{S^{++}S^{--}}}}\right).
\end{equation}%
$\overline{\delta }_{+}$ and $\overline{\delta }_{-}$ are now
easily
calculated from%
\begin{eqnarray}
\overline{\delta }_{-} &=&\frac{1}{2}\arg \left[ \frac{S^{--}}{\cos \left( 2%
\overline{\varepsilon }\right) }\right] \\
\overline{\delta }_{+} &=&\frac{1}{2}\arg \left[ \frac{S^{++}}{\cos \left( 2%
\overline{\varepsilon }\right) }\right]
\end{eqnarray}
\subsubsection*{Blatt and Biedenharn parametrisation}
In the also used Blatt and Biedenharn parametrisation\cite{Blatt},
$S$ is written as
\begin{eqnarray}
S &=&\left(
\begin{array}{cc}
\cos \varepsilon  & \sin \varepsilon  \\
-\sin \varepsilon  & \cos \varepsilon
\end{array}%
\right) ^{-1}\left(
\begin{array}{cc}
e^{i2\delta _{-}} & 0 \\
0 & e^{i2\delta _{+}}%
\end{array}%
\right) \left(
\begin{array}{cc}
\cos \varepsilon  & \sin \varepsilon  \\
-\sin \varepsilon  & \cos \varepsilon
\end{array}%
\right)  \\
&=&\left(
\begin{array}{cc}
\cos ^{2}\left( \varepsilon \right) e^{i2\delta _{-}}+\sin
^{2}\left( \varepsilon \right) e^{i2\delta _{+}} & \sin
\varepsilon \cos \varepsilon
\left( e^{i2\delta _{-}}-e^{i2\delta +}\right)  \\
\sin \varepsilon \cos \varepsilon \left( e^{i2\delta
_{-}}-e^{i2\delta +}\right)  & \sin ^{2}\left( \varepsilon \right)
e^{i2\delta _{-}}+\cos
^{2}\left( \varepsilon \right) e^{i2\delta _{+}}%
\end{array}%
\right).
\end{eqnarray}%
Defining
\begin{equation}
c_{ppmm}=\frac{S_{--}+S_{++}}{2}=\frac{e^{i2\delta _{-}}+e^{i2\delta _{+}}}{2%
}
\end{equation}
and
\begin{eqnarray}
c_{mx} &=&\sqrt{\left( \frac{S_{--}+S_{++}}{2}\right)
^{2}-S_{++}S_{--}+S_{+-}S_{-+}} \\
&=&\sqrt{\left( \frac{e^{i2\delta _{-}}+e^{i2\delta
_{+}}}{2}\right)
^{2}-e^{i2\left( \delta _{+}+\delta _{-}\right) }} \\
&=&\frac{e^{i2\delta _{+}}-e^{i2\delta _{-}}}{2},
\end{eqnarray}
it is straightforward to obtain
\begin{equation}
c_{dm}=c_{ppmm}-c_{mx}=e^{i2\delta _{-}}
\end{equation}%
\begin{equation}
c_{dp}=c_{ppmm}-c_{mx}=e^{i2\delta +},
\end{equation}%
and thus giving for the phase-shifts $\delta _{-}$ and $\delta
_{+}$ the following expressions:
\begin{equation}
tg(2\delta _{-})=\frac{ {\mathrm{Re}}\,{[c}_{{dm}}{]}}{\frac{1}{2}%
- {\mathrm{Im}}\,{[c}_{{dm}}{]}}
\end{equation}
\begin{equation}
tg(2\delta _{+})=\frac{ {\mathrm{Re}}\,{[c}_{{dp}}{]}}{\frac{1}{2}%
- {\mathrm{Im}}\,{[c}_{{dp}}{]}}.
\end{equation}

For both the uncoupled and coupled cases, $S=\boldsymbol{1}+i
\alpha T$.

%% file: Appendix_E.tex
\clearpage\chapter{Partial wave decomposition of the amplitudes
\label{ApPWAmp}}

%
%
%
\section*{Re-scattering diagram \label{tforpwrescatt}}
The re-scattering diagram amplitude involves terms like $\left(
\vec{\sigma}^{\left( i\right) }\cdot\vec {q}\right)$. Since
$\vec{q}$ is the three-momentum of the exchanged particle, it can
be expressed as
\begin{equation}
\vec{q}=A\vec{q}_{af}-B\vec{q}_{bf}+C\vec{q}_{ej}   \label{tresc}
\end{equation}
where $\vec{q}_{bf}\left( \vec{q}_{af}\right) $ is the relative
momentum of the two nucleons before(after) pion emission and
$\vec{q}_{ej}$ is the momentum of the produced pion. $A$, $B$ and
$C$ are numerical coefficients
which depend on the reaction considered\footnote{%
Amplitudes should be defined such as $A=1$. For the cases of $A\neq1$, then $%
x_{1}=\cos\measuredangle \left( \vec{q}_{ej},A\vec{q}_{af}-B\vec{q}%
_{bf}\right) $.}.


With $x_{1}=\cos\measuredangle\left( \vec{q}_{ej},\vec{q}%
_{af}-\vec{q}_{bf}\right) $ and $x_{2}=\cos\measuredangle\left( \vec{q}_{af},%
\vec{q}_{bf}\right) $, the amplitudes can be written as
\begin{equation}
\mathcal{M}_{r}=\mathcal{F}\left( x_{1},x_{2}\right) \left( \vec{\sigma }%
^{\left( i\right) }\cdot\vec{q}\right)   \label{M1}
\end{equation}
where $\mathcal{F}$ is a function depending on the kinematic
variables of the diagram considered.

Using \eq{sigmadot} and \eq{ysoma}, \eq{M1} reads
\begin{eqnarray}
\mathcal{M}_{r}& = & -\sqrt{4\pi }\left[ \sigma^{\left( i\right)
}\otimes
Y^{1}\left( \hat{q}\right) \right] _{0}^{0}\left\vert \vec{q}\right\vert \mathcal{F}\left( x_{1},x_{2}\right) \\
&& =-\left( 4\pi \right) \mathcal{F}\left( x_{1},x_{2}\right) \sum_{l_{%
3}+l_{4}=1}\left[ \sigma^{\left( i\right) }\otimes
\mathcal{Y}_{l_{3}l_{4}}^{1}\left(
\widehat{A\vec{q}_{af}-B\vec{q}_{bf}},\widehat{q_{ej}}\right)
\right]
_{0}^{0} \\
&& F\left( 1;l_{3}l_{4}\right) \left\vert A\vec{q}_{af}-B\vec{q}%
_{bf}\right\vert ^{l_{3}}\left( Cq_{ej}\right) ^{l_{4}}
\end{eqnarray}
where $F\left( l;l_{1}l_{2}\right) \equiv \sqrt{\frac{\left( 2l+1\right) !}{%
\left( 2l_{1}+1\right) !\left( 2l_{2}+1\right) !}}$. The function
$\mathcal{F}\left( x_{1},x_{2}\right) $ is now decomposed using
the Legendre polynomials
\footnote{%
The coefficients of the expansion are $f^{l_{1}}\left(
x_{2}\right) =\int \mathcal{F}\left( x_{1},x_{2}\right)
P_{l_{1}}\left( x_{1}\right) dx_{1}$. }
\begin{equation}
\mathcal{F}\left( x_{1},x_{2}\right) =\sum_{l_{1}}f^{l_{1}}\left(
x_{2}\right) P_{l_{1}}\left( x_{1}\right)
\frac{\hat{l}_{1}^{2}}{2} \hspace{1cm} \text{with } \hat{l}_{1}
\equiv \sqrt{2l_{1}+1},
\end{equation}
leading to
\begin{eqnarray}
\mathcal{M}_{r}& = &-\frac{\left( 4\pi \right)
^{2}}{2}\sum_{l_{3}+l_{4}=1}\sum_{l_{1}}\left( -1\right)
^{l_{1}}f^{l_{1}}\left( x_{2}\right) \hat{l}_{1}\left\vert
A\vec{q}_{af}-B\vec{q}_{bf}\right\vert
^{l_{3}}\left( Cq_{ej}\right) ^{l_{4}}  \label{M1p1} \\
&& F\left( 1;l_{3}l_{4}\right) \left\{ \sigma^{\left( i\right)
}\otimes
\left[ \mathcal{Y}_{l_{3}l_{4}}^{1}\left( \widehat{%
A\vec{q}_{af}-B\vec{q}_{bf}},\widehat{q_{ej}}\right) \otimes \mathcal{Y}%
_{l_{1}l_{1}}^{00}\left( \widehat{A\vec{q}_{af}-B\vec{q}_{bf}},\widehat{%
q_{ej}}\right) \right] ^{1}\right\} _{0}^{0}  \notag
\end{eqnarray}
where \eq{Plegeharmesfe2} was used. Coupling the spherical
harmonics with the same arguments \eq{tenssamearg} and then using
\eq{AASum}, \eq{M1p1} simplifies to
\begin{eqnarray}
\mathcal{M}_{r}& = &-\frac{\left( 4\pi \right) }{2}\sqrt{4\pi }\sum_{l_{%
3}+l_{4}=1}\left\vert A\vec{q}_{af}-B\vec{q}%
_{bf}\right\vert ^{l_{3}}\left( Cq_{ej}\right) ^{l_{4}}F\left(
1;l_{3}l_{4}\right) \widehat{l_{3}}\widehat{l_{4}}  \label{M1p2} \\
&& \sum_{l_{1}}\left( -1\right) ^{l_{1}}f^{l_{1}}\left( x_{2}\right) \widehat{%
l_{1}}^{2}%
\sum_{t_{1}t_{2}}C_{000}^{l_{3}l_{1}t_{1}}C_{000}^{l_{4}l_{1}t_{2}}%
\sum_{f_{1}+f_{2}=t_{1}}F\left( t_{1};f_{1}f_{2}\right)   \nonumber \\
&& \frac{\left( Aq_{af}\right) ^{f_{1}}\left( -Bq_{bf}\right) ^{f_{2}}}{%
\left\vert A\vec{q}_{af}-B\vec{q}_{bf}\right\vert ^{t_{1}}}\left\{
\begin{array}{lll}
t_{1} & t_{2} & 1 \\
l_{3} & l_{4} & l_{1}%
\end{array}%
\right\} \left\{ \sigma^{\left( i\right) }\otimes \left[ \mathcal{Y}%
_{f_{1}f_{2}}^{t_{1}}\left(
\widehat{q_{af}},\widehat{q_{bf}}\right) \otimes Y^{t_{2}}\left(
\widehat{q_{ej}}\right) \right] ^{1}\right\} _{0}^{0}  \nonumber
\end{eqnarray}

The Legendre polynomials are now used to decompose
$f^{l_{1}}\left(
x_{2}\right) \left\vert A\vec{q}_{af}-B\vec{q}_{bf}\right\vert ^{l_{%
3}-t_{1}}$
\begin{eqnarray}
\mathcal{G}^{l_{1}l_{3}t_{1}}\left( x_{2}\right) & = &
f^{l_{1}}\left(
x_{2}\right) \left\vert A\vec{q}_{af}-B\vec{q}_{bf}\right\vert ^{l_{%
3}-t_{1}} \\
& = &\sum_{l_{2}}g^{l_{1}l_{2}l_{3}t_{1}}P_{l_{2}}\left( x_{2}\right) \frac{%
\widehat{l_{2}}}{2}^{2}
\end{eqnarray}
and \eq{M1p2} reads\footnote{%
Since
\begin{align}
 & \left\{
\begin{array}{lll}
l_{3} & l_{4} & 1 \\
l_{1} & l_{1} & 0 \\
t_{1} & t_{2} & 1%
\end{array}
\right\} \underset{(l_{3}+l_{4}=1)}{=}\left\{
\begin{array}{lll}
l_{3} & l_{4} & 1 \\
t_{1} & t_{2} & 1 \\
l_{1} & l_{1} & 0%
\end{array}
\right\} \underset{C_{000}^{l_{3}l_{1}t_{1}}\Rightarrow l_{3}+l_{1}+t_{1}%
\text{ even}}{=}\frac{1}{\hat{1}\widehat{l_{1}}}W\left(
l_{3}l_{4}t_{1}t_{2};1l_{1}\right) \\
& =\frac{1}{\hat{1}\widehat{l_{1}}}W\left(
t_{1}t_{2}l_{3}l_{4};1l_{1}\right)
=\frac{1}{\hat{1}\widehat{l_{1}}}\left\{
\begin{array}{ccc}
t_{1} & t_{2} & 1 \\
l_{4} & l_{3} & l_{1}%
\end{array}
\right\}
\end{align}%
}%
\begin{align}
\mathcal{M}_{r} & =-\frac{\left( 4\pi\right) ^{2}}{4}\sqrt{4\pi}\sum_{l_{%
3}+l_{4}=1}\left( Cq_{ej}\right) ^{l_{4}}F\left(
1;l_{3}l_{4}\right) \widehat{l_{3}}\widehat{l_{4}}%
\sum_{t_{1}t_{2}}C_{000}^{l_{3}l_{1}t_{1}}C_{000}^{l_{4}l_{1}t_{2}} \\
& \left\{
\begin{array}{ccc}
t_{1} & t_{2} & 1 \\
l_{4} & l_{3} & l_{1}%
\end{array}
\right\} \sum_{f_{1}+f_{2}=t_{1}}F\left( t_{1};f_{1}f_{2}\right)
\left( Aq_{af}\right) ^{f_{1}}\left( -Bq_{bf}\right)
^{f_{2}}\sum_{l_{1}l_{2}}\left( -1\right) ^{l_{1}+l_{2}}  \notag \\
& \widehat{l_{1}}^{2}\widehat{l_{2}}g^{l_{1}l_{2}l_{3}t_{1}}\left[
\sigma^{\left( i\right) }\otimes\left\{ \left[ \mathcal{Y}%
_{f_{1}f_{2}}^{t_{1}}\left(
\widehat{q_{af}},\widehat{q_{bf}}\right) \otimes
Y^{t_{2}}\left( \widehat{q_{ej}}\right) \right] ^{1}\otimes\mathcal{Y}%
_{l_{2}l_{2}}^{0}\left( \widehat{q_{af}},\widehat{q_{bf}}\right)
\right\} ^{1}\right] _{0}^{0}  \notag
\end{align}

Using \eq{tenssamearg} and $%
Y^{t_{2}}\left( \widehat{q_{ej}}\right) =\sqrt{4\pi }\mathcal{Y}%
_{t_{2}0}^{t_{2}}\left( \widehat{q_{ej}},\widehat{q_{bf}}\right)
$, one obtains
\begin{eqnarray}
\mathcal{M}_{r}& = & -\frac{\left( 4\pi \right) ^{2}}{4}\sum_{l_{3%
}+l_{4}=1}\left( Cq_{ej}\right) ^{l_{4}}F\left(
1;l_{3}l_{4}\right) \widehat{l_{3}}\widehat{l_{4}}%
\sum_{t_{1}t_{2}}C_{000}^{l_{3}l_{1}t_{1}}C_{000}^{l_{4}l_{1}t_{2}} \\
&& \left\{
\begin{array}{ccc}
t_{1} & t_{2} & 1 \\
l_{4} & l_{3} & l_{1}%
\end{array}%
\right\}
\sum_{f_{1}+f_{2}=t_{1}}\widehat{f_{1}}\widehat{f_{2}}F\left(
t_{1};f_{1}f_{2}\right) \left( Aq_{af}\right) ^{f_{1}}\left(
-Bq_{bf}\right)
^{f_{2}}\sum_{l_{1}l_{2}}\left( -1\right) ^{l_{1}+l_{2}}\widehat{l_{1}}^{2}%
\widehat{l_{2}}  \nonumber \\
&&
g^{l_{1}l_{2}l_{3}t_{1}}%
\sum_{h_{1}h_{2}}C_{000}^{f_{1}l_{2}h_{1}}C_{000}^{f_{2}l_{2}h_{2}}\left\{
\begin{array}{ccc}
h_{1} & h_{2} & t_{1} \\
f_{2} & f_{1} & l_{2}%
\end{array}%
\right\} \left\{ \sigma^{\left( i\right) }\otimes \left[ \mathcal{Y}%
_{t_{2}0}^{t_{2}}\left( \widehat{q_{ej}},\widehat{q_{bf}}\right)
\otimes
\mathcal{Y}_{l_{2}l_{2}}^{0}\left( \widehat{q_{af}},\widehat{q_{bf}}\right) %
\right] ^{1}\right\} _{0}^{0}  \nonumber
\end{eqnarray}
\subsection*{Orbital matrix elements}
With \eq{tensexcharg} and \eq{sphdef}, the orbital part of $\mathcal{M}_{r}$ reads
\begin{eqnarray}
\mathcal{M}_{r}^{o}& = &-\frac{\left( 4\pi \right)
^{2}}{4}\frac{1}{\sqrt{4\pi
}}\sum_{l_{3}+l_{4}=1}\widehat{l_{3}}\widehat{l_{4}%
}F\left( 1;l_{3}l_{4}\right) \left( Cq_{ej}\right) ^{l_{4}}\sum_{t_{1}t_{2}}%
\widehat{t_{1}}C_{000}^{l_{3}l_{1}t_{1}}C_{000}^{l_{4}l_{1}t_{2}} \\
&& \left\{
\begin{array}{ccc}
t_{1} & t_{2} & 1 \\
l_{4} & l_{3} & l_{1}%
\end{array}%
\right\}
\sum_{f_{1}+f_{2}=t_{1}}\widehat{f_{1}}\widehat{f_{2}}F\left(
t_{1};f_{1}f_{2}\right) \left( Aq_{af}\right) ^{f_{1}}\left(
-Bq_{bf}\right)
^{f_{2}}\sum_{l_{1}l_{2}}\left( -1\right) ^{l_{1}+l_{2}}  \nonumber \\
&& \widehat{l_{1}}^{2}\widehat{l_{2}}^{2}g^{l_{1}l_{2}l_{3}t_{1}}%
\sum_{h_{1}h_{2}}C_{000}^{f_{1}l_{2}h_{1}}C_{000}^{f_{2}l_{2}h_{2}}\left\{
\begin{array}{ccc}
h_{1} & h_{2} & t_{1} \\
f_{2} & f_{1} & l_{2}%
\end{array}%
\right\} \sum_{g_{1}}\widehat{g_{1}}\left( -1\right)
^{t_{1}+t_{2}+1}  \nonumber
\\
&& \left\{
\begin{array}{ccc}
t_{1} & 1 & t_{2} \\
g_{1} & h_{1} & h_{2}%
\end{array}%
\right\}
C_{m_{1}m_{2}m}^{g_{1}g_{2}1}\mathcal{Y}_{t_{2}h_{1}}^{g_{1}}\left(
\widehat{q_{ej}},\widehat{q_{af}}\right) Y^{h_{2}}\left( \widehat{q_{bf}}%
\right)   \nonumber
\end{eqnarray}
Thus, the orbital matrix elements given by
\begin{eqnarray}
O & = &\left\langle
q_{af}q_{ej}\mathcal{L}^{\prime}\mathcal{M}^{\prime
}\left\vert \mathcal{M}_{r}^{o}\right\vert q_{bf}L_{i}M\right\rangle \\
& = &\int d\Omega_{q_{bf}}d\Omega_{q_{af}}d\Omega_{q_{ej}}\mathcal{Y}%
_{l_{f}l_{x}}^{\ast\mathcal{L}^{\prime}\mathcal{M}^{\prime}}\left(
\widehat{\vec{q}_{af}},\widehat{\vec{q}_{ej}}\right)
Y^{L_{i}}_{M}\left(
\widehat{\vec{q}_{bf}}\right) \mathcal{M}_{r}^{o} \\
& = &\int d\Omega_{q_{bf}}d\Omega_{q_{af}}d\Omega_{q_{ej}}\mathcal{Y}%
_{L_{f}l_{x}}^{\ast\mathcal{L}^{\prime}\mathcal{M}^{\prime}}\left(
\widehat{\vec{q}_{af}},\widehat{\vec{q}_{ej}}\right) \left[ \left(
-1\right)
^{M}Y^{\ast L_{i}}_{-M}\left( \widehat{\vec{q}_{bf}}\right) %
\right] \mathcal{M}_{r}^{o}\text{,}
\end{eqnarray}
are
\begin{eqnarray}
O & = & -\frac{\left( 4\pi\right)
^{2}}{4}\frac{1}{\sqrt{4\pi}}\left( -1\right)
^{M}\sum_{l_{3}+l_{4}=1}\widehat{l_{3}}%
\widehat{l_{4}}F\left( 1;l_{3}l_{4}\right) \left( Cq_{ej}\right) ^{l_{4}} \\
&& \sum_{l_{1}l_{2}}\left( -1\right) ^{l_{1}+l_{2}}\widehat{l_{1}}^{2}%
\widehat{l_{2}}^{2}C_{000}^{l_{4}l_{1}l^{\prime}}\sum_{t_{1}}\widehat{t_{1}}%
C_{000}^{l_{3}l_{1}t_{1}}g^{l_{1}l_{2}l_{3}t_{1}}\left\{
\begin{array}{ccc}
t_{1} & l_{x} & 1 \\
l_{4} & l_{3} & l_{1}%
\end{array}
\right\}  \nonumber \\
&& \sum_{f_{1}+f_{2}=t_{1}}\widehat{f_{1}}\widehat{f_{2}}F\left(
t_{1};f_{1}f_{2}\right) \left( Aq_{af}\right) ^{f_{1}}\left(
-Bq_{bf}\right)
^{f_{2}}C_{000}^{f_{1}l_{2}L^{\prime}}C_{000}^{f_{2}l_{2}\mathcal{L}}
\nonumber
\\
&& \left\{
\begin{array}{ccc}
L_{f} & L & t_{1} \\
f_{2} & f_{1} & l_{2}%
\end{array}
\right\} \widehat{\mathcal{L}^{\prime}}\left( -1\right)
^{t_{1}+l_{x}+1}\left\{
\begin{array}{ccc}
t_{1} & 1 & l_{x} \\
\mathcal{L}^{\prime} & L_{f} & L_{i}%
\end{array}
\right\} C_{\mathcal{M}^{\prime}-M m}^{\mathcal{L}^{\prime }L_{i}
1} \notag
\end{eqnarray}

Finally, applying the Wigner-Eckart theorem\cite{Wong:1990bk}
\begin{eqnarray}
\left\langle
q_{af}q_{ej}\mathcal{L}^{\prime}\mathcal{M}^{\prime}\left\vert
O_{k_{l}\lambda}\right\vert q_{bf}L_{i} M \right\rangle & =
&C_{\lambda
M \mathcal{M}^{\prime}}^{k_{l} L_{i} \mathcal{L}^{\prime}}\left\langle \mathcal{L}%
^{\prime}\left\Vert O_{k_{l}\lambda}\right\Vert L_{i} \right\rangle \\
& = &\frac{\widehat{\mathcal{L}^{\prime}}}{\hat{1}}\left( -1\right) ^{\mathcal{%
L}^{\prime}-k_{l}+M}C_{\mathcal{M}^{\prime}-M \lambda}^{%
\mathcal{L}^{\prime}L_{i} k_{l}}\left\langle \mathcal{L}%
^{\prime}\left\Vert O_{k_{l}\lambda}\right\Vert L_{i}
\right\rangle ,
\end{eqnarray}
one obtains the reduced matrix elements%
\begin{eqnarray}
\left\langle \mathcal{L}^{\prime}\left\Vert
\mathcal{M}_{r}^{o}\right\Vert
L_{i} \right\rangle & = & -\frac{\left( 4\pi\right) }{4}\sqrt{4\pi }%
\left( -1\right) ^{\mathcal{L}^{\prime}+l-{x}}\hat{1}\sum _{l_{%
3}+l_{4}=1}\widehat{l_{3}}\widehat{l_{4}}F\left(
1;l_{3}l_{4}\right) \left( Cq_{ej}\right) ^{l_{4}}  \label{M1orb} \\
&& \sum_{l_{1}l_{2}}\left( -1\right) ^{l_{1}+l_{2}}\widehat{l_{1}}^{2}%
\widehat{l_{2}}^{2}C_{000}^{l_{4}l_{1}l^{\prime}}\sum_{t_{1}}\widehat{t_{1}}%
C_{000}^{l_{3}l_{1}t_{1}}g^{l_{1}l_{2}l_{3}t_{1}}\left\{
\begin{array}{ccc}
t_{1} & l_{x} & 1 \\
l_{4} & l_{3} & l_{1}%
\end{array}
\right\}  \nonumber \\
&& \sum_{f_{1}+f_{2}=t_{1}}\widehat{f_{1}}\widehat{f_{2}}F\left(
t_{1};f_{1}f_{2}\right) \left( Aq_{af}\right) ^{f_{1}}\left(
-Bq_{bf}\right)
^{f_{2}}  \nonumber \\
&& C_{000}^{f_{1}l_{2}L_{f}}C_{000}^{f_{2}l_{2}L_{i}}\left\{
\begin{array}{ccc}
L_{f} & L_{i} & t_{1} \\
f_{2} & f_{1} & l_{2}%
\end{array}
\right\} \left\{
\begin{array}{ccc}
t_{1} & 1 & l_{x} \\
\mathcal{L}^{\prime} & L_{f} & L_{i}%
\end{array}
\right\}  \nonumber
\end{eqnarray}
\subsection*{Spin matrix elements}

From \rf{Brink:1993bk}, it follows that
\begin{equation}
\left\langle S_{f}\left\Vert \sigma^{\left( 1\right) }\right\Vert
S_{i}\right\rangle =\left( -1\right) ^{s_{1}+s_{2}+S_{f}}2\widehat{S_{i}}\widehat{%
s_{1}}\sqrt{s_{1}\left( s_{1}+1\right) }\left\{
\begin{array}{ccc}
S_{f} & S_{i} & 1 \\
s_{1} & s_{1} & s_{2}%
\end{array}
\right\}   \label{spin1}
\end{equation}
\begin{equation}
\left\langle S_{f}\left\Vert \sigma^{\left( 2\right) }\right\Vert
S_{i}\right\rangle =\left( -1\right) ^{s_{1}+s_{2}-S_{i}}2\widehat{S_{i}}\widehat{s_{2}}%
\sqrt{s_{2}\left( s_{2}+1\right) }\left\{
\begin{array}{ccc}
S_{f} & S_{i} & 1 \\
s_{2} & s_{2} & s_{1}%
\end{array}
\right\}   \label{spin2}
\end{equation}
\section*{Direct-production diagram \label{tforpwimp}}
The direct-production diagram involves terms like
$\vec{\protect\sigma}^{\left( 1\right) }\cdot\left( \vec
{q}_{\protect\pi}+\vec{p}_{1}+\vec{q}_{1}\right)
\protect\delta\left( \vec{q}_{2}-\vec {p}_{2}\right) $, which have
the form of
\begin{equation}
\vec{\sigma}^{\left( i\right) }\cdot \left( A\vec{q}_{ej}+B\vec{q}_{af}+C%
\vec{q}_{bf}\right) \delta \left[ s\left( \vec{q}_{bf}-\vec{q}_{af}\right) -%
\frac{\vec{q}_{ej}}{2}\right]   \label{timp}
\end{equation}
where $\vec{q}_{bf}\left( \vec{q}_{af}\right) $ is an
external(loop) momentum and $\vec{q}_{ej}=\vec{q}_{\pi }$. With
$x=\cos \measuredangle
\left( \vec{q}_{af},\vec{q}_{bf}\right) $, the amplitudes can be written as
\begin{equation}
\mathcal{M}_{i}=\vec{\sigma}^{\left( i\right) }\cdot \left( A\vec{q}_{ej}+B%
\vec{q}_{af}+C\vec{q}_{bf}\right) \delta \left[ s\left( \vec{q}_{bf}-\vec{q}%
_{af}\right) -\frac{\vec{q}_{ej}}{2}\right]   \label{M2}
\end{equation}
Using \eq{sigmadot}, \eq{M2} reads
\begin{eqnarray}
\mathcal{M}_{i}& = &-\sqrt{4\pi }\left\vert A\vec{q}_{ej}+B\vec{q}_{af}+C\vec{q%
}_{bf}\right\vert \delta \left[ s\left( \vec{q}_{bf}-\vec{q}_{af}\right) -%
\frac{\vec{q}_{ej}}{2}\right]  \\
&& \left\{ \sigma ^{\left(i \right)}\otimes Y^{1}\left( \widehat{A\vec{q}_{ej}+B\vec{q}%
_{af}+C\vec{q}_{bf}}\right) \right\} _{0}^{0}  \nonumber \\
&& =-\sqrt{4\pi }\left\vert A\vec{q}_{ej}+B\vec{q}_{af}+C\vec{q}%
_{bf}\right\vert \delta \left[ s\left( \vec{q}_{bf}-\vec{q}_{af}\right) -%
\frac{\vec{q}_{ej}}{2}\right]  \\
&& \left\{ \sigma ^{\left(i \right)}\otimes Y^{1}\left( \widehat{B^{\prime }\vec{q}%
_{af}+C^{\prime }\vec{q}_{bf}}\right) \right\} _{0}^{0},
\nonumber
\end{eqnarray}
and thus, by \eq{ysoma} one has,
\begin{eqnarray}
\mathcal{M}_{i}& = & -\left( 4\pi \right) \delta \left[ s\left( \vec{q}_{bf}-%
\vec{q}_{af}\right) -\frac{\vec{q}_{ej}}{2}\right] \sum_{f_{1}+f_{2}=1}%
\left( B^{\prime }q_{af}\right) ^{f_{2}}\left( C^{\prime
}q_{bf}\right)
^{f_{1}} \\
&& F\left( 1;f_{1}f_{2}\right) \left\{ \sigma ^{\left(i \right)}\otimes \mathcal{Y}%
_{f_{1}f_{2}}^{1}\left( \widehat{q_{bf}},\widehat{q_{af}}\right)
\right\} _{0}^{0}  \nonumber
\end{eqnarray}
\subsection*{Orbital matrix elements}
The orbital matrix elements are calculated from
\begin{eqnarray}
\mathcal{U} & = & \left\langle q_{af}q_{ej}\mathcal{L}^{\prime}\mathcal{M}%
^{\prime}\left\vert O_{k_{l}\lambda}\right\vert q_{bf}L_{i} M%
\right\rangle \\
&& =\int d\Omega_{q_{bf}}d\Omega_{q_{af}}d\Omega_{q_{ej}}\mathcal{Y}%
_{L_{f}l_{x}}^{\ast\mathcal{L}^{\prime}\mathcal{M}^{\prime}}\left(
\widehat{q_{af}},\widehat{q_{ej}}\right) Y^{L_{i}}_{M}\left( \widehat{%
q_{bf}}\right) \mathcal{M}_{i}.
\end{eqnarray}
Using \eq{BBsum} and
\begin{equation}
\int d\Omega_{q_{ej}}\delta\left[ s\left( \vec{q}_{bf}-\vec{q}_{af}\right) -%
\frac{\vec{q}_{ej}}{2}\right]
\underset{s=\pm1}{=}\frac{\delta\left[
\left\vert s\left( \vec{q}_{bf}-\vec{q}_{af}\right) \right\vert -\frac{q_{ej}%
}{2}\right] }{\left\vert \vec{q}_{bf}-\vec{q}_{af}\right\vert
^{2}}\text{,}
\end{equation}
$\mathcal{U}$ reads
\begin{eqnarray}
\mathcal{U} & = & -\left( 4\pi\right) \sqrt{4\pi}\int
d\Omega_{q_{bf}}d\Omega_{q_{af}}\sum_{g_{1}+g_{2}=l_{x}}\frac{\left(
sq_{bf}\right) ^{g_{1}}\left( -sq_{af}\right) ^{g_{2}}}{\left\vert \vec{q}%
_{bf}-\vec {q}_{af}\right\vert ^{l_{x}}}F\left(
l_{x};g_{1}g_{2}\right) \\
&& \frac{\delta\left[ \left\vert s\left(
\vec{q}_{bf}-\vec{q}_{af}\right)
\right\vert -\frac{q_{ej}}{2}\right] }{\left\vert \vec{q}_{bf}-\vec{q}%
_{af}\right\vert ^{2}}\sum_{f_{1}+f_{2}=1}\left(
B^{\prime}q_{af}\right) ^{f_{2}}\left( C^{\prime}q_{bf}\right)
^{f_{1}}F\left( 1;f_{1}f_{2}\right)
\nonumber \\
&& \left\{ \sigma^{\left(i \right)}\otimes\mathcal{Y}_{f_{1}f_{2}}^{1}\left( \widehat{q_{bf}%
},\widehat{q_{af}}\right) \right\} _{0}^{0}\left\{ Y^{\ast
L_{f}}\left( \widehat{q_{af}}\right)
\otimes\mathcal{Y}_{g_{1}g_{2}}^{\ast
l_{x}}\left( \widehat{q_{bf}},\widehat{q_{af}}\right) \right\} _{%
\mathcal{M}^{\prime}}^{\mathcal{L}^{\prime}}Y^{L_{i}}_{M}\left( \widehat{%
q_{bf}}\right).  \nonumber
\end{eqnarray}
Now, the part depending on $x$ is decomposed in Legendre polynomials
\begin{eqnarray}
\mathcal{A}& = &\frac{\left( sq_{bf}\right) ^{g_{1}}\left(
-sq_{af}\right)
^{g_{2}}}{\left\vert \vec{q}_{bf}-\vec{q}_{af}\right\vert ^{l^{\prime }+2}}%
\left( B^{\prime }q_{af}\right) ^{f_{2}}\left( C^{\prime
}q_{bf}\right) ^{f_{1}}\delta \left[ \left\vert s\left(
\vec{q}_{bf}-\vec{q}_{af}\right)
\right\vert -q_{ej}\right]  \\
& = &\left( 4\pi \right) \sum_{l_{1}}f^{l_{1}l^{\prime
}f_{1}g_{1}}\left(
-1\right) ^{l_{1}}\frac{\widehat{l_{1}}}{2}\mathcal{Y}_{l_{1}l_{1}}^{0}%
\left( \widehat{q_{bf}},\widehat{q_{af}}\right)
\end{eqnarray}
where\footnote{%
Since $\left\vert s\right\vert =1$,%
\begin{equation}
f^{l_{1}l^{\prime }f_{1}g_{1}}=\left[ \frac{\left( sq_{bf}\right)
^{g_{1}}\left( -sq_{af}\right) ^{g_{2}}}{\left\vert \frac{q_{ej}}{2}%
\right\vert ^{l^{\prime }+1}}\left( B^{\prime }q_{l}\right)
^{f_{2}}\left( C^{\prime }q_{fx}\right)
^{f_{1}}\frac{1}{q_{l}q_{fx}}P_{l_{1}}\left( x\right) \right]
_{x=x_{0}}
\end{equation}%
with $x_{0}$ being the solution of $
\sqrt{q_{bf}^{2}-2q_{bf}q_{af}x_{0}+q_{af}^{2}}-\frac{q_{ej}}{2}=0$.
}
\begin{equation}
f^{l_{1}l^{\prime }f_{1}g_{1}}=\int dx\mathcal{A}P_{l_{1}}\left(
x\right),
\end{equation}
leading to
\begin{eqnarray}
\mathcal{U}& = &-\frac{\left( 4\pi \right) ^{2}}{2}\sqrt{4\pi
}\int d\Omega _{q_{bf}}d\Omega
_{q_{af}}\sum_{f_{1}+f_{2}=1}F\left( 1;f_{1}f_{2}\right)
\sum_{g_{1}+g_{2}=l_{x}}F\left( l_{x};g_{1}g_{2}\right)  \\
&& \sum_{l_{1}}f^{l_{1}l_{x}f_{1}g_{1}}\left( -1\right) ^{l_{1}}%
\widehat{l_{1}}\left\{ \sigma ^{\left(i \right)}\otimes
\mathcal{Y}_{f_{1}f_{2}}^{1}\left(
\widehat{q_{bf}},\widehat{q_{af}}\right) \right\} _{0}^{0}  \nonumber \\
&& \left\{ Y^{\ast L_{f}}\left( \widehat{q_{af}}\right) \otimes
\mathcal{Y}_{g_{1}g_{2}}^{\ast l_{x}}\left( \widehat{q_{bf}},\widehat{%
q_{af}}\right) \right\} _{\mathcal{M}^{\prime
}}^{\mathcal{L}^{\prime }}Y^{L_{i}}_{M}\left( \widehat{q_{bf}}\right) \mathcal{Y}_{l_{1}l_{1}}^{0}%
\left( \widehat{q_{bf}},\widehat{q_{af}}\right).   \nonumber
\end{eqnarray}
Using \eq{nYsomasoma}, one obtains
\begin{eqnarray}
\mathcal{U}& = &-\frac{\left( 4\pi \right) }{2}\widehat{L_{f}}\widehat{%
l_{x}}\int d\Omega _{q_{bf}}d\Omega _{q_{af}}\sum_{f_{1}+f_{2}=1}%
\widehat{f_{1}}\widehat{f_{2}}F\left( 1;f_{1}f_{2}\right)
\sum_{g_{1}+g_{2}=l_{x}}\widehat{g_{2}} \\
&& F\left( l_{x};g_{1}g_{2}\right) \sum_{l_{1}}\widehat{l_{1}}%
^{2}f^{l_{1}l_{x}f_{1}g_{1}}\left( -1\right)
^{l_{1}}\sum_{c_{2}}C_{000}^{L_{f}g_{2}c_{2}}\left\{
\begin{array}{ccc}
\mathcal{L}^{\prime } & c_{2} & g_{1} \\
g_{2} & l_{x} & L_{f}%
\end{array}%
\right\}
\sum_{d_{1}d_{2}}C_{000}^{f_{1}l_{1}d_{1}}C_{000}^{f_{2}l_{1}d_{2}}
\nonumber \\
&& \left\{
\begin{array}{ccc}
d_{1} & d_{2} & 1 \\
f_{2} & f_{1} & l_{1}%
\end{array}%
\right\} \left\{ \sigma ^{\left(i \right)}\otimes
\mathcal{Y}_{d_{1}d_{2}}^{1}\left(
\widehat{q_{bf}},\widehat{q_{af}}\right) \right\} _{0}^{0}\mathcal{Y}%
_{g_{1}c_{2}}^{\ast \mathcal{L}^{\prime }\mathcal{M}^{\prime
}}\left(
\widehat{q_{bf}},\widehat{q_{af}}\right) Y^{L_{i}}_{M}\left( \widehat{%
q_{bf}}\right).   \nonumber
\end{eqnarray}
Finally, the result of the angular integration (of the orbital part) is%
\footnote{%
Note that%
\begin{equation}
\left\{ \sigma^{\left( i \right)}\otimes\mathcal{Y}_{d_{1}d_{2}}^{1\mu}\left( \widehat{%
q_{fx}},\widehat{q_{l}}\right) \right\} _{0}^{0}=\sum_{\mu}\left(
-1\right)
^{1+\mu}\sigma_{-\mu}^{\left(i \right)}\mathcal{Y}_{d_{1}d_{2}}^{1\mu}\left( \widehat{%
q_{fx}},\widehat{q_{l}}\right)
\end{equation}
and also by \eq{sphdef}
\begin{eqnarray}
\mathcal{B} & = &\mathcal{Y}_{d_{1}d_{2}}^{1\mu}\left( \widehat{q_{fx}},%
\widehat{q_{l}}\right) \mathcal{Y}_{g_{1}c_{2}}^{\ast\mathcal{L}^{\prime }%
\mathcal{M}^{\prime}}\left( \widehat{q_{fx}},\widehat{q_{l}}\right) Y^{%
L_{i}}_{M}\left( \widehat{q_{fx}}\right) \\
& =
&\sum_{m_{1}m_{2}}C_{m_{1}m_{2}\mu}^{d_{1}d_{2}1}Y^{d_{1}}_{m_{1}}\left(
\widehat{q_{fx}}\right) Y^{d_{2}}_{m_{2}}\left( \widehat{q_{l}}\right) \\
&& \sum_{h_{1}h_{2}}C_{h_{1}h_{2}\mathcal{M}^{\prime}}^{g_{1}c_{2}\mathcal{L}%
^{\prime}}Y^{\ast g_{1}}_{h_{1}}\left( \widehat{q_{fx}}\right)
Y^{\ast
c_{2}}_{h_{2}}\left( \widehat{q_{l}}\right) Y^{L_{i}}_{M}\left( \widehat{%
q_{fx}}\right). \nonumber
\end{eqnarray}
}
\begin{eqnarray}
O & = &-\frac{\sqrt{4\pi}}{2}\widehat{L_{f}}\widehat{L_{i}}\widehat{%
l_{x}}\hat{1}\sum_{f_{1}+f_{2}=1}\widehat{f_{1}}\widehat{f_{2}}F\left(
1;f_{1}f_{2}\right) \sum_{g_{1}+g_{2}=l_{x}}\widehat{g_{2}}F\left(
l_{x};g_{1}g_{2}\right) \\
&& \sum_{l_{1}}f^{l_{1}l_{x}f_{1}g_{1}}\widehat{l_{1}}%
^{2}\sum_{c_{2}}C_{000}^{L_{f}g_{2}c_{2}}\left\{
\begin{array}{ccc}
\mathcal{L}^{\prime} & c_{2} & g_{1} \\
g_{2} & l_{x} & L_{f}%
\end{array}
\right\} \sum_{d_{1}}\widehat{d_{1}}%
C_{000}^{f_{1}l_{1}d_{1}}C_{000}^{f_{2}l_{1}c_{2}}  \nonumber \\
&& \left\{
\begin{array}{lll}
d_{1} & c_{2} & 1 \\
f_{2} & f_{1} & l_{1}%
\end{array}
\right\} C_{000}^{L_{i}d_{1}g_{1}}\left\{
\begin{array}{ccc}
L_{i} & d_{1} & g_{1} \\
c_{2} & \mathcal{L}^{\prime} & 1%
\end{array}
\right\} \left( -1\right) ^{d_{1}}C_{\mu M \mathcal{M}^{\prime}}^{1 L_{i} \mathcal{L}%
\prime},  \nonumber
\end{eqnarray}
and, therefore, the reduced matrix elements are
\begin{eqnarray}
O & = &-\frac{\sqrt{4\pi}}{2}\widehat{L_{f}}\widehat{L_{i}}\widehat{%
l_{x}}\hat{1}\sum_{f_{1}+f_{2}=1}\widehat{f_{1}}\widehat{f_{2}}F\left(
1;f_{1}f_{2}\right) \sum_{g_{1}+g_{2}=l_{x}}\widehat{g_{2}}F\left(
l_{x};g_{1}g_{2}\right)  \label{Oimp} \\
&& \sum_{l_{1}}f^{l_{1}l_{x}f_{1}g_{1}}\widehat{l_{1}}%
^{2}\sum_{c_{2}}C_{000}^{L_{f}g_{2}c_{2}}\left\{
\begin{array}{ccc}
\mathcal{L}^{\prime} & c_{2} & g_{1} \\
g_{2} & l_{x} & L_{f}%
\end{array}
\right\} \sum_{d_{1}}\widehat{d_{1}}%
C_{000}^{f_{1}l_{1}d_{1}}C_{000}^{f_{2}l_{1}c_{2}}  \nonumber \\
&& \left\{
\begin{array}{lll}
d_{1} & c_{2} & 1 \\
f_{2} & f_{1} & l_{1}%
\end{array}
\right\} C_{000}^{L_{i}d_{1}g_{1}}\left\{
\begin{array}{ccc}
L_{i} & d_{1} & g_{1} \\
c_{2} & \mathcal{L}^{\prime} & 1%
\end{array}
\right\} \left( -1\right) ^{d_{1}}  \nonumber
\end{eqnarray}

For the simplest case of a term like $\sigma ^{\left( i\right)
}\cdot \left(
\vec{p}_{1}+\vec{q}_{1}\right) $, \eq{Oimp} reduces to
\begin{eqnarray}
&& O\underset{f_{1}=0}{=}\frac{\sqrt{4\pi }}{2}\widehat{L_{f}}\widehat{%
L_{i}}\widehat{l_{x}}\hat{1}\sum_{g_{1}+g_{2}=l_{x}}%
\widehat{g_{2}}F\left( l_{x};g_{1}g_{2}\right)
\sum_{l_{1}}f^{l_{1}l_{x}0g_{1}} \\
&& \widehat{l_{1}}^{2}\sum_{c_{2}}C_{000}^{L_{f}g_{2}c_{2}}\left\{
\begin{array}{ccc}
\mathcal{L}^{\prime } & c_{2} & g_{1} \\
g_{2} & l_{x} & L_{f}%
\end{array}%
\right\} C_{000}^{1l_{1}c_{2}}C_{000}^{L_{i}l_{1}g_{1}}\left\{
\begin{array}{ccc}
L_{i} & l_{1} & g_{1} \\
c_{2} & \mathcal{L}^{\prime } & 1%
\end{array}%
\right\}   \notag
\end{eqnarray}
\subsection*{Spin matrix elements}
The spin matrix elements are given by \eq{spin1} and by
\eq{spin2}.
\section*{Z-diagrams \label{tforpwZ}}
The computation of the amplitude for the Z-diagrams involves the
calculation of terms like $\vec{\sigma}^{\left( 1\right)
}\cdot\left[ \vec {\sigma}^{\left( 2\right) }\times\vec{q}\right]
$. Since
\begin{equation}
\vec{\sigma}^{\left( 1\right) }\cdot\left[ \vec{\sigma}^{\left(
2\right)
}\times\vec{q}\right] =\left[ \vec{\sigma}^{\left( 1\right) }\times \vec{%
\sigma}^{\left( 2\right) }\right] \cdot\vec{q}   \label{tzcross}
\end{equation}
and
\begin{equation}
\left( \vec{\sigma}^{\left( i\right) }\times\vec{\sigma}^{\left(
j\right) }\right) \cdot\vec{q}=i\sqrt{2}\sqrt{4\pi}\left\vert
\vec{q}\right\vert
\left\{ \left[ \sigma^{\left( i\right) }\otimes\sigma^{\left( j\right) }%
\right] ^{1}\otimes Y^{1}\left( \hat{q}\right) \right\} _{0}^{0}
\end{equation}
the matrix elements of the orbital part are thus $i%
\sqrt{2}$ times the matrix elements of \eq{M1orb}.
\subsection*{Spin matrix elements}
The spin matrix elements are
\begin{eqnarray}
\left\langle s_{a}s_{b}S_{f}\left\Vert \left\{ \sigma^{\left(
1\right) }\otimes\sigma^{\left( 2\right) }\right\} ^{j}\right\Vert
s_{1}s_{2}S_{i}\right\rangle & =&\left( -1\right)
^{S_{f}-S_{i}+j}4\widehat{S_{i}}\hat{\jmath}\widehat{s_{1}}\widehat{s_{2}}
\label{spintens1}
\\
&& \sqrt{s_{1}\left( s_{1}+1\right) }\sqrt{s_{2}\left( s_{2}+1\right) }%
\left\{
\begin{array}{ccc}
S_{f} & S_{i} & j \\
s_{1} & s_{1} & 1 \\
s_{2} & s_{2} & 1%
\end{array}
\right\}  \nonumber
\end{eqnarray}
and
\begin{eqnarray}
\left\langle s_{a}s_{b}S_{f}\left\Vert \left\{ \sigma^{\left(
2\right) }\otimes\sigma^{\left( 1\right) }\right\} ^{j}\right\Vert
s_{1}s_{2}S_{i}\right\rangle & = &\left( -1\right) ^{j+2\left(
s_{1}+s_{2}-S_{i}\right)
}4\widehat{S_{i}}\hat{\jmath}\widehat{s_{1}}\widehat{s_{2}}
\label{spintens2} \\
&& \sqrt{s_{1}\left( s_{1}+1\right) }\sqrt{s_{2}\left( s_{2}+1\right) }%
\left\{
\begin{array}{ccc}
S_{f} & S_{i} & j \\
s_{2} & s_{2} & 1 \\
s_{1} & s_{1} & 1%
\end{array}
\right\}  \nonumber
\end{eqnarray}
\section*{$\Delta$ resonance contribution \label{tforpwZ}}

The $\Delta$ contribution diagram orbital and spin matrix elements
are similar to the corresponding ones for the re-scattering
diagram (\eq{M1orb} and \eq{spin1}, respectively).
\section*{Isospin matrix elements}
The isospin matrix elements involved in the calculations are
\begin{align}
\left\langle pp\left\vert \tau_{3}^{\left( 1\right) }\right\vert
pp\right\rangle & =\left\langle pp\left\vert \tau_{3}^{\left(
2\right)
}\right\vert pp\right\rangle =1 \\
\left\langle pp\left\vert \tau_{+}^{\left( 1\right) }\right\vert
np\right\rangle & =\left\langle pp\left\vert \tau_{+}^{\left(
2\right)
}\right\vert pn\right\rangle =2 \\
\left\langle np\left\vert \tau_{-}^{\left( 1\right) }\right\vert
pp\right\rangle & =\left\langle pn\left\vert \tau_{-}^{\left(
2\right) }\right\vert pp\right\rangle =2
\end{align}
with $\tau_{\pm}=\frac{\tau_{1}\pm i\tau_{2}}{2}$, and
\begin{align}
\left\langle pp\left\vert \epsilon_{abc}\tau_{b}^{\left( 1\right)
}\tau
_{c}^{\left( 2\right) }\right\vert np\right\rangle & =-2i \\
\left\langle np\left\vert \epsilon_{abc}\tau_{b}^{\left( 1\right)
}\tau _{c}^{\left( 2\right) }\right\vert pp\right\rangle & =2i
\end{align}
\section*{Coefficients $A$, $B$ and $C$ for the diagrams
considered}
\subsection*{Re-scattering diagram}
The coefficients $A$, $B$ and $C$ of the partial wave
decomposition of the re-scattering diagram, \eq{tresc}, are listed
on \tb{coeffpwrescatt}. The coefficients for the $\Delta$
contribution diagram can be easily inferred from
\tb{coeffpwrescatt}.
\subsection*{Direct-production diagram}
The coefficients of the partial wave decomposition of the
direct-production term of \eq{timp} are listed in \tb{coeffpwimp}.
\subsection*{Z-diagrams}
The coefficients of the partial wave decomposition of the
Z-diagrams are in \tb{coeffpwzs} (for $\sigma$-exchange) and in
\tb{coeffpwzw1}-\tb{coeffpwzw3} (for $\omega$-exchange).
%
%
\begin{table}
\begin{center}
\begin{tabular}
[c]{l|c|c|c|c}\hline\hline
& FSI$^{\text{a)}}$ & FSI$^{\text{b)}}$ & ISI$^{\text{a)}}$ & ISI$^{\text{b)}%
}$\\\hline $\vec{q}$ &
$\frac{\vec{q}_{\pi}}{2}-\vec{p}+\vec{q}_{k}$ & $\frac{\vec
{q}_{\pi}}{2}+\vec{p}-\vec{q}_{k}$ & $\frac{\vec{q}_{\pi}}{2}-\vec{q}_{k}%
+\vec{q}_{u}$ & $\frac{\vec{q}_{\pi}}{2}+\vec{q}_{k}-\vec{q}_{u}$\\
$\vec{q}_{af}$ & $\vec{q}_{k}$ & $\vec{q}_{k}$ & $\vec{q}_{u}$ & $\vec{q}_{u}%
$\\
$\vec{q}_{bf}$ & $\vec{p}$ & $\vec{p}$ & $\vec{q}_{k}$ & $\vec{q}_{k}$\\
$A$ & $1$ & $-1$ & $1$ & $-1$\\
$B$ & $1$ & $-1$ & $1$ & $-1$\\
$C$ & $\frac{1}{2}$ & $\frac{1}{2}$ & $\frac{1}{2}$ & $\frac{1}{2}%
$\\\hline\hline
\end{tabular}
\captions[Coefficients of the partial wave decomposition of the
re-scattering diagram]{Coefficients of the partial wave
decomposition of the re-scattering diagram. The b) superscript
refers to the $1 \leftrightarrow 2$
diagram.\label{coeffpwrescatt}}
\end{center}
\end{table}
\begin{table}
\begin{center}
\begin{tabular}
[c]{l|c|c|c|c}\hline\hline
& FSI$^{\text{a)}}$ & FSI$^{\text{b)}}$ & ISI$^{\text{a)}}$ & ISI$^{\text{b)}%
}$\\\hline $\vec{q}_{bf}$ & $\vec{p}$ & $\vec{p}$ &
$\vec{q}_{i}=\vec{q}_{u}+\frac
{\vec{q}_{\pi}}{2}$ & $\vec{q}_{i}^{\prime}=\vec{q}_{u}-\frac{\vec{q}_{\pi}%
}{2}$\\
$\vec{q}_{af}$ & $\vec{q}_{f}=\vec{p}-\frac{\vec{q}_{\pi}}{2}$ & $\vec{q}%
_{f}^{\prime}=\vec{p}+\frac{\vec{q}_{\pi}}{2}$ & $\vec{q}_{u}$ & $\vec{q}_{u}%
$\\
$A$ & $-\left[  1+\frac{E_{\pi}}{2\left(  E+M\right)  }\right]  $
& $-\left[
1+\frac{E_{\pi}}{2\left(  F+M\right)  }\right]  $ & $-\left[  1+\frac{E_{\pi}%
}{2\left(  F_{1}+M\right)  }\right]  $ & $-\left[
1+\frac{E_{\pi}}{2\left(
F_{2}+M\right)  }\right]  $\\
$B$ & $\frac{E_{\pi}}{E+M}$ & $-\frac{E_{\pi}}{F+M}$ & $\frac{E_{\pi}}%
{F_{1}+M}$ & $-\frac{E_{\pi}}{F_{2}+M}$\\
$C$ & $\frac{E_{\pi}}{E_{1}+M}$ & $-\frac{E_{\pi}}{E_{2}+M}$ & $\frac{E_{\pi}%
}{E^{\prime}+M}$ & $-\frac{E_{\pi}}{F^{\prime}+M}$\\
$s$ & $1$ & $-1$ & $1$ & $1$\\\hline\hline
\end{tabular}
\captions[Coefficients of the partial wave decomposition of the
direct-production diagram]{Coefficients of the partial wave
decomposition of the direct-production diagram. The b) superscript
refers to the $1 \leftrightarrow 2$ diagram.\label{coeffpwimp}}
\end{center}
\end{table}
\begin{table}[h!]
\begin{center}
\begin{tabular}
[c]{l|c|c|c|c}\hline\hline
& FSI$^{\text{a)}}$ & FSI$^{\text{b)}}$ & ISI$^{\text{a)}}$ & ISI$^{\text{b)}%
}$\\\hline $\vec{q}_{i}$ & $\vec{q}_{bf}-\vec{q}_{\pi}$ &
$-\vec{q}_{bf}-\vec{q}_{\pi}$ &
$\vec{q}_{af}+\vec{q}_{\pi}$ & $-\vec{q}_{af}+\vec{q}_{\pi}$\\
$\vec{k}$ & $\vec{q}_{bf}-\vec{q}_{af}-\frac{\vec{q}_{\pi}}{2}$ &
$\vec
{q}_{bf}-\vec{q}_{af}+\frac{\vec{q}_{\pi}}{2}$ & $\vec{q}_{bf}-\vec{q}%
_{af}-\frac{\vec{q}_{\pi}}{2}$ & $\vec{q}_{bf}-\vec{q}_{af}+\frac{\vec{q}%
_{\pi}}{2}$\\
$A$ & $\frac{1}{F_{1}+M}$ & $-\frac{1}{F_{2}+M}$ &
$\frac{1}{M-E^{\prime}}$ &
$-\frac{1}{M-F^{\prime}}$\\
$B$ & $-\frac{1}{M-E}$ & $-\frac{1}{M-F}$ & $-\frac{1}{E_{1}+M}$ &
$\frac
{1}{E_{2}+M}$\\
$C$ & $-\left[  \frac{1}{M-E}+\frac{1}{2\left(  F_{1}+M\right)
}\right]  $ & $-\left[  \frac{1}{M-F}+\frac{1}{2\left(
F_{2}+M\right)  }\right]  $ & $\frac{1}{2\left(
M-E^{\prime}\right)  }$ & $\frac{1}{2\left(  M-F^{\prime }\right)
}$\\\hline\hline
\end{tabular}
\captions[Coefficients of the partial wave decomposition of the
$\sigma$-exchange part of the Z-diagrams]{Coefficients of the
partial wave decomposition of the $\sigma$-exchange part of the
Z-diagrams \eq{tresc}. $\vec{q}_{bf}=\vec{p}$ and
$\vec{q}_{af}=\vec{q}_{u}$. \label{coeffpwzs}}
\end{center}
\end{table}
\begin{table}
\begin{center}
\begin{tabular}
[c]{l|c|c}\hline\hline & FSI$^{\text{a)}}$ &
FSI$^{\text{b)}}$\\\hline $A$ &
$-\frac{1}{F_{1}+M}-\frac{1}{F_{2}+M}-\frac{x_{\omega}}{2M}$ &
$\frac
{1}{F_{1}+M}+\frac{1}{F_{2}+M}+\frac{x_{\omega}}{2M}$\\
$B$ & $-\frac{1}{M-E}+\frac{1}{E_{2}+M}-\frac{x_{\omega}}{2M}$ &
$-\frac
{1}{E_{1}+M}+\frac{1}{M-F}+\frac{x_{\omega}}{2M}$\\
$C$ & $-\frac{1}{M-E}+\frac{1}{2}\left(  \frac{1}{F_{1}+M}-\frac{1}{F_{2}%
+M}\right)  -\frac{x_{\omega}}{4M}$ &
$-\frac{1}{M-F}-\frac{1}{2}\left(
\frac{1}{F_{1}+M}-\frac{1}{F_{2}+M}\right)  -\frac{x_{\omega}}{4M}%
$\\\hline\hline
\end{tabular}
\captions[Coefficients of the partial wave decomposition of the
$\omega$-exchange part of the Z-diagrams for the FSI
case]{Coefficients of the partial wave decomposition of the
$\omega$-exchange part of the Z-diagrams for the FSI case
\eq{tresc}. \label{coeffpwzw1}}
\end{center}
\end{table}
\begin{table}
\begin{center}
\begin{tabular}
[c]{l|c|c}\hline\hline & ISI$^{\text{a)}}$ &
ISI$^{\text{b)}}$\\\hline $A$ &
$\frac{1}{M-E^{\prime}}-\frac{1}{F_{2}+M}+\frac{x_{\omega}}{2M}$ &
$\frac{1}{F_{1}+M}-\frac{1}{M-F^{\prime}}-\frac{x_{\omega}}{2M}$\\
$B$ & $\frac{1}{E_{1}+M}+\frac{1}{E_{2}+M}+\frac{x_{\omega}}{2M}$
& $-\frac
{1}{E_{1}+M}-\frac{1}{E_{2}+M}-\frac{x_{\omega}}{2M}$\\
$C$ & $\frac{1}{2\left(  M-E^{\prime}\right)  }-\frac{1}{2\left(
F_{2}+M\right)  }+\frac{x_{\omega}}{4M}$ & $\frac{1}{2\left(
M-F^{\prime
}\right)  }-\frac{1}{2\left(  F_{1}+M\right)  }+\frac{x_{\omega}}{4M}%
$\\\hline\hline
\end{tabular}
\captions[Coefficients of the partial wave decomposition of the
$\omega$-exchange part of the Z-diagrams for the ISI case]{The
same of \tb{coeffpwzw1} for the ISI case. \label{coeffpwzw2}}
\end{center}
\end{table}
\begin{table}
\begin{center}
\begin{tabular}
[c]{l|c|c|c|c}\hline\hline
& FSI$^{\text{a)}}$ & FSI$^{\text{b)}}$ & ISI$^{\text{a)}}$ & FSI$^{\text{b)}%
}$\\\hline $A$ & $\left(
-\frac{1}{F_{2}+M}+\frac{x_{\omega}}{2M}\right)  i$ & $\left(
\frac{1}{F_{1}+M}-\frac{x_{\omega}}{2M}\right)  i$ & $\left(  -\frac{1}%
{F_{2}+M}+\frac{x_{\omega}}{2M}\right)  i$ & $\left(  \frac{1}{F_{1}+M}%
-\frac{x_{\omega}}{2M}\right)  i$\\
$B$ & $\left(  -\frac{1}{E_{2}+M}+\frac{x_{\omega}}{2M}\right)  i$
& $\left(
\frac{1}{E_{1}+M}-\frac{x_{\omega}}{2M}\right)  i$ & $\left(  -\frac{1}%
{E_{2}+M}+\frac{x_{\omega}}{2M}\right)  i$ & $\left(  \frac{1}{E_{1}+M}%
-\frac{x_{\omega}}{2M}\right)  i$\\
$C$ & $\left[  -\frac{1}{2\left(  F_{2}+M\right)  }+\frac{x_{\omega}}%
{4M}\right]  i$ & $\left[  -\frac{1}{2\left(  F_{1}+M\right)
}+\frac {x_{\omega}}{4M}\right]  i$ & $\left[  -\frac{1}{2\left(
F_{2}+M\right) }+\frac{x_{\omega}}{4M}\right]  i$ & $\left[
-\frac{1}{2\left( F_{1}+M\right)  }+\frac{x_{\omega}}{4M}\right]
i$\\\hline\hline
\end{tabular}
\captions[Coefficients of the partial wave decomposition of the
$\omega$-exchange part of the Z-diagrams]{Coefficients of the
partial wave decomposition of the $\omega$-exchange part of the
Z-diagrams \eq{tzcross}. \label{coeffpwzw3}}
\end{center}
\end{table}

%% file: Appendix_F.tex
\clearpage
\chapter{Numerical evaluation of integrals with pole singularities \label{ApImpI}}
The integrals appearing in usual $T$-matrix equations are of the
form (see \apx{ApTm}):
\begin{equation}
\int_{0}^{\infty}dk^{\prime} k^{\prime 2}\frac{g \left(k^{\prime}
\right)}{k^{2}-k^{\prime 2}+ i \varepsilon}, \label{typeintforT}
\end{equation}
where a pole exists at $k^{\prime}=k$. $g \left( k^{\prime}
\right)$ is a regular function of $k^{\prime}$ (i. e., it has no
poles). The $i \varepsilon$ term of the denominator of
\eq{typeintforT} can be evaluated by using the Cauchy Principal
value theorem\cite{Marsden:1987bk},
\begin{equation}
\int_{0}^{\infty}dk^{\prime}k^{\prime 2} \frac{g \left(k^{\prime}
\right)}{k^{2}-k^{\prime 2}+ i \varepsilon} =
\mathbf{PV}\int_{0}^{\infty}dk^{\prime} k^{\prime 2}\frac{g
\left(k^{\prime} \right)}{k^{2}-k^{\prime 2}}- i \pi
\int_{0}^{\infty}dk^{\prime}k^{\prime 2}\delta
\left(k^{\prime}\right) g \left(k^{2}-k^{\prime 2}\right),
\label{PVtypeintforT}
\end{equation}
where
\begin{equation}
\mathbf{PV}\int_{0}^{\infty}dk^{\prime} k^{\prime 2}\frac{g
\left(k^{\prime} \right)}{k^{2}-k^{\prime 2}} \equiv
\lim_{\varepsilon \rightarrow 0}\left[
\int_{0}^{k-\varepsilon}dk^{\prime} k^{\prime 2}\frac{g
\left(k^{\prime} \right)}{k^{2}-k^{\prime
2}}+\int_{k+\varepsilon}^{\infty}dk^{\prime} k^{\prime 2}\frac{g
\left(k^{\prime} \right)}{k^{2}-k^{\prime 2}} \right]
\end{equation}
The pole at $k^{\prime}=k$ in \eq{PVtypeintforT} has to be treated
carefully. This can be done numerically by using the subtraction
method with regularisation\cite{Haftel:1970ta} to calculate the
integral. The basis of the subtraction method is to treat the pole
in the numerical integral by subtracting (and adding separately)
an analytically defined integral
\begin{eqnarray}
\int_{0}^{\infty}dk^{\prime}k^{\prime2}\frac{g\left(
k^{\prime}\right) }{k^{2}-k^{\prime2}+i\varepsilon}  && =
\mathbf{PV}\int_{0}^{\infty}dk^{\prime
}\left(  k^{\prime2}\frac{g\left(  k^{\prime}\right)  }{k^{2}-k^{\prime2}%
}-k^{2}\frac{g\left(  k\right) }{k^{2}-k^{\prime2}}\right)
\label{RtypeintforT}
\\
&& +\mathbf{PV}\int_{0}^{\infty}dk^{\prime}k^{2}\frac{g\left(
k\right)}{k^{2}-k^{\prime2}} \nonumber
\\
&& -i\pi\int_{0}^{\infty}dk^{\prime}k^{\prime2}\delta\left(
k^{2}-k^{\prime2}\right)  g\left( k^{\prime}\right) \nonumber
\end{eqnarray}
The first integration on the right hand side of \eq{RtypeintforT}
can be performed numerically because the pole no longer exists
(when $k^{\prime} \rightarrow k$, it goes to $\frac{0}{0}$). The
second integration may be solved analytically as
\begin{eqnarray}
I_{1} & \equiv &  \mathbf{PV}\int_{0}^{k_{max}}dk^{\prime}k^{2}\frac{g\left(  k\right) }%
{k^{2}-k^{\prime2}} \\
&  = & \mathbf{PV}\int_{0}^{k_{max}}dk^{\prime} k^{2}g\left(
k\right)\left[ \frac{1}{2k}\left(  \frac{1}{k-k^{\prime}}+\frac
{1}{k+k^{\prime}}\right)  \right]  \\
& = &\lim_{\varepsilon\rightarrow0}\left(  \left[ \ln\left\vert
\frac {k+k^{\prime}}{k^{\prime}-k}\right\vert \right]
_{0}^{k-\varepsilon}+\left[ \ln\left\vert
\frac{k+k^{\prime}}{k-k^{\prime}}\right\vert \right]
_{k+\varepsilon}^{k_{max}}\right)  \frac{k}{2}g\left(  k\right)  \\
&  = &\lim_{\varepsilon\rightarrow0}\left( \ln\left\vert
\frac{2k-\varepsilon }{\varepsilon}\right\vert -\ln\left\vert
\frac{k}{k}\right\vert +\ln\left\vert
\frac{k_{max}+k}{k-k_{max}}\right\vert -\ln\left\vert
\frac{2k+\varepsilon
}{\varepsilon}\right\vert \right)  \frac{k}{2}g\left(  k\right) \\
& = &-\frac{k}{2}g\left(  k\right) \ln\frac{k_{max}-k}{k_{max}+k}
\label{LntypeintforT}%
\end{eqnarray}
In the limit $k_{max} \rightarrow \infty$ the integral of
\eq{LntypeintforT} reduces to zero, and in that case,
\eq{RtypeintforT} reads
\begin{eqnarray}
\hspace{-3.0cm} I_{2} & \equiv &
\int_{0}^{\infty}dk^{\prime}k^{\prime2}\frac{g\left(
k^{\prime}\right) }{k^{2}-k^{\prime2}+i\varepsilon} \\
& = & \hspace{-0.2cm} \mathbf{PV}\int_{0}^{\infty}dk^{\prime
}\left(  k^{\prime2}\frac{g\left(  k^{\prime}\right)  }{k^{2}-k^{\prime2}%
}-k^{2}\frac{g\left(  k\right)  }{k^{2}-k^{\prime2}}\right)
-i\pi\int_{0}^{\infty}dk^{\prime}k^{\prime2}\delta\left(
k^{2}-k^{\prime 2}\right)  g\left(  k^{\prime}\right) \\
&= & \hspace{-0.2cm} \mathbf{PV}\int_{0}^{\infty}dk^{\prime
}\left(  k^{\prime2}\frac{g\left(  k^{\prime}\right)  }{k^{2}-k^{\prime2}%
}-k^{2}\frac{g\left(  k\right)  }{k^{2}-k^{\prime2}}\right)
-i\frac{\pi}{2}kg\left(  k\right). \label{StypeintforT}
\end{eqnarray}
The integral of \eq{StypeintforT} is no longer a numerical
problem. To evaluate it numerically, one simply needs to map a
variable $x$ in a finite interval to
$k^{\prime}$~\cite{{Nr77:1992},{Nr90:1996}}. For instance, using
\begin{equation}
k^{\prime}=c \tan\left(\frac{\pi}{2} x \right)\hspace{1cm}
\text{with } 0 \leq x \leq 1,
\end{equation}
where $c$ is an appropriate constant, or
\begin{equation}
k^{\prime}=\frac{x+1}{1-x}\hspace{1cm} \text{with } -1 \leq x \leq
1.
\end{equation}

%% file: Appendix_G.tex
\clearpage\chapter{Cross section for pion production
\label{Apcrosssection}}

The unpolarised cross section for pion production in
nucleon-nucleon collisions can be calculated
from\cite{Bjorken:1964bk}
\begin{eqnarray}
\sigma & = & \int \frac{\left( 2M_{1}\right) \left( 2M_{2}\right) }{\sqrt{%
\left( P_{1}\cdot P_{2}\right)
^{2}-M_{1}^{2}M_{2}^{2}}}\overline{\left| \mathcal{M}_{fi}\right|
}^{2}\left( 2\pi \right) ^{4}\delta ^{4}\left(
P_{1}+P_{2}-Q_{1}-Q_{2}-Q_{\pi}\right) \label{crossDirac} \\
&& \frac{%
\left( 2M_{1}\right) \left( 2M_{2}\right) d^{3}q_{1}
d^{3}q_{2}d^{3}q_{\pi}}{\left( 2\pi \right)
^{9}q_{1}^{0}q_{2}^{0}2q_{\pi}^{0}}\mathcal{S} \nonumber
\end{eqnarray}
where $\mathcal{S}$ is the symmetry factor which accounts for
different particles in the final state. $P_{i}\left(Q_{i}\right)$
are the nucleons initial(final) four-momenta. Note that $\sigma$
is built up from three factors, namely the flux factor,
\begin{equation}
\left| \vec{v}_{1}-\vec{v}_{2}\right| \frac{E_{1}}{M_{1}}\frac{E_{2}}{M_{2}%
}=\frac{\sqrt{\left( P_{1}\cdot P_{2}\right) ^{2}-M_{1}^{2}M_{2}^{2}}}{%
M_{1}M_{2}},
\end{equation}
the phase-space,
\begin{equation}
\frac{M_{1}d^{3}q_{1}}{q_{1}^{0}\left( 2\pi \right) ^{3}}\frac{%
M_{2}d^{3}q_{2}}{q_{2}^{0}\left( 2\pi \right) ^{3}}\frac{d^{3}q_{\pi }}{%
\left( 2\pi \right) ^{3}2q_{\pi }^{0}}\left( 2\pi \right)
^{4}\delta ^{4}\left( P_{1}+P_{2}-Q_{1}-Q_{2}-Q_{\pi }\right)
\mathcal{S},
\end{equation}
and the amplitude for the process (averaged over the initial spins
and summed in the final spins),
\begin{equation}
\overline{\left| \mathcal{M}_{fi}\right| }^{2}=\frac{1}{4}%
\sum\limits_{spins}\left| \mathcal{M}_{fi}\right| ^{2}.
\end{equation}
Choosing the overall centre-of-mass system, \eq{crossDirac} reads,
\begin{equation}
\sigma =\frac{M^{2}}{4pE}\frac{1}{\left( 2\pi \right) ^{5}}\mathcal{S}\int \frac{M^{2}}{%
E_{\pi }}\dfrac{1}{E_{pp}E_{k}}\overline{\left|
\mathcal{M}_{fi}\right| }^{2}\delta ^{4}\left(
P_{1}+P_{2}-U_{1}-U_{2}-Q_{\pi }\right)
d^{3}q_{pp}d^{3}kd^{3}q_{\pi },  \label{secb}
\end{equation}
where $2E=\sqrt{s}$ is the initial CM energy,
$E_{k}=\sqrt{M^{2}+k^{2}}$ is the energy of each final nucleon
relatively to the CM of two final $NN$ system. $E_{pp}=2E-E_{\pi
}$ and $E_{\pi}=\sqrt{q_{\pi }^{2}+m_{\pi }^{2}}$ are the energy
of the final $NN$ system and the energy of the produced pion
relatively to the CM, respectively.

The integral in $d^{3}q_{pp}$ can be easily calculated using the
$\delta^{3}$-function,
\begin{equation}
\frac{d\sigma }{d\Omega _{k}d\Omega _{\pi }}=\frac{M^{2}}{4pE}\frac{1}{%
\left( 2\pi \right) ^{5}} \mathcal{S}\int \frac{M^{2}}{E_{\pi }}\frac{1}{E_{pp}E_{k}%
}\overline{\left| \mathcal{M}_{fi}\right| }^{2}\delta \left(
2E-Epp-E_{\pi }\right) k^{2}q_{\pi }^{2}dkdq_{\pi }.
\label{ddsigma}
\end{equation}
The remaining $\delta$ function can then be used to perform the
$dk$ integration. From
\begin{equation}
\left\{
\begin{array}{l}
E_{pp}^{2}=\left( 2E-E_{\pi }\right) ^{2} \\
E_{pp}^{2}-q_{\pi }^{2}=\left[ 2\sqrt{M^{2}+k^{2}}\right] ^{2}
\end{array}
\right. \Rightarrow s-2\sqrt{s}E_{\pi }+E_{\pi }^{2}=4\left(
M^{2}+k^{2}\right) +q_{\pi }^{2},
\end{equation}
one concludes that the argument of the $\delta$-function vanishes
for
\begin{equation}
k_{zero}^{2}=\frac{1}{4}\left( s-2\sqrt{s}\sqrt{m_{\pi }^{2}+q_{\pi }^{2}}%
+m_{\pi }^{2}-4M^{2}\right).  \label{kzero}
\end{equation}
Since
\begin{equation}
E_{pp}^{2}=\left[ 2\sqrt{M^{2}+k^{2}}\right] ^{2}+q_{\pi
}^{2}\Rightarrow 2E_{pp}dE_{pp}=8kdk,
\end{equation}
the final expression for the cross section is\footnote{To express
the cross section in \textit{barn}, \eq{diffcross} must be
multiplied by a factor of $\left( \hslash c\right) ^{2}\times
10^{-2} $.}
\begin{eqnarray}
\frac{d\sigma }{d\Omega _{k}d\Omega _{\pi }}&=& \frac{M^{2}}{4pE}\frac{1}{%
\left( 2\pi \right) ^{5}}\mathcal{S}\frac{1}{4}\int \frac{M^{2}}{E_{\pi }}\frac{1}{%
E_{kzero}}\overline{\left| \mathcal{M}_{fi}\right|
}^{2}k_{zero}q_{\pi }^{2}dq_{\pi } \label{diffcross} \\
&=&  \int K_{i}^{2}  \overline{\left| \mathcal{M}_{fi}\right|
}^{2}dq_{\pi },
\end{eqnarray}
where
\begin{equation}
K_{i}^{2}= \frac{M^{2}}{4pE}\frac{1}{%
\left( 2\pi \right) ^{5}}\frac{1}{4}\mathcal{S}\int
\frac{M^{2}}{E_{\pi }} \frac{1}{E_{kzero}}^{2}k_{zero}q_{\pi }^{2}
\label{Kidef}.
\end{equation}
\eq{diffcross} can also be written in terms of an integral over
the pion energy as,
\begin{equation}
\frac{d\sigma }{d\Omega _{k}d\Omega _{\pi }}=\frac{M^{2}}{4pE}\frac{1}{%
\left( 2\pi \right) ^{5}}\mathcal{S} \int_{m_{\pi }}^{E_{\pi }^{\max }}\dfrac{%
M^{2}}{E_{kzero}}\overline{\left| \mathcal{M}_{fi}\right| }^{2}k_{zero}\sqrt{%
E_{\pi }^{2}-m_{\pi }^{2}}dE_{\pi },  \label{diffcrossE}
\end{equation}
with $E_{\pi }^{\max }=m_{\pi}\sqrt{1+\eta^{2}}$.

%% file: Appendix_H.tex
\clearpage\chapter[Clebsch-Gordan coefficients, Six-J and Nine-J
symbols]{Clebsch-Gordan coefficients, Six-J and Nine-J
symbols\label{ApSixNine}}
\vspace{-0.1cm}

This Appendix lists a series of useful relations of tensorial
calculus\cite{{Wong:1990bk},{Brink:1993bk}} employed in the
partial wave decomposition of the amplitudes (see \apx{ApPWAmp}).
\vspace{-0.1cm}
\section*{Legendre Polynomials}
\vspace{-1cm}
\begin{align}
& P_{l}\left( x\right)
=\dfrac{1}{2^{l}l!}\dfrac{d^{l}}{dx^{l}}\left( x^{2}-1\right) ^{l}
\hspace{8.5cm}\\
& \int_{-1}^{1}P_{l}\left( x\right) P_{l^{\prime }}\left( x\right) dx=\dfrac{2%
}{2l+1}\delta _{ll^{\prime }} \\
& \sum\limits_{l=0}^{\infty }\dfrac{2l+1}{2}P_{l}\left( x\right)
P_{l}\left( x^{\prime }\right) =\delta \left( x-x^{\prime
}\right)\\
& f\left( \hat{a}\cdot \hat{b}\right) =\sum\limits_{l}\dfrac{2l+1}{2}%
P_{l}\left( \hat{a}\cdot \hat{b}\right) f_{l} \\
&  f_{l}=\int_{-1}^{1}d\left( \hat{a}\cdot \hat{b}\right) f\left(
\hat{a}\cdot \hat{b}\right) P_{l}\left( \hat{a}\cdot
\hat{b}\right)
\end{align}
\section*{Spherical Harmonics}
\vspace{-1cm}
\begin{align}
& Y_{m}^{l}\left( -\hat{a}\right) =\left( -\right) ^{l}Y_{m}^{l}\left( \hat{a}%
\right) \hspace{9.0cm} \\
& Y_{m}^{l}\left( \pi -\theta ,\pi +\varphi \right) =\left(
-\right)
^{l}Y_{m}^{l}\left( \theta ,\varphi \right)\\
& Y_{m}^{l \ast }\left( \hat{a}\right) =\left( -\right) ^{m}Y_{-m}^{l}\left( \hat{a%
}\right) \\
& Y_{m}^{l}\left( \hat{0}\right) =\sqrt{\dfrac{2l+1}{4\pi }}\delta
_{m0} \\[0.2cm]
& \int d\hat{q}Y_{m}^{l \ast }\left( \hat{q}\right)
Y_{m^{\prime}}^{l^{\prime }}\left( \hat{q}\right) =\delta
_{ll^{\prime }}\delta
_{mm^{\prime }} \\
& \sum\limits_{lm}Y_{m}^{l \ast }\left( \hat{q}\right) Y_{m}^{l}\left( \hat{q}%
^{\prime }\right) =\delta \left( \hat{q}-\hat{q}^{\prime }\right)
\end{align}
\begin{align}
 Y_{m_{1}}^{l_{1}}\left( \hat{p}\right) Y_{m_{2}}^{l_{2}\ast}\left( \hat {p}%
\right)  & =  \left( -1\right) ^{m_{2}}\frac{1}{\sqrt{4\pi}}\sqrt {2l_{1}+1}%
\sqrt{2l_{2}+1}  \label{prodY} \hspace{4.9cm}\\
& \sum_{LM}Y_{M}^{L}\left( \hat{p}\right)
\frac{1}{\sqrt{2L+1}}\left\langle
l_{1}l_{2}m_{1}-m_{2}|LM\right\rangle \left\langle
l_{1}l_{2}00|L0\right\rangle \nonumber
\end{align}
\begin{align}
 \sum\limits_{lm}Y_{m}^{l \ast }\left( \theta ,\varphi \right)
Y_{m}^{l}\left( \theta ^{\prime },\varphi ^{\prime }\right)
&=\delta \left( \cos \theta -\cos \theta ^{\prime }\right) \delta
\left( \varphi -\varphi ^{\prime }\right) \hspace{4.3cm}\\
&=\dfrac{1}{\sin \theta }\delta \left( \theta -\theta ^{\prime
}\right) \delta \left( \varphi -\varphi ^{\prime }\right)
\nonumber
\end{align}
\begin{align}
& \sum\limits_{m}Y_{m}^{l \ast }\left( \hat{q}\right) Y_{m}^{l}\left( \hat{q}%
^{\prime }\right) =\sqrt{\dfrac{2l+1}{4\pi }}Y_{l0}\left( \widehat{\vec{q}%
^{\prime }-\vec{q}}\right) =\dfrac{2l+1}{4\pi }P_{l}\left( \hat{q};\hat{q}%
^{\prime }\right)\\
& \sum\limits_{m=-l}^{l}Y_{m}^{l \ast }\left( \theta ,\varphi
\right)
Y_{m}^{l}\left( \theta ^{\prime },\varphi ^{\prime }\right) =\dfrac{2l+1}{4\pi }%
P_{l}\left( \cos \alpha \right) \hspace{1.1cm}\text{ with }\alpha
=\measuredangle \left[ \left( \theta ,\varphi \right) ;\left(
\theta ^{\prime },\varphi ^{\prime }\right) \right]
\end{align}
\vspace{0.3cm}
\subsection*{Legendre polynomials related to spherical harmonics}
\vspace{-0.6cm}
\begin{align}
& \mathcal{Y}_{ll}^{00}\left( \hat{a},\hat{b}\right) =
\dfrac{\left(
-1\right) ^{l}}{4\pi }\sqrt{2l+1}P_{l}\left( \hat{a}\cdot \hat{b}\right)\label{Plegeharmesfe1} \hspace{7.25cm}\\
& P_{l}\left( \hat{a}\cdot \hat{b}\right) =\dfrac{\left( -1\right)
^{l}}{\sqrt{2l+1}}4\pi \mathcal{Y}_{ll}^{00}\left( \hat{a},\hat{b}\right) \label{Plegeharmesfe2}\\
& \mathcal{Y}_{l_{1}l_{2}}^{lm}\left( \hat{a},\hat{a}\right) = \sqrt{\dfrac{%
\left( 2l_{1}+1\right) \left( 2l_{2}+1\right) }{4\pi \left( 2l+1\right) }}%
\left\langle l_{1}0l_{2}0\left| l0\right. \right\rangle Y_{m}^{l}\left( \hat{a}%
\right) \label{Plegeharmesfe3}
\end{align}
\vspace{0.5cm}
\section*{Coupling of angular momenta}
%
\subsection*{Clebsch-Gordon coefficients}
%
\subsubsection*{Closure relations}
\vspace{-0.9cm}
\begin{align}
& \sum_{m_{1}m_{2}}(-1)^{-2j_{1}+2j_{2}-m-m^{\prime}}\sqrt{\frac{2j+1}%
{2j^{\prime}+1}}\left\langle j_{1}j_{2}m_{1}m_{2}|jm\right\rangle
\left\langle
j_{1}j_{2}m_{1}m_{2}|j^{\prime}m^{\prime}\right\rangle =\delta_{jj^{\prime}%
}\delta_{mm^{\prime}}\label{cleborth1} \hspace{0.25cm}\\
& \sum_{jm}(-1)^{-2j_{1}+2j_{2}+2m}\left\langle j_{1}j_{2}m_{1}m_{2}%
|j-m\right\rangle \left\langle
j_{1}j_{2}m_{1}^{\prime}m_{2}^{\prime
}|j-m\right\rangle =\delta_{m_{1}m_{1}^{\prime}}\delta_{m_{2}m_{2}^{\prime}%
}\label{cleborth2}
\end{align}

\subsubsection*{Symmetries}
\vspace{-0.9cm}
\begin{align}
& \left\langle j_{1}j_{2}m_{1}m_{2}\left\vert jm\right.
\right\rangle
=\left(  -1\right)  ^{j_{1}+j_{2}-j}\left\langle j_{1}j_{2}-m_{1}%
-m_{2}\left\vert j-m\right.  \right\rangle \label{clebsym1} \hspace{4cm}\\
& \left\langle j_{1}j_{2}m_{1}m_{2}\left\vert jm\right.
\right\rangle
=\left(  -1\right)  ^{j_{1}+j_{2}-j}\left\langle j_{2}j_{1}m_{2}%
m_{1}\left\vert jm\right.  \right\rangle \label{clebsym2}
\end{align}
\begin{align}
& \left\langle j_{1}j_{2}m_{1}m_{2}\left\vert jm\right.
\right\rangle =\sqrt{\frac{2j+1}{2j_{2}+1}}\left(  -1\right)
^{j_{1}-m_{1}}\left\langle j_{1}jm_{1}-m\left\vert
j_{2}-m_{2}\right.  \right\rangle \label{clebsym3} \hspace{2.0cm}\\
&
\left\langle j_{1}j_{2}m_{1}m_{2}\left\vert jm\right.
\right\rangle =\sqrt{\frac{2j+1}{2j_{1}+1}}\left(  -1\right)
^{j_{2}+m_{2}}\left\langle
jj_{2}-mm_{2}\left\vert j_{1}-m_{1}\right.  \right\rangle \label{clebsym4}%
\end{align}
\subsubsection*{Special cases}
\vspace{-0.9cm}
\begin{align}
& \left\langle j_{1}j_{2}m_{1}m_{2}\left\vert 00\right.
\right\rangle
=\left(  -1\right)  ^{j_{1}-m_{1}}\sqrt{2j_{1}+1}\delta_{j_{1}j_{2}}%
\delta_{m_{1},-m_{2}}\label{clebsp1} \hspace{4.05cm}\\
& \left\langle j_{1}0m_{1}0\left\vert jm\right.  \right\rangle
=\delta
_{j_{1}j}\delta_{m_{1}m}\label{clebsp2}%
\end{align}

\vspace{0.2cm}
\subsection*{$3j$-symbols}
\vspace{-0.4cm}
\begin{equation}
\left\langle j_{1}j_{2}m_{1}m_{2}\right. \left\vert
jm\right\rangle =\left( -1\right)
^{j_{1}-j_{2}+m}\hat{\jmath}\left(
\begin{array}{ccc}
j_{1} & j_{2} & j \\
m_{1} & m_{2} & -m%
\end{array}%
\right) \hspace{3.55cm}
\end{equation}
\vspace{0.2cm}
\subsection*{$6j$-symbols and Racah coefficients}
\subsubsection*{Symmetry relations}
\vspace{-0.9cm}
\begin{align}
\left\{
\begin{array}{lll}
j_{1} & j_{2} & j_{3} \\
j_{4} & j_{5} & j_{6}%
\end{array}%
\right\} & =\left\{
\begin{array}{lll}
j_{2} & j_{3} & j_{1} \\
j_{5} & j_{6} & j_{4}%
\end{array}%
\right\} =\left\{
\begin{array}{lll}
j_{3} & j_{1} & j_{2} \\
j_{6} & j_{4} & j_{5}%
\end{array}%
\right\} =\left\{
\begin{array}{lll}
j_{2} & j_{1} & j_{3} \\
j_{5} & j_{4} & j_{6}%
\end{array}%
\right\}  \hspace{0.2cm}\\
& =\left\{
\begin{array}{lll}
j_{4} & j_{5} & j_{3} \\
j_{1} & j_{2} & j_{6}%
\end{array}%
\right\}   \notag
\end{align}
\subsubsection*{Orthogonality relation}
\vspace{-0.4cm}
\begin{equation}
\sum_{j}\hat{\jmath}^{2}\left\{
\begin{array}{lll}
j_{1} & j_{2} & j^{\prime } \\
j_{3} & j_{4} & j%
\end{array}%
\right\} \left\{
\begin{array}{lll}
j_{1} & j_{2} & j^{\prime \prime } \\
j_{3} & j_{4} & j%
\end{array}%
\right\} =\frac{\delta _{j^{\prime }j^{\prime \prime }}}{\widehat{j^{\prime }%
}^{2}} \hspace{4.7cm}
\end{equation}
\subsubsection*{Relation to Racah coefficient}
\vspace{-0.4cm}
\begin{equation}
\left\{
\begin{array}{lll}
j_{1} & j_{2} & J_{12} \\
j_{3} & J & J_{23}%
\end{array}%
\right\} =\left( -1\right) ^{j_{1}+j_{2}+j_{3}+J}W\left(
j_{1}j_{2}Jj_{3};J_{12}J_{23}\right) \hspace{3.4cm}
\end{equation}
\subsubsection*{Relation to 3j symbols}
\vspace{-0.9cm}
\begin{align}
W\left(  j_{1}j_{2}Jj_{3};J_{12}J_{23}\right)   & =
\sum_{\alpha\beta \gamma\delta\varepsilon\phi}\left(  2J+1\right)
\left(  -1\right) ^{J_{23}-J_{12}-\alpha-\delta}\left(
\begin{array}
[c]{ccc}%
j_{1} & j_{2} & J_{12}\\
\alpha & \beta & -\varepsilon
\end{array}
\right) \hspace{0.5cm} \\
&  \left(
\begin{array}
[c]{ccc}%
j_{3} & J & J_{12}\\
\delta & \gamma & \varepsilon
\end{array}
\right)  \left(
\begin{array}
[c]{ccc}%
j_{2} & j_{3} & J_{23}\\
\beta & \delta & -\phi
\end{array}
\right)  \left(
\begin{array}
[c]{ccc}%
J & j_{1} & J_{23}\\
\gamma & \alpha & \phi
\end{array}
\right) \nonumber
\end{align}
\subsubsection*{Sum over components}
\vspace{-0.9cm}
\begin{eqnarray}
W\left(  j_{1}j_{2}Jj_{3};J_{12}J_{23}\right)  \left(
\begin{array}
[c]{ccc}%
J & j_{1} & J_{23}\\
\gamma & \alpha & \phi
\end{array}
\right)   &  = & \sum_{\beta\delta\varepsilon}\left(  -1\right)  ^{J_{23}%
-J_{12}-\alpha-\delta}\left(
\begin{array}
[c]{ccc}%
j_{1} & j_{2} & J_{12}\\
\alpha & \beta & -\varepsilon
\end{array}
\right)  \label{racsum} \hspace{1.9cm}\\
&&  \left(
\begin{array}
[c]{ccc}%
j_{3} & J & J_{12}\\
\delta & \gamma & \varepsilon
\end{array}
\right)  \left(
\begin{array}
[c]{ccc}%
j_{2} & j_{3} & J_{23}\\
\beta & \delta & -\phi
\end{array}
\right) \nonumber
\end{eqnarray}
\begin{eqnarray}
W\left(  j_{1}j_{2}Jj_{3};J_{12}J_{23}\right)  \left\langle
Jj_{1}\gamma \alpha|J_{23}-\phi\right\rangle
(-1)^{-2J+3j_{1}+2\phi} &  = & \sum_{\beta
\delta\varepsilon}\frac{\left( -1\right)
^{J_{23}-J_{12}-\delta}}{\left(
2J_{12}+1\right)  }\label{racsumcleb} \hspace{2.6cm}\\
&&  \left\langle
j_{1}j_{2}\alpha\beta|J_{12}\varepsilon\right\rangle
\left\langle j_{3}J\delta\gamma|J_{12}-\varepsilon\right\rangle \nonumber \\
&&  \left\langle j_{2}j_{3}\beta\delta|J_{23}\phi\right\rangle
\nonumber
\end{eqnarray}
%
%
\subsection*{$9j$-symbols}
\subsubsection*{Relation to $6j$-symbols}
\vspace{-0.9cm}
\begin{align}
\left\{
\begin{array}{lll}
j_{1} & j_{2} & J_{12} \\
j_{3} & j_{4} & J_{34} \\
J_{13} & J_{24} & J%
\end{array}%
\right\} & =\sum_{J^{\prime }}\left( -1\right) ^{2J^{\prime }}\widehat{%
J^{\prime }}^{2}\left\{
\begin{array}{lll}
j_{1} & j_{3} & J_{13} \\
J_{24} & J & J^{\prime }%
\end{array}%
\right\} \left\{
\begin{array}{lll}
j_{2} & j_{4} & J_{24} \\
j_{3} & J^{\prime } & J_{34}%
\end{array}%
\right\} \times  \hspace{2.25cm}\\
& \times \left\{
\begin{array}{lll}
J_{12} & J_{34} & J \\
J^{\prime } & j_{1} & j_{2}%
\end{array}%
\right\}   \notag
\end{align}
\subsubsection*{Relation to 3j-symbols}
\vspace{-0.9cm}
\begin{eqnarray}
\left\{
\begin{array}
[c]{ccc}%
j_{1} & j_{2} & J_{12}\\
j_{3} & j_{4} & J_{34}\\
J_{13} & J_{24} & J
\end{array}
\right\}   &  = & \left(  2j_{1}+1\right)  \sum\left(
\begin{array}
[c]{ccc}%
j_{1} & j_{2} & J_{12}\\
\alpha & \beta & \gamma
\end{array}
\right)  \left(
\begin{array}
[c]{ccc}%
j_{2} & j_{4} & J_{24}\\
\beta & \varepsilon & \eta
\end{array}
\right)  \label{9j3j}\\
&&  \left(
\begin{array}
[c]{ccc}%
J_{12} & J_{34} & J\\
\gamma & \phi & \nu
\end{array}
\right)  \left(
\begin{array}
[c]{ccc}%
j_{1} & j_{3} & J_{13}\\
\alpha & \delta & \rho
\end{array}
\right)  \left(
\begin{array}
[c]{ccc}%
j_{3} & j_{4} & J_{34}\\
\delta & \varepsilon & \phi
\end{array}
\right)  \left(
\begin{array}
[c]{ccc}%
J_{13} & J_{24} & J\\
\rho & \eta & \nu
\end{array}
\right) \nonumber
\end{eqnarray}
\subsubsection*{Relation to Clebsch-Gordan coefficients}
\vspace{-0.9cm}
\begin{eqnarray}
\left\{
\begin{array}
[c]{ccc}%
j_{1} & j_{2} & J_{12}\\
j_{3} & j_{4} & J_{34}\\
J_{13} & J_{24} & J
\end{array}
\right\}   &  = & \frac{\left(  2j_{1}+1\right)  (-1)^{-2j_{1}-J_{12}%
+J_{34}+2j_{4}-J_{13}+J_{24}}}{\sqrt{2J_{12}+1}\sqrt{2J_{34}+1}\sqrt
{2J_{13}+1}\sqrt{2J_{24}+1}\left(  2J+1\right)  }\label{9jcleb}\\
&&  \sum(-1)^{-\rho-\phi-2\nu-\gamma-\eta}\left\langle
j_{1}j_{2}\alpha \beta|J_{12}-\gamma\right\rangle \left\langle
j_{2}j_{4}\beta\varepsilon |J_{24}-\eta\right\rangle \nonumber
\\
&& \left\langle J_{12}J_{34}\gamma\phi|J-\nu\right\rangle
\left\langle
j_{1}j_{3}\alpha\delta|J_{13}-\rho\right\rangle \left\langle j_{3}j_{4}%
\delta\varepsilon|J_{34}-\phi\right\rangle \left\langle
J_{13}J_{24}\rho \eta|J-\nu\right\rangle \nonumber
\end{eqnarray}
\subsubsection*{Sum over components}
\vspace{-0.9cm}
\begin{eqnarray}
\left(
\begin{array}
[c]{ccc}%
j_{1} & j_{2} & J_{12}\\
\alpha & \beta & \gamma
\end{array}
\right)  \left\{
\begin{array}
[c]{ccc}%
j_{1} & j_{2} & J_{12}\\
j_{3} & j_{4} & J_{34}\\
J_{13} & J_{24} & J
\end{array}
\right\}   &  = & \sum\left(
\begin{array}
[c]{ccc}%
j_{2} & j_{4} & J_{24}\\
\beta & \varepsilon & \eta
\end{array}
\right)  \left(
\begin{array}
[c]{ccc}%
J_{12} & J_{34} & J\\
\gamma & \phi & \nu
\end{array}
\right)  \label{9jsum}\\
&&  \left(
\begin{array}
[c]{ccc}%
j_{1} & j_{3} & J_{13}\\
\alpha & \delta & \rho
\end{array}
\right)  \left(
\begin{array}
[c]{ccc}%
j_{3} & j_{4} & J_{34}\\
\delta & \varepsilon & \phi
\end{array}
\right)  \left(
\begin{array}
[c]{ccc}%
J_{13} & J_{24} & J\\
\rho & \eta & \nu
\end{array}
\right) \nonumber
\end{eqnarray}
\begin{eqnarray}
\sum(-1)^{\rho-\eta+\phi+2\nu}\left\langle
j_{2}j_{4}\beta\varepsilon |J_{24}-\eta\right\rangle \left\langle
J_{12}J_{34}\gamma\phi|J-\nu \right\rangle \times && \label{9jsumcleb} \\
\left\langle j_{1}j_{3}\alpha\delta|J_{13}-\rho\right\rangle
\left\langle j_{3}j_{4}\delta\varepsilon|J_{34}-\phi\right\rangle
 \left\langle J_{13}J_{24}\rho\eta|J-\nu\right\rangle & = &
  (-1)^{2j_{2}+2j_{4}+J_{12}+\gamma-J_{34}+J_{13}-J_{24}}
\nonumber \\
&& \frac{\sqrt
{2J_{34}+1}\sqrt{2J_{13}+1}\sqrt{2J_{24}+1}\left(  2J+1\right)  }%
{\sqrt{2J_{12}+1}} \nonumber \\
&&  \left\langle j_{1}j_{2}\alpha\beta|J_{12}-\gamma\right\rangle
\left\{
\begin{array}
[c]{ccc}%
j_{1} & j_{2} & J_{12}\\
j_{3} & j_{4} & J_{34}\\
J_{13} & J_{24} & J
\end{array}
\right\} \nonumber
\end{eqnarray}
\subsubsection*{Symmetry relations}
\vspace{-0.9cm}
\begin{align}
\left\{
\begin{array}{lll}
j_{1} & j_{2} & j_{3} \\
j_{4} & j_{5} & j_{6} \\
j_{7} & j_{8} & j_{9}%
\end{array}%
\right\} & =\left\{
\begin{array}{ccc}
j_{1} & j_{4} & j_{7} \\
j_{2} & j_{5} & j_{8} \\
j_{3} & j_{6} & j_{9}%
\end{array}%
\right\} =\left\{
\begin{array}{lll}
j_{7} & j_{8} & j_{9} \\
j_{1} & j_{2} & j_{3} \\
j_{4} & j_{5} & j_{6}%
\end{array}%
\right\} =\left\{
\begin{array}{lll}
j_{4} & j_{5} & j_{6} \\
j_{7} & j_{8} & j_{9} \\
j_{1} & j_{2} & j_{3}%
\end{array}%
\right\}  \hspace{1.1cm} \\
& =\left( -1\right)
^{j_{1}+j_{2}+j_{3}+j_{4}+j_{5}+j_{6}+j_{7}+j_{8}+j_{9}}\left\{
\begin{array}{lll}
j_{4} & j_{5} & j_{6} \\
j_{1} & j_{2} & j_{3} \\
j_{7} & j_{8} & j_{9}%
\end{array}%
\right\}   \notag
\end{align}
\subsubsection*{Orthogonality relation}
\vspace{-0.6cm}
\begin{equation}
\sum_{J_{13}J_{24}}\widehat{J_{13}}^{2}\widehat{J_{24}}^{2}\left\{
\begin{array}{lll}
j_{1} & j_{2} & J_{12} \\
j_{3} & j_{4} & J_{34} \\
J_{13} & J_{24} & J%
\end{array}%
\right\} \left\{
\begin{array}{lll}
j_{1} & j_{2} & J_{12}^{\prime } \\
j_{3} & j_{4} & J_{34}^{\prime } \\
J_{13} & J_{24} & J%
\end{array}%
\right\} =\frac{\delta _{J_{12}J_{12}^{\prime }}\delta
_{J_{34}J_{34}^{\prime
}}}{\widehat{J_{12}}^{2}\widehat{J_{34}}^{2}} \hspace{1.7cm}
\end{equation}
\vspace{0.6cm}
\section*{Rules for tensorial algebra}
\vspace{-0.4cm}
\begin{align}
& a_{\mu }^{\left( 1\right) }=a\sqrt{\frac{4\pi }{3}}Y^{1}_{\mu }\left( \hat{a}%
\right) \hspace{1cm}\text{with}\hspace{1cm}\mu =-1,0,1
\hspace{5cm} \label{vecesf} \\
& \vec{a}\cdot \vec{b}=-\sqrt{3}\left\{ a^{\left( 1\right)
}\otimes b^{\left( 1\right) }\right\} _{0}^{\left( 0\right) } \hspace{4cm}\\
& \vec{\sigma}\cdot \vec{b}=-b\sqrt{4\pi }\left\{ \sigma
^{1}\otimes Y^{1}\left( \hat{b}\right) \right\} _{0}^{\left(
0\right) } \label{sigmadot} \\
& \left( \vec{a}\times \vec{b}\right) _{\mu }^{\left( 1\right) }=-i\sqrt{2}%
\left\{ a^{\left( 1\right) }\otimes b^{\left( 1\right) }\right\}
_{\mu }^{\left( 1\right) }
\end{align}
\begin{align}
& \vec{a}\cdot \left( \vec{b}\cdot \vec{c}\right) =-\frac{1}{\sqrt{3}}%
\sum_{f}\left( -1\right) ^{f}\hat{f}\left\{ \left[ a^{\left(
1\right) }\otimes b^{\left( 1\right) }\right] ^{\left( f\right)
}\otimes c^{\left( 1\right) }\right\} _{0}^{\left( 0\right) }
\hspace{3.8cm} \\
& \vec{a}\cdot \left( \vec{b}\times \vec{c}\right)
=i\sqrt{6}\left\{ a^{\left(
1\right) }\otimes \left[ b^{\left( 1\right) }\otimes c^{\left( 1\right) }%
\right] ^{\left( 1\right) }\right\} _{0}^{\left( 0\right) }
\label{vdotext} \\
& \left( \vec{a}\times \vec{b}\right) \cdot
\vec{c}=i\sqrt{6}\left\{ \left[ a^{\left( 1\right) }\otimes
b^{\left( 1\right) }\right] ^{\left( 1\right) }\otimes c^{\left(
1\right) }\right\} _{0}^{\left( 0\right) } \\[0.3cm]
& \left(
\vec{a}\cdot \vec{b}\right) \left( \vec{c}\cdot \vec{d}\right)
=3\left\{ \left[ a^{\left( 1\right) }\otimes b^{\left( 1\right)
}\right] ^{\left( 0\right) }\otimes \left[ c^{\left( 1\right)
}\otimes d^{\left( 1\right) }\right] ^{\left(
0\right) }\right\} _{0}^{\left( 0\right) } \\
& \left( \vec{a}\times \vec{b}\right) \left( \vec{c}\cdot \vec{d}\right) =i%
\sqrt{6}\left\{ \left[ a^{\left( 1\right) }\otimes b^{\left( 1\right) }%
\right] ^{\left( 1\right) }\otimes \left[ c^{\left( 1\right)
}\otimes d^{\left( 1\right) }\right] ^{\left( 0\right) }\right\}
_{\mu }^{\left( 1\right) } \\
& \left( \vec{a}\times \vec{b}\right) \cdot \left( \vec{c}\times \vec{d}%
\right) =2\sqrt{3}\left\{ \left[ a^{\left( 1\right) }\otimes
b^{\left( 1\right) }\right] ^{\left( 1\right) }\otimes \left[
c^{\left( 1\right) }\otimes d^{\left( 1\right) }\right] ^{\left(
1\right) }\right\} _{0}^{\left( 0\right) }
\end{align}
\vspace{0.4cm}
\subsection*{Relations for tensorial operators}
\vspace{-0.4cm}
\begin{align}
& \left\{ A^{\left( a\right) }\otimes B^{\left( b\right) }\right\}
_{c_{3}}^{c}=\left( -1\right)
^{2c_{3}}\sum_{a_{3}b_{3}}C_{a_{3}b_{3}c_{3}}^{abc}A^{aa_{3}}B^{bb_{3}}
\label{tensprod} \hspace{6cm}\\
&
A^{aa_{3}}B^{bb_{3}}=\sum_{cc_{3}}C_{a_{3}b_{3}c_{3}}^{abc}\left\{
A^{\left( a\right) }\otimes B^{\left( b\right) }\right\}
_{c_{3}}^{c} \\
& \left\{ A^{\left( a\right) }\otimes B^{\left( b\right) }\right\}
_{c_{3}}^{c}=\left( -1\right) ^{a+b+c}\left\{ B^{\left( b\right)
}\otimes
A^{\left( a\right) }\right\} _{c_{3}}^{c}\hspace{1cm}\text{if }\left[ A,B%
\right] =0
\end{align}
\begin{align}
 \left\{ A^{\left( a\right) }\otimes \left[ B^{\left( b\right)
}\otimes C^{\left( c\right) }\right] ^{\left( d\right) }\right\}
^{\left( e\right) } &=\left( -1\right)
^{a+b+c+e}\sum_{f}\hat{d}\hat{f}\left\{
\begin{array}{ccc}
b & c & d \\
e & a & f%
\end{array}%
\right\}
\label{tenspermthree1} \hspace{2.7cm} \\
& \left\{ \left[ A^{\left( a\right) }\otimes B^{\left( b\right)
}\right] ^{\left( f\right) }\otimes C^{\left( c\right) }\right\}
^{\left( e\right) } \notag
\end{align}
\begin{align}
\left\{ \left[ A^{\left( a\right) }\otimes B^{\left( b\right)
}\right] ^{\left( f\right) }\otimes C^{\left( c\right) }\right\}
^{\left( e\right) } &=\left( -1\right)
^{a+b+c+e}\sum_{d}\hat{d}\hat{f}\left\{
\begin{array}{ccc}
b & a & f \\
e & c & d%
\end{array}%
\right\}  \hspace{2.7cm}\\
&\left\{ A^{\left( a\right) }\otimes \left[ B^{\left( b\right)
}\otimes C^{\left( c\right) }\right] ^{\left( d\right) }\right\}
^{\left( e\right) } \notag
\end{align}%
\begin{align}
\left\{ \left[ A^{\left( a\right) }\otimes B^{\left( b\right)
}\right] ^{\left( c\right) }\otimes \left[ D^{\left( d\right)
}\otimes E^{\left( e\right) }\right] ^{\left( f\right) }\right\}
^{\left( g\right) }
&=\sum_{jk}\hat{c}\hat{f}\hat{\jmath}\hat{k}\left\{
\begin{array}{ccc}
a & b & c \\
d & e & f \\
j & k & g%
\end{array}%
\right\}   \label{tenspermfour}  \hspace{2cm}\\
&\left\{ \left[ A^{\left( a\right) }\otimes D^{\left( d\right)
}\right] ^{\left( j\right) }\otimes \left[ B^{\left( b\right)
}\otimes E^{\left( e\right) }\right] ^{\left( k\right) }\right\}
^{\left( g\right) }  \notag
\end{align}
\subsection*{Spherical (tensor) harmonics}
\vspace{-0.4cm}
\begin{align}
& \mathcal{Y}_{j_{1}j_{2}}^{jm}\left( \hat{a},\hat{b}\right)
=\sum_{m_{1}m_{2}}C_{m_{1}m_{2}m}^{j_{1}j_{2}j}Y^{j_{1}}_{m_{1}}\left( \hat{a}%
\right) Y^{j_{2}}_{m_{2}}\left( \hat{b}\right)   \label{sphdef}
\hspace{6.0cm} \\
& \int d\hat{a}d\hat{b}\mathcal{Y}_{j_{1}j_{2}}^{jm}\left(
\hat{a},b\right) \mathcal{Y}_{j_{1}^{\prime }j_{2}^{\prime
}}^{\mathcal{\ast }j^{\prime }m^{\prime }}\left(
\hat{a},\hat{b}\right) =\delta _{jj^{\prime }}\delta _{mm^{\prime
}}\delta _{j_{1}j_{1}^{\prime }}\delta _{j_{2}j_{2}^{\prime }}
\label{tensinteg} \\[0.4cm]
& Y^{l}_{m}\left( \widehat{\vec{a}+\vec{b}}\right) =\sum_{l_{1}+l_{2}=l}\frac{%
a^{l_{1}}b^{l_{2}}}{\left\vert \vec{a}+\vec{b}\right\vert ^{l}}\sqrt{\frac{%
4\pi \left( 2l+1\right) !}{\left( 2l_{1}+1\right) !\left( 2l_{2}+1\right) !}}%
\mathcal{Y}_{l_{1}l_{2}}^{lm}\left( \hat{a},\hat{b}\right)
\label{ysoma} \\[0.4cm]
& \mathcal{Y}_{ll}^{00}\left( \hat{a},\hat{b}\right)
=\dfrac{\left( -1\right) ^{l}}{4\pi }\sqrt{2l+1}P_{l}\left(
\hat{a}\cdot \hat{b}\right) \\
& P_{l}\left( \hat{a}\cdot \hat{b}\right) =\dfrac{\left( -1\right) ^{l}}{\sqrt{%
2l+1}}4\pi \mathcal{Y}_{ll}^{00}\left( \hat{a},\hat{b}\right) \\
&
\mathcal{Y}_{l_{1}l_{2}}^{lm}\left( \hat{a},\hat{a}\right) =\sqrt{\dfrac{%
\left( 2l_{1}+1\right) \left( 2l_{2}+1\right) }{4\pi \left( 2l+1\right) }}%
C_{000}^{l_{1}l_{2}l}Y^{l}_{m}\left( \hat{a}\right)
\end{align}
\vspace{0.4cm}
\begin{align}
\mathcal{Y}_{fg}^{lm}\left( \widehat{\vec{a}+\vec{b}},\widehat{\vec{c}+\vec{d%
}}\right) & =\sum_{\substack{ f_{1}+f_{2}=f \\ g_{1}+g_{2}=g}}\frac{%
a^{f_{1}}b^{f_{2}}}{\left\vert \vec{a}+\vec{b}\right\vert ^{f}}\frac{%
c^{g_{1}}d^{g_{2}}}{\left\vert \vec{c}+\vec{d}\right\vert ^{g}}
\sqrt{\frac{4\pi \left( 2f+1\right) !}{\left( 2f_{1}+1\right)
!\left( 2f_{2}+1\right) !}} \hspace{2.5cm}\label{nYsomasoma} \\
& \sqrt{\frac{4\pi \left( 2g+1\right) !}{\left( 2g_{1}+1\right)
!\left( 2g_{2}+1\right) !}}  \left\{
\mathcal{Y}_{f_{1}f_{2}}^{f}\left( \hat{a},\hat{b}\right) \otimes
\mathcal{Y}_{g_{1}g_{2}}^{g}\left( \hat{c},\hat{d}\right) \right\}
_{m}^{\left( l\right) }  \notag
\end{align}
%
%
\vspace{0.6cm}
\begin{align}
\mathcal{Y}_{fg}^{lm}\left( \widehat{\vec{a}+\vec{b}},\widehat{\vec{c}}%
\right)  &=\sum_{f_{1}+f_{2}=f}\frac{a^{f_{1}}b^{f_{2}}}{\left\vert \vec{a}+%
\vec{b}\right\vert ^{f}}\sqrt{\frac{4\pi \left( 2f+1\right)
!}{\left(
2f_{1}+1\right) !\left( 2f_{2}+1\right) !}}  \label{AASum} \hspace{3.3cm}\\
&\left\{ \mathcal{Y}_{f_{1}f_{2}}^{f}\left( \hat{a},\hat{b}\right)
\otimes Y^{g}\left( \hat{c}\right) \right\} _{m}^{\left( l\right)
}  \notag
\end{align}
\begin{align}
\mathcal{Y}_{fg}^{lm}\left( \widehat{\vec{a}},\widehat{\vec{c}+\vec{d}}%
\right)  &=\sum_{g_{1}+g_{2}=g}\frac{c^{g_{1}}d^{g_{2}}}{\left\vert \vec{c}+%
\vec{d}\right\vert ^{g}}\sqrt{\frac{4\pi \left( 2g+1\right)
!}{\left(
2g_{1}+1\right) !\left( 2g_{2}+1\right) !}}  \label{BBsum} \hspace{3.3cm}\\
&\left\{ Y^{f}\left( \hat{a}\right) \otimes \mathcal{Y}_{g_{1}g_{2}}^{g}%
\left( \hat{c},\hat{d}\right) \right\} _{m}^{\left( l\right) }
\notag
\end{align}
\begin{align}
\left\{ \mathcal{Y}_{L_{1}L_{2}}^{L}\left( \hat{p},\hat{q}\right)
\otimes
\mathcal{Y}_{l_{1}l_{2}}^{l}\left( \hat{p},\hat{q}\right) \right\} _{%
\mathcal{M}}^{\mathcal{L}}& =\sum_{f_{1}f_{2}}\tfrac{1}{4\pi }\hat{L}\hat{l}%
\widehat{L_{1}}\widehat{l_{1}}\widehat{L_{2}}\widehat{l_{2}}%
C_{000}^{L_{1}l_{1}f_{1}}C_{000}^{L_{2}l_{2}f_{2}}\times
\label{tenssamearg} \hspace{2.2cm}\\
& \times \left\{
\begin{array}{lll}
L_{1} & L_{2} & L \\
l_{1} & l_{2} & l \\
f_{1} & f_{2} & \mathcal{L}%
\end{array}%
\right\} \mathcal{Y}_{f_{1}f_{2}}^{\mathcal{LM}}\left( \hat{p},\hat{q}%
\right)   \notag
\end{align}
\begin{align}
\left[ \mathcal{Y}_{j_{1}j_{2}}^{j}\left( \hat{a},\hat{b}\right)
\otimes
\mathcal{Y}_{k_{1}k_{2}}^{k}\left( \widehat{d},\widehat{e}\right) \right] _{%
\mathcal{M}}^{\mathcal{L}}& =\sum_{f_{1}f_{2}}\widehat{j}\widehat{k}\widehat{%
f_{1}}\widehat{f_{2}}\left\{
\begin{array}{lll}
j_{1} & j_{2} & j \\
k_{1} & k_{2} & k \\
f_{1} & f_{2} & \mathcal{L}%
\end{array}%
\right\} \times   \label{tensexcharg} \hspace{2.2cm} \\
& \times \left[ \mathcal{Y}_{j_{1}k_{1}}^{f_{1}}\left( \widehat{a},\widehat{d%
}\right) \otimes \mathcal{Y}_{j_{2}k_{2}}^{f_{2}}\left( \widehat{b},\widehat{%
e}\right) \right] _{\mathcal{M}}^{\mathcal{L}}  \notag
\end{align}
\vspace{0.5cm}
\section*{Wigner-Eckart theorem for the matrix elements of \\
spherical tensors}
\begin{equation}
\left\langle J^{\prime}M^{\prime}\left\vert T_{KQ}\right\vert
JM\right\rangle =C_{QMM^{\prime}}^{KJJ^{\prime}}\left\langle
J^{\prime}\left\Vert T_{K}\right\Vert J\right\rangle
\end{equation}
with\footnote{%
Since $\left\langle J^{\prime}M^{\prime}\left\vert
T_{KQ}\right\vert
JM\right\rangle^{\text{\rf{Brink:1993bk}}}=\left( -1\right)
^{2K}C_{MQM^{\prime}}^{JKJ^{\prime}}\left\langle
J^{\prime}\left\Vert T_{K}\right\Vert
J\right\rangle^{\text{\rf{Brink:1993bk}}}
$.%
}%
\begin{equation}
\left\langle J^{\prime}\left\Vert T_{K}\right\Vert
J\right\rangle^{\text{\rf{Wong:1990bk}}} =\left( -1\right)
^{2K}\left( -1\right) ^{J^{\prime}-\left( J+K\right) }\left\langle
J^{\prime}\left\Vert T_{K}\right\Vert
J\right\rangle^{\text{\rf{Brink:1993bk}}}
\end{equation}